%% file: Thesis.tex
\documentclass[a4paper,oneside,openright,12pt]{book}

\usepackage{customisations}
\usepackage{booktabs}
\usepackage{amsmath}
\usepackage{textcomp}
\usepackage{tikz}

\title{On-Shell Physics of Black Holes}
\author{Ben Maybee}
\date{March 2021}

\begin{document}

\singlespacing
\frontmatter
\eighteenptleading
\input{thesis-frontmatter/thesis-frontmatter}

\mainmatter
\eighteenptleading
\include{chapter1/chapter1}
\include{chapter2/chapter2}
\include{chapter3/chapter3}
\include{chapter4/chapter4}
\include{chapter5/chapter5}
\include{chapter6/chapter6}
\include{chapter7/chapter7}

\appendix
\include{appendices/appendix1}
\include{appendices/appendix2}

\providecommand{\href}[2]{#2}\begingroup\raggedright\endgroup

\end{document}

%% file: thesis-frontmatter/thesis-frontmatter.tex
\singlespacing
\maketitlepage
\frontmatter
\eighteenptleading
\input{thesis-frontmatter/abstract}

\cleardoublepage
\phantomsection
\makedeclaration

\input{thesis-frontmatter/acknowledgements}

\cleardoublepage
\phantomsection
\addcontentsline{toc}{chapter}{\contentsname}
\setcounter{tocdepth}{2}
\tableofcontents

\cleardoublepage
\phantomsection
\addcontentsline{toc}{chapter}{\listfigurename}
\listoffigures


%% file: thesis-frontmatter/abstract.tex
\chapter{Lay summary}

Black holes are some of the most beguiling objects in theoretical physics. Their interiors, shrouded by the cloak of the inescapable event horizon, will always remain one of the
least understood environments in modern science. Yet, from the perspective of a (very) safely removed external observer, black holes are remarkably simple objects. So simple in fact that any classical black hole in our leading theory of gravity, general relativity, is exactly specified by just three parameters: its mass, charge and angular momentum. This sparsity of information leads physicists to say that black holes ``\textit{have no hair}'', in contrast to ``hairy'' bodies, such as stars and PhD students, which unfortunately cannot be exactly identified by three numbers.

Black holes' ``no hair'' constraints are maybe more reminiscent of the description of fundamental particles, which are catalogued by their mass, charge and quantum spin, than extended bodies with complicated physiognomies. Yet astronomers readily observe influences on the propagation of light, the motion of stars and the structure of galaxies that are precisely predicted by black hole solutions. Moreover, general relativity predicts that black holes can interact and merge with each other. Such interactions take the form of a long courtship, performing a well separated inspiral that steadily, but inescapably, leads to a final unifying merger under immense gravitational forces. These interactions do not happen in isolation from the rest of the universe. They are violent, explosive events which release enormous amounts of energy, exclusively in the form of gravitational waves.

The frequent experimental detection of these signals from distant black hole mergers marks the beginning of a new era of observational astronomy. Yet despite the enormous total energy released, gravitational wave signals observable on Earth are very, very faint. In addition to advanced instrumentation, extracting these clean whispers from the background hubbub of our messy world requires precise predictions from general relativity, so as to filter out meaningful signals from noise. The difficulty of performing these calculations means that they are rapidly approaching the point of becoming the largest source of errors in experiments; promoting gravitational wave astronomy to a precision science now depends on improving theoretical predictions.

This has motivated the development of novel ways of looking at gravitational dynamics, and one such route has been to adopt methods from particle physics. Modern particle collider programmes have necessitated the development of an enormous host of theoretical techniques, targeted towards experimental predictions, which have enabled precision measurements to be extracted from swathes of messy data. These techniques focus on calculating \textit{scattering amplitudes}: probabilities for how particles will interact quantum mechanically. These techniques can be applied to gravity provided that the interactions are weak, and astonishingly theorists have found that they can facilitate precise predictions for the initial inspiral phase of black hole coalescence. This is a relatively new line of research, and one which is growing rapidly.

This thesis is a small part of this endeavour. It is focussed on obtaining \textit{on-shell} observables, relevant for black hole physics, from quantum gravity. An on-shell quantity is one that is physical and directly measurable by an experimentalist, such as the change in a body's momentum. Scattering amplitudes are also on-shell, in the sense that they underpin measurable probabilities. We will be interested in direct routes from amplitudes to observables, which seems like an obvious path to choose. However, many of the techniques in a theorist's toolbox instead involve going via an unmeasurable, mathematical midstep which contains all physical data needed to make measurable predictions. One example is the gravitational potential: one cannot measure the potential directly, but it contains everything needed to make concrete predictions of classical dynamics.

Much of the power of modern scattering amplitudes comes from their reliance on physical, on-shell data. This thesis will show that a direct map to observables opens new insights into the physics of classical black holes. In particular, we can make the apparent similarity between fundamental particles and black holes concrete: certain scattering amplitudes are the on-shell incarnation of the special ``no hair'' constraint for black holes. This relationship is beautiful and provocative. It opens new practical treatments of spinning black holes in particular, and to this end we will use unique complex number properties of these solutions, combined with quantum mechanics, to obtain powerful classical descriptions of how spinning black holes interact with one another.

\cleardoublepage
\phantomsection
\chapter{Abstract}

\noindent On-shell methods are a key component of the modern amplitudes programme. By utilising the power of generalised unitarity cuts, and focusing on gauge invariant quantities, enormous progress has been made in the calculation of amplitudes required for theoretical input into experiments such as the LHC. Recently, a new experimental context has emerged in which scattering amplitudes can be of great utility: gravitational wave astronomy. Indeed, developing new theoretical techniques for tackling the two-body problem in general relativity is essential for future precision measurements. Scattering amplitudes have already contributed new state of the art calculations of post--Minkowskian (PM) corrections to the classical gravitational potential.

The gravitational potential is an unphysical, gauge dependent quantity. This thesis seeks to apply the advances of modern amplitudes to classical gravitational physics by constructing physical, on-shell observables applicable to black hole scattering, but valid in any quantum field theory. We will derive formulae for the impulse (change in momentum), total radiated momentum, and angular impulse (change in spin vector) from basic principles, directly in terms of scattering amplitudes.

By undertaking a careful analysis of the classical region of these observables, we derive from explicit wavepackets how to take the classical limit of the associated amplitudes. These methods are then applied to examples in both QED and QCD, through which we obtain new theoretical results; however, the main focus is on black hole physics. We exploit the double copy relationship between gravity and gauge theory to calculate amplitudes in perturbative quantum gravity, from whose classical limits we derive results in the PM approximation of general relativity.

Applying amplitudes to black hole physics offers more than computational power: in this thesis we will show that the observables we have constructed provide particularly clear evidence that massive, spinning particles are the on-shell avatar of the no-hair theorem. Building on these results, we will furthermore show that the classically obscure Newman--Janis shift property of the exact Kerr solution can be interpreted in terms of a worldsheet effective action. At the level of equations of motion, we show that the Newman--Janis shift holds also for the leading interactions of the Kerr black hole. These leading interactions will be conveniently described using chiral classical equations of motion with the help of the spinor helicity method familiar from scattering amplitudes, providing a powerful and purely classical method for computing on-shell black hole observables.

%% file: thesis-frontmatter/acknowledgements.tex
\chapter{Acknowledgements}

\noindent

\normalsize

My foremost thanks go to my supervisor, Donal O'Connell. I will be forever grateful to you for helping me learn to navigate the dense jungle of theoretical physics; your capacity for novel ideas has been truly inspirational, and working with you has been a fantastic experience. Thank you for placing your trust in me as a collaborator these last few years, and also for your sage mountaineering advice: don't fall off.

Thank you also to my fellow collaborators: Leonardo de la Cruz, Alfredo Guevara, David Kosower, Dhritiman Nandan, Alex Ochirov, Alasdair Ross, Matteo Sergola, and Justin Vines. I am very proud of the science we have done together. I am especially grateful to Justin, for his hospitality at the Albert--Einstein institute, insights into the world of general relativity, and all-round kindness. My thanks also to Andr\'{e}s Luna for his friendship and warmth; Chris White, for his enthusiastic introduction to the eikonal; and Yu-Tin Huang, Paolo Pichini and Oliver Schlotterer for the successful joint conquest of the great Bavarian mountain Kaiserschmarrn. I'm still full.

I am extremely grateful for the additional hospitality of Daniel Jenkins at Regensburg; Thibault Damour at the IHES; and the organisers of the MIAPP workshop ``Precision Gravity''. These visits, coupled with the QCD meets Gravity conferences, have been highlights of my PhD. Thank you to the wider community for so many exciting scientific discussions in a welcoming environment, and to David, Alessandra Buonanno and Henrik Johansson for offering me the opportunity to continue my tenure; I am sorry that I have not done so.

The PPT group at Edinburgh has been a great place to work, and one I have sorely missed following the virtual transition. Thank you to all my fellow PhD students and TA's for your support and knowledge, especially Christian Br\o{}nnum--Hansen, Tomasso Giani, Michael Marshall, Calum Milloy, Izzy Nicholson, Rosalyn Pearson and Saad Nabeebaccus. Teaching in particular has been a great joy throughout; thank you to Lucile Cangemi and Maria Derda for working with me on your MPhys projects, and to JC Denis for encouraging my outreach activities.

Research is not always a smooth path, but one which is certainly made easier with the support of friends and family: thank you especially to Rosemary Green for her continual love and encouragement. Thank you to Mike Daniels, Laura Glover, Kirsty McIntyre and Sophie Walker for (literally) accommodating my poor decision making and keeping me going, even if you didn't know it! Meanwhile colloquial debates at dinner club tables were more taxing than any calculation, and highlights of my week; I miss them dearly, thank you all. And thank you to everyone I have tied onto a rope with for unwittingly heeding Donal's advice while sharing many fun adventures at the same time. Here's to many more.

Finally, I could not have completed this thesis without the true love of my life, Ruth Green. Thank you for encouraging me to pursue my work in Edinburgh in the first place, for putting up with my geographical superposition from our nest, and for always being there for me.

I acknowledge financial support under STFC studentship ST/R504737/1.

%% file: chapter1/chapter1.tex
\chapter{Introduction}
\label{chap:intro}

%
The dawn of gravitational wave astronomy, heralded by the binary black hole and neutron star mergers detected by the LIGO and VIRGO collaborations~\cite{Abbott:2016blz,Abbott:2016nmj,Abbott:2017oio,Abbott:2017vtc,TheLIGOScientific:2017qsa}, has opened a new observational window on the universe. Future experiments offer the tantalising prospect of unprecedented insights into the physics of black holes, as well as neutron star structure, extreme nuclear matter and general relativity itself. Theorists have a critical role to play in this endeavour: to access such insights, an extensive bank of theoretical waveform templates are required for both event detection and parameter extraction \cite{Buonanno:2014aza}.

The vast majority of accessible data in a gravitational wave signal lies in the inspiral regime. This is the phase preceding the dramatic merger, in which the inspiralling pair coalesce and begin to influence each other's motion. The two bodies remain well separated, and one therefore can tackle their dynamics perturbatively: we can begin by treating them as point particles, and then increase precision by calculating corrections at higher orders in a given approximation. For an inspiral with roughly equal mass black holes the most directly applicable perturbative series is the non-relativistic \textit{post--Newtonian} (PN) expansion, where one expands in powers of the bodies' velocities $v$. Meanwhile the \textit{post--Minkowskian} (PM) expansion in powers of Newton's constant $G$ is fully relativistic, and thus more naturally suited to scattering interactions; however, it makes crucial contributions to precision inspiral calculations \cite{Antonelli:2019ytb}. Finally, when one black hole is far heavier than the other a \textit{self-force} expansion can be taken about the test body limit, expanding in the mass ratio of the black holes but keeping $v$ and $G$ to all orders. 

Although simple in concept, the inherent non-linearity of general relativity (GR) makes working even with these approximations an extremely difficult task. Yet future prospects in gravitational wave astronomy require perturbative calculations at very high precision~\cite{Babak:2017tow}. This has spawned interest in new techniques for solving the two-body problem in gravity and generating the required waveforms. Such techniques would complement methods based on the `traditional' Arnowitt--Deser--Misner Hamiltonian formalism~\cite{Deser:1959zza,Arnowitt:1960es,Arnowitt:1962hi,Schafer:2018kuf}, direct post--Newtonian solutions in harmonic gauge \cite{Blanchet:2013haa}, long-established effective-one-body (EOB) methods introduced by Buonnano and Damour~\cite{Buonanno:1998gg,Buonanno:2000ef,Damour:2000we,Damour:2001tu}, numerical-relativity approaches~\cite{Pretorius:2005gq,Pretorius:2007nq}, and the effective field theory approach pioneered by Goldberger and Rothstein~\cite{Goldberger:2004jt,Porto:2016pyg,Levi:2018nxp}.

Remarkably, ideas and methods from quantum field theory (QFT) offer a particularly promising avenue of investigation. Here, interactions are encoded by scattering amplitudes. Utilising amplitudes allows a powerful armoury of modern on-shell methods\footnote{See \cite{Cheung:2017pzi,Elvang:2015rqa} for an introduction.} to be applied to a problem, drawing on the success of the NLO (next-to-leading order) revolution in particle phenomenology. An appropriate method for extracting observables relevant to the problem at hand is also required. The relevance of a scattering amplitude --- in particular, a loop amplitude --- to the classical potential, for example, is well understood from work on gravity as an effective field theory~\cite{Iwasaki:1971,Duff:1973zz,Donoghue:1993eb,Donoghue:1994dn,Donoghue:1996mt,Donoghue:2001qc,BjerrumBohr:2002ks,BjerrumBohr:2002kt,Khriplovich:2004cx,Holstein:2004dn,Holstein:2008sx}. There now exists a panoply of techniques for applying modern amplitudes methods to the computation of the classical gravitational potential \cite{Neill:2013wsa,Bjerrum-Bohr:2013bxa,Bjerrum-Bohr:2014lea,Bjerrum-Bohr:2014zsa,Bjerrum-Bohr:2016hpa,Bjerrum-Bohr:2017dxw,Cachazo:2017jef,Cheung:2018wkq,Caron-Huot:2018ape,Cristofoli:2019neg,Bjerrum-Bohr:2019kec,Cristofoli:2020uzm,Kalin:2020mvi,Cheung:2020gbf}, generating results directly applicable to gravitational wave physics \cite{Damour:2016gwp,Damour:2017zjx,Bjerrum-Bohr:2018xdl,Bern:2019nnu,Brandhuber:2019qpg,Bern:2019crd,Huber:2019ugz,Cheung:2020gyp,AccettulliHuber:2020oou,Cheung:2020sdj,Kalin:2020fhe,Haddad:2020que,Kalin:2020lmz,Bern:2020uwk,Huber:2020xny,Bern:2021dqo}. By far the most natural relativistic expansion from an amplitudes perspective is the post--Minkowskian expansion (in the coupling constant): indeed, amplitudes methods have achieved the first calculations of the 3PM~\cite{Bern:2019nnu} and 4PM~\cite{Bern:2021dqo} potential. Furthermore, it is possible to use analytic continuation to obtain bound state observables directly from the scattering problem \cite{Kalin:2019rwq,Kalin:2019inp}. 

The gravitational potential is a versatile tool; however it is also coordinate, and thus gauge, dependent. The conservative potential also neglects the radiation emitted from interactions, leading to complications at higher orders. Amplitudes and physical observables, meanwhile, are on-shell and gauge-invariant, and should naturally capture all the physics of the problem. Direct maps between amplitudes and classical physics are well known to hold in certain regimes: for example, the eikonal exponentation of amplitudes in the extreme high energy limit has long been used to derive scattering angles \cite{Amati:1987wq,tHooft:1987vrq,Muzinich:1987in,Amati:1987uf,Amati:1990xe,Amati:1992zb,Kabat:1992tb,Amati:1993tb,Muzinich:1995uj,DAppollonio:2010krb,Melville:2013qca,Akhoury:2013yua,DAppollonio:2015fly,Ciafaloni:2015vsa,DAppollonio:2015oag,Ciafaloni:2015xsr,Luna:2016idw,Collado:2018isu,KoemansCollado:2019ggb,DiVecchia:2020ymx,DiVecchia:2021ndb}. Furthermore, calculating amplitudes in the high energy regime exposes striking universal features in gravitational scattering \cite{Bern:2020gjj,Parra-Martinez:2020dzs,DiVecchia:2020ymx}. Meanwhile, a careful analysis of soft limits of amplitudes with massless particles can extract data about both classical radiation \cite{Laddha:2018rle,Laddha:2018myi,Sahoo:2018lxl,Laddha:2018vbn,Laddha:2019yaj,A:2020lub,Sahoo:2020ryf,Bonocore:2020xuj} and radiation reaction effects \cite{DiVecchia:2021ndb}. Calculations with radiation can also be accomplished in the eikonal formalism \cite{Amati:1990xe}, but have proven particularly natural in classical worldline approaches, whereby one applies perturbation theory directly to point-particle worldlines rather than quantum states \cite{Goldberger:2016iau,Goldberger:2017frp, Goldberger:2017vcg,Goldberger:2017ogt,Chester:2017vcz,Li:2018qap, Shen:2018ebu,Plefka:2018dpa,Plefka:2019hmz,PV:2019uuv,Almeida:2020mrg,Prabhu:2020avf,Mougiakakos:2021ckm}. Using path integrals to develop a worldline QFT enables access to on-shell amplitudes techniques in this context \cite{Mogull:2020sak,Jakobsen:2021smu}, and this method has been used to calculate the NLO current due to Schwarzschild black hole bremsstrahlung.

We know how to extract information about classical scattering and radiation from quantum amplitudes in a gauge invariant manner --- but only in specific regimes. It is therefore natural to seek a more generally applicable, on-shell mapping between amplitudes and classical observables: this will form the topic of the first part of this thesis. We will construct general formulae for a variety of on-shell observables, valid in any quantum field theory and for any two-body scattering event. In this context we will also systematically study how to extract the classical limit of an amplitude, developing in the process a precise understanding of how to use quantum amplitudes to calculate observables for classical point-particles. We will show that by studying appropriate observables, expressed directly in terms of amplitudes, we can handle the nuances of the classical relationship between conservative and dissipative physics in a single, systematic approach, avoiding the difficulties surrounding the Abraham--Lorentz--Dirac radiation reaction force in electrodynamics \cite{Lorentz,Abraham:1903,Abraham:1904a,Abraham:1904b,Dirac:1938nz}. First presented in ref.~\cite{Kosower:2018adc}, the formalism we will develop has proven particularly useful for calculating on-shell observables for black hole processes involving classical radiation \cite{Luna:2017dtq,Bautista:2019tdr,Cristofoli:2020hnk,A:2020lub,delaCruz:2020bbn,Mogull:2020sak,Gonzo:2020xza,Herrmann:2021lqe} and spin \cite{Maybee:2019jus,Guevara:2019fsj,Arkani-Hamed:2019ymq,Moynihan:2019bor,Huang:2019cja,Bern:2020buy,Emond:2020lwi,Monteiro:2020plf}. Wider applications also exist to other aspects of classical physics, such as the Yang--Mills--Wong equations \cite{delaCruz:2020bbn,Wong:1970fu} and hard thermal loops \cite{delaCruz:2020cpc}.

Even the most powerful QFT techniques require a precise understanding of how to handle the numerous subtleties involved in taking the classical limit and accurately calculating observables. The reader may therefore wonder whether the philosophy of applying quantum amplitudes to classical physics really offers any fundamental improvement --- after all,  there are many concurrent advances in our understanding of the two-body problem arising from alternative calculational methods. For example, information from the self-force approximation can provide extraordinary simplifications directly at the level of classical calculations \cite{Bini:2020flp,Bini:2020hmy,Bini:2020nsb,Bini:2020uiq,Bini:2020wpo,Damour:2020tta,Bini:2020rzn}. Aside from the fact that amplitudes methods have achieved state-of-the-art precision in the PM approximation \cite{Bern:2019nnu,Bern:2021dqo,Herrmann:2021lqe}, such a sweeping judgement would be premature, as we have still yet to encounter two unique facets of the amplitudes programme: the double copy, and the treatment of spin effects.

\subsection{The double copy}

An important insight arising from the study of scattering amplitudes is that amplitudes in perturbative quantum gravity are far simpler than one would expect, and in particular are closely connected to the amplitudes of Yang--Mills (YM) theory. This connection is called the double copy, because gravitational amplitudes are obtained as a product of two Yang--Mills quantities. One can implement this double copy in a variety of ways: the original statement, by Kawai, Lewellen and Tye~\cite{Kawai:1985xq} presents a tree-level gravitational (closed string) amplitude as a sum over terms, each of which is a product of two tree-level colour-ordered Yang--Mills (open string) amplitudes, multiplied by appropriate Mandelstam invariants. More recently, Bern, Carrasco and Johansson~\cite{Bern:2008qj,Bern:2010ue} demonstrated that the double copy can be understood very simply in terms of a diagrammatic expansion of a scattering amplitude. They noted that any tree-level $m$-point amplitude in Yang--Mills theory could be expressed as a sum over the set of cubic diagrams $\Gamma$,
\begin{equation}
\mathcal{A}_{m} = g^{m-2}\sum_{\Gamma}\frac{n_i c_i}{\Delta_i}\,,
\end{equation}
where $\Delta_i$ are the propagators, $n_i$ are gauge-dependent kinematic numerators, and $c_i$ are colour factors which are related in overlapping sets of three by Jacobi identities,
\begin{equation}
c_\alpha \pm c_\beta \pm c_\gamma = 0\,.
\end{equation}
The colour factors are single trace products of $SU(N)$ generators $T^a$, normalised such that $\textrm{tr}(T^aT^b) = \delta^{ab}$. Remarkably, BCJ found that gauge freedom makes it possible to always choose numerators satisfying the same Jacobi identities \cite{Bern:2010yg}. This fundamental property is called \textit{colour-kinematics duality}, and has been proven to hold for tree-level Yang--Mills theories \cite{Bern:2010yg}.

When colour-kinematics duality holds, the double copy then tells us that
\begin{equation}
\mathcal{M}_m = \left(\frac{\kappa}{2}\right)^{m-2}\sum_{\Gamma}\frac{n_i\tilde{n}_i}{\Delta_i}\,
\end{equation}
is the corresponding $m$-point gravity amplitude, obtained by the replacements
\begin{equation}
g\mapsto\frac{\kappa}{2}\,,\quad c_i\mapsto \tilde{n}_i\,.
\end{equation}
Here $\kappa = \sqrt{32\pi G}$ is the appropriate gravitational coupling, and $\tilde{n}_i$ is a distinct second set of numerators satisfying colour-kinematics duality. The choice of numerator determines the resulting gravity theory. To obtain gravity the original numerators are chosen, and thus amplitude numerators for gravity are simply the square of kinematic numerators in Yang--Mills theory, provided that colour-kinematics duality holds. 

One complication is that regardless of the choice of $\tilde{n}_i$, the result is not a pure theory of gravitons, but instead is a factorisable graviton multiplet. This can easily be seen in pure Yang--Mills theory, where the tensor product $A^\mu \otimes A^\nu \sim \phi_\textrm{d} \oplus B^{\mu\nu} \oplus h^{\mu\nu}$ leads to a scalar dilaton field, antisymmetric Kalb--Ramond axion and traceless, symmetric graviton respectively. In 4 dimensions the axion has only one degree of freedom, so its field strength $H^{\mu\nu\rho} = \partial^{[\mu}B^{\nu\rho]}$ can be written as
\begin{equation}
H^{\mu\nu\rho}=\frac{1}{2}\epsilon^{\mu\nu\rho\sigma}\partial_\sigma \zeta,\label{eqn:axionscalar}
\end{equation}
with $\zeta$ representing the single propagating pseudoscalar degree of freedom. There any many possible ways to deal with the unphysical axion and dilaton modes and isolate the graviton degrees of freedom \cite{Bern:2019prr} --- we will see some such methods in the course of the thesis. However, the main point here is that Einstein gravity amplitudes can be determined exclusively by gauge theory data.

This modern formulation of the double copy is particularly exciting as it has a clear generalisation to loop level; one simply includes integrals over loop momentum and appropriate symmetry factors. A wealth of non-trivial evidence  supports this conjecture --- for reviews, see \cite{Carrasco:2015iwa,Bern:2019prr}. The work of BCJ suggests that gravity may be simpler than it seems, and also more closely connected to Yang--Mills theory than one would guess after inspecting their Lagrangians. Here our simple presentation of colour-kinematics duality was only for tree level, massless gauge theory. However, the double copy can be applied far more generally: it forms bridges between a veritable web of theories, for both massless and massive states \cite{Johansson:2014zca,Johansson:2015oia,Johansson:2019dnu,Haddad:2020tvs}.

Since perturbation theory is far simpler in Yang--Mills theory than in standard approaches to gravity, the double copy has revolutionary potential for gravitational physics. Indeed, it has proven to be the key tool enabling state-of-the-art calculations of the PM potential from amplitudes \cite{Bern:2019crd,Bern:2019nnu,Bern:2021dqo}. Furthermore, it has also raised the provocative question of whether exact solutions in general relativity satisfy similar simple relationships to their classical Yang--Mills counterparts, extending the relationship beyond perturbation theory. First explored in \cite{Monteiro:2014cda}, many exact classical double copy maps are now known to hold between classical solutions of gauge theory and gravity \cite{Luna:2015paa,Luna:2016hge,Adamo:2017nia,Bahjat-Abbas:2017htu,Carrillo-Gonzalez:2017iyj,Lee:2018gxc,Berman:2018hwd,Carrillo-Gonzalez:2018pjk,Adamo:2018mpq,Luna:2018dpt,CarrilloGonzalez:2019gof,Cho:2019ype,Carrillo-Gonzalez:2019aao,Bah:2019sda,Huang:2019cja,Alawadhi:2019urr,Borsten:2019prq,Kim:2019jwm,Banerjee:2019saj,Bahjat-Abbas:2020cyb,Moynihan:2020gxj,Adamo:2020syc,Alfonsi:2020lub,Luna:2020adi,Keeler:2020rcv,Elor:2020nqe,Alawadhi:2020jrv,Casali:2020vuy,Adamo:2020qru,Easson:2020esh,Chacon:2020fmr,Emond:2020lwi,White:2020sfn,Monteiro:2020plf,Lescano:2021ooe}, even when there is gravitational radiation present~\cite{Luna:2016due}. 

To emphasise that the classical double copy has not been found simply for esoteric exact solutions, let us briefly consider the original Kerr--Schild map constructed in \cite{Monteiro:2014cda}. Kerr--Schild spacetimes are a particularly special class of solutions possessing sufficient symmetry that their metrics can be written
\begin{equation}
g_{\mu\nu} = \eta_{\mu\nu} + \varphi k_\mu k_\nu\,,\label{eqn:KSmetric}
\end{equation}
where $\varphi$ is a scalar function and $k_\mu$ is null with respect to both the background and full metric, and satisfies the background geodesic equation:
\begin{equation}
g^{\mu\nu} k_\mu k_\nu = \eta^{\mu\nu} k_\mu k_\nu = 0\,, \qquad k\cdot\partial k_\mu = 0\,.\label{eqn:KSvector}
\end{equation}
The symmetries of this class of spacetime ensure that the (mixed index placement) Ricci tensor is linearised. It was proposed in \cite{Monteiro:2014cda} that for such spacetimes there then exists a single copy gauge theory solution,
\begin{equation}
A^a_\mu = \varphi\, c^a k_\mu\,.\label{eqn:singleKScopy}
\end{equation}
where $c^a$ is a classical colour charge. This incarnation of the double copy is therefore enacted by replacing copies of the classical colour with the null vector $k_\mu$, in analogue to the BCJ amplitude replacement rules.

The crucial importance of the Kerr--Schild double copy is that it encompasses both Schwarzschild and Kerr black holes. Both (exterior) spacetime metrics can be written in the compact Kerr--Schild form, with respective data \cite{Monteiro:2014cda}
\begin{equation}
\varphi_\textrm{Schwz}(r) = \frac{2GM}{r}\,, \quad k^\mu = \left(1, \frac{\v{x}}{r}\right)\\
\end{equation}
for Schwarzschild, where $r^2 = \v{x}^2$; and
\begin{equation}
\varphi_\textrm{Kerr}(\tilde r, \theta) = \frac{2GM\tilde r}{\tilde r^2 + a^2 \cos^2\theta}\,, \quad k^\mu = \left(1,\frac{\tilde r x + ay}{\tilde r^2 + a^2}, \frac{\tilde ry - ax}{\tilde r^2 + a^2},\frac{z}{r}\right)\label{eqn:blackholesKSforms}
\end{equation}
for Kerr, where the parameter $a$ is the radius of the Kerr singularity about the $z$ axis. This key parameter is the norm of a pseudovector $a^\mu$ which fully encodes the spin of the black hole, the \textit{spin vector}. It is important to note that in the Kerr case $(\tilde r, \theta)$ are not the usual polar coordinates, instead satisfying
\begin{equation}
\frac{x^2 +y^2}{\tilde r^2 + a^2} + \frac{z^2}{\tilde{r}^2} = 1\label{eqn:KerrKSradial}
\end{equation}
and $z = \tilde r \cos\theta$. The corresponding gauge theory single copies are then given by \eqn~\eqref{eqn:singleKScopy}. The Schwarzchild single copy is simply a Coulomb charge. The Kerr single copy meanwhile is a disk of uniform charge rotating about the $z$, axis whose mass distribution exhibits a singularity at $x^2 + y^2 = a^2$ \cite{Monteiro:2014cda}. We will refer to this unique charged particle by its modern name, $\rootKerr$ \cite{Arkani-Hamed:2019ymq}.

The $\rootKerr$ solution was first explored by Israel in \cite{Israel:1970kp}, and will be of great interest for us in the second part of the thesis: its double copy relation to Kerr ensures that the structure and dynamics of Kerr in gravity are precisely mirrored by the behaviour of $\rootKerr$ in gauge theory, where calculations are often simpler. This is particularly important in the context of the second key area in which applying amplitudes ideas to  black hole interactions can offer a significant computational and conceptual advantage: spin.


\subsection{Spin}

The astrophysical bodies observed in gravitational wave experiments spin. The spins of the individual bodies in a compact binary coalescence event influence the details of the outgoing gravitational radiation \cite{Buonanno:2014aza}, and moreover contain information on the poorly-understood formation channels of the binaries \cite{Mandel:2018hfr}. Measurement of spin is therefore one of the primary physics outputs of gravitational wave observations.

Any stationary axisymmetric extended body has an infinite tower of mass-multipole moments $\mathcal{I}_\ell$ and current-multipole moments $\mathcal{J}_\ell$, which generally depend intricately on its internal structure and composition. In the point-particle limit it is thus the multipole structure of the body which accurately identifies to an observer whether that object is a neutron star, black hole or other entity. Incorporating spin multipoles into the major theoretical platform for these experiments, the EOB formalism \cite{Buonanno:1998gg,Buonanno:2000ef}, is well established in the PN approximation \cite{Damour:2001tu,Damour:2008qf,Barausse:2009aa,Barausse:2009xi,Barausse:2011ys,Damour:2014sva,Bini:2017wfr,Khalil:2020mmr}, and has also been extended to the PM approximation by means of a gauge-invariant spin holonomy \cite{Bini:2017xzy}. This has been used to compute the dipole (or spin-orbit) contribution to the conservative potential for two spinning bodies through 2PM order \cite{Bini:2018ywr}. Calculating higher-order PN spin corrections has been a particular strength of the effective field theory treatment of PN dynamics \cite{Goldberger:2004jt,Porto:2005ac,Porto:2006bt,Porto:2008tb,Levi:2011eq,Levi:2015msa,Levi:2016ofk,Levi:2020kvb,Levi:2020uwu,Levi:2020lfn}, while self-force data has also driven independent progress in this approximation \cite{Siemonsen:2019dsu,Antonelli:2020aeb,Antonelli:2020ybz}. EFT progress in the handling of spin has also recently been extended to the PM series, yielding the first calculation of finite-size effects beyond leading-order \cite{Liu:2021zxr}; moreover, these results can be mapped to bound observables by analytic continuation \cite{Kalin:2019rwq,Kalin:2019inp}.

A common feature of all of these calculations is that they are significantly more complicated than the spinless examples considered previously, and moreover are nearly unanimously restricted to the special case where the spins of the bodies are aligned with each other.

The black hole case is special. For a Kerr black hole, every multipole is determined by only the mass $m$ and spin vector $a^\mu$, through the simple relation due to Hansen \cite{Hansen:1974zz},
\begin{equation}
\mathcal I_\ell+i\mathcal J_\ell = m\left(ia\right)^\ell \,.\label{eqn:multipoles}
\end{equation}
This distinctive behaviour is a precise reflection of the \textit{no-hair theorem} \cite{Israel:1967wq,Israel:1967za,Carter:1971zc}, which ensures that higher multipoles are constrained by the dipole. This simple multipole structure is also reflected in the dynamics of spinning black holes --- for example, remarkable all-spin results are known for black hole scattering at leading order in both the PN and PM approximations \cite{Vines:2016qwa,Siemonsen:2017yux,Vines:2017hyw}; aligned-spin black hole scattering was also considered at 2PM order for low multipoles in \cite{Vines:2018gqi}.

Moreover, over the last few years it has become increasingly apparent that an on-shell expression of the no-hair theorem is that black holes correspond to \textit{minimal coupling} in classical limits of quantum scattering amplitudes for massive spin~$s$ particles and gravitons.  Amplitudes for long-range gravitational scattering of spin 1/2 and spin 1 particles were found in \cite{Ross:2007zza,Holstein:2008sx} to give the universal spin-orbit (pole-dipole level) couplings in the post-Newtonian corrections to the gravitational potential.  Further similar work in \cite{Vaidya:2014kza}, up to spin 2, suggested that the black hole multipoles \eqref{eqn:multipoles} up to order $\ell=2s$ are faithfully reproduced from tree-level amplitudes for minimally coupled spin~$s$ particles.

Such amplitudes for arbitrary spin $s$ were computed in \cite{Guevara:2017csg}, by adopting the representation of minimal coupling for arbitrary spins presented in \cite{Arkani-Hamed:2017jhn} using the massive spinor-helicity formalism---see also~\cite{Conde:2016vxs,Conde:2016izb}. Those amplitudes were shown in \cite{Guevara:2018wpp,Bautista:2019tdr} to lead in the limit $s\to\infty$ to the two-black-hole aligned-spin scattering angle found in \cite{Vines:2017hyw} at first post--Minkowskian order and to all orders in the spin-multipole expansion, while in \cite{Chung:2018kqs} they were shown to yield the contributions to the interaction potential (for arbitrary spin orientations) at the leading post--Newtonian orders at each order in spin. Meanwhile in \cite{Bern:2020buy,Kosmopoulos:2021zoq} amplitudes for arbitrary spin fields were combined with the powerful effective theory matching techniques of \cite{Cheung:2018wkq} to yield the first dipole-quadrapole coupling calculation at 2PM order. Methods from heavy quark effective theory \cite{Damgaard:2019lfh,Aoude:2020onz,Haddad:2020tvs} and quantum information \cite{Aoude:2020mlg} have also proven applicable to spinning black hole scattering, the former leading to the first amplitudes treatment of tidal effects on spinning particles \cite{Aoude:2020ygw}.

To replicate the behaviour of Kerr black holes, the massive spin $s$ states in amplitudes must be minimally coupled to the graviton field, by which we mean that the high energy limit is dominated by the corresponding helicity configuration of massless particles \cite{Arkani-Hamed:2017jhn}. This has been especially emphasised in \cite{Chung:2018kqs}, where, by matching at tree-level to the classical effective action of Levi and Steinhoff \cite{Levi:2015msa}, it was shown that the theory which reproduces the infinite-spin limit of minimally coupled graviton amplitudes is an effective field theory (EFT) of spinning black holes. It can be explicitly shown that any deviation from minimal coupling adds further internal structure to the effective theory \cite{Chung:2019duq,Chung:2019yfs,Chung:2020rrz}, departing the special black hole case. 

Applying amplitudes methods to the scattering of any spinning object, black hole or otherwise, we face the familiar problem of requiring an appropriate observable and precise understanding of the classical limit. When the spins of scattering objects are not aligned there no longer exists a well defined scattering plane, and thus the most common observable calculated from classical potentials, the scattering angle, becomes meaningless. We shall therefore apply the methods developed in Part~\ref{part:observables} to quantum field theories of particles with spin, setting up observables in terms of scattering amplitudes which can fully specify the dynamics of spinning black holes. When the spins are large these methods are known to exactly reproduce established 1PM results \cite{Guevara:2019fsj,Vines:2017hyw}. 

After systematically dealing with the classical limit of quantum spinning particles, we will apply insights from amplitudes to the classical dynamics of Kerr and its single copy, $\rootKerr$. We will utilise the fact that the beautiful relationship between Kerr black holes and minimally coupled amplitudes goes far deeper than simply being a powerful calculational tool. Amplitudes can explain and reveal structures in general relativity that are obscured by geometrical perspectives: for example, the double copy. 

Another key example in the context of spin is the fact, first noted by Newman and Janis in \cite{Newman:1965tw}, that the Kerr metric can be obtained from Schwarzschild by means of a complex coordinate transformation. This is easy to see when the metrics are in Kerr--Schild form: take the data for Scwharzschild in \eqn~\eqref{eqn:blackholesKSforms}. Under the transformation $z \rightarrow z + ia$,
\[
r^2 \rightarrow &\, r^2 + 2iaz - a^2 \\ &\equiv  \tilde r^2 - \frac{a^2 z^2}{\tilde r^2} + 2ia\tilde{r}\cos\theta = (\tilde r +i a\cos\theta)^2\,,
\]
where the Kerr radial coordinate $\tilde r$ is defined in \eqn~\eqref{eqn:KerrKSradial}. Hence under $z \rightarrow z + ia$ we have that $r \rightarrow \tilde r + ia\cos\theta$, and moreover,
\[
\varphi_{\rm Schwz}(r) \rightarrow &\, 2GM\Re\left\{\frac1{r}\right\}\bigg|_{r\rightarrow \tilde r + ia\cos\theta}\\ &= \frac{2GM \tilde r}{(\tilde r^2 + a^2 \cos^2\theta)} \equiv \varphi_{\rm Kerr}(\tilde r,\theta)\,.
\]
In other words, the Kerr solution looks like a complex translation of the Schwarzschild solution \cite{Newman:2002mk}. Clearly the same shift holds in the gauge theory single copies. This is but one example of complex maps between spacetimes; further examples were derived by Talbot \cite{Talbot:1969bpa}, encompassing the Kerr--Newman and Taub--NUT solutions.

A closely related way to understand these properties of classical solutions is to consider their Weyl curvature spinor $\Psi$. For example, with appropriate coordinates the NJ shift applies exactly to the spinor:
\[
\Psi^\text{Kerr}(x) = \Psi^\text{Schwarzschild}(x + i a) \,,\label{eqn:NJshift}
\]
Similarly, in the electromagnetic $\rootKerr$~case~\cite{Newman:1965tw} it is the Maxwell spinor $\maxwell$ that undergoes a shift:
\[
\maxwell^{\sqrt{\text{Kerr}}}(x) = \maxwell^\text{Coulomb}(x + i a) \,.
\]
Therefore, $\rootKerr$~is a kind of complex translation of the Coulomb solution. 

Although these complex maps are established classically, there is no geometric understanding for \textit{why} such a complex map holds. However, this is not the case from the perspective of amplitudes --- it was explicitly shown in \cite{Arkani-Hamed:2019ymq} that the Newman--Janis shift is a simple consequence of the the exponentiation of minimally coupled amplitudes in the large spin limit. The simplicity of minimally coupled amplitudes has since been utilised to explain a wider range of complex mappings between spacetime and gauge theory solutions \cite{Moynihan:2019bor,Huang:2019cja,Moynihan:2020gxj,Kim:2020cvf}, culminating in a precise network of relationships constructed from the double copy, Newman--Janis shifts and electric-magnetic duality \cite{Emond:2020lwi}. These investigations have relied on the on-shell observables we will introduce in Part~\ref{part:observables} \cite{Kosower:2018adc}.

Inspired by the insights offered by amplitudes, we will adopt the complex Newman--Janis shift as a starting point for investigating the classical dynamics of these unique spinning objects. In particular, we will show that interacting effective actions for Kerr and $\rootKerr$, in the vein of Levi and Steinhoff \cite{Levi:2015msa}, can be interpreted as actions for a complex worldsheet. We will also apply the power of the massive spinor helicity representations of ref.~\cite{Arkani-Hamed:2017jhn} to classical dynamics, rapidly deriving on-shell scattering observables for Kerr and $\rootKerr$ from spinor equations of motion. Although working purely classically, our methodology and philosophy will be entirely drawn from amplitudes-based investigations.

\section{Summary}

To summarise, the structure of this thesis is as follows. Part~\ref{part:observables} is dedicated to the construction of on-shell observables which are well defined in both classical and quantum field theory. We begin in chapter~\ref{chap:pointParticles} by setting up single particle quantum wavepackets which describe charged scalar point-particles in the classical limit. In chapter~\ref{chap:impulse} we then turn to descriptions of point-particle scattering by considering our first observable, the impulse, or the total change in the momentum of a scattering particle. We derive general expressions for this observable in terms of amplitudes, and undertake a careful examination of the classical limit, extracting the rules needed to pass from the quantum to the classical regime. In chapter~\ref{chap:radiation} we introduce the total radiated momentum and demonstrate momentum conservation and the automatic handling of radiation reaction effects in our formalism. We introduce spin in part~\ref{part:spin}, which is concerned with spinning black holes. In chapter~\ref{chap:spin} we construct on-shell observables in QFT for spinning particles, reproducing results for Kerr black holes after considering in detail the classical limit of amplitudes with finite spin. We then return to classical dynamics in chapter~\ref{chap:worldsheet}, using insights from structures in on-shell amplitudes to uncover worldsheet effective actions for $\rootKerr$ and Kerr particles. We finish by discussing our results in~\ref{chap:conclusions}. Results in chapters~\ref{chap:pointParticles}, \ref{chap:impulse} and~\ref{chap:radiation} were published in refs.~\cite{Kosower:2018adc,delaCruz:2020bbn}, chapter~\ref{chap:spin} is based on \cite{Maybee:2019jus}, and chapter~\ref{chap:worldsheet} appeared in \cite{Guevara:2020xjx}. 

\subsection{Conventions}

In all the work that follows, our conventions for Fourier transforms are
\begin{equation}
f(x) = 	\int\!\frac{\d^n q}{(2\pi)^n}\, \tilde{f}(q) e^{-i q\cdot x}\,, \qquad \tilde{f}(q) = \int\! d^4x\, f(x) e^{i q\cdot x}\,.
\end{equation}
We will consistently work in relativistically natural units where $c=1$, however we will always treat $\hbar$ as being dimensionful. We work in the mostly minus metric signature $(+,-,-,-)$, where we choose $\epsilon_{0123} = +1$ for the Levi--Civita tensor. We will occasionally find it convenient to separate a Lorentz vector $x^\mu$ into its time component $x^0$ and its spatial components $\v{x}$, so that $x^\mu = (x^0,x^i) = (x^0, \v{x})$, where $i=1,2,3$. 

For a given tensor $X$ of higher rank, total symmmetrisation and antisymmetrisation respectively of tensor indices are represented as usual by
\begin{equation}
\begin{aligned}
X^{(\mu_1} \dots X^{\mu_n)} &= \frac1{n!}\left(X^{\mu_1} X^{\mu_2} \dots X^{\mu_n} + X^{\mu_2} X^{\mu_1} \dots X^{\mu_n} + \cdots\right)\\
X^{[\mu_1} \dots X^{\mu_n]} &= \frac1{n!}\left(X^{\mu_1} X^{\mu_2} \dots X^{\mu_n} - X^{\mu_2} X^{\mu_1} \dots X^{\mu_n} + \cdots\right).
\end{aligned}
\end{equation}
Finally, our definition of the amplitude will consistently differ by a phase factor relative to the standard definition used for the double copy. Here, in either gauge theory or gravity
\begin{equation}
i\mathcal{A}(p_1, p_2 \rightarrow p_1 + q, p_2 - q) = \sum \left(\text{Feynman diagrams}\right)\,,
\end{equation}
whereas in the convention used in the original work of BCJ~\cite{Bern:2008qj,Bern:2010ue} the entire left hand side is defined as the amplitude.

%% file: chapter2/chapter2.tex
\part{Classical observables from quantum field theory}	
\label{part:observables}

\chapter{Point-particles}
\label{chap:pointParticles}

On-shell amplitudes in quantum field theory are typically calculated on a basis of plane wave states: the context of the calculation is the physics of states with definite momenta, but indefinite positions. However, to capture the physics of a black hole in quantum mechanics, or indeed any other classical point-particle, this is clearly not sufficient: we need states which are well localised. We also need quantum states that accurately correspond to point-particles when the ``classical limit'' is taken. These requirements motivate the goals of this first chapter. We will precisely specify what we mean by the classical limit, and explicitly construct localised wavepackets which describe single, non-spinning point-particles in this limit. The technology that we develop will provide the foundations for our construction of on-shell observables for interacting black holes in later chapters.

To ensure full generality we will consider charged particles, studying the classical limits of states in an $SU(N)$ gauge group representation. Such point-particles are described by the Yang--Mills--Wong equations in the classical regime~\cite{Wong:1970fu}. For gravitational physics one could have in mind an Einstein--Yang--Mills black hole, but there are more interesting perspectives available. YM theory, treated as a classical field theory, shares many of the important physical features of gravity, including non-linearity and a subtle gauge structure. In this respect the YM case has always served as an excellent toy model for gravitational dynamics. But, as we discussed in the previous chapter, our developing understanding of the double copy has taught us that the connection between Yang--Mills theory and gravity is deeper than this; detailed aspects
of the perturbative dynamics of gravity, including gravitational radiation, can be deduced from Yang--Mills theory and the double copy. Understanding non--trivial gauge, or \textit{colour}, representation states will thus play a key role in our later calculations of black hole observables. 

This chapter is based on work published in refs.~\cite{Kosower:2018adc,delaCruz:2020bbn}, in collaboration with Leonardo de la Cruz, David Kosower, Donal O'Connell and Alasdair Ross.

\section{Restoring $\hbar$}
\label{sec:RestoringHBar}

To extract the classical limit of a quantum mechanical system describing the physics of point-particles we are of course going to need to be careful in our treatment of Planck's constant, $\hbar$. A straightforward and pragmatic approach to restoring all factors of $\hbar$ in an expression is dimensional analysis: we denote the dimensions of mass and length by $[M]$ and $[L]$ respectively.

We may choose the dimensions of an $n$-point scattering amplitude in four dimensions to be $[M]^{4-n}$ even when $\hbar \neq 1$. This is consistent with choosing the dimensions of creation and annihilation operators so that
\begin{equation}
[a_i(p), a^{\dagger j}(p')] = 2E_p (2\pi)^3 \delta^{(3)}(\v{p} - \v{p}')\,\delta_i{ }^j\,,\label{eqn:ladderCommutator}
\end{equation}
Here the indices label the representation $R$ of any Lie group. We define single-particle momentum eigenstates in this representation by
\begin{equation}
|p^i \rangle = a^{\dagger i}(p) |0\rangle\,.\label{eqn:singleParticleStateDef}
\end{equation}
Since the vacuum state is taken to be dimensionless, the dimension of $|p^i\rangle$ is thus $[M]^{-1}$. We further define $n$-particle asymptotic states as tensor products of these normalised single-particle states. In order to avoid an unsightly splatter of factors of $2\pi$, it is convenient to define
\begin{equation}
\del^{(n)}(p) \equiv (2\pi)^n \delta^{(n)}(p)
\label{eqn:delDefinition}
\end{equation}
for the $n$-fold Dirac $\delta$ distribution. With these conventions the state normalisation is
\begin{equation}
\langle p'_i | p^j \rangle = 2 E_p \, \del^{(3)} (\v{p}-\v{p}') \delta_i{ }^j\,.
\label{eqn:MomentumStateNormalization}
\end{equation}
We define the amplitudes in four dimensions on this plane wave basis by
\begin{multline}
\langle p'_1 \cdots p'_m  | T | p_1 \cdots p_n \rangle = \Ampl(p_1 \cdots p_n \rightarrow p'_1 \cdots p'_m) \\ \times \del^{(4)}(p_1 + \cdots p_n - p'_1 - \cdots - p'_m)\,.
\label{eqn:amplitudeDef}
\end{multline}
The scattering matrix $S$ and the transition matrix $T$ are both dimensionless, leading to the initially advertised dimensions for amplitudes.

Let us now imagine restoring the $\hbar$'s in a given amplitude. When $\hbar = 1$, the amplitude has dimensions of $[M]^{4-n}$.  When $\hbar \neq 1$, the dimensions of the momenta and masses in the amplitude are unchanged. Similarly there is no change to the dimensions of polarisation vectors. However, we must remember that the dimensionless coupling in electrodynamics is $e/\sqrt{\hbar}$. Similarly, in gravity a factor of $1/\sqrt{\hbar}$ appears, as the appropriate coupling with dimensions of inverse mass is $\kappa = \sqrt{32 \pi G/ \hbar}$. We will see shortly that the situation is a little more intricate in Yang-Mills theory, as the colour factors can carry dimensions of $\hbar$. However we will establish conventions such that the coupling has the same scaling as the QED/gravity case.  The algorithm to restore the dimensions of any amplitude in electrodynamics, chromodynamics or gravity is then simple: each factor of a coupling is multiplied by an additional factor of $1/\sqrt{\hbar}$. For example, an $n$-point, $L$-loop amplitude in scalar QED is proportional to $\hbar^{1-n/2-L}$. 

This conclusion, though well-known, may be surprising in the present context because it seems na\"{i}vely that as $\hbar \rightarrow 0$, higher multiplicities and higher loop orders are \textit{more\/} important. However, when restoring powers of $\hbar$ one must distinguish between the momentum $p^\mu$ of a particle and its wavenumber, which has dimensions of $[L]^{-1}$. This distinction will be important for us, so we introduce a notation for the wavenumber $\barp$ associated with a momentum $p$:
\begin{align}
\wn p \equiv p / \hbar.
\label{eqn:notationWavenumber}
\end{align}
In the course of restoring powers of $\hbar$ by dimensional analysis, we will first treat the momenta of all particles as genuine momenta. We will also treat any mass as a mass, rather than the associated (reduced) Compton wavelength $\ell_c = \hbar/m$.

As we will, the approach to the classical limit --- for observables that make sense classically --- effectively forces the wavenumber scaling upon certain momenta. Examples include the momenta of massless particles, such as photons or gravitons.  In putting the factors of $\hbar$ back into the couplings, we have therefore not yet made manifest all of the physically relevant factors of $\hbar$ in an amplitude. This provides one motivation for this part of the thesis: we wish to construct on-shell observables which are both classically and quantum-mechanically sensible. 

\section{Single particle states}
\label{sec:stateSetup}

Let us take a generic single particle state expanded on the plane wave basis of \eqref{eqn:singleParticleStateDef}:
\begin{equation}
|\psi\rangle = \sum_i \int\! \dd^4 p \, \delp(p^2 - m^2) \, \psi_i(p) \, |p^i\rangle\,,\label{eqn:InitialState}
\end{equation}
Here $\dd p$ absorbs a factor of $2 \pi$; more generally $\dd^n p$ is defined by
\begin{equation}
\dd^n p \equiv \frac{d^n p}{(2 \pi)^n}\,.
\label{eqn:ddxDefinition}
\end{equation}
We restrict the integration to positive-energy solutions of the delta functions of $p^2-m^2$, as indicated by the $(+)$ superscript in $\delp$, as well as absorbing a factor of $2\pi$, just as for $\del(p)$:
\begin{equation}
\delp(p^2-m^2) \equiv 2\pi\Theta(p^0)\delta(p^2-m^2)\,.
\label{eqn:delpDefinition}
\end{equation}
We will find it convenient to further abbreviate the notation for on-shell integrals (over Lorentz-invariant phase space), defining
\begin{equation}
\df(p) \equiv \dd^4 p \, \delp(p^2-m^2)\,.
\label{eqn:dfDefinition}
\end{equation}
We will generally leave the mass implicit, along with the designation of the integration variable as the first summand when the argument is a sum. Note that the right-hand side of~\eqref{eqn:MomentumStateNormalization} is the appropriately normalised delta function for this measure, 
\begin{equation}
\label{eqn:norm1}
\int \df(p') \, 2 E_{p'} \, \del^{(3)} (\v{p}-\v{p}') f(p') = f(p)\,.
\end{equation}
Thus for any function $f(p_1')$, we define
\begin{equation}
\Del(p-p') \equiv  2 E_{p'} \del^{(3)} (\v{p}-\v{p}')\,.
\end{equation}
The argument on the left-hand side is understood as a function of four-vectors. This leads to a notationally clearer version of \eqn~\eqref{eqn:norm1}:
\begin{equation}
\int \df(p') \, \Del(p - p') f(p') = f(p)\,,
\end{equation}
and of \eqn~\eqref{eqn:MomentumStateNormalization}:
\begin{equation}
\langle p'_i | p^j \rangle = \Del(p-p') \delta_i{ }^j\,.
\end{equation}

The full state $|\psi\rangle$ is a non-trivial representation, of a Lie group associated with symmetries which constrain the description of our particle. The kinematic data, however, should be independent of these symmetries, and thus a singlet of $R$. The full state is thus a tensor product of momentum and representation states: 
\begin{align}
\ket{\psi}= \sum \ket{\psi_{\text{mom}}} \otimes  \ket{\psi_{R}}.
\end{align}
We will make this explicit by splitting the wavefunctions $\psi_i(p)$, writing
\begin{equation}
\sum_i\psi_i(p)| p^i\rangle = \sum_i \varphi(p) \chi_i |p^i\rangle = \varphi(p) |p\, \chi\rangle\,.\label{eqn:wavefunctionSplit}
\end{equation}

In these conventions, \eqn~\eqref{eqn:InitialState} becomes
\begin{equation}
| \psi \rangle = \int \! \df(p)\;
\varphi(p) | p \, \chi \rangle\,.
\label{eqn:InitialStateSimple}
\end{equation}
Using this simplified notation, the normalisation condition is
\begin{equation}
\begin{aligned}
1 &= \langle \psi | \psi \rangle \\
&= \sum_{i,j} \! \int \! \df(p) \df(p') \varphi^*(p_1') \varphi(p_1) \chi^{*i} \chi_j\, \Del(p_1 - p_1') \, \delta_i{ } ^j\\
&= \sum_i \! \int \! \df(p)\; |\varphi(p)|^2 |\chi_i|^2\,.
\end{aligned}
\end{equation}
We can obtain this normalisation by requiring that both wavefunctions $\phi(p)$ and $\chi_i$ be normalised to unity:
\begin{equation}
\int \! \df(p)\; |\varphi(p)|^2 = 1\,, \qquad \sum_i \chi^{i*} \chi_i = 1\,.
\label{eqn:WavefunctionNormalization}
\end{equation}

Since $|\psi\rangle$ is expanded on a basis of momentum eigenstates, it is trivial to measure the momentum of the state with the momentum operator $\mathbb P^\mu$:
\begin{equation}
\langle \psi|\mathbb{P}^\mu |\psi\rangle = \int \! \df(p)\; p^\mu |\varphi(p)|^2\label{eqn:momentumExp}
\end{equation} 
But how do we measure the physical charges associated with the representation states $|\chi\rangle$?

\subsection{Review of the theory of colour}
\label{sec:setup}

For a physical particle state, there are two distinct interpretations for the representation $R$: it could be the irreducible representation of the little group for a particle of non-zero spin; or it could be the representation of an internal symmetry group of the theory. In this first part of the thesis we are only interested in scalar particle states. We will therefore restrict to the second option for the time being, returning to little group representations in part~\ref{part:spin}. 

Our ultimate goal is to extract, from QFT, long-range interactions between point particles, mediated by a classical field. We will therefore take $R$ to be any representation of an $SU(N)$ gauge group. The classical dynamics of the corresponding Yang--Mills field $A_\mu = A_\mu^a T^a$, coupled to several classical point-like particles, are then described by the Yang--Mills--Wong equations:. 
\begin{subequations}
	\label{eqn:classicalWong}
	\begin{gather}
	\frac{\d p_\alpha^\mu }{\d \tau_\alpha} = g\, c^a_\alpha(\tau_\alpha)\, F^{a\,\mu\nu}\!(x_\alpha(\tau_\alpha))\,  v_{\alpha\, \nu}(\tau_\alpha)\,,  \label{eqn:Wong-momentum} 
	\\
	\frac{\d c^a_\alpha}{\d \tau_\alpha}= g f^{abc} v^\mu_\alpha(\tau_\alpha) A_\mu^b(x_\alpha(\tau_\alpha))\,c^c_\alpha(\tau_\alpha)\,,
	\label{eqn:Wong-color}
	\\
	D^\mu F_{\mu\nu}^a(x) = J^a_\nu(x) = g \sum\limits_{\alpha= 1}^N \int\!\d\tau_\alpha\,  c^a_\alpha(\tau_\alpha)  v^\mu(\tau_\alpha)\, \delta^{(4)}(x-x(\tau_\alpha))\,, \label{eqn:YangMillsEOM}
	\end{gather}
\end{subequations}
These equations describe particles, following worldlines $x_\alpha(\tau_\alpha)$ and with velocities $v_\alpha$, that each carry colour charges $c^a$ which are time-dependent vectors in the adjoint-representation of the gauge group. 

Let us review the emergence of of these non-Abelian colour charges from quantum field theory by restricting our attention to scalars $\pi_\alpha$ in any representation $R_\alpha$ of the gauge group, coupled to the Yang--Mills field. The action is
\begin{equation}
S = \int\!\d^4x\, \left(\sum_{\alpha}\left[ (D_\mu \pi_\alpha)^\dagger D^\mu \pi_\alpha - \frac{m_\alpha^2}{\hbar^2} \pi_\alpha^\dagger \pi_\alpha\right] - \frac14 F^a_{\mu\nu} F^{a\,\mu\nu}\right), \label{eqn:scalarAction}
\end{equation}
where  $D_\mu = \partial_\mu + i g A_\mu^a T^a_R$. The generator matrices (in a representation $R$) are $T^a_R = (T_R^a)_i{ }^j$, and satisfy the Lie algebra 
$[T_R^a, T_R^b]_i{ }^j = if^{abc} (T_R^c)_i{ }^j$. 

Let us consider only a single massive scalar. At the classical level, the colour charge can be obtained from the Noether current $j^a_\mu$ associated with the global part of the gauge symmetry. The colour charge is explicitly given by
\begin{equation}
\int\!\d^3x\, j^a_0(t,\v{x}) = i\!\int\! \d^3x\, \Big(\pi^\dagger T^a_R\, \partial_0 \pi - (\partial_0 \pi^\dagger) T^a_R\, \pi\Big)\,.  \label{eqn:colourNoetherCharge}
\end{equation}
Notice that a direct application of the Noether procedure has led to a colour charge with dimensions of action, or equivalently, of angular momentum. It is now worth dwelling on dimensional analysis in the context of the Wong equations~\eqref{eqn:classicalWong}, since they motivate us to make certain choices which may, at first, seem surprising. The Yang--Mills field strength
\begin{equation}
F^a_{\mu\nu}  =\partial_\mu A^a_\nu - \partial_\nu A^a_\mu - gf^{abc} A^b_\mu A^c_\nu\,\label{eqn:fieldStrength}
\end{equation}
is obviously an important actor in these classical equations. Classical equations should contain no factors\footnote{An equivalent point of view is that any factors of $\hbar$ appearing in an equation which has classical meaning should be absorbed into parameters of the classical theory.} of $\hbar$, so we choose to maintain this precise expression for the field strength when $\hbar \neq 1$. By inspection it follows that $[g A^{a}_\mu] = L^{-1}$. We can develop this further; since the action of \eqn~\eqref{eqn:scalarAction} has dimensions of angular momentum, the Yang--Mills field strength must have dimensions of $\sqrt{M/L^3}$. Thus, from \eqn~\eqref{eqn:fieldStrength},
\begin{equation}
[A^a_\mu] = \sqrt{\frac{M}{L}}\,, \qquad [g] = \frac1{\sqrt{ML}}\,.\label{eqn:YMdims}
\end{equation}
This conclusion about the dimensions of $g$ is in contrast to the situation in electrodynamics, where $[e] = \sqrt{ML}$. Put another way, in electrodynamics the dimensionless fine structure constant is $e^2 / 4\pi \hbar$ while in our conventions the Yang--Mills analogue is $\hbar g^2 / 4\pi\,$! It is possible to arrange matters such that the YM and EM cases are more similar, but we find the present conventions to be convenient in perturbative calculations.

Continuing with our discussion of dimensions, note that the Yang--Mills version of the Lorentz force, \eqn~\eqref{eqn:Wong-momentum}, demonstrates that the quantity $g c^a$ must have the same dimension as the electric charge. This is consistent with our observation above that the colour has dimensions of angular momentum.

At first our assignment of dimensions of $g$ may seem troubling; the fact that $g$ has dimensions of $1/\sqrt{ML}$ implies that the dimensionless coupling at each vertex is $g \sqrt \hbar$, so factors of $\hbar$ associated with the coupling appear with the opposite power to the case of electrodynamics (and gravity). However, because the colour charges are dimensionful the net power of $\hbar$ turns out to be the same. The classical limit of this aspect of the theory is clarified by the dimensionful nature of the colour --- to see how this works we must quantise. 

Dimensional analysis demonstrates that the field $\pi$ has dimensions of $\sqrt{M/L}$, so its mode expansion is
\begin{equation}
{\pi}_i(x) = \frac1{\sqrt{\hbar}} \int\!\df(p)\, \left(a_i(p) e^{-ip\cdot x/\hbar} + b_i^\dagger(p) e^{ip\cdot x/\hbar}\right).
\end{equation}
The ladder operators are normalised as in equation~\eqref{eqn:ladderCommutator}, with the index $i$ again labelling the representation $R$. After quantisation, the colour charge of \eqn~\eqref{eqn:colourNoetherCharge} becomes a Hilbert space operator,
\begin{equation}
\begin{aligned}
\C^a &= i\!\int\! \d^3x\, \Big(\pi^\dagger T^a_R\, \partial_0 \pi - (\partial_0 \pi^\dagger) T^a_R\, \pi\Big) \\
&= \hbar\int\!\df(p)\, \left( a^\dagger(p) \,T^a_R \, a(p) + b^\dagger(p)\, T^a_{\bar R} \, b(p)\right),\label{eqn:colourOp}
\end{aligned}
\end{equation}
where we have used that the generators of the conjugate representation $\bar R$ satisfy $T^a_{\bar R} = - T^a_R$. The overall $\hbar$ factor guarantees that the colour has dimensions of angular momentum, as we require. It is important to note that these global colour operators inherit the usual Lie algebra of the generators, modified by factors of $\hbar$, so that
\begin{equation}
[\C^a, \C^b] = i\hbar f^{abc} \C^c\,.\label{eqn:chargeLieAlgebra}
\end{equation}

Acting with the colour charge operator of \eqn~\eqref{eqn:colourOp} on momentum eigenstates (as defined in \eqn~\eqref{eqn:singleParticleStateDef}), we immediately see that
\begin{equation}
\C^a|p^i\rangle = \hbar\, (T^a_R)_j{ }^i|p^j\rangle\,, \qquad \langle p_i|\C^a = \hbar\, \langle p_j|(T^a_R)_i{ }^j\,.
\end{equation}
Thus inner products yield generators scaled by $\hbar$:
\begin{equation}
\langle p_i|\C^a|p^j\rangle \equiv (\newT^a)_i{ }^j = \hbar\, (T^a_R)_i{ }^j\,.
\end{equation}
The $(C^a)_i{ }^j$ are simply rescalings of the usual generators $T^a_R$ by a factor of $\hbar$, and thus satisfy the rescaled Lie algebra in \eqn~\eqref{eqn:chargeLieAlgebra}; since this rescaling is important for us, it is useful to make the distinction between the two. 

We can now finally act with the colour operator on the single particle state of equation~\eqref{eqn:InitialStateSimple}:
\begin{equation}
\C^a|\psi\rangle = \int\!\df(p)\, (\newT^a)_i{ }^j\, \varphi(p) \chi_j|p^i\rangle\,,
\end{equation}
allowing us to define the colour charge of the particle as
\begin{equation}
\langle \psi |\C^a| \psi \rangle =  \chi^{i*} (\newT^a)_{i}{ }^{j}\, \chi_j\,. \label{eqn:colourCharge}
\end{equation}

As a final remark on the rescaled generators, let us write out the covariant derivative in the representation $R$. In terms of $\newT^a$, 
the $\hbar$ scaling of interactions is precisely the same as in QED (and in perturbative gravity):
\begin{equation}
D_\mu = \partial_\mu + i \, g A^a_\mu T^a = \partial_\mu + \frac{ig}{\hbar}\, A^a_\mu \newT^a \,;\label{eqn:covDerivative}
\end{equation}
for comparison, the covariant derivative in QED consistent with our discussion in section~\ref{sec:RestoringHBar} is $\partial_\mu + i e A_\mu/ \hbar$. Thus we have arranged that factors of $\hbar$ appear in the same place in YM theory as in electrodynamics, provided that the colour is measured by $C^a$. This ensures that the basic rules for obtaining the classical limits of amplitudes will be the same; in practical calculations one restores $\hbar$'s in colour factors and works using $C^a$'s everywhere. However, it is worth emphasising that unlike classical colour charges, the factors $C^a$ do not commute.

\section{Classical point-particles}
\label{sec:PointParticleLimit}

For the states in equation~\eqref{eqn:InitialStateSimple} to have a well defined point-particle limit, for any operator $\mathbb{O}$ they must, at a bare minimum, satisfy the following two constraints in the classical limit \cite{Yaffe:1981vf}:
\[
\langle \psi |\mathbb{O} | \psi \rangle &= \textrm{finite} \,, \\
\langle \psi |\mathbb{O} \, \mathbb{O}| \psi \rangle &= \langle \psi |\mathbb{O} | \psi \rangle\langle \psi | \mathbb{O}| \psi \rangle + \textrm{negligible} \,.\label{eqn:classicalConstraints}
\]
Furthermore the classical limit is not necessarily injective: distinct quantum states $|\psi\rangle$ and $|\psi'\rangle$ may yield the same classical limit. Classical physics should of course be independent of the details of quantum states, and therefore we also require that in the limit, the overlap
\begin{equation}
\langle \psi' | \psi \rangle = \langle \psi | \psi \rangle + \textrm{negligible}\,.\label{eqn:classicalOverlap}
\end{equation}
Similarly, the expectation values above should remain unchanged in the limit if taken over distinct but classicaly equivalent states \cite{Yaffe:1981vf}.

Our goal in this section is to choose suitable momentum and colour wavefunctions, $\varphi(p)$ and $\chi$ respectively, which ensure that the observables in equations~\eqref{eqn:momentumExp} and~\eqref{eqn:colourCharge} meet these crucial requirements.

\subsection{Wavepackets}
\label{subsec:Wavefunctions}

Classical point particles have well defined positions and momenta. Heuristically, we therefore require well localised quantum states. We will take the momentum space wavefunctions $\varphi(p)$ to be wavepackets, characterised by a smearing or spread in momenta\footnote{Evaluating positions and uncertainties therein in	relativistic field theory is a bit delicate, and we will not consider the question in this thesis.}.   

Let us ground our intuition by first examining nonrelativistic wavefunctions.  An example of a minimum-uncertainty wavefunction in momentum space (ignoring normalisation) for a particle of mass $m$ growing sharper in the $\hbar\rightarrow 0$ limit has the form
\begin{equation}
\exp\left( -\frac{\v{p}\mskip1mu{}^2}{2 \hbar m \lcomp/ \lpack^2}\right)
= \exp\left( -\frac{\v{p}\mskip1mu{}^2}{2m^2 \lcomp^2/\lpack^2}\right),
\label{eqn:NonrelativisticMomentumSpaceWavefunction}
\end{equation}
where $\lcomp$ is the particle's Compton wavelength, and where $\lpack$ is an additional parameter with dimensions of length. We can obtain the conjugate in position space by Fourier transforming:
\begin{equation}
\exp\left( -\frac{(\v{x}-\v{x}_0)^2}{2 \lpack^2}\right).
\end{equation}
The precision with which we know the particle's location is given by $\lpack$, which we could take as an intrinsic measure of the wavefunction's spread.

This suggests that in considering relativistic wavefunctions, we should also take the dimensionless parameter controlling the approach to the classical limit in momentum space to be the square of the ratio of the Compton wavelength $\lcomp$ to the intrinsic spread~$\lpack$,
\begin{equation}
\xi \equiv \biggl(\frac{\lcomp}{\lpack}\biggr){\vphantom{\frac{\lcomp}{\lpack}}}^2\,.\label{eqn:defOfXi}
\end{equation}
We therefore obtain the classical result by studying the behaviour of expectation values as $\xi\rightarrow 0$; or alternatively, in the region where
\begin{equation}
\ell_c \ll \ell_w\,.\label{eqn:ComptonConstraint1}
\end{equation}
Towards the limit, the wavefunctions must be sharply peaked around the classical value for the momenta, $\pcl = m \ucl$, with the classical four-velocity $\ucl$ normalised to $\ucl^2 = 1$.  We can express this requirement through the conditions\footnote{The integration measure for $p$ enforces $\langle p^2\rangle = m^2$.}
\begin{equation}
\begin{aligned}
\langle p^\mu\rangle &= \int \df(p)\; p^\mu\, |\varphi(p)|^2 = 
m \uapprox^\mu f_{p}(\xi)\,,
\\ f_{p}(\xi) &= 1+\Ord(\xi^{\beta'})\,,
\\ \uapprox\cdot \ucl &= 1+\Ord(\xi^{\beta''})\,,
\\ \spread(p)/m^2 &=
\langle \bigl(p-\langle p\rangle\bigr){}^2\rangle/m^2
\\&= \bigl(\langle p^2\rangle-\langle p\rangle{}^2\bigr)/m^2
= c_\Delta \xi^\beta\,,
\end{aligned}
\label{eqn:expectations}
\end{equation}
where $c_\Delta$ is a constant of order unity, and the $\beta$'s are simple rational exponents. For the simplest wavefunctions, $\beta=1$. This spread around the classical value is not necessarily positive, as the difference $p^\mu-\langle p^\mu\rangle$ may be spacelike, and the expectation of its Lorentz square possibly negative. For that reason, we should resist the usual temptation of taking its square root to obtain a variance.

These constraints are the specific statement of those in equation~\eqref{eqn:classicalConstraints} for momentum space wavepackets. What about the vanishing overlap between classically equivalent states,~\eqref{eqn:classicalOverlap}? To determine a constraint on the wavepackets we need a little more detail of their functional form. Now, because of the on-shell condition $p^2=m^2$ imposed by the phase-space integral over the wavepacket's momenta, the only Lorentz invariant built out of the momentum is constant, and so the wavefunction cannot usefully depend on it. This means the wavefunction must depend on at least one four-vector parameter. The simplest wavefunctions will depend on exactly one four-vector, which we can think of as the (classical) 4-velocity $\ucl$ of the corresponding particle.  It can depend only on the dimensionless combination $p\cdot \ucl/m$ in addition to the parameter $\xi$.  The simplest form will be a function of these two in the combination $p\cdot\ucl/(m\xi)$, so that large deviations from $m \ucl$ will be suppressed in a classical quantity. The wavefunction will have additional dependence on $\xi$ in its normalisation.

The difference between two classically equivalent wavepackets must therefore come down to a characteristic mismatch $q_0$ of their momentum arguments --- without loss of generality, classically equivalent wavepackets are then specified by wavefunctions $\varphi(p)$ and $\varphi(p + q_0)$. As one nears the classical limit, both wavefunctions must represent the particle: that is they should be sharply peaked, and in addition their overlap should be $\Ord(1)$, up to corrections of $\Ord(\xi)$. Requiring the overlap to be $\Ord(1)$ is equivalent to requiring that $\varphi(p+q_0)$ does not differ much from $\varphi(p)$, which in turn requires that the derivative at $p$ is small, or that
\begin{equation}
\frac{q_0\cdot\ucl}{m\xi} \ll 1\,.
\label{eqn:qConstraint1}
\end{equation}
If we scale $q$ by $1/\hbar$, this constraint takes the following form:
\begin{equation}
\qb_0\cdot\ucl\,\lpack \ll \sqrt{\xi}.
\label{eqn:qbConstraint1}
\end{equation}
We have replaced the momentum by a wavenumber. We will see in the next chapter that this constraint is the fundamental constraint forcing the classical scaling advertised in equation~\eqref{eqn:notationWavenumber} upon certain momenta in scattering amplitudes.

Let us remain in the single particle case and finally examine an explicit example wavefunction satisfying our constraints. We will take a linear exponential,
\begin{equation}
\varphi(p) = \Norm m^{-1}\exp\biggl[-\frac{p\cdot u}{\hbar\lcomp/\lpack^2}\biggr]
= \Norm m^{-1}\exp\biggl[-\frac{p\cdot u}{m\xi}\biggr]\,,
\label{eqn:LinearExponential}
\end{equation}
which shares some features with relativistic wavefunctions discussed in ref.~\cite{AlHashimi:2009bb}. In spite of the linearity of the exponent in $p$, this function gives rise to the Gaussian of \eqn~\eqref{eqn:NonrelativisticMomentumSpaceWavefunction} in the nonrelativistic limit (in the rest frame of $u$).

The normalisation condition~(\ref{eqn:WavefunctionNormalization}) requires
\begin{equation}
\Norm = \frac{2\sqrt2\pi}{\xi^{1/2} K_1^{1/2}(2/\xi)}
\,,
\end{equation}
where $K_1$ is a modified Bessel function of the second kind. For details of this computation and following ones, see appendix~\ref{app:wavefunctions}.  An immediate corollary is that the overlap
\begin{equation}
\int \! \df(p) \, \varphi^*(p + q_0) \varphi(p) = \exp\left[-\frac{u\cdot q_0}{m\xi}\right] \equiv \eta_1(q_0;p)\,.\label{eqn:wavefunctionOverlap}
\end{equation}
Clearly, for this result to vanish in the limit $\xi=0$ we must rescale $q_0 = \hbar \wn q_0$, which then explicitly recovers the constraint~\eqref{eqn:qbConstraint1}.

We can compute the momentum expectation value of the wavepacket straightforwardly, obtaining
\begin{equation}
\langle p^\mu\rangle = m u^\mu \frac{K_2(2/\xi)}{K_1(2/\xi)}\,.
\end{equation}
As we approach the classical region, where $\xi\rightarrow 0$, the wavefunction indeed becomes sharply peaked, as
\begin{equation}
\langle p^\mu \rangle \rightarrow m u^\mu \left(1 + \frac34 \xi\right) + \Ord(\xi^2)\,.
\end{equation}
Moreover, the spread of the wavepacket
\begin{equation}
\frac{\sigma^2(p)}{\langle p^2\rangle} = 1 - \frac{K_2^2(2/\xi)}{K_1^2(2/\xi)} \rightarrow -\frac32 \xi + \Ord(\xi^2)\,.
\end{equation}

Finally, a similar calculation yields
\begin{equation}
\langle p^\mu p^\nu \rangle = m^2 u^\mu u^\nu \left(1 +  \frac{2\xi \, K_2(2/\xi)}{K_1(2/\xi)}\right)  - \frac{m^2}{2}  \frac{\xi\, K_2(2/\xi)}{K_1(2/\xi)}\, \eta^{\mu\nu}\,,\label{eqn:doubleMomExp}
\end{equation}
so in the classical region our wavepackets explicitly satisfy
\[
\langle p^\mu p^\nu \rangle &\rightarrow m^2 u^\mu u^\nu \left(1 + 2 \xi\right)  - \frac{m^2}{2} \xi\, \eta^{\mu\nu}  + \mathcal{O}(\xi^2)\\ 
&= \langle p^\mu \rangle \langle p^\nu \rangle + \mathcal{O}(\xi)\,.
\]
From these results, we see that the conditions in equation~\eqref{eqn:expectations} are explicitly satisfied, with $c_\Delta = -3/2$ and rational exponents $\beta = \beta' = \beta'' = 1$.

\subsection{Coherent colour states}
\label{sec:classicalSingleParticleColour}

We have seen that the classical point-particle picture emerges from sharply peaked quantum wavepackets. To understand colour, governed by the Yang--Mills--Wong equations in the classical arena, a similar picture should emerge for our quantum colour operator in \eqn~\eqref{eqn:colourOp}. We define the classical limit of the colour charge in equation~\eqref{eqn:colourCharge} to be
\begin{equation}
c^a \equiv \langle \psi |\C^a| \psi \rangle\,.
\end{equation}
Since the colour operator in \eqref{eqn:colourOp} explicitly involves a factor of $\hbar$, another parameter must be large so that the colour expectation $\langle \psi |\C^a | \psi \rangle$ is much bigger than $\hbar$ in the classical region. For states in irreducible representation $R$ the only new dimensionless parameter available is the size of the representation, $n$, and indeed we will see explicitly in the case of $SU(3)$ that we indeed need $n$ large in this limit.

Coherent states are the key to the classical limit very generally~\cite{Yaffe:1981vf}, and we will choose a coherent state to describe the colour of our particle. The states adopted previously to describe momenta can themselves be  understood as coherent states for a ``first-quantised'' particle --- more specifically they are states for the restricted Poincar\'e group \cite{Kaiser:1977ys, TwarequeAli:1988tvp, Kowalski:2018xsw}. By ``coherent" we mean in the sense of the definition introduced by Perelomov \cite{perelomov:1972}, which formalises the notion of coherent state for any Lie group and hence can be utilised for both the kinematic and the colour parts.

To construct explicit colour states we will use the Schwinger boson formalism.  For $SU(2)$, constructing irreducible representations from Schwinger bosons is a standard textbook exercise \cite{Sakurai:2011zz}. One simply introduces the Schwinger bosons --- that is, creation $a^{\dagger i}$ and annihilation $a_i$ operators, transforming in the fundamental two-dimensional representation so that $i = 1,2$. The irreducible representations of $SU(2)$ are all symmetrised tensor powers of the fundamental, so the state
\[
a^{\dagger i_1} a^{\dagger i_2} \cdots a^{\dagger i_{2j}} \ket{0} ,
\]
which is automatically symmetric in all its indices, transforms in the spin $j$ representation. 

For groups larger than $SU(2)$, the situation is a little more complicated because the construction of a general irreducible representation requires both symmetrisation and 
antisymmetrisation over appropriate sets of indices. This leads to expressions which are involved already for $SU(3)$ \cite{Mathur:2000sv,Mathur:2002mx}. We content 
ourselves with a  brief discussion of the $SU(3)$ case, which captures all of the interesting features of the general case.

One can construct all irreducible representations from tensor products only of fundamentals \cite{Mathur:2010wc,Mathur:2010ey}; however, for our treatment of $SU(3)$ it is helpful to instead make use of the fundamental and antifundamental, and tensor these together to generate representations.  Following \cite{Mathur:2000sv}, we introduce two sets of ladder operators $a_i$ and $b^i$ , $i=1, 2, 3$, which transform in the $\mathbf{3}$ and $\mathbf{3}^*$ respectively.  The colour operator can then be written as
\begin{equation}
\C^e= \hbar \left( a^\dagger \frac{\lambda^e}{2} a -
b^\dagger \frac{\bar{\lambda}^e}{2} b  \right), \quad e=1, \dots, 8\,, \label{eqn:charge-SU3}
\end{equation}
where $\lambda^e$ are the Gell--Mann matrices and $\bar\lambda^e$ are their conjugates. The operators $a$ and $b$ satisfy the usual commutation relations
\begin{align}
[a_i, a^{\dagger j}]= \delta_{i}{ }^{j}\,, \quad   [b^i, b^{\dagger}_j]= \delta^{i}{ }_{j}\,, \quad 
[a_i, b^j]= 0\,, \quad  [a^{\dagger i}, b^{\dagger }_j]= 0\,.
\end{align}
By virtue of these commutators, the colour operator \eqref{eqn:charge-SU3} obeys the commutation relation \eqref{eqn:chargeLieAlgebra}.

There are two Casimir operators given by the number operators\footnote{Here we define $a^{\dagger}  \cdot a \equiv \sum_{i=1}^3 a^{\dagger i} a_i$ and
	$|\varsigma|^2\equiv \sum_{i=1}^3 |\varsigma_i|^2$.}
\begin{equation}
\mathcal{N}_1\equiv a^\dagger \cdot a\,, \qquad  \mathcal{N}_2\equiv b^\dagger \cdot b\,,
\end{equation}
with eigenvalues $n_1$ and $n_2$ respectively, so we label irreducible representations by $[n_1, n_2]$. 
Na\"ively, the states we are looking for are constructed by acting on the vacuum state as follows:
\begin{align}
\left(a^{\dagger i_1} \cdots  a^{\dagger i_{n_1}} \right)
\left(b_{j_1}^{\dagger} \cdots  b_{j_{n_2}}^{\dagger} \right)
\ket{0}.\label{eqn:states-reducible}
\end{align}
However, these states are $SU(3)$ reducible and thus cannot be used in our construction of coherent states. We write the irreducible states schematically by acting with a Young projector $\mathcal{P}$ which appropriately (anti-) symmetrises upper and lower indices, thereby subtracting traces:
\begin{align}
\ket{\psi}_{[n_1, n_2]} \equiv \mathcal{P} \left( \left(a^{\dagger i_1} \cdots  a^{\dagger i_{n_1}} \right)
\left(b^{\dagger}_{j_1} \cdots  b_{j_{n_2}}^{\dagger} \right)\ket{0} \right).
\label{eqn:YPstate}
\end{align}
In general these operations will lead to involved expressions for the states, but we can understand them from their associated Young tableaux (Fig.~\ref{fig:SU3-YT}). Each double box column represents an operator $b_i^{\dagger}$ and each single column box represents the operator $a^{\dagger i}$, and thus for a mixed representation we have $n_2$ double columns and $n_1$ single columns.

\begin{figure}
	\centering 
	\begin{ytableau}
		j_1 & j_2  & \dots  &j_{n_2} & i_1 & i_2 & \cdots & i_{n_1}  \cr    &  &  &  
	\end{ytableau}
	\caption{Young tableau of $SU(3)$.}
	\label{fig:SU3-YT}
\end{figure}

Having constructed the irreducible states, one can define a coherent state parametrised by two triplets of complex numbers $\varsigma_i$ and $ \varrho^i$, $i=1,2, 3$.  These are normalised according to
\begin{equation}
|\varsigma|^2 = |\varrho|^2 = 1\,, \qquad \varsigma \cdot \varrho=0\,.
\end{equation}
We won't require fully general coherent states, but instead their projections onto the $[n_1,n_2]$ representation, which are
\begin{equation}
\ket{\varsigma \,\varrho}_{[n_1,n_2]}\equiv  \frac{1}{\sqrt{(n_1! n_2!)}} \left( \varrho \cdot b^\dagger\right)^{n_2} \left(\varsigma \cdot a^\dagger \right)^{n_1} \ket{0}.
\label{eqn:restricted-coherent}
\end{equation}

The square roots ensure that the states are normalised to unity\footnote{Note that the Young projector in equation~\eqref{eqn:YPstate} is no longer necessary since the constraint $\xi \cdot \zeta = 0$ removes all the unwanted traces.}. With this normalisation we can write the identity operator as  
\begin{equation}
\mathbb{I}_{[n_1,n_2]} = \int \d \mu(\varsigma,\varrho) \Big(\ket{\varsigma \,\varrho}\bra{\varsigma \,\varrho}\Big)_{[n_1,n_2]},\label{eqn:Haar}
\end{equation}
where $\int \d \mu(\varsigma,\varrho)$ is the $SU(3)$ Haar measure, normalised such that $\int \d \mu(\varsigma,\varrho)=1$. Its precise form is irrelevant for our purposes.

With the states in hand, we can return to the expectation value of the colour operator $\C^a$ in \eqn~\eqref{eqn:colourOp}. The size of the representation, that is $n_1$ and $n_2$, must be large compared to $\hbar$ in the classical regime so that the final result is finite. To see this let us compute this expectation value explicitly.
By definition we have 
\begin{equation}
\langle \varsigma \,\varrho|\mathbb{C}^e|\varsigma \,\varrho \rangle_{[n_1,n_2]} =  \frac{\hbar}{2} \left(\langle \varsigma \,\varrho|a^\dagger \lambda^e a|\varsigma \,\varrho \rangle_{[n_1,n_2]} -
\langle \varsigma \,\varrho|b^\dagger \bar{\lambda}^e b|\varsigma \,\varrho\rangle_{[n_1,n_2]} \right).
\end{equation}
After a little algebra we find that 
\begin{equation}
\langle \varsigma \,\varrho|\C^e|\varsigma \,\varrho \rangle = \frac{\hbar}{2} \left( n_1   \varsigma^{*} \lambda^e \varsigma - n_2    \varrho^*\bar \lambda^e \varrho\right).
\end{equation}
We see that a finite charge requires a scaling limit in which we take $n_1$, $n_2$ large  as $\hbar \to 0$, keeping the product $\hbar n_\alpha$  fixed for at least one value of $\alpha=1,2$. The classical charge is therefore the finite c-number
\begin{equation}
c^a =  \langle \varsigma \,\varrho|\C^a|\varsigma \,\varrho \rangle_{[n_1, n_2]}  = \frac{\hbar}{2} \left( n_1   \varsigma^{*}\lambda^a \varsigma - n_2 \varrho^{*} \bar\lambda^a \varrho\right). \label{eqn:clas-charge-SU3}
\end{equation}

The other feature we must check is the expectation value of products. Using the result above, a similar calculation for two pairs of charge operators in a large representation leads to
\begin{multline}
\langle \varsigma\,\varrho|\C^a\C^b | \varsigma\,\varrho\rangle_{[n_1, n_2]}  = \langle \varsigma\,\varrho|\C^a| \varsigma\,\varrho\rangle_{[n_1,n_2]}  \langle \varsigma\,\varrho|\C^b | \varsigma\,\varrho\rangle_{[n_1,n_2]} \\ + \hbar \left( \hbar n_1 \, \varsigma^*\lambda^a\cdot \lambda ^b \varsigma - \hbar n_2\, \varrho^* \bar\lambda^a\cdot \bar\lambda^b \varrho \right).
\end{multline}
The finite quantity in the classical limit $\hbar \to 0, \, n_\alpha \to \infty$ is the product $\hbar n_\alpha$. The term inside the brackets on the second line is itself finite, but comes with a lone $\hbar$ coefficient, and thus vanishes in the classical limit. Thus,
\[
\langle \varsigma\,\varrho|\C^a \C^b|\varsigma\,\varrho \rangle _{[n_1,n_2]} = c^a c^b + \mathcal O(\hbar)\,. \label{eqn:factorisation-charges}
\]
This is in fact a special case of a more general construction discussed in detail by Yaffe~\cite{Yaffe:1981vf}. Similar calculations can also be used to demonstrate  that the overlap $\langle \chi' | \chi \rangle$ is very strongly peaked about $\chi = \chi'$, as required by equation~\eqref{eqn:classicalOverlap}. We have thus constructed explicit colour states which ensure the correct classical behaviour of the colour charges. 

For the remainder of the thesis we will only need to make use of the finiteness and factorisation properties, so we will avoid further use of the explicit form of the representation states. Henceforth we write $\chi$ for the parameters of a general colour state $\ket{\chi}$ with these properties, and $\d \mu(\chi)$ for the Haar measure of the $SU(N)$ colour group.

\section{Multi-particle wavepackets}

Having set up appropriate wavepackets for a single particle, we can now consider multiple particles, and thus generalise the generic states adopted in equation~\eqref{eqn:InitialStateSimple}. We will take two distinguishable scalar particles, associated with distinct quantum fields $\pi_\alpha$ with $\alpha = 1, 2$. The action is therefore as given in \eqn~\eqref{eqn:scalarAction}. Both fields $\pi_\alpha$ must be in representations $R_\alpha$ which are large, so that a classical limit is available for the individual colours.

In anticipation of considering scattering processes in the next section, we will now take our state to be at some initial time in the far past, where we assume that our two particles both have well-defined positions, momenta and colours. In other words, particle $\alpha$ has a wavepacket $\varphi_\alpha(p_\alpha)$ describing its momentum-space distribution, and a coherent colour wavefunction $\chi_\alpha$, as described in the previous section. Then the appropriate generalisation of the multi-particle state is
\[
|\Psi\rangle &= \int\!\df(p_1)\df(p_2)\, \varphi_1(p_1) \varphi_2(p_2)\, e^{ib\cdot p_1/\hbar}\, |{p_1}\, \chi_1 ; \, {p_2}\, \chi_2 \rangle \\
&= \int\!\df(p_1)\df(p_2) \, \varphi_1(p_1) \varphi_2(p_2)\, e^{ib\cdot p_1/\hbar}\, \chi_{1i}\, \chi_{2j} |{p_1}^i ; \, {p_2}^j \rangle\,,\label{eqn:inState}
\]
where the displacement operator insertion accounts for the particles' spatial separation.

We measure observables for multi-particle states by acting with operators which are simply the sum of the individual operators for each of the scalar fields. For example, acting with the colour operator~\eqref{eqn:colourOp} on the state $|{p_1}\, \chi_1 ; \, {p_2}\, \chi_2 \rangle$ we have
\[
\C^a |{p_1}\,\chi_1 ;&\, {p_2}\, \chi_2 \rangle = |{p_1}^{i'} \, {p_2}^{j'} \rangle \, \left( (C^a_{1})_{i'}{}^i \delta_{j'}{}^j + \delta_{i'}{}^i (C^a_{2})_{j'}{}^j \right) \chi_{1i}\, \chi_{2j} \\
&= \int \! d\mu(\chi'_1) d\mu(\chi'_2) \, \ket{{p_1} \, \chi'_1 ; \, {p_2}\, \chi'_2} \,\langle \chi'_1\, \chi'_2| C^a_{1} \otimes 1 + 1 \otimes C^a_{2} |\chi_1\, \chi_2 \rangle\\
&= \int \! d\mu(\chi'_1) d\mu(\chi'_2) \, \ket{{p_1} \, \chi'_1 ; \, {p_2}\, \chi'_2} \,\langle\chi'_1\, \chi'_2| C^a_{1+2}  |\chi_1\, \chi_2\rangle\,,\label{eqn:charge2particleAction}
\]
where $C^a_\alpha$ is the colour in representation $R_\alpha$ and we have written $C^a_{1+2}$ for the colour operator on the tensor product of representations $R_1$ and $R_2$.
In the classical regime, using the property that the overlap between states sets $\chi'_\alpha=\chi_\alpha $ in the classical limit, it follows that
\[
\bra{p_1\,\chi_1; \, p_2 \,\chi_2} C^a_{1+2} \ket{p_1\, \chi_1 ; \, p_2\, \chi_2 } = c_1^a + c_2^a \,,
\]
so the colours simply add. A trivial similar result holds for the momenta of the two particles.

Suppose that at some later time the two particles described by our initial state interact --- for example, two black holes scattering elastically. When does a point-particle description  remain appropriate? This will be the crucial topic of the next chapter. We will take the initial separation $b^\mu$ to be the transverse impact parameter for the scattering of two point-like objects with momenta $p_{1,2}$. (The impact parameter is transverse in the sense that $p_\alpha \cdot b = 0$ for $\alpha = 1, 2$.) At the quantum level, the particles are individually described by the wavefunctions in section~\ref{sec:PointParticleLimit}. We would expect the point-particle description to be valid when the separation of the two scattering particles is always very large compared to their (reduced) Compton wavelengths, so the point-particle description will be accurate provided that
\begin{equation}
\sqrt{-b^2} \gg \lcomp^{(1,2)}\,.
\end{equation}
The impact parameter and the Compton wavelengths are not the only scales we must consider, however --- the spread of the wavepackets, $\lpack$, is another intrinsic scale. As we will discuss, the quantum-mechanical expectation values of observables are well approximated by the corresponding classical ones when the packet spreads are in the `Goldilocks' zone, $\lcomp\ll \lpack\ll \sqrt{-b^2}$. These inequalities will have powerful ramifications on the behaviour of scattering amplitudes in the classical limit. To see this however, we need an on-shell scattering observable.

%% file: chapter3/chapter3.tex
\chapter{The impulse}
\label{chap:impulse}

At a gravitational wave observatory we are of course interested in the gravitational radiation emitted by the source of interest. However, gravitational waves also carry information about the potential experienced by, for example, a black hole binary system. This observation motivates our interest in an on-shell observable related to the potential. We choose to explore the \textit{impulse\/} on a particle during a scattering event: at the classical level, this is simply the total change in the momentum of one of the particles --- say particle~1 --- during the collision.

In this chapter we will begin by examining the change in momentum during a scattering event, without accompanying radiation, extracting the classical values from a fully relativistic quantum-mechanical computation. We examine scattering events in which two widely separated particles are prepared in the state~\eqref{eqn:inState} at $t \rightarrow -\infty$, and then shot at each other with impact parameter $b^\mu$. We will use this observable as a laboratory to explore certain conceptual and practical issues in approaching the classical limit. Using the explicit wavepackets we have constructed in chapter~\ref{chap:pointParticles}, we will carefully analyse the small-$\hbar$ region to understand how scattering amplitudes encode classical physics. We will see that the appropriate treatment is one where point-particles have momenta which are fixed as we take $\hbar$ to zero, whereas for massless particles and momentum transfers between massive particles, it is the wavenumber which we should treat as fixed in the limit.

Our formalism is quite general, applying in both gauge theory and gravity; for simplicity, we will nonetheless continue to focus on the scattering of two massive, stable quanta of scalar fields described by the Lagrangian in equation~\eqref{eqn:scalarAction}. We will generalise to higher spin fields in part~\ref{part:spin}. We will always restrict our attention to scattering processes in which quanta of fields 1 and 2 are both present in the final state. This will happen, for example, if the particles have separately conserved quantum numbers.  We also always assume that no new quanta of fields 1 and 2 can be produced during the collision, for example because the centre-of-mass energy is too small.

In the first section of this chapter we construct expressions for the impulse in terms of on-shell scattering amplitudes, providing a formal definition of the momentum transfer to a particle in quantum field theory. In \sect{sec:classicalLimit}, we derive the Goldilocks zone in which the point-particle limit of our wavepackets remains valid, deriving from first principles the behaviour of scattering amplitudes in the classical limit. In \sect{sec:examples} we apply our formalism explicitly, deriving the NLO impulse in scalar Yang--Mills theory, a result which is analagous to well-established post--Minkowskian results for the scattering of Schwarzschild black holes \cite{Portilla:1979xx,Portilla:1980uz,Westpfahl:1979gu,Westpfahl:1985tsl}.

This chapter continues to be based on work published in refs.~\cite{Kosower:2018adc,delaCruz:2020bbn}.

\section{Impulse in quantum field theory}
\label{sec:QFTsetup}

To define the observable, we place detectors at asymptotically large distances pointing at the collision region. The detectors measure only the momentum of particle 1. We assume that these detectors cover all possible scattering angles. Let $\mathbb{P}_\alpha^\mu$ be the momentum operator for particle~$\alpha$; the expectation of the first particle's outgoing momentum $\outp1^\mu$ is then
\begin{equation}
\begin{aligned}
\langle \outp1^\mu \rangle &= 
{}_\textrm{out}{\langle}\Psi|  \mathbb{P}_1^\mu |\Psi\rangle_\textrm{out} \\
&= {}_\textrm{out}{\langle}\Psi|  \mathbb{P}_1^\mu U(\infty,-\infty)\,
|\Psi\rangle_\textrm{in} \\
&= {}_\textrm{in}{\langle} \Psi | \, U(\infty, -\infty)^\dagger   \mathbb{P}_1^\mu U(\infty, -\infty) \, | \Psi \rangle_\textrm{in}\,,
\end{aligned}
\end{equation}
where $U(\infty, -\infty)$ is the time evolution operator from the far past to the far future.  This evolution operator is just the $S$ matrix, so the expectation value is simply
\begin{equation}
\begin{aligned}
\langle \outp1^\mu \rangle &= 
{}_\textrm{in}{\langle} \Psi | S^\dagger   \mathbb{P}_1^\mu S\, 
| \Psi \rangle_\textrm{in}\,.
\end{aligned}
\end{equation}
We can insert a complete set of states and rewrite the expectation value as
\begin{equation}
\begin{aligned}
\langle \outp1^\mu \rangle 
&=\sum_X \int \df(\finalk_1)\, \df(\finalk_2) \, d\mu(\zeta_1) \, d\mu(\zeta_2)\; \finalk_1^\mu 
\; \bigl| \langle \finalk_1\, \zeta_1; \finalk_2\, \zeta_2; X |S| \Psi\rangle\bigr|^2\,,
\end{aligned}
\label{eqn:p1Expectation}
\end{equation}
where we can think of the inserted states as the final state of a scattering process. In this equation, $X$ refers to any other particles which may be created. The intermediate state containing $X$ also necessarily contains exactly one particle each corresponding to fields~1 and~2. Their momenta are denoted by $\finalk_{1,2}$ respectively, while $\d\mu(\zeta_\alpha)$ is the $SU(N)$ Haar measure for their coherent colour states, as introduced in~\eqref{eqn:Haar}. The sum over $X$ is a sum over all states, including $X$ empty, and includes phase-space integrals for $X$ non-empty. The expression~(\ref{eqn:p1Expectation}) already hints at the possibility of evaluating the momentum in terms of on-shell scattering amplitudes.

The physically interesting quantity is rather the change of momentum of the particle during the scattering, so we define
\begin{equation}
\langle \Delta p_1^\mu \rangle =  \langle \Psi |S^\dagger \,   \mathbb{P}^\mu_1 \, S |\Psi\rangle - \langle \Psi | \,   \mathbb{P}^\mu_1 \,  |\Psi\rangle\,.
\end{equation}
This impulse is the difference between the expected outgoing and
the incoming momenta of particle 1. It is an on--shell observable, defined in both the quantum and the classical theories. Similarly, we can measure the impulse imparted to particle 2. In terms of the momentum operator, $\mathbb{P}_2^\mu$, of quantum field 2, this impulse is evidently
\begin{equation}
\langle \Delta p_2^\mu \rangle =  \langle \Psi |S^\dagger \,  \mathbb{P}^\mu_2 \, S |\Psi\rangle - \langle \Psi | \,   \mathbb{P}^\mu_2 \,  |\Psi\rangle.
\end{equation}

Returning to the impulse on particle 1, we proceed by writing the scattering matrix in terms of the transition matrix $T$ via $S = 1 + i T$, in order to make contact with the usual scattering amplitudes. The no-scattering (unity) part of the $S$ matrix cancels in the impulse, leaving behind only delta functions that identify the final-state momenta with the initial-state ones in the wavefunction or its conjugate. Using unitarity we obtain the result
\begin{equation}
\langle \Delta p_1^\mu \rangle 
= \langle \Psi | \, i [   \mathbb{P}_1^\mu, T ] \, | \Psi \rangle + \langle \Psi | \, T^\dagger [   \mathbb{P}_1^\mu, T] \, |\Psi \rangle\,.
\label{eqn:defl1}
\end{equation}

\subsection{Impulse in terms of amplitudes}

Having established a general expression for the impulse, we turn to expressing it in terms of scattering amplitudes. It is convenient to work on the two terms in equation~\eqref{eqn:defl1} separately. For ease of discussion, we define
\begin{equation}
\begin{aligned}
\ImpA \equiv \langle \Psi | \, i [ \mathbb{P}_1^\mu, T ] \, | \Psi \rangle\,, \qquad
\ImpB \equiv \langle \Psi | \, T^\dagger [ \mathbb{P}_1^\mu, T] \, |\Psi \rangle\,,
\end{aligned}
\end{equation}
so that the impulse is $\langle \Delta p_1^\mu \rangle = \ImpA + \ImpB$. Using equation~\eqref{eqn:inState} to expand the wavepacket in the first term, $\ImpA$, we find
\[
\hspace*{-4mm}\ImpA &= 
\int \! \df(\initialk_1)\df(\initialk_2)
\df(\initialkc_1)\df(\initialkc_2)\;
e^{i b \cdot (\initialk_1 - \initialkc_1)/\hbar} \, 
\varphi_1(\initialk_1) \varphi_1^*(\initialkc_1) 
\varphi_2(\initialk_2) \varphi_2^*(\initialkc_2) 
\\ &\hspace{40mm}
\times i (\initialkc_1\!{}^\mu - \initialk_1^\mu) \, 
\langle \initialkc_1\, \chi_1'; \initialkc_2 \, \chi_2'| \,T\, |
\initialk_1 \, \chi_1; \initialk_2\, \chi_2 \rangle
\\&= \int \! \df(\initialk_1)\df(\initialk_2)
\df(\initialkc_1)\df(\initialkc_2)\;
e^{i b \cdot (\initialk_1 - \initialkc_1)/\hbar} \, 
\varphi_1(\initialk_1) \varphi_1^*(\initialkc_1) 
\varphi_2(\initialk_2) \varphi_2^*(\initialkc_2) 
\\ &\qquad\qquad
\times i \int \df(\finalk_1)\df(\finalk_2) \d\mu(\zeta_1) \d\mu(\zeta_2)\;(\finalk_1^\mu-\initialk_1^\mu)
\\ &\qquad\qquad\qquad\qquad 
\times \langle \initialkc_1 \, \chi_1'; \initialkc_2 \, \chi_2'| \finalk_1 \, \zeta_1 ;\finalk_2\, \zeta_2 \rangle
\langle \finalk_1 \, \zeta_1; \finalk_2\, \zeta_2 | \,T\, |\initialk_1 \, \chi_1; \initialk_2 \, \chi_2\rangle
\,,
\label{eqn:defl2}
\]
where in the second equality we have re-inserted the final-state momenta $\finalk_\alpha$ in
order to make manifest the phase independence of the result. We label the states in the incoming wavefunction by $\initialk_{1,2}$, those in the conjugate ones by $\initialkc_{1,2}$. Let us now introduce the momentum shifts $q_\alpha = \initialkc_\alpha-\initialk_\alpha$, and then change variables in the integration from the $p_\alpha'$ to the $q_\alpha$. In these variables, the matrix element is
\begin{equation}
\begin{aligned}
\langle p_1'\, \chi'_1;\,p_2'\, \chi'_2|T| p_1\,\chi_1;\,p_2\,\chi_2\rangle &= \langle \chi'_1\, \chi'_2|\mathcal{A}(p_1,p_2 \rightarrow p_1',p_2')|\chi_1\,\chi_2\rangle\\ &\qquad\qquad\qquad\qquad \times
\del^{(4)}(\initialkc_1+\initialkc_2-\initialk_1-\initialk_2) 
\\&\equiv
\langle \Ampl(\initialk_1 \initialk_2 \rightarrow \initialk_1 + q_1\,, \initialk_2 + q_2)\rangle
\del^{(4)}(q_1 + q_2)\,,\label{eqn:defOfAmplitude}
\end{aligned}
\end{equation}
yielding
\begin{equation}
\begin{aligned}
\ImpA &= \int \! \df(\initialk_1) \df(\initialk_2)
\df(q_1+\initialk_1)\df(q_2+\initialk_2)\;
\\&\qquad\times 
\varphi_1(\initialk_1) \varphi_1^*(\initialk_1 + q_1)
\varphi_2(\initialk_2) \varphi_2^*(\initialk_2+q_2) 
\, \del^{(4)}(q_1 + q_2)
\\&\qquad\times 
\, e^{-i b \cdot q_1/\hbar} 
\,i q_1^\mu  \, \langle\Ampl(\initialk_1 \initialk_2 \rightarrow 
\initialk_1 + q_1, \initialk_2 + q_2)\rangle\,
\,,
\end{aligned}
\label{eqn:impulseGeneralTerm1a}
\end{equation}
where the remaining expectation value is solely over the representation states $\chi_\alpha$. Note that we are implicitly using the condition~\eqref{eqn:classicalOverlap}, in anticipation of the classical limit, for clarity of presentation. Now, recall the shorthand notation introduced earlier for the phase-space measure,
\begin{equation}
\df(q_1+p_1) = \dd^4 q_1\; \del\bigl((p_1 + q_1)^2 - m_1^2\bigr)
\Theta(p_1^0 + q_1^0)\,.
\end{equation} 
We can perform the integral over $q_2$ in \eqn~\eqref{eqn:impulseGeneralTerm1a} using the four-fold delta function. Further relabeling $q_1 \rightarrow q$, we obtain
\[
\ImpA&= \int \! \df(\initialk_1)\df(\initialk_2) \dd^4 q  \; 
\del(2\initialk_1 \cdot q + q^2) \del(2 \initialk_2 \cdot q - q^2) \, e^{-i b \cdot q/\hbar}\\
&\qquad \times \Theta(\initialk_1^0+q^0) \Theta(\initialk_2^0-q^0)\, \varphi_1(\initialk_1) \varphi_1^*(\initialk_1 + q)
\varphi_2(\initialk_2) \varphi_2^*(\initialk_2-q)
\\& \qquad\qquad \times  
\,  i q^\mu  \, \langle\Ampl(\initialk_1 \initialk_2 \rightarrow 
\initialk_1 + q, \initialk_2 - q)\rangle\,.
\label{eqn:impulseGeneralTerm1}
\]
Unusually for a physical observable, this contribution is linear in the amplitude. We emphasise that the incoming and outgoing momenta of this amplitude do \textit{not\/} correspond to the initial- and final-state momenta of the scattering process, but rather both correspond to the initial-state momenta, as they appear in the wavefunction and in its conjugate.  The momentum $q$ looks like a momentum transfer if we examine the amplitude alone, but for the physical scattering process it represents a difference between the momentum within the wavefunction and that in the conjugate.  Inspired by our discussion in section~\ref{sec:PointParticleLimit}, we will refer to it as a `momentum mismatch'.  As indicated on the first line of \eqn~\eqref{eqn:defl2}, we should think of this term as an interference of a standard amplitude with an interactionless forward scattering. Recalling that in equation~\eqref{eqn:wavefunctionSplit} we defined ${\psi_i}_\alpha(p_\alpha) = \varphi_\alpha(p_\alpha) {\chi_i}_\alpha$, we can write this diagrammatically as
\begin{equation}
\begin{aligned}
\ImpA & = 
\int \! \df(\initialk_1)\df(\initialk_2) \dd^4 q \, 
\del(2\initialk_1 \cdot q + q^2) \del(2 \initialk_2 \cdot q - q^2) \\
& \qquad \times  \Theta(\initialk_1^0+q^0)\Theta(\initialk_2^0-q^0) \, e^{-i b \cdot q/\hbar}  \,  iq^\mu \!\!\!\!
\begin{tikzpicture}[scale=1.0, 
baseline={([yshift=-\the\dimexpr\fontdimen22\textfont2\relax]
	current bounding box.center)},
] 
\begin{feynman}

\vertex (b) ;
\vertex [above left=1 and 0.66 of b] (i1) {$\psi_1(p_1)$};
\vertex [above right=1 and 0.33 of b] (o1) {$\psi_1^*(p_1+q)$};
\vertex [below left=1 and 0.66 of b] (i2) {$\psi_2(p_2)$};
\vertex [below right=1 and 0.33 of b] (o2) {$\psi_2^*(p_2-q)$};

\begin{scope}[decoration={
	markings,
	mark=at position 0.7 with {\arrow{Stealth}}}] 
\draw[postaction={decorate}] (b) -- (o2);
\draw[postaction={decorate}] (b) -- (o1);
\end{scope}
\begin{scope}[decoration={
	markings,
	mark=at position 0.4 with {\arrow{Stealth}}}] 
\draw[postaction={decorate}] (i1) -- (b);
\draw[postaction={decorate}] (i2) -- (b);
\end{scope}

\filldraw [color=white] (b) circle [radius=10pt];
\filldraw [fill=allOrderBlue] (b) circle [radius=10pt];
\end{feynman}
\end{tikzpicture} 
\!\!\!\!.
\end{aligned}
\end{equation}

Turning to the second term, $\ImpB$, in the impulse, we again introduce a complete set of states labelled by momenta $\finalk_1$, $\finalk_2$ and $X$ so that
\begin{equation}
\begin{aligned}
\ImpB &= \langle \Psi | \, T^\dagger [ \mathbb{P}_1^\mu, T] \, |\Psi \rangle
\\&= \sum_X \int \! \df(\finalk_1) \df(\finalk_2) \d\mu(\zeta_1) \d\mu(\zeta_2) 
\\& \qquad\qquad \times\langle \Psi | \, T^\dagger | \finalk_1 \, \zeta_1; \finalk_2 \, \zeta_2; X \rangle 
\langle \finalk_1  \, \zeta_1; \finalk_2 \, \zeta_2; X| [  \mathbb{P}_1^\mu, T] \, |\Psi \rangle\,.
\end{aligned}
\end{equation}
As above, we can now expand the wavepackets. We again label the momenta in the incoming wavefunction by $\initialk_{1,2}$, and those in the conjugate ones by $\initialkc_{1,2}$:
\begin{equation}
\begin{aligned}
\ImpB
&=
\sum_X \int \!\prod_{\alpha = 1, 2}  \df(\finalk_\alpha)  \df(\initialk_\alpha) \df(\initialkc_\alpha)
\; \varphi_\alpha(\initialk_\alpha) \varphi^*_\alpha(\initialkc_\alpha) 
e^{i b \cdot (\initialk_1 - \initialkc_1) / \hbar}
(\finalk_1^\mu - \initialk_1^\mu) \\
&\hspace*{5mm}\times \del^{(4)}(\initialk_1 + \initialk_2 
- \finalk_1 - \finalk_2 -\finalk_X) 
\del^{(4)}(\initialkc_1+\initialkc_2 - \finalk_1 - \finalk_2 -\finalk_X)\\
&\hspace*{15mm}\times \langle \Ampl^*(\initialkc_1\,, \initialkc_2 \rightarrow \finalk_1 \,, \finalk_2 \,, \finalk_X) 
\Ampl(\initialk_1\,,\initialk_2 \rightarrow 
\finalk_1\,, \finalk_2 \,, \finalk_X)\rangle
\,.
\end{aligned}
\label{eqn:forcedef2}
\end{equation}
In this expression we again absorb the representation states $\chi_\alpha$ into an expectation value over the amplitudes, while $\finalk_X$ denotes the total momentum carried by particles in $X$. The second term in the impulse can thus be interpreted as a weighted cut of an amplitude; the lowest order contribution is a weighted two-particle cut of a one-loop amplitude. 

In order to simplify $\ImpB$, let us again define the momentum shifts $q_\alpha = \initialkc_\alpha-\initialk_\alpha$, and change variables in the integration from the $\initialkc_\alpha$ to the $q_\alpha$, so that
\begin{equation}
\begin{aligned}
\ImpB
&=
\sum_X  \int \!\prod_{\alpha = 1,2} \df(\finalk_\alpha)  \df(\initialk_\alpha) 
\df(q_\alpha+\initialk_\alpha)
\; \varphi_\alpha(\initialk_\alpha) \varphi^*_\alpha(\initialk_\alpha+q_\alpha) \\
&\hspace*{5mm}\times \del^{(4)}(\initialk_1 + \initialk_2 
- \finalk_1 - \finalk_2 -\finalk_X)\, 
\del^{(4)}(q_1+q_2)\, e^{-i b \cdot q_1 / \hbar} (\finalk_1^\mu - \initialk_1^\mu)\\
&\hspace*{5mm}\times \langle\Ampl^*(\initialk_1+q_1\,, \initialk_2+q_2 \rightarrow \finalk_1 \,,\finalk_2 \,, \finalk_X)
\Ampl(\initialk_1\,, \initialk_2 \rightarrow \finalk_1\,, \finalk_2 \,, \finalk_X)
\rangle
\,.
\end{aligned} \label{eqn:impulseGeneralTerm2a}
\end{equation}
We can again perform the integral over $q_2$ using the four-fold delta function, and relabel $q_1 \rightarrow q$ to obtain
\begin{equation}
\begin{aligned}
\ImpB
&=
\sum_X  \int \!\prod_{\alpha = 1,2} \df(\finalk_\alpha)  \df(\initialk_\alpha) 
\dd^4 q\;
\del(2\initialk_1 \cdot q + q^2) \del(2 \initialk_2 \cdot q - q^2) \\
&\hspace*{5mm}\times \Theta(\initialk_1^0+q^0)\Theta(\initialk_2^0-q^0)\, \varphi_1(\initialk_1) \varphi_2(\initialk_2)\, \varphi^*_1(\initialk_1+q) \varphi^*_2(\initialk_2-q) \\
&\hspace*{5mm}\times \del^{(4)}(\initialk_1 + \initialk_2 
- \finalk_1 - \finalk_2 -\finalk_X) \, e^{-i b \cdot q / \hbar} (\finalk_1^\mu - \initialk_1^\mu)
\\ &\hspace*{5mm}\times \langle \Ampl^*(\initialk_1+q\,, \initialk_2-q \rightarrow \finalk_1 \,,\finalk_2 \,, \finalk_X)
\Ampl(\initialk_1\,, \initialk_2 \rightarrow 
\finalk_1\,, \finalk_2 \,, \finalk_X)\rangle
\,.
\end{aligned} 
\label{eqn:impulseGeneralTerm2b}
\end{equation}
The momentum $q$ is again a momentum mismatch. The momentum transfers $\xfer_\alpha\equiv r_\alpha-p_\alpha$ will play an important role in analysing the classical limit, so it is convenient to change variables to them from the final-state momenta $\finalk_\alpha$,
\begin{equation}
\begin{aligned}
\ImpB &= \sum_X  \int \!\prod_{\alpha = 1,2}  \df(\initialk_i) \dd^4 \xfer_\alpha
\dd^4 q\;
\del(2p_\alpha\cdot \xfer_\alpha+\xfer_\alpha^2)\Theta(p_\alpha^0+\xfer_\alpha^0)
\\&\qquad\times
\del(2\initialk_1 \cdot q + q^2) \del(2 \initialk_2 \cdot q - q^2) 
\Theta(\initialk_1^0+q^0)\Theta(\initialk_2^0-q^0)\, \varphi_1(\initialk_1) \varphi_2(\initialk_2)\\
&\qquad \times\varphi^*_1(\initialk_1+q) \varphi^*_2(\initialk_2-q) e^{-i b \cdot q / \hbar}\,\xfer_1^\mu \; \del^{(4)}(\xfer_1+\xfer_2+\finalk_X) 
\\ &\qquad\qquad\times 
\langle \Ampl^*(\initialk_1+q, \initialk_2-q \rightarrow 
\initialk_1+\xfer_1 \,,\initialk_2+\xfer_2 \,, \finalk_X)
\\ &\qquad\qquad\qquad\qquad\qquad \times 
\Ampl(\initialk_1\,, \initialk_2 \rightarrow 
\initialk_1+\xfer_1\,, \initialk_2+\xfer_2 \,, \finalk_X) \rangle\,.
\end{aligned} 
\label{eqn:impulseGeneralTerm2}
\end{equation}
Diagrammatically, this second contribution to the impulse is
\begin{equation}
\begin{aligned}
\usetikzlibrary{decorations.markings}
\ImpB&= 
\sum_X {\int} \! \prod_{\alpha = 1,2}  \df(\initialk_\alpha) \dd^4 \xfer_\alpha
\dd^4 q\; \del(2p_\alpha\cdot \xfer_\alpha+\xfer_\alpha^2)\Theta(p_\alpha^0+\xfer_\alpha^0)\,e^{-i b \cdot q / \hbar}\,\xfer_1^\mu
\\&\qquad\times
\del(2\initialk_1 \cdot q + q^2) \del(2 \initialk_2 \cdot q - q^2) 
\Theta(\initialk_1^0+q^0)\Theta(\initialk_2^0-q^0)\\
& \hspace*{10mm} \times \del^{(4)}(\xfer_1+\xfer_2+\finalk_X) \!\!\!\!\!
\begin{tikzpicture}[scale=1.0, 
baseline={([yshift=-\the\dimexpr\fontdimen22\textfont2\relax]
	current bounding box.center)},
] 
\begin{feynman}
\begin{scope}
\vertex (ip1) ;
\vertex [right=3 of ip1] (ip2);
\node [] (X) at ($ (ip1)!.5!(ip2) $) {};
\begin{scope}[even odd rule]

\vertex [above left=0.66 and 0.5 of ip1] (q1) {$ \psi_1(p_1)$};
\vertex [above right=0.66 and 0.33 of ip2] (qp1) {$ \psi^*_1(p_1 + q)$};
\vertex [below left=0.66 and 0.5 of ip1] (q2) {$ \psi_2(p_2)$};
\vertex [below right=0.66 and 0.33 of ip2] (qp2) {$ \psi^*_2(p_2 - q)$};
\diagram* {
	(ip1) -- [photon]  (ip2)
};
\begin{scope}[decoration={
	markings,
	mark=at position 0.4 with {\arrow{Stealth}}}] 
\draw[postaction={decorate}] (q1) -- (ip1);
\draw[postaction={decorate}] (q2) -- (ip1);
\end{scope}
\begin{scope}[decoration={
	markings,
	mark=at position 0.7 with {\arrow{Stealth}}}] 
\draw[postaction={decorate}] (ip2) -- (qp1);
\draw[postaction={decorate}] (ip2) -- (qp2);
\end{scope}
\begin{scope}[decoration={
	markings,
	mark=at position 0.34 with {\arrow{Stealth}},
	mark=at position 0.7 with {\arrow{Stealth}}}] 
\draw[postaction={decorate}] (ip1) to [out=90, in=90,looseness=1.2] node[above left] {{$p_1 + w_1$}} (ip2);
\draw[postaction={decorate}] (ip1) to [out=270, in=270,looseness=1.2]node[below left] {$p_2 + w_2$} (ip2);
\end{scope}

\node [] (Y) at ($(X) + (0,1.4)$) {};
\node [] (Z) at ($(X) - (0,1.4)$) {};
\node [] (k) at ($ (X) - (0.65,-0.25) $) {$\finalk_X$};

\filldraw [color=white] ($ (ip1)$) circle [radius=8pt];
\filldraw  [fill=allOrderBlue] ($ (ip1) $) circle [radius=8pt];

\filldraw [color=white] ($ (ip2) $) circle [radius=8pt];
\filldraw  [fill=allOrderBlue] ($ (ip2) $) circle [radius=8pt];

\end{scope}
\end{scope}
\filldraw [color=white] ($  (Y) - (3pt, 0) $) rectangle ($ (Z) + (3pt,0) $) ;
\draw [dashed] (Y) to (Z);
\end{feynman}
\end{tikzpicture} .
\end{aligned}
\end{equation}

\section{Point-particle scattering}
\label{sec:classicalLimit}

The observable we have discussed  --- the impulse --- is designed to be well-defined in both the quantum and the classical theories. As we approach the classical limit, the quantum expectation values should reduce to the classical impulse, ensuring that we are able to explore the $\hbar \rightarrow 0$ limit. Here we explore this limit, and its ramifications on scattering amplitudes, in detail.

\subsection{The Goldilocks inequalities}

We have already discussed in section~\ref{sec:RestoringHBar} how to make explicit the factors of $\hbar$ in the observables, and in section~\ref{sec:PointParticleLimit} we selected wavefunctions which have the desired classical point-particle limit, which we established was the region where
\begin{equation}
\ell_c \ll \ell_w\,.\label{eqn:comptonConstraint}
\end{equation}
At this point, we could in principle perform the full quantum calculation, using the specific wavefunctions we chose, and expand in the $\xi\!\rightarrow\! 0$ limit at the end.  However, having established previously the detailed properties of our wavefunctions $\varphi_\alpha$, it is far more efficient to neglect the details and simply use the fact that they allow us to approach the limit as early as possible in calculations.  This will lead us to impose stronger constraints on our choice than the mere existence of a suitable classical limit.

Heuristically, the wavefunctions for the scattered particles must satisfy two separate conditions. As discussed in the single particle case, the details of the wavepacket should not be sensitive to quantum effects. At the same time, now that we aim to describe the scattering of point-particles the spread of the wavefunctions should not be too large, so that the interaction with the other particle cannot peer into the details of the quantum wavepacket.

To quantify this discussion let us examine $\ImpA$ in~(\ref{eqn:impulseGeneralTerm1}) more closely. It has the form of an amplitude integrated over the on-shell phase space for both of the incoming momenta, subject to additional $\delta$ function constraints --- and then weighted by a phase $e^{-ib\cdot q/\hbar}$ dependent on the momentum mismatch $q$, and finally integrated over all $q$. As one nears the classical limit~\eqref{eqn:comptonConstraint}, the wavefunction and its conjugate should both represent the particle. The amplitude will vary slowly on the scale of the wavefunction when one is close to the limit. This is therefore precisely the same constraint as we had in the single particle case, and we immediately have
\begin{equation}
\qb\cdot\ucl_\alpha\,\lpack \ll \sqrt{\xi}\,,
\label{eqn:qbConstraint}
\end{equation}
where we have scaled $q$ by $1/\hbar$, replacing the momentum by a wavenumber.

We next examine another rapidly varying factor that appears in all our integrands, the delta functions in $q$ arising from the on-shell constraints on the conjugate momenta $\initialkc_\alpha$. These delta functions, appearing in equations~(\ref{eqn:impulseGeneralTerm1} and~\ref{eqn:impulseGeneralTerm2}), take the form
\begin{equation}
\del(2p_\alpha\cdot q+q^2) = \frac1{\hbar m_\alpha}\del(2\qb\cdot u_\alpha+\lcomp \qb^2)\,.
\label{eqn:universalDeltaFunction}
\end{equation}
The integration over the initial momenta $\initialk_\alpha$ and the initial wavefunctions will smear out these delta functions to sharply peaked functions whose scale is of the same order as the original wavefunctions.  As $\xi$ gets smaller, this function will turn back into a delta function imposed on the $\qb$ integration. To see this, let us consider an explicit wavefunction integral similar to $\ImpA$, but with a simpler integrand:
\def\IntOne{T_1}
\begin{equation}
\IntOne = \int \df(p_1)\,\varphi(p_1)\varphi^*(p_1+q)\,\del(2 p_1\cdot q+q^2)\,.\label{eqn:deltaFunctionIntegral}
\end{equation}
With $\varphi$ chosen to be the linear exponential~(\ref{eqn:LinearExponential}), this integral simplifies to
\begin{equation}
\IntOne = \frac{1}{\hbar m_1} \eta_1(\qb;p_1)\,\int \df(p_1)\,
\del(2 p_1\cdot\qb/m_1+\hbar\qb^2/m_1)\,|\varphi(p_1)|^2\,,
\end{equation}
where $\eta_1(\qb;p_1)$ is the overlap defined in equation~\eqref{eqn:wavefunctionOverlap} and we have also replaced $q\rightarrow\hbar \qb$.

The remaining integrations in $\IntOne$ are relegated in appendix~\ref{app:wavefunctions}, but yield\footnote{The wavenumber transfer is necessarily spacelike.}
\begin{multline}
\IntOne = \frac1{4 \hbar m_1\sqrt{(\qb\cdot u)^2-\qb^2}\,K_1(2/\xi)} 
\\\times \exp\biggl[-\frac2{\xi}\frac{\sqrt{(\qb\cdot u)^2-\qb^2}}{\sqrt{-\qb^2}}
\sqrt{1-\hbar^2\qb^2/(4m_1^2)}\biggr]
\,.\label{eqn:TIntegral}
\end{multline}
Notice that our result depends on two dimensionless ratios in addition to its dependence on $\xi$,
\begin{equation}
\lcomp \sqrt{-\qb^2}
\qquad
\textrm{and}\qquad
\frac{\qb\cdot u}{\sqrt{-\qb^2}}\,.
\end{equation}
Let us call $1/\sqrt{-\qb^2}$ a `scattering length' $\lscatt$.  In terms of this length, our two dimensionless ratios are
\begin{equation}
\frac{\lcomp}{\lscatt}
\qquad
\textrm{and}\qquad
{\qb\cdot u}\,\lscatt\,.\label{eqn:dimensionlessRatios}
\end{equation}

As we approach the $\hbar,\xi\rightarrow 0$ limit, we may expect $\IntOne$ to be concentrated in a small region in $\qb$.  Towards the limit, the dependence on the magnitude is just given by the prefactor.  To understand the behaviour in the boost and angular degrees of freedom, we may note that 
\begin{equation}
\frac1{K_1(2/\xi)} \sim \frac2{\sqrt{\pi}\sqrt{\xi}} \exp\biggl[\frac2{\xi}\biggr]\,,\label{eqn:BesselLimit}
\end{equation}
and that $\hbar\sqrt{\xi}$ is of order $\xi$, so that overall $\IntOne$ has the form
\begin{equation}
\frac1{\xi}\exp\biggl[-\frac{f(\qb)}{\xi}\biggr]\,.
\label{eqn:LimitForm}
\end{equation}
This will yield a delta function so long as $f(\qb)$ is positive. To figure out its argument, we recall that $\qb^2<0$, and parametrise the wavenumber as
\begin{equation}
\qb^\mu = \Eqb\bigl(\sinh\zeta,\,\cosh\zeta\sin\theta\cos\phi,
\,\cosh\zeta\sin\theta\sin\phi,
\,\cosh\zeta\cos\theta\bigr)\,,\label{eqn:qbRapidityParametrisation}
\end{equation}
with rapidity $\zeta$ running over $[0,\infty]$, $\theta$ over $[0,\pi]$, and $\phi$ over $[0,2\pi]$. Working in the rest frame of $u^\mu$, the exponent in \eqn~\eqref{eqn:TIntegral} (including the term from \eqn~\eqref{eqn:BesselLimit}) is
\begin{equation}
-\frac2{\xi}\Bigl(\cosh\zeta\sqrt{1+\hbar^2 \Eqb^2/(4m^2)}-1\Bigr)\,,
\end{equation}
so that the delta function will ultimately localise 
\begin{equation}
\cosh\zeta \rightarrow \frac1{\sqrt{1+\hbar^2 \Eqb^2/(4m^2)}} = 
1-\frac{\hbar^2 \Eqb^2}{8m^2}+\Ord(\hbar^4)
\end{equation}
to zero. Thus in terms of the Lorentz--invariant dimensionless ratios in equation~\eqref{eqn:dimensionlessRatios}, we find that the delta function is
\begin{equation}
\delta\!\left(\qb\cdot u\, \lscatt + \frac{\ell_c^2}{4 \lscatt^2} \, \frac1{\qb\cdot u\, \lscatt}\right). \label{eqn:deltaArgument}
\end{equation}
The direction-averaging implicit in the integration over $\initialk_1$ has led to a constraint on two positive quantities built out of the ratios.

Recall that to arrive at this expression we absorbed a factor of $\hbar\sim\sqrt{\xi}$. Since the argument of the delta function is a polynomial in the dimensionless ratios, both must be independently constrained to be of this order:
\begin{subequations}
\begin{align}
{\qb\cdot u}\,\lscatt &\lesssim \sqrt{\xi}\,,\label{eqn:deltaConstraint1}
\\\frac{\lcomp}{\lscatt} &\lesssim \sqrt{\xi}\,.\label{eqn:deltaConstraint2}
\end{align}
\end{subequations}
If we had not already scaled out a factor of $\hbar$ from $q$, these constraints would make it natural to do so.  
%

Combining constraint~\eqref{eqn:deltaConstraint1} with that in \eqn~\eqref{eqn:qbConstraint}, we obtain the constraint $\lpack\ll\lscatt$. Then including constraint~\eqref{eqn:ComptonConstraint1}, or $\xi\ll 1$, we obtain our first version of the `Goldilocks' inequalities,
\begin{equation}
\lcomp \ll \lpack \ll \lscatt\,.
\label{eqn:Goldilocks1}
\end{equation}
As we shall see later in the explicit evaluation of $\ImpA$, $\lscatt\sim \sqrt{-b^2}$; this follows on dimensional grounds.  This gives us the second version of the `Goldilocks' requirement,
\begin{equation}
\lcomp\ll \lpack\ll \sqrt{-b^2}\,. 
\label{eqn:Goldilocks2}
\end{equation}

Note that the constraint following from~\eqref{eqn:deltaConstraint2} is weaker, $\lpack\lesssim\lscatt$. Indeed, we should not expect a similar strengthening of this restriction; the sharp peaking of the wavefunctions alone will not force the left-hand side to be much smaller than the right-hand side. This means that we should expect $\qb\cdot u$ to be smaller than, but still of order, $\sqrt{\xi}/\lscatt$.  If we compare the two terms in the argument to the delta function~(\ref{eqn:universalDeltaFunction}), we see that the second term
\begin{equation}
\lcomp \qb^2 \sim \frac{\lcomp}{\lscatt} \frac1{\lscatt} \ll \frac{\sqrt{\xi}}{\lscatt}\,,\label{eqn:neglectq2}
\end{equation}
so that $\lcomp \qb^2 \ll \qb\cdot u_\alpha$, and the second term should be negligible. In our evaluation of $\IntOne$, we see that the integral is sharply peaked about the delta function
\begin{equation}
\delta(\wn q\cdot u)\,.\label{eqn:universalDeltaFunction2}
\end{equation}
We are thus free to drop the $\wn q^2$ correction in the classical limit. There is one important caveat to this simplification, which we will mention below.
\begin{figure}[t]
	\center
	\includegraphics[width = 0.75\textwidth]{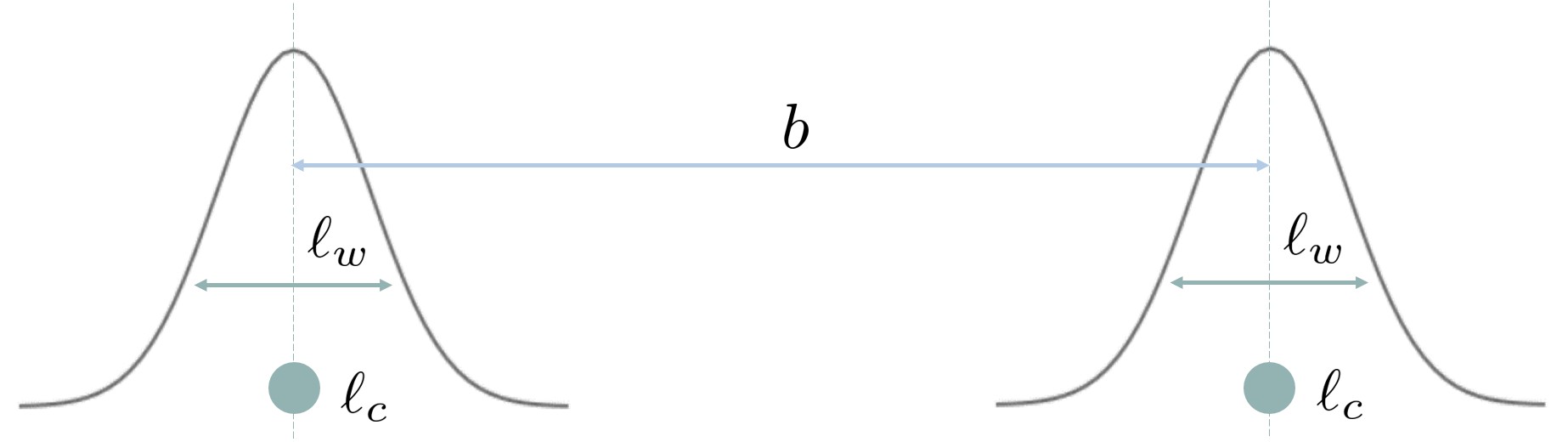}
	\vspace{-3pt}
	\caption{Heuristic depiction of the Goldilocks inequalities.\label{Goldilocks}}
\end{figure}

\subsection{Taking the limit of observables}

In computing the classical observable, we cannot simply set $\xi=0$. Indeed, we don't even want to fully take the $\xi\rightarrow 0$ limit. Rather, we want to take the leading term in that limit. This term may in fact be proportional to a power of $\xi$.  To understand this, we should take note of one additional length scale in the problem, namely the classical radius of the point particle.  In electrodynamics, this is $\lclass=e^2/(4\pi m)$. However,
\begin{equation}
\lclass = \frac{\hbar e^2}{4\pi\hbar m} = \alpha\lcomp\,,
\end{equation}
where $\alpha$ is the usual, dimensionless, electromagnetic coupling. Dimensionless ratios of $\lclass$ to other length scales will be the expansion parameters in classical observables; but as this relation shows, they too will vanish in the $\xi\rightarrow 0$ limit. There are really three dimensionless parameters we must consider: $\xi$; $\lpack/\lscatt$; and $\lclass/\lscatt$.  We want to retain the full dependence on the latter, while considering only effects independent of the first two.

Under the influence of a perturbatively weak interaction (such as electrodynamics or gravity) below the particle-creation threshold, we expect a wavepacket's shape to be distorted slightly, but not radically changed by the scattering.  We would expect the outgoing particles to be characterised by wavepackets similar to those of the incoming particles. However, using a wavepacket basis of states for the state sums in \sect{sec:QFTsetup} would be cumbersome, inconvenient, and computationally less efficient than the plane-wave states we used.  We expect the narrow peaking of the wavefunction to impose constraints on the momentum transfers as they appear in higher-order corrections to the impulse $\ImpB$ in equation~\eqref{eqn:impulseGeneralTerm2}; but we will need to see this narrowness indirectly, via assessments of the spread as in \eqn~\eqref{eqn:expectations}, rather than directly through the presence of wavefunction (or wavefunction mismatch) factors in our observables. We can estimate the spread $\spread(\finalk_\alpha)$ in a final-state momentum $\finalk_\alpha$ as follows:
\begin{equation}
\begin{aligned}
\spread(\finalk_\alpha)/m_\alpha^2 &= 
\langle\bigl(\finalk_\alpha-\langle\finalk_\alpha\rangle\bigr)^2\rangle/m_\alpha^2
\\&= \bigl(\langle \finalk_\alpha^2\rangle-\langle\finalk_\alpha\rangle{}^2\bigr)/m_\alpha^2
\\&= 1-\bigl(\langle\initialk_\alpha\rangle+\expchange\bigr){}^2/m_\alpha^2
\\&= \spread(p_\alpha)/m_\alpha^2 -\langle\Delta p_\alpha\rangle\cdot 
\bigl(2\langle\initialk_\alpha\rangle+\expchange\bigr)/m_\alpha^2\,.
\end{aligned}
\end{equation}
So long as $\expchange/m_\alpha\lesssim\spread(\initialk_\alpha)/m_\alpha^2$, the second term will not greatly increase the result, and the spread in the final-state momentum will be of the same order as that in the initial-state momentum.   Whether this condition holds depends on the details of the wavefunction. Even if it is violated, so long as $\expchange/m_\alpha \lesssim c'_\Delta \xi^{\beta'''}$ with $c'_\Delta$ a constant of $\Ord(1)$, then the final-state momentum will have a narrow spread towards the limit.  (It would be broader than the initial-state momentum spread, but that does not affect the applicability of our results.)

The magnitude of $\expchange$ can be determined perturbatively.  The leading-order value comes from $\ImpA$, with $\ImpB$ contributing yet-smaller corrections.  As we shall see, these computations reveal $\expchange/m_\alpha$ to scale like $\hbar$, or $\sqrt{\xi}$, and be numerically much smaller.

This in turn implies that for perturbative consistency, the `characteristic' values of momentum transfers $w_\alpha$ inside the definition of $I^\mu_{(2)}$ must also be very small compared to $m_\alpha\sqrt{\xi}$. This constraint is in fact much weaker than implied by the leading-order value of $\expchange$. Just as for $q_0$ in \eqn~\eqref{eqn:qConstraint1}, we should scale these momentum transfers by $1/\hbar$, replacing them by wavenumbers $\xferb_\alpha$.  The corresponding scattering lengths $\tilde\lscatt^\alpha = \sqrt{-w_{\alpha}^2}$ must again satisfy $\tilde\lscatt^\alpha \gg \lpack$. If we now examine the energy-momentum-conserving delta function 
in \eqn~\eqref{eqn:impulseGeneralTerm2},
\begin{equation}
\del^{(4)}(w_1+w_2 + \finalk_X)\,,\label{eqn:radiationScalingDeltaFunction}
\end{equation}
we see that all other transferred momenta $\finalk_\alpha$ must likewise be small compared to $m_\alpha\sqrt{\xi}$: all their energy components must be positive and hence no cancellations are possible inside the delta function. The typical values of these momenta should again by scaled by $1/\hbar$ and replaced by wavenumbers. We will see in the next chapter that $I_{(2)}^\mu$ encodes radiative effects, and the same constraint will force these momenta for emitted radiation to scale as wavenumbers. 

What about loop integrations?  As we integrate the loop  momentum over all values, it is a matter of taste how we scale it.  If it is the momentum of a (virtual) massless line, however, unitarity considerations suggest that as the natural scaling is to remove a factor of $\hbar$ in the real contributions to the cut in equation~\eqref{eqn:impulseGeneralTerm2}, we should likewise do so for virtual lines.  More generally, we should scale those differences of the loop momentum with external legs that correspond to massless particles, and replace them by wavenumbers.  Moreover, unitarity considerations also suggest that we should choose the loop momentum to be that of a massless line in the loop, if there is one.

In general, we may not be able to approach the $\hbar\rightarrow0$ limit of each contribution to an observable separately, because they may contain terms which are singular, having too many inverse powers of $\hbar$.  We find that such singular terms meet one of two fates: they are multiplied by functions which vanish in the regime of validity of the limit; or they cancel in the sum over all contributions.  We cannot yet offer a general argument that such troublesome terms necessarily disappear in one of these two
manners.  We can treat independently contributions whose singular terms ultimately cancel in the sum, so long as we expand each contribution in a Laurent series in $\hbar$.

For theories with non-trivial internal symmetries, when identifying singular terms (in both parts of the impulse) it is essential not to forget factors coming from the classical limit of the representation states $\chi_\alpha$. Since the scattering particles remain well separated at all times there is no change to the story in chapter~\ref{chap:pointParticles}. However, it is important to keep in mind that in our conventions colour factors in scattering amplitudes carry dimensions of $\hbar$, and in particular satisfy the Lie algebra in equation~\eqref{eqn:chargeLieAlgebra}. There is subsequently an independent Laurent series available when evaluating colour factors, and this can remove terms which appear singular in the kinematic expansion.

Full impulse integrand factors that appear uniformly in all contributions --- that is, factors which appear directly in a final expression after cancellation of terms singular in the $\hbar\rightarrow 0$ limit --- can benefit from applying two simplifications to the integrand: setting $p_\alpha$ to $m_\alpha\ucl_\alpha$, as prescribed by equation~\eqref{eqn:universalDeltaFunction2}, and truncating at the lowest order in $\hbar$ or $\xi$.  For other factors, we must be careful to expand in a Laurent series.  As mentioned above, a consequence of equation~\eqref{eqn:neglectq2} is that inside the on-shell delta functions $\del(2p_\alpha\cdot \qb\pm \hbar \qb^2)$ we can neglect the $\hbar \qb^2$ term; this is true so long as the factors multiplying these delta functions are not singular in $\hbar$. If they are indeed nonsingular (after summing over terms), we can safely neglect the second term inside such delta functions, and replace them by $\del(2p_\alpha\cdot \qb)$.  A similar argument allows us to neglect the $\hbar \qb^0$ term inside the positive-energy theta functions; the $\qb$ integration then becomes independent of them.  Similar arguments, and caveats, apply to the squared momentum-transfer terms $\hbar \xferb_\alpha^2$ appearing inside on-shell delta functions in higher-order contributions, along with the energy components $\xferb_\alpha^0$ appearing inside positive-energy theta functions.  They can be neglected so long as the accompanying factors are not singular in $\hbar$.  If accompanying factors \textit{are} singular as $\hbar\rightarrow 0$, then we may need to retain such formally suppressed $\hbar \qb^2$ or $\hbar \xferb_\alpha^2$ terms inside delta functions.
We will see an example of this in the calculation of the NLO contributions to the impulse in section~\ref{sec:examples}.

\subsubsection{Summary}

For ease of future reference, let us collect the rules we have derived for calculating classical scattering observables from quantum field theory. We have all together established that, in the classical limit, we must apply the following constraints when evaluating amplitudes in explicit calculations:
\begin{itemize}
	\item The momentum mismatch $q = p'_1 - p_1$ scales as a wavenumber, $q = \hbar \wn q$.
	\item The momentum transfers $w_\alpha$ in $I_{(2)}^\mu$ scale as wavenumbers.
	\item Massless loop momenta scale as wavenumbers.
	\item The $\hbar \wn q^2$ factors in on--shell delta functions can be dropped, but only when there are no terms singular in $\hbar$.
	\item Any amplitude colour factors are evaluated using the commutation relation~\eqref{eqn:chargeLieAlgebra}.
\end{itemize}
We derived these rules using the impulse, but they hold for any on-shell observable constructed in the manner of section~\ref{sec:QFTsetup}. Furthermore, it will be convenient to introduce a notation to allow us to manipulate integrands under the eventual approach to the $\hbar\rightarrow0$ limit; we will use large angle brackets for the purpose,
\begin{multline}
\Lexp f(p_1,p_2,\cdots) \Rexp = \int\! \df(p_1) \df(p_2) |\varphi(p_1)|^2 |\varphi(p_2)|^2\\ \times \langle \chi_1\,\chi_2|f(p_1,p_2,\cdots)|\chi_1\,\chi_2\rangle\,,
\label{eqn:angleBrackets}
\end{multline}
where the integration over both $\initialk_1$ and $\initialk_2$ is implicit.  Within the angle brackets, we have approximated $\varphi_\alpha(p\pm q)\simeq \varphi_\alpha(p)$ and $\chi'_\alpha \simeq  \chi_\alpha$. Then, relying on our detailed study of the momentum and colour wavefunctions in sections~\ref{sec:PointParticleLimit} and~\ref{sec:classicalLimit}, to evaluate the integrals and representation expectation values implicit in the large angle brackets we can simply set $p_\alpha\simeq m_\alpha \ucl_\alpha$, and replace quantum colour charges $C_\alpha^a$ with their (commuting) classical counterparts $c^a_\alpha$.

\subsection{The classical impulse}
\label{sec:classicalImpulse}

We have written the impulse in terms of two terms, $\langle \Delta p_1^\mu \rangle = \ImpA + \ImpB$, and expanded these in terms of wavefunctions in equations~\eqref{eqn:impulseGeneralTerm1} and ~\eqref{eqn:impulseGeneralTerm2}. We will now discuss the classical limit of these terms in detail, applying the rules gathered above.

We begin with the first and simplest term in the impulse, $\ImpA$, given in \eqn~\eqref{eqn:impulseGeneralTerm1}, and here recast in the notation of \eqn~\eqref{eqn:angleBrackets} in preparation:
\begin{multline}
\ImpAcl =  \Lexp i\!\int \!\dd^4 q  \; 
\del(2\initialk_1 \cdot q + q^2) \del(2 \initialk_2 \cdot q - q^2) 
\Theta(\initialk_1^0+q^0)\Theta(\initialk_2^0-q^0)\\
 \times  e^{-i b \cdot q/\hbar} 
\, q^\mu  \, \Ampl(\initialk_1, \initialk_2 \rightarrow 
\initialk_1 + q, \initialk_2 - q)\,\Rexp\,.
\label{eqn:impulseGeneralTerm1recast}
\end{multline}
Rescale $q \rightarrow \hbar\qb$; drop the $q^2$ inside the on-shell delta functions;
and also remove the overall factor of $\tilde g^2$ and accompanying $\hbar$'s from the amplitude, to obtain the leading-order (LO) contribution to the classical impulse,
\begin{multline}
\DeltaPlo \equiv \ImpAclsup{(0)} =  \frac{i\tilde g^2}{4} \Lexp \hbar^2\! \int \!\dd^4 \qb  \; 
\del(\qb\cdot p_1) \del(\qb\cdot p_2) 
\\\times  
e^{-i b \cdot \qb} 
\, \qb^\mu  \, \AmplB^{(0)}(p_1,\,p_2 \rightarrow 
p_1 + \hbar\qb, p_2 - \hbar\qb)\,\Rexp\,.
\label{eqn:impulseGeneralTerm1classicalLO}
\end{multline}
We denote by $\AmplB^{(L)}$ the reduced $L$-loop amplitude, that is the $L$-loop amplitude with a factor of the (generic) coupling $\tilde g/\sqrt{\hbar}$ removed for every interaction: in the gauge theory case, this removes a factor of $g/\sqrt{\hbar}$, while in the gravitational case, we would remove a factor of $\kappa/\sqrt{\hbar}$.  In general, this rescaled fixed-order amplitude depends only on $\hbar$-free ratios of couplings; in pure electrodynamics or gravitational theory, it is independent of couplings.  In pure electrodynamics, it depends on the charges of the scattering particles. While it is free of the powers of $\hbar$ discussed in section~\ref{sec:RestoringHBar}, it will in general still scale with an overall power of $\hbar$ thanks to dependence on momentum mismatches or transfers. As we shall see in the next section, additional inverse powers of $\hbar$ emerging from $\AmplB$ will cancel the $\hbar^2$ prefactor and yield a nonvanishing result.

As a reminder, while this contribution to a physical observable is linear in an amplitude, it arises from an expression involving wavefunctions multiplied by their conjugates.  This is reflected in the fact that both the `incoming' and `outgoing' momenta in the amplitude here are in fact initial-state momenta. Any phase which could be introduced by hand in the initial state would thus cancel out of the observable.

The LO classical impulse is special in that only the first term~(\ref{eqn:impulseGeneralTerm1}) contributes. In general however, it is only the sum of the two terms in \eqn~\eqref{eqn:defl1} that has a well-defined classical limit.  We may write this sum as
\begin{multline}
\Delta p_1^\mu = 
\Lexp i\hbar^{-2}\!
\int \!\dd^4 q  \; \del(2\initialk_1 \cdot q + q^2) \del(2 \initialk_2 \cdot q - q^2)\\ \times 
\Theta(\initialk_1^0+q^0)\Theta(\initialk_2^0-q^0) \; e^{-i b \cdot q/\hbar} \; \impKer \Rexp \,,
\label{eqn:partialClassicalLimitNLO}
\end{multline}
where the \textit{impulse kernel\/} $\impKer$ is defined as
\begin{equation}
\begin{aligned}
\impKer \equiv&\,  \hbar^2 q^\mu \, \Ampl(\initialk_1 \initialk_2 \rightarrow 
\initialk_1 + q, \initialk_2 - q)
\\& -i \hbar^2 \sum_X \int \!\prod_{\alpha = 1,2}  \dd^4 \xfer_\alpha
\del(2p_\alpha\cdot \xfer_\alpha+\xfer_\alpha^2)\Theta(p_\alpha^0+\xfer_\alpha^0)
\\&\hphantom{-} \times \xfer_1^\mu\, \del^{(4)}(\xfer_1+\xfer_2+\finalk_X) \Ampl(\initialk_1 \initialk_2 \rightarrow \initialk_1+\xfer_1\,, \initialk_2+\xfer_2 \,, \finalk_X)
\\ &\hspace*{25mm}\times \Ampl^*(\initialk_1+q, \initialk_2-q \rightarrow \initialk_1+\xfer_1 \,,\initialk_2+\xfer_2 \,, \finalk_X)\,.
\end{aligned}
\label{eqn:FullImpulse}
\end{equation}
The prefactor in \eqn~\eqref{eqn:partialClassicalLimitNLO} and the normalization of $\impKer$ are chosen so that the latter is $\Ord(\hbar^0)$ in the classical limit. At leading order, 
the only contribution comes from the tree-level four-point amplitude in the first term, and after passing to the classical limit, we recover \eqn~\eqref{eqn:impulseGeneralTerm1classicalLO} as expected. At next-to-leading order (NLO), both terms contribute. The contribution from
the first term is from the one-loop amplitude, while that from the second term has $X=\emptyset$, so that both the amplitude and conjugate inside the integral are tree level four-point amplitudes.

Focus on the NLO contributions, and pass to the classical limit. As discussed in section~\ref{subsec:Wavefunctions} we may neglect the $q^2$ terms in the delta functions present in \eqn~\eqref{eqn:partialClassicalLimitNLO} so long as any singular terms in the impulse
kernel cancel. We then rescale $q \rightarrow \hbar\qb$; and remove an overall factor of $\tilde g^4$ and accompanying $\hbar$''s from the amplitudes. In addition, we may rescale $\xfer\rightarrow \hbar\xferb$. However, since singular terms may be present in the individual summands of the impulse kernel --- in general, they will cancel against singular terms emerging from the loop integration in the first term in \eqn~\eqref{eqn:FullImpulse} ---  
we are not entitled to drop the $w^2$ inside the on-shell delta functions. We obtain
\begin{equation}
\DeltaPnlo= \frac{i\tilde g^4}{4}\Lexp \int \!\dd^4 \qb  \; \del(\initialk_1 \cdot \qb ) 
\del(\initialk_2 \cdot \qb) 
\; e^{-i b \cdot \qb} \;  \impKerCl \Rexp \,,
\label{eqn:classicalLimitNLO}
\end{equation}
where
\begin{equation}
\begin{aligned}
\impKerCl &= \hbar \qb^\mu \, \AmplB^{(1)}(\initialk_1 \initialk_2 \rightarrow 
\initialk_1 + \hbar\qb  , \initialk_2 -  \hbar\qb)
\\&\hphantom{=} 
-i \hbar^3 \int \! \dd^4 \xferb \; 
\del(2p_1\cdot \xferb+ \hbar\xferb^2)\del(2p_2\cdot \xferb- \hbar\xferb^2)  \; \xferb^\mu \;
\\&\hspace*{15mm}\times 
\AmplB^{(0)}(\initialk_1\,, \initialk_2 \rightarrow 
\initialk_1+ \hbar \xferb\,, \initialk_2- \hbar\xferb)
\\&\hspace*{15mm}\times 
\AmplB^{(0)*}(\initialk_1+ \hbar\qb\,, \initialk_2-  \hbar\qb \rightarrow 
\initialk_1+\hbar\xferb  \,,\initialk_2- \hbar\xferb) \,.
\label{eqn:impKerClDef}
\end{aligned}
\end{equation}
Once again, we will see in the next section that additional inverse powers of $\hbar$ will arise from the amplitudes, and will yield a finite and nonvanishing answer
in the classical limit.

\section{Examples}
\label{sec:examples}
\newcommand{\colStructure}{\mathcal{C}}

To build confidence in the formalism we have developed, let us use it to conduct explicit calculations of the classical impulse. We will work in the context of scalar Yang--Mills theory, as defined by the Lagrangian in equation~\eqref{eqn:scalarAction}, using the double copy where our interest is in perturbative gravity.

Before we begin to study the impulse at leading and next-to-leading order, note that it is frequently convenient to write amplitudes in Yang--Mills theory in colour-ordered form; for example, see~\cite{Ochirov:2019mtf} for an application to amplitudes with multiple different external particles. The full amplitude $\mathcal{A}$ is decomposed onto a basis of colour factors times partial amplitudes $A$. The colour factors are associated with some set of Feynman topologies. Once a basis of independent colour structures is chosen, the corresponding partial amplitudes must be gauge invariant. Thus,
\[
\mathcal{A}(p_1,p_2 \rightarrow p_1',p_2') = \sum_D \colStructure(D)\, A_D(p_1,p_2 \rightarrow p_1',p_2')\,,\label{eqn:colourStripping}
\]
where $\colStructure(D)$ is the colour factor of diagram $D$ and $A_D$ is the associated partial amplitude. Expectation values of the representation states $\chi_\alpha$ can now be taken as being purely over the colour structures.

\subsection{Leading-order impulse}
\label{sec:LOimpulse}
\newcommand{\tree}{\begin{tikzpicture}[thick, baseline={([yshift=-\the\dimexpr\fontdimen22\textfont2\relax] current bounding box.center)}, decoration={markings,mark=at position 0.6 with {\arrow{Stealth}}}]
	\begin{feynman}
	\vertex (v1);
	\vertex [below = 0.3 of v1] (v2);
	\vertex [above right = .125 and .275 of v1] (o1);
	\vertex [below right = .125 and .275 of v2] (o2);
	\vertex [above left = .125 and .275 = of v1] (i1);
	\vertex [below left = .125 and .275 of v2] (i2);
	\draw (i1) -- (v1);
	\draw (v1) -- (o1);
	\draw (i2) -- (v2);
	\draw (v2) -- (o2);
	\draw (v1) -- (v2);
	\end{feynman}
	\end{tikzpicture}}
\subsubsection{Gauge theory}

We begin by computing in YM theory the impulse, $\DeltaPlo$, on particle 1 at leading order. At this order, only $\ImpA$ contributes, as expressed in \eqn~\eqref{eqn:impulseGeneralTerm1classicalLO}. To evaluate the impulse, we must first compute the $2\rightarrow 2$ tree-level scattering amplitude. The reduced amplitude $\AmplB^{(0)}$ is
\begin{equation}
i\AmplB^{(0)}(p_1, p_2 \rightarrow p_1+\hbar\qb\,, p_2-\hbar\qb) = \!\!\!\!
\begin{tikzpicture}[scale=1.0, baseline={([yshift=-\the\dimexpr\fontdimen22\textfont2\relax] current bounding box.center)}, decoration={markings,mark=at position 0.6 with {\arrow{Stealth}}}]
\begin{feynman}
\vertex (v1);
\vertex [below = 0.97 of v1] (v2);
\vertex [above left=0.5 and 0.66 of v1] (i1) {$p_1$};
\vertex [above right=0.5 and 0.33 of v1] (o1) {$p_1+\hbar \qb$};
\vertex [below left=0.5 and 0.66 of v2] (i2) {$p_2$};
\vertex [below right=0.5 and 0.33 of v2] (o2) {$p_2-\hbar \qb$};
\draw [postaction={decorate}] (i1) -- (v1);
\draw [postaction={decorate}] (v1) -- (o1);
\draw [postaction={decorate}] (i2) -- (v2);
\draw [postaction={decorate}] (v2) -- (o2);
\diagram*{(v1) -- [gluon] (v2)};
\end{feynman}	
\end{tikzpicture}
\!\!\!\!\!\!\!\!\!\!= i \newT_1\cdot\newT_2 \frac{4 p_1 \cdot p_2 + \hbar^2 \qb^2}
{\hbar^2 \qb^2}\,.
\label{eqn:ReducedAmplitude1}
\end{equation}
Clearly, the colour decomposition of the amplitude is trivial:
\begin{equation}\label{eqn:treeamp}
\bar{A}_{\scalebox{0.5}{\tree}} =  \frac{4 p_1\cdot p_2 +\hbar \barq^2}{\hbar^2 \barq^2}\,, \qquad \colStructure\!\left(\tree\right) = \newT_1\cdot\newT_2\,.
\end{equation}
We can neglect the second term in the numerator, which is subleading in the classical limit.

Substituting this expression into \eqn~\eqref{eqn:impulseGeneralTerm1classicalLO}, we obtain
\begin{equation}
\DeltaPlo = i g^2 \Lexp \int \!\dd^4 \qb  \; \del(\qb\cdot p_1) \del(\qb\cdot p_2)\, e^{-i b \cdot \qb} \newT_1\cdot\newT_2 \frac{p_1 \cdot p_2}{\qb^2}\, \qb^\mu\,\Rexp\,.
\label{eqn:impulseClassicalLOa}
\end{equation}
As promised, the leading-order expression is independent of $\hbar$. Evaluating the $p_{1,2}$ integrals, in the process applying the simplifications explained in section~\ref{sec:classicalLimit}, namely replacing $p_\alpha\rightarrow m_\alpha\ucl_\alpha$, we find that
\begin{equation}
\DeltaPlo = i g^2 c_1\cdot c_2 \int \!\dd^4 \qb  \; \del(\qb\cdot \ucl_1) \del(\qb\cdot \ucl_2) 
e^{-i b \cdot \qb} \frac{\ucl_1 \cdot \ucl_2}{\qb^2} \, \qb^\mu\,.
\label{eqn:impulseClassicalLO}
\end{equation}
Note that evaluating the double angle brackets has also replaced quantum colour factors with classical colour charges. Replacing the classical colour with electric charges $Q_\alpha$ yields the result for QED; this expression then has intriguing similarities to quantities that arise in the high-energy
limit of two-body scattering~\cite{Amati:1987wq,tHooft:1987vrq,Muzinich:1987in,Amati:1987uf,Amati:1990xe,Amati:1992zb,Kabat:1992tb,Amati:1993tb,Muzinich:1995uj,DAppollonio:2010krb,Melville:2013qca,Akhoury:2013yua,DAppollonio:2015fly,Ciafaloni:2015vsa,DAppollonio:2015oag,Ciafaloni:2015xsr,Luna:2016idw,Collado:2018isu}. The eikonal approximation used there is known to
exponentiate, and it would be interesting to explore this connection further. 

Note that it is natural that the Yang--Mills LO impulse is a simple colour dressing of its QED counterpart, since at leading order the gluons do not self interact.

It is straightforward to perform the integral over $\qb$ in \eqn~\eqref{eqn:impulseClassicalLO} to obtain an explicit expression for the leading order impulse. To do so, we work in the rest frame of particle 1, so that $\ucl_1 = (1, 0, 0, 0)$. Without loss of generality we can orientate the spatial coordinates in this frame so that particle 2 is moving along the $z$ axis, with proper velocity $\ucl_2 = (\gamma, 0, 0, \gamma \beta)$. We have introduced the standard Lorentz gamma factor $\gamma = \ucl_1 \cdot \ucl_2$ and the velocity parameter $\beta$ satisfying $\gamma^2 ( 1- \beta^2 ) = 1$. In terms of these variables, the impulse is
\begin{equation}
\begin{aligned}
\DeltaPlo &= i g^2 c_1\cdot c_2 \int \!\dd^4 \qb  \;
\del(\qb^0) \del(\gamma \qb^0 - \gamma \beta \qb^3) \;
e^{-i b \cdot \qb} \frac{\gamma}{\qb^2}
\, \qb^\mu \\
&= -i \frac{g^2 c_1 \cdot c_2 }{4\pi^2 |\beta|}\int \! d^2 \qb  \;
e^{i \v{b} \cdot \v{\qb}_\perp} \frac{1}{\v \qb_\perp^2}
\, \qb^\mu \, ,
\end{aligned}
\end{equation}
where $\qb^0 = \qb^3 = 0$ and the non-vanishing components of $\qb^\mu$ in the $xy$ plane of our corrdinate system are $\v \qb_\perp$. It remains to perform the two dimensional integral over $\v \qb_\perp$, which is easily done using polar coordinates. Let the magnitude of $\v \qb_\perp$ be $\chi$ and orient the $x$ and $y$ axes so that $\v b \cdot \v \qb_\perp = | \v b| \chi \cos \theta$. Then the non-vanishing components of $\qb^\mu$ are $\qb^\mu = (0, \chi \cos \theta, \chi \sin \theta, 0)$ and the impulse is
\begin{equation}
\begin{aligned}
\DeltaPlo &= -i \frac{g^2 c_1\cdot c_2 }{4\pi^2 |\beta|}\int_0^\infty d \chi \; \chi \int_{-\pi}^\pi d \theta   \;
e^{i | \v b| \chi \cos \theta} \frac{1}{\chi^2}
\, (0, \chi \cos \theta, \chi \sin \theta, 0) \\
&= -i \frac{g^2 c_1\cdot c_2 }{4\pi^2 |\beta|}\int_0^\infty d \chi \; \int_{-\pi}^\pi d \theta   \;
e^{i | \v b| \chi \cos \theta} 
\, (0, \cos \theta, \sin \theta, 0) \\
&= \frac{g^2 c_1\cdot c_2 }{2\pi |\beta|}\int_0^\infty d \chi \; 
J_1 ( |\v b| \chi) \; \hat{\v b} \\ 
&= \frac{g^2 c_1\cdot c_2 }{2\pi |\beta|} \; \frac{\hat{\v b}}{| \v b|} \, ,\label{eqn:LOimpulseIntegral}
\end{aligned}
\end{equation}
where $\hat {\v b}$ is the spatial unit vector in the direction of the impact parameter. To restore manifest Lorentz invariance, note that
\begin{equation}
\frac{1}{| \beta|} = \frac{\gamma}{\sqrt{\gamma^2 - 1}}\,, 
\quad \frac{\hat{\v b}}{|\v b|} = - \frac{b^\mu}{b^2}\,.
\end{equation}
(Recall that $b^\mu$ is spacelike, so $-b^2>0$.) With this input, we may write the impulse as
\begin{equation}
\DeltaPlo  
= -\frac{g^2 c_1\cdot c_2}{2\pi} \frac{\gamma}{\sqrt{\gamma^2 - 1}} \frac{b^\mu}{b^2}\,.
\end{equation}

Stripping away the colour and adopting the QED coupling $e$, in the non-relativistic limit this should match a familiar formula: the expansion of the Rutherford scattering angle $\theta(b)$ as a function of the impact parameter. To keep things simple, we consider Rutherford scattering of a light particle (for example, an electron) off a heavy particle (a nucleus). Taking particle 1 to be the moving light particle, particle 2 is very heavy and we work in its rest frame. Expanding the textbook Rutherford result to order $e^2$, we find
\begin{equation}
\theta(b) = 2 \tan^{-1} \frac{e^2}{4 \pi m v^2 b} \simeq \frac{e^2}{2 \pi m v^2 b},
\end{equation}
where $v$ is the non-relativistic velocity of the particle. To recover this simple result from equation~\eqref{eqn:impulseClassicalLO}, recall that in the non-relativistic limit $\gamma \simeq 1 + v^2/2$. The scattering angle, at this order, is simply $\Delta v/v$.  We will make use of this frame in later sections as well.

We note in passing that the second term in the numerator \eqn~\eqref{eqn:ReducedAmplitude1} is a quantum correction.  It will ultimately be suppressed by $\lcomp^2/b^2$, and in addition would contribute only a contact interaction, as it leads to a $\delta^{(2)}(b)$ term in the impulse.

\subsubsection{Gravity}

Rather than compute gravity amplitudes using the Feynman rules associated with the Einstein--Hilbert action, we can easily just apply the double copy where we have knowledge of their gauge theory counterparts. The generalisation of the traditional BCJ gauge theory replacement rules \cite{Bern:2008qj,Bern:2010ue} to massive matter states was developed by Johansson and Ochirov \cite{Johansson:2014zca}.  In our context the colour-kinematics replacement is simple: the amplitude only has a $t$-channel diagram, making the Jacobi identity trivial. Thus by replacing the colour factor with the desired numerator we are guaranteed to land on a gravity amplitude, provided we replace $g\rightarrow\frac{\kappa}{2}$, where $\kappa = \sqrt{32\pi G}$ is the coupling in the Einstein--Hilbert Lagrangian.

A minor point before double-copying is to further rescale\footnote{We choose this normalisation as it simplifies the colour replacements in the double copy.} the (dimensionful) colour factors as $\tilde{\newT}^a$ = $\sqrt{2}\newT^a$, such that
\begin{equation}
\mathcal{A}^{(0)}(p_1, p_2 \rightarrow p_1+\hbar\qb\,, p_2-\hbar\qb) = \frac{g^2}{\hbar^3}\frac{2p_1 \cdot p_2 + \mathcal{O}(\hbar)}{\wn q^2} \tilde{\newT}_1 \cdot \tilde{\newT}_2\,.\label{eqn:scalarYMamp}
\end{equation}
Then replacing the colour factor with the (rescaled) scalar numerator from equation~\eqref{eqn:treeamp}, we immediately obtain the gravity tree amplitude\footnote{The overall sign is consistent with the replacements in \cite{Bern:2008qj,Bern:2010ue} for our amplitudes' conventions.}
\begin{equation}
\mathcal{M}^{(0)}(p_1, p_2 \rightarrow p_1+\hbar\qb\,, p_2-\hbar\qb) = -\frac{4}{\hbar^3}\left(\frac{\kappa}{2}\right)\frac{(p_1 \cdot p_2)^2 + \mathcal{O}(\hbar)}{\wn q^2} \,.
\end{equation}
This is not quite an amplitude in Einstein gravity: the interactions suffer from dilaton pollution, as can immediately be seen by examining the amplitude's factorisation channels:
\[
\lim\limits_{\wn q^2 \rightarrow 0} \left(\wn q^2 \hbar^3 \mathcal{M}^{(0)}\right) &= -4\left(\frac{\kappa}{2}\right)^2\, p_1^\mu p_1^{\tilde{\mu}} \left(\mathcal{P}^{(4)}_{\mu\tilde{\mu}\nu\tilde{\nu}} + \mathcal{D}^{(4)}_{\mu\tilde{\mu}\nu\tilde{\nu}}\right) p_2^\nu p_2^{\tilde{\nu}}\,,
\]
where
\begin{equation}
\mathcal{P}^{(D)}_{\mu\tilde{\mu}\nu\tilde{\nu}} = \eta_{\mu(\nu}\eta_{\tilde{\nu})\tilde{\mu}} - \frac{1}{D-2}\eta_{\mu\tilde{\mu}}\eta_{\nu\tilde{\nu}} \qquad \text{and} \qquad
\mathcal{D}^{(D)}_{\mu\tilde{\mu}\nu\tilde{\nu}} = \frac{1}{D-2}\eta_{\mu\tilde{\mu}}\eta_{\nu\tilde{\nu}}\label{eqn:gravityProjectors}
\end{equation}
are the $D$-dimensional de-Donder gauge graviton and dilaton projectors respectively. The pure Einstein gravity amplitude can now just be read off as the part of the amplitude contracted with the graviton projector. We find that
\begin{equation}
\mathcal{M}^{(0)}_{\rm GR}(p_1, p_2 \rightarrow p_1+\hbar\qb\,, p_2-\hbar\qb) = -\left(\frac{\kappa}{2}\right)^2 \frac{4}{\hbar^3\,\wn q^2} \left((p_1\cdot p_2)^2 - \frac12 m_1^2 m_2^2\right).
\end{equation}
Following the same steps as those before~\eqref{eqn:impulseClassicalLO}, we find that the LO impulse for massive scalar point-particles in general relativity (such as Schwarzschild black holes) is
\begin{equation}
\Delta p_1^{\mu,(0)} = -2i m_1 m_2 \left(\frac{\kappa}{2}\right)^2\! \int \!\dd^4 \qb  \; \del(2\qb\cdot \ucl_1) \del(2\qb\cdot \ucl_2) 
e^{-i b \cdot \qb} \frac{(2\gamma^2 - 1)}{\qb^2} \, \qb^\mu\,.
\end{equation}
Integrating as in equation~\eqref{eqn:LOimpulseIntegral} yields the well known 1PM result~\cite{Westpfahl:1979gu,Portilla:1980uz}
\begin{equation}
\Delta p_1^{\mu,(0)} = \frac{2G m_1 m_2}{\sqrt{\gamma^2 - 1}} (2\gamma^2 - 1) \frac{{b}^\mu}{b^2}\,.
\end{equation}

\subsection{Next-to-leading order impulse}
\label{sec:nloQimpulse}

At the next order in perturbation theory, a well-defined classical impulse is only obtained by combining all terms in the impulse $\langle \Delta p_1^\mu \rangle$ of order $\tilde g^4$. As we discussed in section~\ref{sec:classicalImpulse}, both $\ImpA$ and $\ImpB$ contribute. We found in \eqn~\eqref{eqn:classicalLimitNLO} that the impulse is a simple integral over an impulse kernel $\impKerCl$, defined in \eqn~\eqref{eqn:impKerClDef}, which has a well-defined classical limit. 

The determination of the impulse kernel at this order requires us to compute the four-point one-loop amplitude along with a cut amplitude; that is, an integral over a term quadratic in the tree amplitude. We will compute the NLO impulse in scalar Yang--Mills theory. As the one-loop amplitude in gauge theory is simple, we compute using on-shell renormalised perturbation theory in Feynman gauge.

\subsubsection{Purely Quantum Contributions}
\label{sec:PurelyQuantum}

The contributions to the impulse in the quantum theory can be divided into three classes, according to the prefactor in the charges they carry. For simplicity of counting, let us momentarily restrict to Abelian gauge theory, with charges $Q_\alpha$. There are then three classes of diagrams: $\Gamma_1$, those proportional to $Q_1^3 Q_2$; $\Gamma_2$, those to $Q_1^2 Q_2^2$; and $\Gamma_3$, those to $Q_1 Q_2^3$. The first class can be further subdivided into $\Gamma_{1a}$, terms which would be proportional to $Q_1 (Q_1^2+n_s Q_3^2) Q_2$ were we to add $n_s$ species of a third scalar with charge $Q_3$, and into $\Gamma_{1b}$, terms which would retain the simple $Q_1^3 Q_2$ prefactor.  Likewise, the last class can be further subdivided into $\Gamma_{3a}$, terms which would be proportional to $Q_1 (Q_2^2+n_s Q_3^2) Q_2$, and into $\Gamma_{3b}$, those whose prefactor would remain simply $Q_1 Q_2^3$.

Classes $\Gamma_{1a}$ and $\Gamma_{3a}$ consist of gauge boson self-energy corrections along with renormalisation counterterms.  They appear only in the 1-loop corrections to the four-point amplitude, in the first term in the impulse kernel $\impKerCl$. As one may intuitively expect, they give no contribution in the classical limit. Consider, for example, the self-energy terms, focussing on internal scalars of mass $m$ and charge $Q_i$. We define the self-energy via
\begin{equation}
\hspace{-10pt}Q_i^2 \Pi(q^2) \left( q^2 \eta^{\mu\nu} - q^\mu q^\nu \right) \equiv \scalebox{0.9}{\feynmandiagram [inline = (a.base), horizontal=a to b, horizontal=c to d] { a -- [photon, momentum'=\(q\)] b -- [fermion, half left] c -- [fermion, half left] b -- [draw = none] c -- [photon] d};
\,  + \!\!\!\! \feynmandiagram[inline = (a.base), horizontal=a to b]{a -- [photon, momentum'=\(q\)] c -- [out=45, in=135, loop, min distance=2cm]c -- [photon] b};
\!\!\!\! + \, \feynmandiagram[inline = (a.base), layered layout, horizontal=a to b] { a -- [photon, momentum'=\(q\)] b [crossed dot] -- [photon] c};} \,,
\label{eqn:SelfEnergyContributions}
\end{equation}
where we have made the projector required by gauge invariance manifest, but have not included factors of the coupling. We have extracted the charges $Q_i$ for later convenience. The contribution of the photon self-energy to the reduced 4-point amplitude is
\begin{equation}
\AmplB_\Pi = {Q_1 Q_2 Q_i^2} \frac{(2p_1 + \hbar \qb) \cdot (2p_2 -\hbar \qb)}
{\hbar^2 \qb^2} \Pi(\hbar^2 \qb^2)\,.
\end{equation}
The counterterm is adjusted to impose the renormalisation condition that $\Pi(0) = 0$, 
required in order to match the identification of the gauge coupling with its classical counterpart. As a power series in the dimensionless ratio $q^2 / m^2 = \hbar^2 \qb^2 / m^2$, which is of order $\lcomp^2 / b^2$,
\begin{equation}
\Pi(q^2) = \hbar^2 \Pi'(0) \frac{\qb^2}{m^2} 
+ \mathcal{O}\biggl(\frac{\lcomp^4}{b^4} \biggr)\,.
\end{equation}
The renormalisation condition is essential in eliminating possible contributions of $\Ord(\hbar^0)$. One way to see that $\AmplB_\Pi$ is a purely quantum correction is to follow the powers of $\hbar$. As $\Pi(q^2)$ is of order $\hbar^2$, $\AmplB_\Pi$ is of order $\hbar^0$. This gives a contribution of $\Ord(\hbar)$ to the impulse kernel~(\ref{eqn:impKerClDef}), which in turn gives a contribution of $\Ord(\hbar)$ to the impulse, as can be seen in \eqn~\eqref{eqn:classicalLimitNLO}.

Alternatively, one can consider the contribution of these graphs to $\Delta p / p$. Counting each factor of $\qb$ as of order $b$, and using $\Pi(q^2) \sim \lcomp^2 / b^2$, it is easy to see that these self-energy graphs yield a contribution to $\Delta p / p$ of order $\alpha^2 \hbar^3 / (mb)^3 \sim (\lclass^2 / b^2) \,( \lcomp / b)$.

The renormalisation of the vertex is similarly a purely quantum effect. Since the classes $\Gamma_{1b}$ and $\Gamma_{3b}$ consisted of vertex corrections, wavefunction renormalisation, and their counterterms, they too give no contribution in the classical limit.

These conclusions continue to hold in the non--Abelian theory, with charges promoted to colour factors $C_\alpha$. The different colour structures present in each class of diagram introduces a further splitting of topologies, but one that does not disrupt our identification of quantum effects.

\subsubsection{Classical colour basis}
\label{sec:colourDecomp}
\newcommand{\boxy}{\begin{tikzpicture}[thick, baseline={([yshift=-\the\dimexpr\fontdimen22\textfont2\relax] current bounding box.center)}, decoration={markings,mark=at position 0.6 with {\arrow{Stealth}}}]
	\begin{feynman}
	\vertex (v1);
	\vertex [right = 0.25 of v1] (v2);
	\vertex [above = 0.25 of v1] (v3);
	\vertex [right = 0.25 of v3] (v4);
	\vertex [above left = 0.15 and 0.15 of v3] (o1);
	\vertex [below left = 0.15 and 0.15 of v1] (i1);
	\vertex [above right = 0.15 and 0.15 of v4] (o2);
	\vertex [below right = 0.15 and 0.15 of v2] (i2);
	\draw (i1) -- (v1);
	\draw (v1) -- (v3);
	\draw (v3) -- (o1);
	\draw (i2) -- (v2);
	\draw (v2) -- (v4);
	\draw (v4) -- (o2);
	\draw (v1) -- (v2);
	\draw (v3) -- (v4);
	\end{feynman}	
	\end{tikzpicture}}
\newcommand{\crossbox}{\begin{tikzpicture}[thick, baseline={([yshift=-\the\dimexpr\fontdimen22\textfont2\relax] current bounding box.center)}, decoration={markings,mark=at position 0.6 with {\arrow{Stealth}}}]
	\begin{feynman}
	\vertex (v1);
	\vertex [right = 0.3 of v1] (v2);
	\vertex [above = 0.3 of v1] (v3);
	\vertex [right = 0.3 of v3] (v4);
	\vertex [above left = 0.125 and 0.125 of v3] (o1);
	\vertex [below left = 0.125 and 0.125 of v1] (i1);
	\vertex [above right = 0.125 and 0.125 of v4] (o2);
	\vertex [below right = 0.125 and 0.125 of v2] (i2);
	\vertex [above right = 0.1 and 0.1 of v1] (g1);
	\vertex [below left = 0.1 and 0.1 of v4] (g2);
	\draw (i1) -- (v1);
	\draw (v1) -- (v2);
	\draw (v3) -- (o1);
	\draw (i2) -- (v2);
	\draw (v3) -- (v4);
	\draw (v4) -- (o2);
	\draw (v4) -- (g2);
	\draw (g1) -- (v1);
	\draw (v2) -- (v3);
	\end{feynman}
	\end{tikzpicture}}
\newcommand{\triR}{\begin{tikzpicture}[thick, baseline={([yshift=-\the\dimexpr\fontdimen22\textfont2\relax] current bounding box.center)}, decoration={markings,mark=at position 0.6 with {\arrow{Stealth}}}]
	\begin{feynman}
	\vertex (v1);
	\vertex [above left = 0.25 and 0.17of v1] (v2);
	\vertex [above right = 0.25 and 0.17 of v1] (v3);
	\vertex [below right = 0.2 and 0.275 of v1] (o1);
	\vertex [below left = 0.2 and 0.275 of v1] (i1);
	\vertex [above right = 0.1 and 0.15 of v3] (o2);
	\vertex [above left = 0.1 and 0.15 of v2] (i2);
	\draw (i1) -- (v1);
	\draw (v1) -- (o1);
	\draw (i2) -- (v2);
	\draw (v2) -- (v3);
	\draw (v3) -- (o2);
	\draw (v2) -- (v1);
	\draw (v3) -- (v1);
	\end{feynman}
	\end{tikzpicture}}
\newcommand{\triL}{\begin{tikzpicture}[thick, baseline={([yshift=-\the\dimexpr\fontdimen22\textfont2\relax] current bounding box.center)}, decoration={markings,mark=at position 0.6 with {\arrow{Stealth}}}]
	\begin{feynman}
	\vertex (v1);
	\vertex [below left = 0.25 and 0.17of v1] (v2);
	\vertex [below right = 0.25 and 0.17 of v1] (v3);
	\vertex [above right = 0.2 and 0.275 of v1] (o1);
	\vertex [above left = 0.2 and 0.275 of v1] (i1);
	\vertex [below right = 0.1 and 0.15 of v3] (o2);
	\vertex [below left = 0.1 and 0.15 of v2] (i2);
	\draw (i1) -- (v1);
	\draw (v1) -- (o1);
	\draw (i2) -- (v2);
	\draw (v2) -- (v3);
	\draw (v3) -- (o2);
	\draw (v2) -- (v1);
	\draw (v3) -- (v1);
	\end{feynman}	
	\end{tikzpicture}}
\newcommand{\nonAbL}{\begin{tikzpicture}[thick, baseline={([yshift=-\the\dimexpr\fontdimen22\textfont2\relax] current bounding box.center)}, decoration={markings,mark=at position 0.6 with {\arrow{Stealth}}}]
	\begin{feynman}
	\vertex (v1);
	\vertex [below = 0.15 of v1] (g1);
	\vertex [below left = 0.2 and 0.175 of g1] (v2);
	\vertex [below right = 0.2 and 0.175 of g1] (v3);
	\vertex [above right = 0.2 and 0.275 of v1] (o1);
	\vertex [above left = 0.2 and 0.275 of v1] (i1);
	\vertex [below right = 0.1 and 0.15 of v3] (o2);
	\vertex [below left = 0.1 and 0.15 of v2] (i2);
	\draw (i1) -- (v1);
	\draw (v1) -- (o1);
	\draw (i2) -- (v2);
	\draw (v2) -- (v3);
	\draw (v3) -- (o2);
	\draw (v1) -- (g1);
	\draw (v2) -- (g1);
	\draw (v3) -- (g1);
	\end{feynman}	
	\end{tikzpicture}}
\newcommand{\nonAbR}{\begin{tikzpicture}[thick, baseline={([yshift=-\the\dimexpr\fontdimen22\textfont2\relax] current bounding box.center)}, decoration={markings,mark=at position 0.6 with {\arrow{Stealth}}}]
	\begin{feynman}
	\vertex (v1);
	\vertex [above = 0.15 of v1] (g1);
	\vertex [above left = 0.2 and 0.175 of g1] (v2);
	\vertex [above right = 0.2 and 0.175 of g1] (v3);
	\vertex [below right = 0.2 and 0.275 of v1] (o1);
	\vertex [below left = 0.2 and 0.275 of v1] (i1);
	\vertex [above right = 0.1 and 0.15 of v3] (o2);
	\vertex [above left = 0.1 and 0.15 of v2] (i2);
	\draw (i1) -- (v1);
	\draw (v1) -- (o1);
	\draw (i2) -- (v2);
	\draw (v2) -- (v3);
	\draw (v3) -- (o2);
	\draw (v1) -- (g1);
	\draw (v2) -- (g1);
	\draw (v3) -- (g1);
	\end{feynman}	
	\end{tikzpicture}}

This leaves us with contributions of class $\Gamma_2$; these appear in both terms in the impulse kernel. These contributions to the 1-loop amplitude in the first term take the form
\begin{equation}
\begin{aligned}
i \AmplB^{(1)}(p_1,p_2 \rightarrow p_1', p_2') &= \begin{tikzpicture}[scale=1.0, baseline={([yshift=-\the\dimexpr\fontdimen22\textfont2\relax] current bounding box.center)}] 
\begin{feynman}
\vertex (b) ;
\vertex [above left=1 and 0.66 of b] (i1) {$p_1$};
\vertex [above right=1 and 0.33 of b] (o1) {$p_1+q$};
\vertex [below left=1 and 0.66 of b] (i2) {$p_2$};
\vertex [below right=1 and 0.33 of b] (o2) {$p_2-q$};
\begin{scope}[decoration={
	markings,
	mark=at position 0.7 with {\arrow{Stealth}}}] 
\draw[postaction={decorate}] (b) -- (o2);
\draw[postaction={decorate}] (b) -- (o1);
\end{scope}
\begin{scope}[decoration={
	markings,
	mark=at position 0.4 with {\arrow{Stealth}}}] 
\draw[postaction={decorate}] (i1) -- (b);
\draw[postaction={decorate}] (i2) -- (b);
\end{scope}	
\filldraw [color=white] (b) circle [radius=10pt];
\draw [pattern=north west lines, pattern color=patternBlue] (b) circle [radius=10pt];
\filldraw [fill=white] (b) circle [radius=6pt];
\end{feynman}
\end{tikzpicture} 
\\
&\hspace*{-30mm}=
\scalebox{1.1}{
	\begin{tikzpicture}[baseline={([yshift=-\the\dimexpr\fontdimen22\textfont2\relax] current bounding box.center)}, decoration={markings,mark=at position 0.6 with {\arrow{Stealth}}}]
	\begin{feynman}
	\vertex (v1);
	\vertex [right = 0.9 of v1] (v2);
	\vertex [above = 0.97 of v1] (v3);
	\vertex [right = 0.9 of v3] (v4);
	\vertex [above left = 0.5 and 0.5 of v3] (i2);
	\vertex [below left = 0.5 and 0.5 of v1] (i1);
	\vertex [above right = 0.5 and 0.5 of v4] (o2);
	\vertex [below right = 0.5 and 0.5 of v2] (o1);
	\draw [postaction={decorate}] (i1) -- (v1);
	\draw [postaction={decorate}] (v1) -- (v2);
	\draw [postaction={decorate}] (v2) -- (o1);
	\draw [postaction={decorate}] (i2) -- (v3);
	\draw [postaction={decorate}] (v3) -- (v4);
	\draw [postaction={decorate}] (v4) -- (o2);
	\diagram*{(v3) -- [gluon] (v1)};
	\diagram*{(v2) -- [gluon] (v4)};
	\end{feynman}	
	\end{tikzpicture} + \begin{tikzpicture}[baseline={([yshift=-\the\dimexpr\fontdimen22\textfont2\relax] current bounding box.center)}, decoration={markings,mark=at position 0.6 with {\arrow{Stealth}}}]
	\begin{feynman}
	\vertex (v1);
	\vertex [right = 0.9 of v1] (v2);
	\vertex [above = 0.97 of v1] (v3);
	\vertex [right = 0.9 of v3] (v4);
	\vertex [above left = 0.5 and 0.5 of v3] (i2);
	\vertex [below left = 0.5 and 0.5 of v1] (i1);
	\vertex [above right = 0.5 and 0.5 of v4] (o2);
	\vertex [below right = 0.5 and 0.5 of v2] (o1);
	\vertex [above right = 0.45 and 0.485 of v1] (g1);
	\vertex [below left = 0.4 and 0.4 of v4] (g2);
	\draw [postaction={decorate}] (i1) -- (v1);
	\draw [postaction={decorate}] (v1) -- (v2);
	\draw [postaction={decorate}] (v2) -- (o1);
	\draw [postaction={decorate}] (i2) -- (v3);
	\draw [postaction={decorate}] (v3) -- (v4);
	\draw [postaction={decorate}] (v4) -- (o2);
	\diagram*{(v4) -- [gluon] (v1)};
	\filldraw [color=white] (g1) circle [radius=6.4pt];
	\diagram*{(v2) -- [gluon] (v3)};
	\end{feynman}
	\end{tikzpicture} + \begin{tikzpicture}[baseline={([yshift=-\the\dimexpr\fontdimen22\textfont2\relax] current bounding box.center)}, decoration={markings,mark=at position 0.6 with {\arrow{Stealth}}}]
	\begin{feynman}
	\vertex (v1);
	\vertex [above left = 0.94 and 0.45 of v1] (v2);
	\vertex [above right = 0.94 and 0.45 of v1] (v3);
	\vertex [below right = 0.5 and 0.95 of v1] (o1);
	\vertex [below left = 0.5 and 0.95 of v1] (i1);
	\vertex [above right = 0.4 and 0.5 of v3] (o2);
	\vertex [above left = 0.4 and 0.5 of v2] (i2);
	\draw [postaction={decorate}] (i1) -- (v1);
	\draw [postaction={decorate}] (v1) -- (o1);
	\draw [postaction={decorate}] (i2) -- (v2);
	\draw [postaction={decorate}] (v2) -- (v3);
	\draw [postaction={decorate}] (v3) -- (o2);
	\diagram*{(v1) -- [gluon] (v2)};
	\diagram*{(v1) -- [gluon] (v3)};
	\end{feynman}	
	\end{tikzpicture} + \begin{tikzpicture}[baseline={([yshift=-\the\dimexpr\fontdimen22\textfont2\relax] current bounding box.center)}, decoration={markings,mark=at position 0.6 with {\arrow{Stealth}}}]
	\begin{feynman}
	\vertex (v1);
	\vertex [below left = 0.94 and 0.45 of v1] (v2);
	\vertex [below right = 0.94 and 0.45 of v1] (v3);
	\vertex [above right = 0.5 and 0.95 of v1] (o1);
	\vertex [above left = 0.5 and 0.95 of v1] (i1);
	\vertex [below right = 0.4 and 0.5 of v3] (o2);
	\vertex [below left = 0.4 and 0.5 of v2] (i2);
	\draw [postaction={decorate}] (i1) -- (v1);
	\draw [postaction={decorate}] (v1) -- (o1);
	\draw [postaction={decorate}] (i2) -- (v2);
	\draw [postaction={decorate}] (v2) -- (v3);
	\draw [postaction={decorate}] (v3) -- (o2);
	\diagram*{(v2) -- [gluon] (v1)};
	\diagram*{(v3) -- [gluon] (v1)};
	\end{feynman}	
	\end{tikzpicture} } \\ &\hspace{-10mm}+\scalebox{1.1}{\begin{tikzpicture}[baseline={([yshift=-\the\dimexpr\fontdimen22\textfont2\relax] current bounding box.center)}, decoration={markings,mark=at position 0.6 with {\arrow{Stealth}}}]
	\begin{feynman}
	\vertex (v1);
	\vertex [above = 0.55 of v1] (g1);
	\vertex [above left = 0.45 and 0.45 of g1] (v2);
	\vertex [above right = 0.45 and 0.45 of g1] (v3);
	\vertex [below right = 0.5 and 0.95 of v1] (o1);
	\vertex [below left = 0.5 and 0.95 of v1] (i1);
	\vertex [above right = 0.4 and 0.5 of v3] (o2);
	\vertex [above left = 0.4 and 0.5 of v2] (i2);
	\draw [postaction={decorate}] (i1) -- (v1);
	\draw [postaction={decorate}] (v1) -- (o1);
	\draw [postaction={decorate}] (i2) -- (v2);
	\draw [postaction={decorate}] (v2) -- (v3);
	\draw [postaction={decorate}] (v3) -- (o2);
	\diagram*{(g1) -- [gluon] (v1)};
	\diagram*{(v2) -- [gluon] (g1)};
	\diagram*{(g1) -- [gluon] (v3)};
	\end{feynman}	
	\end{tikzpicture} +	\begin{tikzpicture}[baseline={([yshift=-\the\dimexpr\fontdimen22\textfont2\relax] current bounding box.center)}, decoration={markings,mark=at position 0.6 with {\arrow{Stealth}}}]
	\begin{feynman}
	\vertex (v1);
	\vertex [below = 0.55 of v1] (g1);
	\vertex [below left = 0.45 and 0.45 of g1] (v2);
	\vertex [below right = 0.45 and 0.45 of g1] (v3);
	\vertex [above right = 0.5 and 0.95 of v1] (o1);
	\vertex [above left = 0.5 and 0.95 of v1] (i1);
	\vertex [below right = 0.4 and 0.5 of v3] (o2);
	\vertex [below left = 0.4 and 0.5 of v2] (i2);
	\draw [postaction={decorate}] (i1) -- (v1);
	\draw [postaction={decorate}] (v1) -- (o1);
	\draw [postaction={decorate}] (i2) -- (v2);
	\draw [postaction={decorate}] (v2) -- (v3);
	\draw [postaction={decorate}] (v3) -- (o2);
	\diagram*{(g1) -- [gluon] (v1)};
	\diagram*{(g1) -- [gluon] (v2)};
	\diagram*{(v3) -- [gluon] (g1)};
	\end{feynman}	
	\end{tikzpicture} + \begin{tikzpicture}[scale=1.0, baseline={([yshift=-\the\dimexpr\fontdimen22\textfont2\relax] current bounding box.center)}, decoration={markings,mark=at position 0.6 with {\arrow{Stealth}}}] 
	\begin{feynman}
	\vertex (v1) ;
	\vertex [above left= 0.6 and 1 of v1] (i1);
	\vertex [above right= 0.6 and 1 of v1] (o1);
	\vertex [below = 0.7 of v1] (v2);
	\vertex [below left= 0.6 and 1 of v2] (i2);
	\vertex [below right= 0.6 and 1 of v2] (o2);
	\draw [postaction={decorate}] (i1) -- (v1);
	\draw [postaction={decorate}] (v1) -- (o1);
	\draw [postaction={decorate}] (i2) -- (v2);
	\draw [postaction={decorate}] (v2) -- (o2);
	\diagram*{(v1) -- [gluon, half left] (v2)};
	\diagram*{(v2) -- [gluon, half left] (v1)};
	\end{feynman}
	\end{tikzpicture}}.
\end{aligned}
\end{equation}
In each contribution, we count powers of $\hbar$ following the rules in section~\ref{subsec:Wavefunctions}, replacing $\ell\rightarrow\hbar\ellb$ and $q\rightarrow\hbar \qb$.  In the final double-seagull contribution, we will get four powers from the loop measure, and four inverse powers from the two photon propagators.  Overall, we will not get enough inverse powers to compensate the power in front of the integral in \eqn~\eqref{eqn:impKerClDef}, and thus the seagull will die in the classical limit. 

We will refer to the remaining topologies as the box $B$, cross box $C$, triangles $T_{\alpha\beta}$, and non-Abelian diagrams $Y_{\alpha\beta}$, respectively. Applying the colour decomposition of equation~\eqref{eqn:colourStripping}, the 1-loop amplitude contributing classically to the linear part of the impulse is
\begin{multline}
\AmplB^{(1)}(p_1,p_2 \rightarrow p_1', p_2') = \colStructure\!\left(\boxy \right) B + \colStructure\!\left(\crossbox \right) C + \colStructure\!\left(\triR \right) T_{12} \\ + \colStructure\!\left(\triL \right) T_{21} + \colStructure\!\left(\nonAbR \right) Y_{12} + \colStructure\!\left(\nonAbL \right) Y_{21}\,.
\end{multline}
A first task is to choose a basis of independent colour structures. The complete set of colour factors can easily be calculated:
\begin{equation}
\begin{gathered}
\colStructure\!\left(\boxy \right) = \newT_1^a \newT_2^a \newT_1^b \newT_2^b\,, \qquad
\colStructure\!\left(\crossbox \right) = \newT_1^a \newT_2^b \newT_1^b \newT^a_2\,,\\
\colStructure\!\left(\nonAbL \right) = \hbar\, \newT_1^a f^{abc} \newT_2^b \newT_2^c\,, \qquad 
\colStructure\!\left(\nonAbR \right) = \hbar\, \newT_1^a \newT_1^b f^{abc} \newT_2^c\,,\\
\colStructure\!\left(\triL \right) = \frac12\, \colStructure\!\left(\boxy \right) + \frac12\, \colStructure\!\left(\crossbox \right) = \colStructure\!\left(\triR \right).
\end{gathered}
\end{equation}
At first sight, we appear to have a basis of four independent colour factors: the box, cross box and the two non-Abelian triangles. However, it is very simple to see that the latter are in fact both proportional to the tree colour factor of \eqn~\eqref{eqn:treeamp}; for example, 
\[
\colStructure\!\left(\nonAbL \right) = \frac{\hbar}{2}\, \newT_1^a f^{abc} [\newT_2^b, \newT_2^c] &= \frac{i\hbar^2}{2} f^{abc} f^{bcd} \newT_1^a \newT_2^d\\
& = \frac{i\hbar^2}{2}\, \colStructure\!\left(\tree \right),
\]
where we have used \eqn~\eqref{eqn:chargeLieAlgebra}. Moreover, similar manipulations demonstrate that the cross-box colour factor is not in fact linearly independent:
\[
\colStructure\!\left(\crossbox \right) &= \newT_1^a \newT_1^b \left( \newT_2^a \newT_2^b - i\hbar f^{abc} \newT_2^c\right)\\
&= (\newT_1 \cdot \newT_2) (\newT_1 \cdot \newT_2 ) - \frac{i\hbar}{2} [\newT_1^a, \newT_2^b] f^{abc} \newT_2^c\\
& = \colStructure\!\left(\boxy \right) + \frac{\hbar^2}{2}\, \colStructure\!\left(\tree \right).
\]
Thus at 1-loop the classically significant part of the amplitude has a basis of two colour structures: the box and tree. Hence the decomposition of the 1-loop amplitude into partial amplitudes and colour structures is
\begin{multline}
\AmplB^{(1)}(p_1,p_2 \rightarrow p_1', p_2') =  \colStructure\!\left(\boxy \right) \bigg[B + C + T_{12} + T_{21}\bigg] \\ + \frac{\hbar^2}{2}\, \colStructure\!\left(\tree \right) \bigg[C + \frac{ T_{12}}{2} + \frac{T_{21}}{2} + iY_{12} + iY_{21}\bigg]\,.\label{eqn:1loopDecomposition}
\end{multline}
This expression for the amplitude is particularly useful when taking the classical limit. The second term is proportional to two powers of $\hbar$, while the only possible singularity in $\hbar$ at one loop order is a factor $1/\hbar$ in the evaluation of the kinematic parts of the diagrams. Thus, it is clear that the second line of the expression must be a quantum 
correction, and can be dropped in calculating the classical impulse. Perhaps surprisingly, these terms include the sole contribution from the non-Abelian triangles $Y_{\alpha\beta}$, and thus we will not need to calculate these diagrams. We learn that classically, the 1-loop scalar YM amplitude has a basis of only one colour factor:
\[
\AmplB^{(1)}(p_1,p_2 \rightarrow p_1', p_2') &=  \colStructure\!\left(\boxy \right) \bigg[B + C + T_{12} + T_{21}\bigg] + \mathcal{O}(\hbar)\,.\label{eqn:OneLoopImpulse}
\]
Moreover, the impulse depends on precisely the same topologies as in QED \cite{Kosower:2018adc}.

\subsubsection{Triangles}
\label{sec:Triangles}

Let us first examine the two (colour stripped) triangle diagrams in \eqn~\eqref{eqn:OneLoopImpulse}.  They are related by swapping particles~1 and~2. The first diagram is
\begin{equation}
i T_{12} = 
\begin{tikzpicture}[scale=1.0, baseline={([yshift=-\the\dimexpr\fontdimen22\textfont2\relax] current bounding box.center)}, decoration={markings,mark=at position 0.6 with {\arrow{Stealth}}}] 
\begin{feynman}
\vertex (i1) {$p_1$};
\vertex [right=2.5 of i1] (i2) {$p_1 + q$};
\vertex [below=2.5 of i1] (o1) {$p_2$};
\vertex [below=2.5 of i2] (o2) {$p_2 - q$};

\vertex [below right=1.1 of i1] (v1);
\vertex [below left=1.1 of i2] (v2);
\vertex [above right=1.1 and 1.25 of o1] (v3);

\draw [postaction={decorate}] (i1) -- (v1);
\draw [postaction={decorate}] (v1) -- (v2);
\draw [postaction={decorate}] (v2) -- (i2);
\draw [postaction={decorate}] (o1) -- (v3);
\draw [postaction={decorate}] (v3) -- (o2);

\diagram*{
	(v3) -- [gluon, momentum=\(\ell\)] (v1);
	(v2) -- [gluon] (v3);
};
\end{feynman}
\end{tikzpicture}
= -2 \!\int \!\dd^D \ell\, \frac{(2p_1 + \ell) \cdot (2 p_1 + q + \ell)}
{\ell^2 (\ell - q)^2 (2p_1 \cdot \ell + \ell^2 + i \epsilon)}\,.
\end{equation}
In this integral, we use a dimensional regulator in a standard way ($D=4-2\varepsilon$)
in order to regulate potential divergences. We have retained an explicit $i \epsilon$ in the massive scalar propagator, because it will play an important role below.

To extract the classical contribution of this integral to the amplitude, we recall from section~\ref{subsec:Wavefunctions} that we should set $q =  \hbar\qb$ and $\ell =  \hbar\ellb$, and therefore that the components of $q$ and $\ell$ are all small compared 
to $m$. Consequently, the triangle simplifies to
\begin{equation}
T_{12} = \frac{4 i m_1^2}{\hbar} \!\int\! \dd^4 \bar \ell \, \frac{1}{\bar \ell^2 (\bar \ell - \bar q)^2 (p_1 \cdot \bar \ell + i \epsilon)}\,.
\label{eqn:triangleIntermediate1}
\end{equation}
Here, we have taken the limit $D\rightarrow 4$, as the integral is now free of divergences.
Notice that we have exposed one additional inverse power of $\hbar$. Comparing to the definition of $\impKerCl$ in \eqn~\eqref{eqn:impKerClDef}, we see that this inverse power of $\hbar$ will cancel against the explicit factor of $\hbar$ in $\ImpAclsup{(1)}$, signalling a classical contribution to the impulse.

At this point we employ a simple trick which simplifies the loop integral appearing in \eqn~\eqref{eqn:triangleIntermediate1}, and which will be of great help in simplifying the more complicated box topologies below. The on-shell condition for the outgoing particle 1 requires that $p_1 \cdot \qb = - \hbar \qb^2/2$, so replace $\ellb \rightarrow \ellb' = \qb - \ellb$ in $T_{12}$:
\begin{equation}
\begin{aligned}
T_{12} &= -\frac{4 i m_1^2}{\hbar}\! \int\! \dd^4 \bar \ell'\, \frac{1}{\bar \ell'^2 (\bar \ell' - \bar q)^2 (p_1 \cdot \bar \ell' + \hbar \qb^2 - i \epsilon)} \\
&= -\frac{4 i m_1^2}{\hbar} \!\int \!\dd^4 \bar \ell' \,\frac{1}{\bar \ell'^2 (\bar \ell' - \bar q)^2 (p_1 \cdot \bar \ell' - i \epsilon)} + \mathcal{O}(\hbar^0)\,,
\end{aligned}
\end{equation}
Because of the linear power of $\hbar$ appearing in \eqn~\eqref{eqn:impKerClDef}, the second term
is in fact a quantum correction. We therefore neglect it, and write
\begin{equation}
T_{12}= -\frac{4 i m_1^2}{\hbar}\! \int\! \dd^4 \bar \ell\, \frac{1}{\bar \ell^2 (\bar \ell - \bar q)^2 (p_1 \cdot \bar \ell - i \epsilon)} \, ,
\end{equation}
where we have dropped the prime on the loop momentum: $\ell' \rightarrow \ell$. Comparing with our previous expression, \eqn~\eqref{eqn:triangleIntermediate1}, for the triangle, the net result of these replacements has simply been to introduce an overall sign while, crucially, also switching the sign of the $i \epsilon$ term. Symmetrising over the two expressions for $T_{12}$, we learn that
\begin{equation}
T_{12}= \frac{2 m_1^2}{\hbar}\! \int\! \dd^4 \bar \ell\, \frac{\del(p_1 \cdot \bar \ell)}{\bar \ell^2 (\bar \ell - \bar q)^2} \,,
\end{equation}
using the identity
\begin{equation}
\frac{1}{x-i \epsilon} - \frac{1}{x+i \epsilon} = i \del(x)\,.
\label{eqn:deltaPoles}
\end{equation}

The second triangle contributing to the amplitude, $T_{21}$, can be obtained from $T_{12}$ simply by interchanging the labels 1 and 2:
\begin{equation}
T_{21} = \frac{2 m_2^2}{\hbar}\! \int \! \dd^4 \bar \ell \,
\frac{\del(p_2 \cdot \bar \ell)}{\bar \ell^2 (\bar \ell - \bar q)^2}\,.
\end{equation}
These triangles contribute to the impulse kernel via
\begin{equation}
\begin{aligned}
\impKerCl \big|_\mathrm{triangle} &= \hbar\qb^\mu\, \colStructure\!\left(\boxy \right) (T_{12} + T_{21}) 
\\&= 2\left(\newT_1\cdot \newT_2 \right)^2 \bar q^\mu\! \int \! \frac{\dd^4 \bar \ell}{\bar \ell^2 (\bar \ell - \bar q)^2} \left(m_1^2\del(p_1 \cdot \bar \ell) + m_2^2\del(p_2 \cdot \bar \ell) \right).
\end{aligned}
\end{equation}
Recall that we must integrate over the wavefunctions in order to obtain the classical impulse from the impulse kernel. As we have discussed in section~\ref{subsec:Wavefunctions}, because the inverse power of $\hbar$ here is cancelled by the linear power present explicitly in \eqn~\eqref{eqn:classicalLimitNLO}, we may evaluate the wavefunction integrals by replacing the $p_\alpha$ by their classical values $m_\alpha \ucl_\alpha$. The result for the contribution to the kernel is
\begin{equation}
\impKerTerm1 \equiv
{2 (c_1\cdot c_2)^2} \qb^\mu \!\int \! \dd^4 \ellb\;
\frac{1}{\ellb^2 (\ellb - \qb)^2} 
\biggl(m_1{\del(\ucl_1 \cdot \ellb)} 
+ m_2{\del(\ucl_2 \cdot \ellb)} \biggr)\,.
\label{eqn:TriangleContribution}
\end{equation}
One must still integrate this expression over $\qb$, as in \eqn~\eqref{eqn:classicalLimitNLO}, to
obtain the contribution to the impulse.

\subsubsection{Boxes}
\label{sec:Boxes}

The one-loop amplitude also includes boxes and crossed boxes, and the NLO contribution to the impulse includes as well a term quadratic in the tree amplitude which we can think of as the cut of a one-loop box. Because of the power of $\hbar$ in front of the first term in \eqn~\eqref{eqn:impKerClDef}, we need to extract the contributions of all of these quantities at order $1/\hbar$. However, as we will see, each individual diagram also contains singular terms of order $1/\hbar^2$.  We might fear that these terms pose an obstruction to the 
very existence of a classical limit of the observable in which we are interested. As we will see, this fear is misplaced, as these singular terms cancel completely, leaving a well-defined classical result. It is straightforward to evaluate the individual contributions, but making the cancellation explicit requires some care. We begin with the colour-stripped box:
\begin{equation}
\begin{aligned}
\hspace*{-7mm}i B &= \hspace*{-2mm}
\begin{tikzpicture}[scale=1.0, baseline={([yshift=-\the\dimexpr\fontdimen22\textfont2\relax] current bounding box.center)}, decoration={markings,mark=at position 0.6 with {\arrow{Stealth}}}] 
\begin{feynman}
\vertex (i1) {$p_1$};
\vertex [right=2.5 of i1] (o1) {$p_1 + q$};
\vertex [below=2.5 of i1] (i2) {$p_2$};
\vertex [below=2.5 of o1] (o2) {$p_2 - q$};
\vertex [below right=1.1 of i1] (v1);
\vertex [below left=1.1 of o1] (v2);
\vertex [above right=1.1 of i2] (v3);
\vertex [above left=1.1 of o2] (v4);
\draw [postaction={decorate}] (i1) -- (v1);
\draw [postaction={decorate}] (v1) -- (v2);
\draw [postaction={decorate}] (v2) -- (o1);
\draw [postaction={decorate}] (i2) -- (v3);
\draw [postaction={decorate}] (v3) -- (v4);
\draw [postaction={decorate}] (v4) -- (o2);
\diagram*{
	(v3) -- [gluon, momentum=\(\ell\)] (v1);
	(v2) -- [gluon] (v4);
};
\end{feynman}
\end{tikzpicture} 
\hspace*{-9mm}
= \int \! \dd^D \ell \;
\frac{(2 p_1 \tp \ell) \td (2p_2 \tm \ell)\,(2 p_1 \tp q\tp \ell) \td (2 p_2\tm q\tm \ell)}
{\ell^2 (\ell \tm q)^2 (2 p_1 \cdot \ell \tp \ell^2 \tp i \epsilon)
	(-2p_2 \cdot \ell \tp \ell^2 \tp i \epsilon)}
=\hspace*{-8mm}
\\[-2mm]
\\&\hspace*{-5mm} \frac{1}{\hbar^{2+2\varepsilon}} \!\!\int \! \dd^D \ellb\;
\frac{\bigl[4 p_1\td p_2\tm 2\hbar(p_1\tm p_2)\td\ellb\tm \hbar^2\ellb^2\bigr]
	\bigl[4 p_1\td p_2\tm 2\hbar(p_1\tm p_2)\td(\ellb\tp \qb)\tm \hbar^2(\ellb\tp \qb)^2\bigr]}
{\ellb^2 (\ellb - \qb)^2 (2 p_1 \cdot \ellb + \hbar\ellb^2 + i \epsilon)
	(-2p_2 \cdot \ellb + \hbar\ellb^2 + i \epsilon)}\,,\hspace*{-8mm}
\end{aligned}
\end{equation}
where as usual, we have set $q = \hbar \qb$, $\ell = \hbar \ellb$. We get four powers of $\hbar$ from changing variables in the measure, but six inverse powers from the propagators\footnote{We omit fractional powers of $\hbar$ in this counting as they will disappear when we take $D \rightarrow 4$.}. We thus encounter an apparently singular $1/\hbar^2$ leading behaviour.  We must extract both this singular, $\Ord(1/\hbar^2)$, term 
as well as the terms contributing in the classical limit, which here are $\Ord(1/\hbar)$.
Consequently, we must take care to remember that the on-shell delta functions enforce $\qb \cdot p_1 = - \hbar \qb^2 / 2$ and $\qb \cdot p_2 = \hbar \qb^2 / 2$. 

Performing a Laurent expansion in $\hbar$, truncating after order $1/\hbar$, and separating different orders in $\hbar$, we find that the box's leading terms are given by
\begin{equation}
\begin{aligned}
B &= B_{-1}+B_0\,,
\\ B_{-1} &= \frac{4 i}{\hbar^{2+2\varepsilon}} (p_1 \cdot p_2)^2
\int \frac{\dd^D \ellb}{\ellb^2 (\ellb - \qb)^2
	(p_1 \cdot \ellb + i \epsilon)(p_2 \cdot \ellb - i \epsilon)} \,,
\\ B_{0} &= -\frac{2i }{\hbar^{1+2\varepsilon}} p_1 \cdot p_2 
\int \frac{\dd^D \ellb}{\ellb^2 (\ellb - \qb)^2
	(p_1 \cdot \ellb + i \epsilon)(p_2 \cdot \ellb - i \epsilon)}
\\& \hspace*{30mm}\times
\biggl[ 2{(p_1 - p_2)\cdot \ellb}
+ \frac{ (p_1 \cdot p_2) \ellb^2}{(p_1 \cdot \ellb + i \epsilon)} 
- \frac{(p_1 \cdot p_2) \ellb^2}{(p_2 \cdot \ellb - i \epsilon)}\biggl]\,.
\end{aligned}
\label{eqn:BoxExpansion}
\end{equation}
Note that pulling out a sign from one of the denominators has given the appearance of
flipping the sign of one of the denominator $i\epsilon$ terms.  We must also bear in mind that the integral in $B_{-1}$ is itself \textit{not\/} $\hbar$-independent, so that we will later need to expand it as well.

Similarly, the crossed box is
\begin{align}
i C &= 
\begin{tikzpicture}[scale=1.0, baseline={([yshift=-\the\dimexpr\fontdimen22\textfont2\relax] current bounding box.center)}, decoration={markings,mark=at position 0.6 with {\arrow{Stealth}}}] 
\begin{feynman}
\vertex (i1) {$p_1$};
\vertex [right=2.5 of i1] (o1) {$p_1 + q$};
\vertex [below=2.5 of i1] (i2) {$p_2$};
\vertex [below=2.5 of o1] (o2) {$p_2 - q$};
\vertex [below right=1.1 of i1] (v1);
\vertex [below left=1.1 of o1] (v2);
\vertex [above right=1.1 of i2] (v3);
\vertex [above left=1.1 of o2] (v4);
\draw [postaction={decorate}] (i1) -- (v1);
\draw [postaction={decorate}] (v1) -- node [above] {$\scriptstyle{p_1 + \ell}$} (v2);
\draw [postaction={decorate}] (v2) -- (o1);
\draw [postaction={decorate}] (i2) -- (v3);
\draw [postaction={decorate}] (v3) -- (v4);
\draw [postaction={decorate}] (v4) -- (o2);
\diagram*{(v3) -- [gluon] (v2);};
\filldraw [color=white] ($ (v3) !.5! (v2) $) circle [radius = 4.3pt];
\diagram*{	(v1) -- [gluon] (v4);};
\end{feynman}
\end{tikzpicture} 
\\
&=  \int \! \dd^D \ell\, \frac{(2 p_1 + \ell) \cdot (2p_2 - 2q + \ell)(2 p_1 +q+ \ell) \cdot (2 p_2 -q + \ell)}{\ell^2 (\ell - q)^2 (2 p_1 \cdot \ell + \ell^2 + i \epsilon)(2p_2 \cdot (\ell-q) + (\ell-q)^2 + i \epsilon)}\nonumber
\\
&= \frac{1}{\hbar^{2+2\varepsilon}} \!\! \int \! \dd^D \ellb\, 
\frac{(2 p_1 + \hbar\ellb) \cdot (2p_2 - 2\hbar\qb + \hbar\ellb)\,
	(2 p_1 +\hbar\qb+ \hbar\ellb) \cdot (2 p_2 -\hbar\qb + \hbar\ellb)}
{\ellb^2 (\ellb - \qb)^2 (2 p_1 \cdot \ellb + \hbar\ellb^2 + i \epsilon)
	(2p_2 \cdot (\ellb-\qb) + \hbar(\ellb-\qb)^2 + i \epsilon)}\,.\nonumber
\end{align}
Using the on-shell conditions to simplify $p_\alpha\cdot \qb$ terms in the denominator
and numerator, and once again expanding in powers of $\hbar$, truncating after order $1/\hbar$, and separating different orders in $\hbar$,  we find
\begin{equation}
\begin{aligned}
C &= C_{-1}+C_0\,,
\\ C_{-1} &= -\frac{4i}{\hbar^{2+2\varepsilon}} (p_1 \cdot p_2)^2
\!\int \!\frac{\dd^D \ellb}{\ellb^2(\ellb - \qb)^2} 
\frac{1}
{(p_1 \cdot \ellb + i \epsilon)(p_2 \cdot \ellb + i \epsilon)} 
\\ C_{0} &= -\frac{2i}{\hbar^{1+2\varepsilon}} p_1 \cdot p_2
\!\int \!\frac{\dd^D \ellb}{\ellb^2(\ellb - \qb)^2
	(p_1 \cdot \ellb + i \epsilon)(p_2 \cdot \ellb + i \epsilon)} 
\\& \qquad \times
\biggl[2  (p_1 + p_2) \cdot \ellb
- \frac{(p_1 \cdot p_2) \ellb^2}{(p_1 \cdot \bar \ell + i \epsilon)} 
- \frac{(p_1 \cdot p_2) [(\ellb - \qb)^2 - \qb^2]}
{(p_2 \cdot \ellb + i \epsilon)}\biggr]\,.\hspace*{-20mm}
\end{aligned}
\label{eqn:CrossedBoxExpansion}
\end{equation}
Comparing the expressions for the $\Ord(1/\hbar^2)$ terms in the box and the crossed box, 
$B_{-1}$ and $C_{-1}$ respectively, we see that there is only a partial cancellation of 
the singular, $\mathcal{O}(1/\hbar^2)$, term in the reduced amplitude $\AmplB^{(1)}$. The impulse kernel, \eqn~\eqref{eqn:impKerClDef}, does contain another term, which is quadratic in the tree-level reduced amplitude  $\AmplB^{(0)}$.  We will see below that taking this additional contribution into account leads to a complete cancellation of the singular term; 
but the classical limit does not exist for each of these terms separately.

\subsubsection{Cut Box}
\label{sec:CutBoxes}

In order to see the cancellation of the singular term we must incorporate the term in the impulse kernel which is quadratic in tree amplitudes. As with the previous loop diagrams, let us begin by splitting the colour and kinematic information as in equation~\eqref{eqn:colourStripping}. Then the quadratic term in~\eqref{eqn:impKerClDef} can be written as
\begin{equation}
\impKerCl \big|_\textrm{non-lin} = \colStructure\!\left({\scalebox{1}{\tree}} \right)^\dagger \colStructure\!\left({\scalebox{1}{\tree}}\right) \cutbox^\mu\,,\label{eqn:cutBoxColDecomp}
\end{equation}
where the kinematic data $\cutbox^\mu$ can be viewed as proportional to the cut of the one-loop box, weighted by the loop momentum $\hbar \xferb^\mu$:
\begin{equation}
\cutbox^\mu = -i\hbar^2\int \! \dd^4 \xferb \, \xferb^\mu \, \del(2 p_1 \cdot \xferb + \hbar \xferb^2) \del(2p_2 \cdot \xferb - \hbar \xferb^2) \times
\begin{tikzpicture}[scale=1.0, baseline={([yshift=-\the\dimexpr\fontdimen22\textfont2\relax] current bounding box.center)}, decoration={markings,mark=at position 0.6 with {\arrow{Stealth}}}] 
\begin{feynman}
\vertex (i1) {$p_1$};
\vertex [right=2.5 of i1] (o1) {$p_1 + \hbar \qb$};
\vertex [below=2.5 of i1] (i2) {$p_2$};
\vertex [below=2.5 of o1] (o2) {$p_2 - \hbar\qb$};
\node [] (cutTop) at ($ (i1)!.5!(o1) $) {};
\node [] (cutBottom) at ($ (i2)!.5!(o2) $) {};
\vertex [below right=1.1 of i1] (v1);
\vertex [below left=1.1 of o1] (v2);
\vertex [above right=1.1 of i2] (v3);
\vertex [above left=1.1 of o2] (v4);
\draw [postaction={decorate}] (i1) -- (v1);
\draw (v1) -- (v2);
\draw [postaction={decorate}] (v2) -- (o1);
\draw [postaction={decorate}] (i2) -- (v3);
\draw (v3) -- (v4);
\draw [postaction={decorate}] (v4) -- (o2);
\filldraw [color=white] ($  (cutTop) - (3pt, 0) $) rectangle ($ (cutBottom) + (3pt,0) $) ;
\draw [dashed] (cutTop) -- (cutBottom);
\diagram*{
	(v3) -- [gluon, momentum=\(\hbar\xferb\)] (v1);
	(v2) -- [gluon] (v4);
};
\end{feynman}
\end{tikzpicture}\!\!\!\!\!\!\!\!\!\!\!\!;.
\end{equation}
Note that an additional factor of $\hbar$ in the second term of \eqn~\eqref{eqn:impKerClDef} will be multiplied into \eqn~\eqref{CombiningBoxes} below, as it parallels the factor in the first term of \eqn~\eqref{eqn:impKerClDef}. Evaluating the Feynman diagrams, we obtain
\begin{multline}
\cutbox^\mu = -i\frac{1}{\hbar^2}\! \int \! \dd^4 \xferb \, 
\del(2 p_1 \cdot \xferb + \hbar \xferb^2) \del(2p_2 \cdot \xferb - \hbar \xferb^2) \,
\frac{\xferb^\mu}{\xferb^2 (\xferb - \qb)^2} \\
\quad \times (2 p_1 + \hbar\xferb ) \cdot (2p_2 - \xferb \hbar)\,
(2 p_1 + \hbar\qb + \hbar\xferb ) \cdot (2 p_2 - \hbar\qb  -  \hbar\xferb)\, .\label{eqn:cutBoxFull}
\end{multline}
As in the previous subsection, expand in $\hbar$, and truncate after order $1/\hbar$,
so that
\begin{align}
\cutbox^\mu &= \cutbox_{-1}^\mu + \cutbox_{0}^\mu\,,\nonumber
\\ \cutbox_{-1}^\mu &=  -\frac{4i}{\hbar^2} (p_1 \cdot p_2)^2  
\!\int\! \frac{\dd^4 \ellb \; \ellb^\mu}{\ellb^2 (\ellb - \qb)^2}
\del(p_1 \cdot \ellb) \del(p_2 \cdot \ellb) \,, \label{eqn:CutBoxExpansion}
\\ \cutbox_{0}^\mu &=  -\frac{2i}{\hbar} (p_1 \cdot p_2)^2  
\!\int\! \frac{\dd^4 \ellb \; \ellb^\mu}{\ellb^2 (\ellb - \qb)^2}\,
{\ellb^2} \Big(\del'(p_1 \cdot \ellb) \del(p_2 \cdot \ellb) 
- \del(p_1 \cdot \ellb) \del'(p_2 \cdot \ellb) \Big)\,.\nonumber
\end{align}
We have relabelled $\xferb\rightarrow\ellb$ in order to line up terms more transparently with corresponding ones in the box and crossed box contributions.

Finally, it is easy to see that the cut box colour factor in~\eqref{eqn:cutBoxColDecomp} is simply
\begin{equation}
\colStructure\!\left(\tree \right)^\dagger \colStructure\!\left(\tree \right) = (\newT_2\cdot \newT_1) (\newT_1 \cdot \newT_2) = \colStructure\!\left(\boxy \right)\,.
\end{equation}
Thus there is only one relevant colour structure in the NLO momentum impulse, that of the box. This will be important in the following.

\subsubsection{Combining Contributions}
\label{sec:CombiningTerms}

We are now in a position to assemble the elements computed in the three previous subsections in order to obtain the NLO contributions to the impulse kernel $\impKerCl$, and thence the NLO contributions to the impulse using \eqn~\eqref{eqn:classicalLimitNLO}. Let us begin by examining the singular terms. We must combine the terms from the box, crossed box, and cut box. We can simplify the cut-box contribution $\cutbox_{-1}^\mu$ by exploiting the linear change of variable $\ellb' = \qb - \ellb$:
\begin{align}
\cutbox_{-1}^\mu 
&= -\frac{4i}{\hbar^2} (p_1 \cdot p_2)^2  \nonumber
\!\int\! \dd^4 \ellb' \;\frac{ (\qb^\mu - \ellb'^\mu)}{\ellb'^2 (\ellb' - \qb)^2}
\del(p_1 \cdot \ellb'-p_1\cdot\qb)
\del(p_2 \cdot \ellb'-p_2\cdot \qb)
\\&= -\frac{4i}{\hbar^2} (p_1 \cdot p_2)^2  
\!\int\! \dd^4 \ellb' \;\frac{ (\qb^\mu - \ellb'^\mu)}{\ellb'^2 (\ellb' - \qb)^2}
\del(p_1 \cdot \ellb'+\hbar\qb^2/2) 
\del(p_2 \cdot \ellb'-\hbar\qb^2/2)\nonumber
\\&= -\frac{2i}{\hbar^2} (p_1 \cdot p_2)^2 \qb^\mu
\! \int\! \dd^4 \ellb \;\frac{\del(p_1 \cdot \ellb) \del(p_2 \cdot \ellb)}
{\ellb^2 (\ellb - \qb)^2} +\Ord(1/\hbar)\,,\label{eqn:CutSingular}
\end{align}
where we have used the on-shell conditions to replace $p_1\cdot\qb\rightarrow -\hbar\qb^2/2$
and $p_2\cdot\qb\rightarrow \hbar\qb^2/2$, and where the last line arises from averaging over the two equivalent expressions for $\cutbox_{-1}^\mu$.

We may similarly simplify the singular terms from the box and cross box. Indeed, using the identity~\eqref{eqn:deltaPoles} followed by the linear change of variable, we have
\begin{equation}
\begin{aligned}
B_{-1} + C_{-1} &= 
-\frac{4 }{\hbar^{2+2\varepsilon}} (p_1 \cdot p_2)^2
\!\int \! \frac{\dd^D \ellb}{\ellb^2 (\bar \ell - \bar q)^2} 
\frac{1}{(p_1 \cdot \ellb + i \epsilon)} \del(p_2 \cdot \ellb)
\\&= \frac{4 Q_1^2 Q_2^2}{\hbar^{2+2\varepsilon}} (p_1 \cdot p_2)^2
\!\int\! \frac{\dd^D \ellb'}{\ellb'^2 (\ellb' - \qb)^2}
\frac{\del(p_2 \cdot \ellb'-\hbar\qb^2/2) }{(p_1 \cdot \ellb'+\hbar\qb^2/2 - i \epsilon)} 
\\&= \frac{2i}{\hbar^2} (p_1 \cdot p_2)^2
\!\int\! \frac{\dd^4 \ellb\;\del(p_1 \cdot \ellb)\del(p_2 \cdot \ellb) }
{\ellb^2 (\ellb - \qb)^2} + \Ord(1/\hbar)\,,
\end{aligned}
\label{eqn:BoxSingular}
\end{equation}
where we have averaged over equivalent forms, and then used \eqn~\eqref{eqn:deltaPoles} a second time in obtaining the last line. At the very end, we took $D\rightarrow 4$.

Combining \eqns{eqn:CutSingular}{eqn:BoxSingular}, we find that the potentially singular contributions to the impulse kernel in the classical limit are
\begin{equation}
\begin{aligned}
\impKerCl &\big|_\textrm{singular} =
\hbar\qb^\mu\, \colStructure\!\left(\boxy \right) (B_{-1}+C_{-1}) +\hbar\, \colStructure\!\left(\boxy \right) \cutbox^\mu_{-1} 
\\&\hspace*{-6mm} = \frac{2i}{\hbar} (p_1 \cdot p_2)^2\qb^\mu \left(\newT_1\cdot \newT_2 \right)^2  \Bigg[
\int \frac{\dd^4 \ellb\;\del(p_1 \cdot \ellb)\del(p_2 \cdot \ellb) }
{\ellb^2 (\ellb - \qb)^2}
\\ &\hspace{50mm} - \!\int \dd^4 \ellb \;\frac{\del(p_1 \cdot \ellb) \del(p_2 \cdot \ellb)}
{\ellb^2 (\ellb - \qb)^2}\Bigg]  +\Ord(\hbar^0)
\\&\hspace*{-6mm}=\Ord(\hbar^0)\,.
\end{aligned}
\label{CombiningBoxes}
\end{equation}
Since all terms have common colour factors the dangerous terms cancel, leaving only well-defined contributions.

\def\Cancelling{Z}
It remains to extract the $\Ord(1/\hbar)$ terms from the box, crossed box, and cut box contributions, and to combine them with the triangles~(\ref{eqn:TriangleContribution}), which are of this order.  In addition to $B_0$ from \eqn~\eqref{eqn:BoxExpansion}, $C_0$ from \eqn~\eqref{eqn:CrossedBoxExpansion}, and $\cutbox^\mu_0$ from \eqn~\eqref{eqn:CutBoxExpansion}, we must also include the $\Ord(1/\hbar)$ terms left implicit in \eqns{eqn:CutSingular}{eqn:BoxSingular}.  In the former contributions, we can now set $p_\alpha\cdot \qb = 0$, as the $\hbar$ terms in the on-shell delta functions would give rise to contributions of $\Ord(\hbar^0)$ to the impulse kernel, which in turn will give contributions of $\Ord(\hbar)$ to the impulse.  In combining all these terms, we make use of summing over an expression and the expression after the linear change of variables;
the identity~(\ref{eqn:deltaPoles}); and the identity
\begin{equation}
\del'(x) = \frac{i}{(x-i\epsilon)^2} - \frac{i}{(x+i\epsilon)^2}\,.
\end{equation}
One finds that
\begin{equation}
\begin{aligned}
&\hbar\qb^\mu (B_0 + C_0) +\bigl[\hbar\qb^\mu (B_{-1} + C_{-1})\bigr]\big|_{\Ord(\hbar^0)}
= \Cancelling^\mu
\\& + 2  (p_1 \cdot p_2)^2 \qb^\mu
\!\int \!\frac{\dd^4 \ellb}{\ellb^2 (\ellb \tm \qb)^2}  
\biggl(\del(p_2 \td \ellb) 
\frac{\ellb \td (\ellb \tm\qb) }{(p_1 \td \ellb \tp i \epsilon )^2} 
+ \del(p_1 \td \ellb) 
\frac{\ellb \td (\ellb \tm\qb) }{(p_2 \td \ellb \tm i \epsilon )^2}\biggr) 
\,,
\\ &\hbar\cutbox^\mu_0 +\bigl[\hbar\cutbox^\mu_{-1}\bigr]\big|_{\Ord(\hbar^0)}
= - \Cancelling^\mu 
\\& -\!2 i (p_1 \cdot p_2)^2 
\! \int\! \frac{\dd^4 \ellb}{\ellb^2 (\ellb \tm \qb)^2}  
\ellb^\mu \, \ellb \td (\ellb \tm \qb)\,
\bigl( \del^\prime(p_1 \td \ellb) \del(p_2 \td \ellb) - \del^\prime(p_2 \td \ellb)
\del(p_1 \td \ellb)\bigr) \,,
\end{aligned}
\end{equation}
where we have now taken $D\rightarrow4$, and where the quantity $\Cancelling^\mu$ is
\begin{multline}
\Cancelling^\mu = i (p_1 \cdot p_2)^2 \qb^\mu 
\!\int \! \frac{\dd^4 \ellb}{\ellb^2 (\ellb \tm \qb)^2} 
\;(2 \ellb \cdot \qb - \ellb^2 )
\bigl( \del^\prime(p_1 \td \ellb) \del(p_2 \td \ellb) \\ - \del^\prime(p_2 \td \ellb)
\del(p_1 \td \ellb)\bigr) \, .
\end{multline}

Finally, we integrate over the external wavefunctions. The possible singularity in $\hbar$ has cancelled, so as discussed in section~\ref{subsec:Wavefunctions}, we perform the integrals by replacing the momenta $p_\alpha$ with their classical values $m_\alpha \ucl_\alpha$, and replace the quantum colour factors with classical colour charges. The box-derived contribution is therefore
\begin{multline}
\impKerTerm2 \equiv
2 (c_1\cdot c_2)^2 \gamma^2 \qb^\mu
\! \int \!\frac{\dd^4 \ellb}{\ellb^2 (\ellb \tm \qb)^2}  
\biggl(m_2\del(\ucl_2 \td \ellb) 
\frac{\ellb \td (\ellb \tm\qb) }{(\ucl_1 \td \ellb \tp i \epsilon )^2} 
\\ + m_1\del(\ucl_1 \td \ellb) 
\frac{\ellb \td (\ellb \tm\qb) }{(\ucl_2 \td \ellb \tm i \epsilon )^2}\biggr) \,,
\end{multline}
while that from the cut box is
\begin{multline}
\impKerTerm3 \equiv
-2 i (c_1\cdot c_2)^2 \, \gamma^2 
\! \int\! \frac{\dd^4 \ellb}{\ellb^2 (\ellb \tm \qb)^2}  
\ellb^\mu \, \ellb \td (\ellb \tm \qb)\\
\times\Big( m_2\del^\prime(\ucl_1 \td \ellb)  \del(\ucl_2 \td \ellb) - m_1\del^\prime(\ucl_2 \td \ellb) \del(\ucl_1 \td \ellb)\Big)\,.
\end{multline}
In both contributions we have dropped the $\Cancelling^\mu$ term which cancels
between the two. The full impulse kernel is given by the sum $\impKerTerm1+\impKerTerm2+\impKerTerm3$, and the impulse by
\begin{equation}
\begin{aligned}
\DeltaPnlo &= \frac{i g^4}{4}\hbar\! \int \! \dd^4 \qb \, \del(\qb \cdot \ucl_1) \del(\qb \cdot \ucl_2) 
e^{-i \qb\cdot b} 
\left( \impKerTerm1 + \impKerTerm2 + \impKerTerm3 \right)
\\&= \frac{ig^4}{2} (c_1\cdot c_2)^2\!
\int \! \frac{\dd^4 \ellb}{\ellb^2 (\ellb - \qb)^2}\dd^4 \qb \, \del(\qb \cdot \ucl_1) \del(\qb \cdot \ucl_2) e^{-i \qb\cdot b}
\\&\hphantom{=}\times\biggl[
\qb^\mu \biggl( \frac{\del(\ucl_1 \cdot \ellb)}{m_2}
+ \frac{\del(\ucl_2 \cdot \ellb)}{m_1} \biggr)
\\&\hphantom{=} \hphantom{\times\biggl[}
+\gamma^2\qb^\mu \biggl(\frac{\del(\ucl_2 \td \ellb)}{m_1}
\frac{\ellb \td (\ellb -\qb) }{(\ucl_1 \td \ellb + i \epsilon )^2} 
+ \frac{\del(\ucl_1 \td \ellb)}{m_2}
\frac{\ellb \td (\ellb -\qb) }{(\ucl_2 \td \ellb - i \epsilon )^2}\biggr) 
\\&\hphantom{=} \hphantom{\times\biggl[} -
i \gamma^2\ellb^\mu \, \ellb \td (\ellb - \qb)\,
\biggl( \frac{\del^\prime(\ucl_1 \td \ellb) \del(\ucl_2 \td \ellb)}{m_1}
- \frac{\del^\prime(\ucl_2 \td \ellb) \del(\ucl_1 \td \ellb)}{m_2}\biggr)\biggr]\,.
\end{aligned}
\label{eqn:NLOImpulse}
\end{equation}
It was shown in \cite{delaCruz:2020bbn} that this result is precisely reproduced by applying worldline perturbation theory to iteratively solve the Yang--Mills--Wong equations in equation~\eqref{eqn:classicalWong}. Moreover, our the final result for the impulse in non-Abelian gauge theory is in fact identical to QED \cite{Kosower:2018adc} (in which context this calculation was first performed), but with the charge to colour replacement $Q_1 Q_2 \rightarrow c_1 \cdot c_2$. This is a little peculiar, as it is natural to expect the non-linearity of the Yang--Mills field to enter at this order (and it does so in the quantum theory). The origin of the result is the colour basis decomposition in equation~\eqref{eqn:1loopDecomposition}, and in particular the fact that the non-Abelian triangle diagrams only contribute to the $\hbar^2$ suppressed second colour structure.

With the relevant Yang--Mills amplitude at hand, one may of course wonder about the prospect of double copying to obtain the NLO impulse in gravity. The construction of colour-kinematics dual numerators at loop level following our methods is highly non-trivial; however, recent progress with massive particles may now make this problem tractable \cite{Carrasco:2015iwa}. It is also interesting to compare our methods to those of Shen~\cite{Shen:2018ebu}, who implemented the double copy at NLO wholly within the classical worldline formalism following ground-breaking work of Goldberger and Ridgway~\cite{Goldberger:2016iau}. Shen found it necessary to include vanishing terms involving structure constants in his work. Similarly, in our context, some colour factors are paired with kinematic numerators proportional to $\hbar$. It would be interesting to use the tools developed in these chapters to explore the double copy construction of Shen~\cite{Shen:2018ebu} from the perspective of amplitudes.

The agreement of~\eqref{eqn:NLOImpulse} with worldline perturbation theory offers a strong check on our formalism, and is of greater importance than the evaluation of the remaining integrals, which also arise in the classical theory. Their evaluation is surprisingly intricate; however, we can gain some interesting insights into the physics just from considering momentum conservation.

\subsubsection{On-Shell Cross Check}

As we have seen, careful inclusion of boxes, crossed boxes as well as cut boxes are necessary to determine the impulse in the classical regime. This may seem to be at odds with other work on the classical limit of amplitudes, which often emphasises the particular importance of triangle diagrams to the classical potential at next to leading order. However, in the context of the potential, the partial cancellation between boxes and crossed boxes is well-understood~\cite{Donoghue:1996mt}, and it is because of this fact that triangle diagrams are particularly important. The residual phase is known to exponentiate so that it does not effect classical physics. Meanwhile, the relevance of the subtraction of iterated (cut) diagrams has long been a topic of discussion~\cite{Sucher:1994qe,BjerrumBohr:2002ks,Neill:2013wsa}.

Nevertheless, in the case of the impulse it may seem that the various boxes play a more significant role, as they certainly contribute to the classical result for the impulse. In fact, it is easy to see that these terms must be included to recover a physically sensible result. The key observation is that the final momentum, $\finalk_1^\mu$, of the outgoing particle after a classical scattering process must be on shell, $\finalk_1^2 = m_1^2$.

We may express the final momentum in terms of the initial momentum and the impulse, so that
\begin{equation}
\finalk_1^\mu = p_1^\mu + \Delta p_1^\mu\,.
\end{equation}
The on-shell condition is then
\begin{equation}
(\Delta p_1)^2 + 2 p_1 \cdot \Delta p_1 = 0\,.
\end{equation}
At order $g^2$, this requirement is satisfied trivially. At this order $(\Delta p_1)^2$ is negligible, while
\begin{equation}
p_1 \cdot \Delta p_1 = i m_1 g^2 c_1\cdot c_2 \!\int \! \dd^4 \qb \,
\del(\qb \cdot \ucl_1) \del(\qb \cdot \ucl_2) 
\, e^{-i \qb \cdot b} \, \qb \cdot \ucl_1 \frac{\ucl_1 \cdot \ucl_2 }{\qb^2} = 0\,,
\end{equation}
using our result for the LO impulse in \eqn~\eqref{eqn:impulseClassicalLO}.

The situation is less trivial at order $g^4$, as neither $p_1 \cdot \Delta p_1$ nor $(\Delta p_1)^2$ vanish. In fact, at this order we may use \eqn~\eqref{eqn:impulseClassicalLO} once again to find that
\begin{multline}
(\Delta p_1)^2 = - g^4  (c_1\cdot c_2)^2 \, (\ucl_1 \cdot \ucl_2)^2 \\
\times \int \! \dd^4 \qb \,\dd^4 \qb' \, \del(\qb \cdot \ucl_1) 
\del(\qb \cdot \ucl_2) \del(\qb' \cdot \ucl_1) \del(\qb' \cdot \ucl_2) 
\, e^{-i (\qb + \qb') \cdot b} \,  \frac{\qb \cdot \qb'}{\qb^2 \, \qb'^2}\,.
\label{eqn:DeltaPsquared}
\end{multline}
Meanwhile, to evaluate $p_1 \cdot \Delta p_1$ we must turn to our NLO result for the impulse, \eqn~\eqref{eqn:NLOImpulse}. Thanks to the delta functions present in the impulse, we 
find a simple expression:
\begin{multline}
2 p_1 \cdot \Delta p_1 = g^4 (c_1\cdot c_2)^2 \, (\ucl_1 \cdot \ucl_2)^2\! \int \! \dd^4 \qb \, \del(\qb \cdot \ucl_1) \del(\qb \cdot \ucl_2) \, e^{-i \qb\cdot b} \\
\times \int \! \dd^4 \ellb\; \ellb \cdot \ucl_1 \, \del'(\ellb \cdot \ucl_1) \del(\ellb \cdot \ucl_2) \, \frac{\ellb \cdot (\ellb - \qb)}{\ellb^2 (\ellb - \qb)^2}\,.
\label{eqn:pDotDeltaP}
\end{multline}
To simplify this expression, it may be helpful to imagine working in the restframe of the timelike vector $u_1$. Then, the $\ellb$ integral involves the distribution $\ellb_0 \, \del'(\ellb_0)$, while $\qb_0 = 0$. Thus the $\ellb_0$ integral has the form
\begin{equation}
\int \! \dd \ellb_0 \, \ellb_0 \, \del'(\ellb_0) \, f(\ellb_0{}^2) = -\!\int \! \dd \ellb_0 \, \del(\ellb_0) \, f(\ellb_0{}^2)\,.
\end{equation}
Using this observation, we may simplify equation~\eqref{eqn:pDotDeltaP} to find
\begin{equation}
\begin{aligned}
\vspace{-2mm}2 p_1 \cdot \Delta p_1 &= -g^4 (c_1\cdot c_2)^2 \, (\ucl_1 \cdot \ucl_2)^2 \\
& \,\,\, \times 
\int \! \dd^4 \qb\, \dd^4 \ellb \; \del(\qb \cdot \ucl_1) \del(\qb \cdot \ucl_2) 
\del(\ellb \cdot \ucl_1) \del(\ellb \cdot \ucl_2)  
e^{-i \qb\cdot b}  \frac{\ellb \cdot (\ellb - \qb)}{\ellb^2 (\ellb - \qb)^2}  \\
&=  g^4 (c_1\cdot c_2)^2 \, (\ucl_1 \cdot \ucl_2)^2 \\
& \,\,\,\times
\int \! \dd^4 \ellb \, \dd^4 \qb' \;
\del(\ellb \cdot \ucl_1) \del(\ellb \cdot \ucl_2) \del(\qb' \cdot \ucl_1) 
\del(\qb' \cdot \ucl_2) \, e^{-i (\ellb+\qb')\cdot b} 
\frac{\ellb \cdot \qb'}{\ellb^2\, \qb'^2}\,,
\end{aligned}
\end{equation}
where in the last line we set $\qb' = \qb - \ellb$. This expression is equal but opposite to \eqn~\eqref{eqn:DeltaPsquared}, and so the final momentum is on-shell as it must be.

It is worth remarking that the part of the NLO impulse that is relevant in this cancellation arises solely from the cut boxes. One can therefore view this phenomenon as an analogue of the removal of iterations of the tree in the potential.

\section{Beyond next-to-leading-order}
\label{sec:NNLO}

We have worked in this chapter under the premise of studying conservative scattering. Yet the LO and NLO impulse are only conservative in the sense that momentum is simply exchanged from particle 1 to particle 2 at these orders. However, beyond these lowest orders in perturbation theory physics does not clearly distinguish between conservative and dissipative behaviour: we will see shortly that at NNLO momentum can be radiated away, and moreover back-reacts on the impulse. To complete our on-shell formalism we must therefore incorporate radiation --- the interplay between the impulse and the radiated momentum forms the subject of our next chapter.

%% file: chapter4/chapter4.tex
\chapter{Radiation: emission and reaction}
\label{chap:radiation}

\section{Introduction}

Gravitational wave astronomy relies on extracting measurable data from radiation. In this chapter we will therefore apply the methods developed for the impulse to construct a second on-shell and quantum-mechanical observable, the total emitted radiation.

These two observables are not independent. Indeed, the relation between them goes to the heart of one of the difficulties in traditional approaches to classical field theory with point sources. In two-particle scattering in classical electrodynamics, for example, momentum is transferred from one particle to the other via the electromagnetic field, as described by the Lorentz force. But the energy-momentum lost by point-particles to radiation is not accounted for by the Lorentz force. Conservation of momentum is restored by taking into account an additional force, the Abraham--Lorentz--Dirac (ALD) force~\cite{Lorentz,Abraham:1903,Abraham:1904a,Abraham:1904b,Dirac:1938nz,LandauLifshitz}; see e.g. refs.~\cite{Higuchi:2002qc,Galley:2006gs,Galley:2010es,Birnholtz:2013nta,Birnholtz:2014fwa,Birnholtz:2014gna} for more recent treatments. Inclusion of this radiation reaction force is not without cost: rather, it leads to the celebrated issues of runaway solutions or causality violations in the classical electrodynamics of point sources.

Using quantum mechanics to describe charged-particle scattering in should cure these ills. Indeed, we will see explicitly that a quantum-mechanical description will conserve energy and momentum in particle scattering automatically. First, in section~\ref{sec:radiatedmomentum} we will set up expressions for the total radiated momentum in quantum field theory, and show that when combined with the impulse of the previous chapter, momentum is automatically conserved to all orders in perturbation theory. We will apply our previous investigation of the classical limit in section~\ref{sec:classicalradiation}, introducing the radiation kernel and discussing how it relates to objects familiar from classical field theory. In section~\ref{sec:LOradiation} we explicitly compute the radiation kernel at leading order in gauge and gravitational theories, and using its form in QED explicitly show that our impulse formalism from chapter~\ref{chap:impulse} reproduces the predictions of the classical Abraham--Lorentz--Dirac force. We discuss the results of this and the previous chapter in section~\ref{sec:KMOCdiscussion}.

This chapter continues to be based on work published in refs.~\cite{Kosower:2018adc,delaCruz:2020bbn}.

\section{The momentum radiated during a collision}
\label{sec:radiatedmomentum}

A familiar classical observable is the energy radiated by an accelerating particle, for example during a scattering process. More generally we can compute the four-momentum radiated. In quantum mechanics there is no precise prediction for the energy or the momentum radiated by localised particles; we obtain a continuous spectrum if we measure a large number of events. However we can compute the expectation value of the four-momentum radiated during a scattering process. This is a well-defined observable, and as we will see it is on-shell in the sense that it can be expressed in terms of on-shell amplitudes.

To define the observable, let us again surround the collision with detectors which measure outgoing radiation of some type. We will call the radiated quanta `messengers'. Let $\mathbb{K}^\mu$ be the momentum operator for whatever field is radiated; then the expectation of the radiated momentum is
\begin{equation}
\begin{aligned}
\langle k^\mu \rangle = {}_\textrm{out}{\langle} \Psi |  \mathbb{K}^\mu S \, | \Psi \rangle_\textrm{in} = {}_\textrm{in}{\langle} \Psi | \, S^\dagger   \mathbb{K}^\mu S\, | \Psi \rangle_\textrm{in}\,,
\end{aligned}
\end{equation}
where $|\Psi\rangle_\textrm{in}$ is again taken as the wavepacket in equation~\eqref{eqn:inState}. Once again we can anticipate that the radiation will be expressed in terms of amplitudes. Rewriting $S = 1 + i T$, the expectation value becomes
\begin{align}
\Rad^\mu \equiv \langle k^\mu \rangle &= {}_\textrm{in}{\langle} \Psi | \, S^\dagger   \mathbb{K}^\mu S \, | \Psi \rangle_\textrm{in}
= {}_\textrm{in}\langle \Psi | \, T^\dagger   \mathbb{K}^\mu T \, | \Psi \rangle_\textrm{in}\,,
\end{align}
because $ \mathbb{K}^\mu |\Psi \rangle_\textrm{in} = 0$ since there are no quanta of radiation in the incoming state. 

We can insert a complete set of states $|X ; k; \finalk_1 \, \zeta_1; \finalk_2 \, \zeta_2\rangle$ containing at least one radiated messenger of momentum $k$, and write the expectation value of the radiated momentum as follows:
\begin{equation}
\begin{aligned}
\hspace{-2mm}\Rad^\mu = \sum_X \int\! \df(k) \df(\finalk_1) \df(\finalk_2) \d\mu(\zeta_1) \d\mu(\zeta_2)\;
k_X^\mu \bigl|\langle k; \finalk_1 \, \zeta_1; \finalk_2 \, \zeta_2; X \,| \, T \,
| \Psi \rangle\bigr|^2\,.
\end{aligned}
\label{eqn:radiationTform}
\end{equation}
In this expression, $X$ can again be empty, and $k_X^\mu$ is the sum of the explicit messenger momentum $k^\mu$ and the momenta of any messengers in the state $X$. Notice that we are including explicit integrals for particles 1 and 2, consistent with our assumption that the number of these particles is conserved during the process. The state $| k \rangle$ describes a radiated messenger; the phase space integral over $k$ implicitly includes a sum over its helicity.

Expanding the initial state, we find that the expectation value of the radiated momentum is given by
\begin{equation}
\begin{aligned}
\Rad^\mu &= \sum_X \int\! \df(k) \df(\finalk_1) \df(\finalk_2) \d\mu(\zeta_1) \d\mu(\zeta_2)\,
k_X^\mu \\ 
& \times\bigg| \int\! \df(\initialk_1)\df(\initialk_2)  e^{i b \cdot \initialk_1/\hbar} \varphi_1(\initialk_1) \varphi_2(\initialk_2) \del^{(4)}(\initialk_1 + \initialk_2 - \finalk_1 - \finalk_2 - k - \finalk_X) \\
& \hspace{40mm} \times \langle \zeta_1 \, \zeta_2 \, X|\Ampl(\initialk_1\,, \initialk_2 \rightarrow 
\finalk_1\,, \finalk_2\,, k\,, \finalk_X) |\chi_1\, \chi_2 \rangle \bigg|^2
\,,
\label{eqn:ExpectedMomentum}
\end{aligned}
\end{equation}
where we have accounted for any representation states in $X$, with the appropriate Haar measure implicity contained in the external sum. We can again introduce momentum transfers, $q_\alpha=\initialkc_\alpha-\initialk_\alpha$, and trade the integrals over $\initialkc_\alpha$ for integrals over the $q_\alpha$.  One of the four-fold $\delta$ functions will again become $\del^{(4)}(q_1+q_2)$, and we can use it to perform the $q_2$ integrations.  We again relabel $q_1\rightarrow q$. The integration leaves behind a pair of on-shell $\delta$ functions and positive-energy $\Theta$ functions, just as in \eqns{eqn:impulseGeneralTerm1}{eqn:impulseGeneralTerm2}:
\begin{equation}
\begin{aligned}
\hspace{-8pt}\Rad^\mu =&\, \sum_X \int\! \df(k) \prod_{\alpha=1,2}\df(\finalk_\alpha) \df(\initialk_\alpha) \dd^4 q\;
\varphi_1(\initialk_1) \varphi_2(\initialk_2) 
\varphi_1^*(\initialk_1+q) \varphi_2^*(\initialk_2-q) \,
\\&\times 
\del(2\initialk_1 \cdot q + q^2) \del(2 \initialk_2 \cdot q - q^2) 
\Theta(\initialk_1{}^0+q^0)\Theta(\initialk_2{}^0-q^0)
\\&\times 
k_X^\mu \, e^{-i b \cdot q/\hbar} 
\,\del^{(4)}(\initialk_1 + \initialk_2 
- \finalk_1 - \finalk_2 - k - \finalk_X)%
\\&\times 
\langle \Ampl^*(\initialk_1+q\,, \initialk_2-q \rightarrow \finalk_1\,, \finalk_2\,, k\,, \finalk_X)
 \Ampl(\initialk_1\,, \initialk_2 \rightarrow \finalk_1\,, \finalk_2\,, k\,, \finalk_X)
\rangle
\,.
\end{aligned}
\label{eqn:ExpectedMomentum2b}
\end{equation}
Representation states have been absorbed into an expectation value as in equation~\eqref{eqn:defOfAmplitude}. We emphasise that this is an all-orders expression: the amplitude $\Ampl(\initialk_1,\squeeze \initialk_2 \squeeze\rightarrow \squeeze \finalk_1,\squeeze \finalk_2,\squeeze k,\squeeze \finalk_X)$ includes all loop corrections, though of course it can be expanded in perturbation theory. The corresponding real-emission contributions are present in the sum over states $X$.   If we truncate the amplitude at a fixed order in perturbation theory, we should similarly truncate the sum over states. Given that the expectation value is expressed in terms of an on-shell amplitude, it is also appropriate to regard this observable as a fully on-shell quantity.

It can be useful to represent the observables diagrammatically. Two equivalent expressions for the radiated momentum are helpful:
\begin{multline}
\usetikzlibrary{decorations.markings}
\usetikzlibrary{positioning}
\Rad^\mu =
\sum_X  \mathlarger{\int}\! \df(k)\df(\finalk_1)\df(\finalk_2)\;
k_X^\mu 
\\ \times\left| \mathlarger{\int}\! \df(\initialk_1) \df(\initialk_2)\; 
e^{i b \cdot \initialk_1/\hbar} \, 
\del^{(4)}\!\left(\sum p\right) \hspace*{-13mm}
\begin{tikzpicture}[scale=1.0, 
baseline={([xshift=-5cm,yshift=-\the\dimexpr\fontdimen22\textfont2\relax]
	current bounding box.center)},
] 
\begin{feynman}
\vertex (b) ;
\vertex [above left=of b] (i1) {$\psi_1(\initialk_1)$};
\vertex [above right=of b] (o1) {$\finalk_1$};
\vertex [above right =0.2 and 1.4 of b] (k) {$k$};
\vertex [below right =0.2 and 1.4 of b] (X) {$\finalk_X$};
\vertex [below left=1 and 1 of b] (i2) {${\psi_2(\initialk_2)}$};
\vertex [below right=1 and 1 of b] (o2) {$\finalk_2$};
\diagram* {(b) -- [photon, photonRed] (k)};
\diagram*{(b) -- [boson] (X)};
\begin{scope}[decoration={
	markings,
	mark=at position 0.7 with {\arrow{Stealth}}}] 
\draw[postaction={decorate}] (b) -- node [right=4pt] {}(o2);
\draw[postaction={decorate}] (b) -- node [left=4pt] {} (o1);
\end{scope}
\begin{scope}[decoration={
	markings,
	mark=at position 0.42 with {\arrow{Stealth}}}] 
\draw[postaction={decorate}] (i1) -- node [left=4pt] {} (b);
\draw[postaction={decorate}] (i2) --node [right=4pt] {} (b);
\end{scope}	
\filldraw [color=white] (b) circle [radius=10pt];
\filldraw [fill=allOrderBlue] (b) circle [radius=10pt];	
\end{feynman}
\end{tikzpicture}
\right|^2,
\label{eqn:RadiationPerfectSquare}
\end{multline}
which is a direct pictorial interpretation of equation~\eqref{eqn:ExpectedMomentum}, and
\begin{equation}
\usetikzlibrary{decorations.markings}
\usetikzlibrary{positioning}
\begin{aligned}
\Rad^\mu &=
\sum_X \mathlarger{\int}\! \df(k) \prod_{\alpha = 1, 2} \df(\finalk_\alpha)
 \df(\initialk_\alpha) \df(\initialkc_\alpha)\; k_X^\mu  \, e^{i b \cdot (\initialk_1 - \initialkc_1)/\hbar}   \\
& \times \del^{(4)}(\initialk_1 + \initialk_2 - \finalk_1 - \finalk_2 - k - \finalk_X)\, \del^{(4)}(\initialkc_1 + \initialkc_2 - \finalk_1 - \finalk_2 - k - \finalk_X) \\
& \hspace{50mm}\times
\begin{tikzpicture}[scale=1.0, 
baseline={([yshift=-\the\dimexpr\fontdimen22\textfont2\relax]
	current bounding box.center)},
] 
\begin{feynman}
\begin{scope}
\vertex (ip1) ;
\vertex [right=2 of ip1] (ip2);
\node [] (X) at ($ (ip1)!.5!(ip2) $) {};
\begin{scope}[even odd rule]
\begin{pgfinterruptboundingbox} 
\path[invclip] ($  (X) - (4pt, 30pt) $) rectangle ($ (X) + (4pt,30pt) $) ;
\end{pgfinterruptboundingbox} 

\vertex [above left=0.66 and 0.33 of ip1] (q1) {$ \psi_1(\initialk_1)$};
\vertex [above right=0.66 and 0.33 of ip2] (qp1) {$ \psi^*_1(\initialkc_1)$};
\vertex [below left=0.66 and 0.33 of ip1] (q2) {$ \psi_2(\initialk_2)$};
\vertex [below right=0.66 and 0.33 of ip2] (qp2) {$ \psi^*_2(\initialkc_2)$};

\diagram* {(ip1) -- [photon, out=30, in=150, photonRed]  (ip2)};
\diagram*{(ip1) -- [photon, out=330, in=210]  (ip2)};
\begin{scope}[decoration={
	markings,
	mark=at position 0.4 with {\arrow{Stealth}}}] 
\draw[postaction={decorate}] (q1) -- (ip1);
\draw[postaction={decorate}] (q2) -- (ip1);
\end{scope}
\begin{scope}[decoration={
	markings,
	mark=at position 0.7 with {\arrow{Stealth}}}] 
\draw[postaction={decorate}] (ip2) -- (qp1);
\draw[postaction={decorate}] (ip2) -- (qp2);
\end{scope}
\begin{scope}[decoration={
	markings,
	mark=at position 0.38 with {\arrow{Stealth}},
	mark=at position 0.74 with {\arrow{Stealth}}}] 
\draw[postaction={decorate}] (ip1) to [out=90, in=90,looseness=1.7] node[above left] {{$ \finalk_1$}} (ip2);
\draw[postaction={decorate}] (ip1) to [out=270, in=270,looseness=1.7]node[below left] {${\finalk_2}$} (ip2);
\end{scope}

\node [] (Y) at ($(X) + (0,1.5)$) {};
\node [] (Z) at ($(X) - (0,1.5)$) {};
\node [] (k) at ($ (X) - (0.35,-0.55) $) {$k$};
\node [] (x) at ($ (X) - (0.35,0.55) $) {$\finalk_X$};

\filldraw [color=white] ($ (ip1)$) circle [radius=8pt];
\filldraw  [fill=allOrderBlue] ($ (ip1) $) circle [radius=8pt];

\filldraw [color=white] ($ (ip2) $) circle [radius=8pt];
\filldraw  [fill=allOrderBlue] ($ (ip2) $) circle [radius=8pt];

\end{scope} 
\end{scope}
\draw [dashed] (Y) to (Z);
\end{feynman}
\end{tikzpicture},
\end{aligned}
\end{equation}
which demonstrates that we can think of the expectation value as the weighted cut of a loop amplitude. As $X$ can be empty, the lowest-order contribution arises from the weighted cut of a two-loop amplitude.

\subsection{Conservation of momentum}
\label{sect:allOrderConservation}

The expectation of the radiated momentum is not independent of the impulse. In fact the relation between these quantities is physically rich. In the classical electrodynamics of point~particles, for example, the impulse is due to a total time integral of the usual Lorentz force,~\eqref{eqn:Wong-momentum}. However, when the particles emit radiation the point-particle approximation leads to well-known issues. This is a celebrated problem in classical field theory.  Problems arise because of the singular nature of the point-particle source. In particular, the electromagnetic field at the position of a point charge is infinite, so to make sense of the Lorentz force acting on the particle the traditional route is to subtract the particle's own field from the full electromagnetic field in the force law. The result is a well-defined force, but conservation of momentum is lost.

Conservation of momentum is restored by including another force, the Abraham--Lorentz--Dirac (ALD) force~\cite{Lorentz,Abraham:1903,Abraham:1904a,Abraham:1904b,Dirac:1938nz}, acting on the particles. This gives rise to an impulse on particle 1 in addition to the impulse due to the Lorentz force. The Lorentz force exchanges momentum between particles 1 and 2, while the radiation reaction impulse,
\begin{equation}
\Delta {p^\mu_1}_{\rm ALD} = \frac{e^2 Q_1^2}{6\pi m_1}\int_{-\infty}^\infty\! d\tau \left(\frac{d^2p_1^\mu}{d\tau^2} + \frac{p_1^\mu}{m_1^2}\frac{dp_1}{d\tau}\cdot\frac{dp_1}{d\tau}\right),
\label{eqn:ALDclass}
\end{equation}
accounts for the irreversible loss of momentum due to radiation. Of course, the ALD force is a notably subtle issue in the classical theory.

In the quantum theory of electrodynamics there can be no question of violating conservation of momentum, so the quantum observables we have defined must already include all the effects which would classically be attributed to both the Lorentz and ALD forces. This must also hold for the counterparts of these forces in any other theory. In particular, it must be the case that our definitions respect conservation of momentum; it is easy to demonstrate this formally to all orders using our definitions. Later, in section~\ref{sec:ALD}, we will indicate how the radiation reaction is included in the impulse more explicitly.

Our scattering processes involve two incoming particles. Consider, then,
\begin{equation}
\begin{aligned}
\langle \Delta p_1^\mu \rangle + \langle \Delta p_2^\mu \rangle &= 
\langle \Psi | i [   \mathbb{P}_1^\mu +   \mathbb{P}_2^\mu, T ] | \Psi \rangle 
+ \langle \Psi | T^\dagger [   \mathbb{P}_1^\mu +   \mathbb{P}_2^\mu, T ] | \Psi \rangle \\
&= \bigl\langle \Psi \big| i \bigl[ \textstyle{\sum_\alpha}   \mathbb{P}_\alpha^\mu, T \bigr] 
\big| \Psi \bigr\rangle 
+ \langle \Psi | T^\dagger [  \mathbb{P}_1^\mu +   \mathbb{P}_2^\mu, T ] | \Psi \rangle\,,
\end{aligned}
\end{equation}
where the sum $\sum   \mathbb{P}_\alpha^\mu$ is now over all momentum operators in the theory, not just those for the two initial particles. The second equality above holds because $  \mathbb{P}_\alpha^\mu | \Psi \rangle = 0$ for $\alpha \neq 1,2$; only quanta of fields 1 and 2 are present in the incoming state. Next, we use the fact that the total momentum is time independent, or in other words
\begin{equation}
\Bigl[ \sum  \mathbb{P}_\alpha^\mu, T \Bigr] = 0\,,
\end{equation}
where the sum extends over all fields. Consequently,
\begin{equation}
\langle \Psi | i [   \mathbb{P}_1^\mu +   \mathbb{P}_2^\mu, T ] | \Psi \rangle =  
\bigl\langle \Psi \big| i \bigl[ \textstyle{\sum_\alpha}   \mathbb{P}_\alpha^\mu, T \bigr] \big| 
\Psi \bigr\rangle = 0\,.
\label{eqn:commutatorVanishes}
\end{equation}
Thus the first term $\langle \Psi | i [   \mathbb{P}_1^\mu, T ] | \Psi \rangle$ in the impulse~(\ref{eqn:defl1}) describes only the exchange of momentum between particles~1 and~2; in this sense it is associated with the classical Lorentz force (which shares this property) rather than with the classical ALD force (which does not). The second term in the impulse, on the other hand, includes radiation. To make the situation as clear as possible, let us restrict attention to the case where the only other momentum operator is $  \mathbb{K}^\mu$, the momentum operator for the messenger field. Then we know that $[   \mathbb{P}_1^\mu +   \mathbb{P}_2^\mu +   \mathbb{K}^\mu, T] = 0$, and conservation of momentum at the level of expectation values is easy to demonstrate:
\begin{equation}
\langle \Delta p_1^\mu \rangle + \langle \Delta p_2^\mu \rangle = 
- \langle \Psi | T^\dagger [   \mathbb{K}^\mu, T ] | \Psi \rangle = 
- \langle \Psi | T^\dagger    \mathbb{K}^\mu T  | \Psi \rangle = 
- \langle k^\mu \rangle = - \Rad^\mu\,,
\end{equation}
once again using the fact that there are no messengers in the incoming state.

In the classical theory, radiation reaction is a subleading effect, entering for two-body scattering at order $e^6$ in perturbation theory in electrodynamics. This is also the case in the quantum theory. To see why, we again expand the operator product in the second term of \eqn~\eqref{eqn:defl1} using a complete set of states:
\begin{multline}
\langle \Psi | \, T^\dagger [   \mathbb{P}_1^\mu, T] \, |\Psi \rangle = \sum_X \int \! \df(\finalk_1)\df(\finalk_2) \d\mu(\zeta_1) \d\mu(\zeta_2)\;
\\ \times\langle \Psi | \, T^\dagger | \finalk_1 \, \zeta_1; \finalk_2 \, \zeta_2; X \rangle 
\langle \finalk_1 \, \zeta_1; \finalk_2 \, \zeta_2; X | [  \mathbb{P}_1^\mu, T] \, |\Psi \rangle\,.
\end{multline}
The sum over $X$ is over all states, including an implicit integral over their momenta and a sum over any other quantum numbers. The inserted-state momenta of particles 1 and~2 (necessarily present) are labeled by $\finalk_\alpha$, and the corresponding integrations over these momenta by $\df(\finalk_\alpha)$.  These will ultimately become integrations over the final-state momenta in the scattering. To make the loss of momentum due to radiation explicit at this level, we note that
\begin{multline}
\langle \Psi | \, T^\dagger [   \mathbb{P}_1^\mu +    \mathbb{P}_2^\mu, T] \, |\Psi \rangle 
= -\sum_X \int \! \df(\finalk_1)\df(\finalk_2) \d\mu(\zeta_1) \d\mu(\zeta_2)\;
\\\times\langle \Psi | \, T^\dagger | \finalk_1 \, \zeta_1; \finalk_2 \, \zeta_2; X\rangle  
\langle \finalk_1 \, \zeta_1; \finalk_2 \, \zeta_2; X | \,   \mathbb{P}_X^\mu T \,  |\Psi \rangle\,,
\end{multline}
where $ \mathbb{P}_X$ is the sum over momentum operators of all quantum fields other than the scalars~1 and 2. The sum over all states $X$ will contain, for example, terms where the state $X$ includes messengers of momentum $k^\mu$ along with other massless particles. We can further restrict attention to the contributions of the messenger's momentum to $\mathbb{P}_X^\mu$. This contribution produces a net change of momentum of particle 1 given by
\begin{multline}
-\sum_X \int \!  \df(k) \df(\finalk_1)\df(\finalk_2)\d\mu(\zeta_1) \d\mu(\zeta_2)\; k^\mu \, 
\\\times\langle \Psi | \, T^\dagger | k ; \finalk_1 \, \zeta_1; \finalk_2 \, \zeta_2; X\rangle
\langle k; \finalk_1 \, \zeta_1; \finalk_2 \, \zeta_2; X| \, T \,  |\Psi \rangle 
= - \langle k^\mu \rangle\,,
\end{multline} 
with the help of equation~\eqref{eqn:radiationTform}. Thus we explicitly see the net loss of momentum due to radiating messengers. In any theory this quantity is suppressed by factors of the coupling $\tilde g$ because of the additional state. The lowest order case corresponds to $X = \emptyset$; as there are two quanta in $|\psi \rangle$, we must compute the modulus squared of a five-point tree amplitude. The term is proportional to $\tilde g^6$, where $\tilde g$ is the coupling of an elementary three-point amplitude; as far as the impulse is concerned, it is a next-to-next-to-leading order (NNLO) effect. Other particles in the state $X$, and other contributions to its momentum, describe higher-order effects.

\section{Classical radiation}
\label{sec:classicalradiation}
\def\Radcl{K^\mu}

Following our intensive study of the classical limit of the impulse in the previous chapter, the avenue leading to the classical limit of $R^\mu$ is clear: provided we work with the wavefunctions of chapter~\ref{chap:pointParticles} in the the Goldilocks zone $\ell_c \ll \ell_w \ll \lscatt$, we can simply adopt the rules of section~\ref{sec:classicalLimit}. In particular the radiated momentum $k$ will scale as a wavenumber in the classical region. This is enforced by the energy-momentum-conserving delta function in \eqn~\eqref{eqn:ExpectedMomentum2b}, rewritten in terms of momentum transfers $w_\alpha = r_\alpha - p_\alpha$:
\begin{equation}
\del^{(4)}(w_1+w_2 + k + \finalk_X)\,.
\end{equation}
The arguments given after equation~\eqref{eqn:radiationScalingDeltaFunction} then ensure that the typical values of all momenta in the argument should again by scaled by $1/\hbar$ and replaced by wavenumbers.

With no new work required on the formalities of the classical limit, let us turn to explicit expressions for the classical radiated momentum in terms of amplitudes. Recall that our expressions for the total emitted radiation in section~\ref{sec:radiatedmomentum} depended on $q$, which represents a momentum mismatch rather than a momentum transfer. However, we expect the momentum transfers to play an important role in the classical limit, and so it is convenient to change variables from the $r_\alpha$ to make use of them:
\begin{equation}
\begin{aligned}
\Rad^\mu &= \sum_X \int\! \df(k) \prod_{\alpha=1,2} \df(\initialk_\alpha) \dd^4\xfer_\alpha\dd^4 q\;
\del(2p_\alpha\cdot \xfer_\alpha+\xfer_\alpha^2)\Theta(p_\alpha^0+\xfer_\alpha^0)
\\&\times 
\del(2\initialk_1 \cdot q + q^2) \del(2 \initialk_2 \cdot q - q^2) 
\Theta(\initialk_1{}^0+q^0)\Theta(\initialk_2{}^0-q^0)\,\varphi_1(\initialk_1) \varphi_2(\initialk_2) 
\\&\qquad\times \varphi_1^*(\initialk_1+q) \varphi_2^*(\initialk_2-q) \, k_X^\mu \, e^{-i b \cdot q/\hbar}  \del^{(4)}(\xfer_1+\xfer_2+ k+ \finalk_X)
\\&\qquad\qquad\times 
\langle\Ampl^*(\initialk_1+q\,, \initialk_2-q \rightarrow 
\initialk_1+\xfer_1\,, \initialk_2+\xfer_2\,, k\,, \finalk_X)
\\&\qquad\qquad\qquad\times \Ampl(\initialk_1\,, \initialk_2 \rightarrow 
\initialk_1+\xfer_1\,, \initialk_2+\xfer_2\,, k\,, \finalk_X)
\rangle
\,.
\end{aligned}
\label{eqn:ExpectedMomentum2}
\end{equation}
We can now recast this expression in the notation of \eqn~\eqref{eqn:angleBrackets}:
\begin{equation}
\begin{aligned}
\Rad^\mu_\class &= \sum_X\, \Lexp \int\! \df(k) \prod_{\alpha=1,2} \dd^4\xfer_\alpha\,\dd^4 q\;
\del(2p_\alpha\cdot \xfer_\alpha+\xfer_\alpha^2)\Theta(p_\alpha^0+\xfer_\alpha^0)\,  k_X^\mu 
\\&\times 
\del(2\initialk_1 \cdot q + q^2) \del(2 \initialk_2 \cdot q - q^2) 
\del^{(4)}(\xfer_1+\xfer_2+ k+ \finalk_X) \Theta(\initialk_1{}^0+q^0)%
\\& \times \Theta(\initialk_2{}^0-q^0)\, e^{-i b \cdot q/\hbar} \, \Ampl^*(\initialk_1+q, \initialk_2-q \rightarrow 
\initialk_1+\xfer_1\,, \initialk_2+\xfer_2\,, k\,, \finalk_X)
\\&\qquad\qquad\qquad\times 
\Ampl(\initialk_1, \initialk_2 \rightarrow 
\initialk_1+\xfer_1\,, \initialk_2+\xfer_2\,, k\,, \finalk_X)\,\Rexp
\,.
\end{aligned}
\label{eqn:ExpectedMomentum2recast}
\end{equation}
We will determine the classical limit of this expression using precisely the same logic as in the preceding chapter. Let us again focus on the leading contribution, with $X=\emptyset$.  Once again, rescale $q \rightarrow \hbar\qb$, and drop the $q^2$ inside the on-shell delta functions. Here, remove an overall factor of $\tilde g^6$ and accompanying $\hbar$'s from the amplitude and its conjugate. In addition, rescale the momentum transfers $\xfer\rightarrow \hbar\xferb$ and the radiation momenta, $k\rightarrow\hbar\wn k$. At leading order there is no sum, so there will be no hidden cancellations, and we may drop the $\xfer_\alpha^2$ inside the on-shell delta functions to obtain
\def\kb{\bar k}
\begin{equation}
\begin{aligned}
\Rad^{\mu,(0)}_\class &= 
\tilde g^6 \Lexp \hbar^4\! \int\! \df(\kb) \prod_{\alpha=1,2} \dd^4\xferb_\alpha\dd^4 \qb\, \del(2\xferb_\alpha\cdot p_\alpha)
\del(2\qb\cdot p_1) \del(2\qb\cdot p_2) \, e^{-i b \cdot \qb}
\\& \qquad \times \kb^\mu \, \AmplB^{(0)*}(\initialk_1+\hbar \qb, \initialk_2-\hbar \qb \rightarrow 
\initialk_1+\hbar\xferb_1\,, \initialk_2+\hbar\xferb_2\,, \hbar\kb)
\\& \qquad\times 
\AmplB^{(0)}(\initialk_1, \initialk_2 \rightarrow 
\initialk_1+\hbar\xferb_1\,, \initialk_2+\hbar\xferb_2\,, \hbar\kb)\,\del^{(4)}(\xferb_1+\xferb_2+ \kb)\,\Rexp
\,.
\end{aligned}
\label{eqn:ExpectedMomentum2classicalLO}
\end{equation}
We will make use of this expression below to verify that momentum is conserved as expected.

One disadvantage of this expression for the leading order radiated momentum is that it is no longer in a form of an 
integral over a perfect square, such as shown in \eqn~\eqref{eqn:RadiationPerfectSquare}. Nevertheless we can recast \eqn~\eqref{eqn:ExpectedMomentum2recast} in such a form.
To do so, perform a change of variable, including in the (momentum space) wavefunctions. To begin, it is helpful to write \eqn~\eqref{eqn:ExpectedMomentum2recast} as
\begin{equation}
\begin{aligned}
\Rad^\mu_\class =&\, \sum_X \prod_{\alpha=1,2}  \int \! \df(\initialk_\alpha)\,  |\varphi_\alpha(\initialk_\alpha)|^2 \int \! \df(k) \df( \xfer_\alpha+\initialk_\alpha) \df(q_\alpha+\initialk_\alpha) \; 
\\& \times \del^{(4)}(\xfer_1+\xfer_2+ k+ \finalk_X) \del^{(4)}(q_1 + q_2) \, e^{-i b \cdot q_1/\hbar} \, k_X^\mu \,   %
\\&\qquad\times 
\langle \Ampl^*(\initialk_1+q_1\,, \initialk_2 + q_2 \rightarrow 
\initialk_1+\xfer_1\,, \initialk_2+\xfer_2\,, k\,, \finalk_X)
\\&\qquad\qquad\times \Ampl(\initialk_1\,, \initialk_2 \rightarrow 
\initialk_1+\xfer_1\,, \initialk_2+\xfer_2\,, k\,, \finalk_X)
\rangle\,
\,.
\end{aligned}
\end{equation}
\def\tinitialk{\tilde\initialk}
\def\txfer{\tilde\xfer}
\def\tq{\tilde q}
\noindent We will now re-order the integration and perform a change of variables. Let us define $\tinitialk_\alpha=\initialk_\alpha - \txfer_\alpha$, $\tq_\alpha = q_\alpha + \txfer_\alpha$, and $\txfer_\alpha = - \xfer_\alpha$, changing variables from $\initialk_\alpha$ to $\tinitialk_\alpha$, from $q_\alpha$ to $\tq_\alpha$, and from $\xfer_\alpha$ to $\txfer_\alpha$:
\begin{equation}
\begin{aligned}
\Rad^\mu_\class =&\, \sum_X \prod_{\alpha=1,2}  \int \! \df(\tinitialk_\alpha)  \df(k) \df(\txfer_\alpha+\tinitialk_\alpha) \df (\tq_\alpha+\tinitialk_\alpha) |\varphi_\alpha(\tinitialk_\alpha+\txfer_\alpha)|^2 \; 
\\& \times \del^{(4)}(\tilde \xfer_1+ \tilde \xfer_2- k- \finalk_X) \del^{(4)}(\tq_1 + \tq_2 - k - \finalk_X)\, e^{-i b \cdot (\tq_1 - \txfer_1)/\hbar} \, k_X^\mu
\\&\qquad\times 
\langle\Ampl^*(\tinitialk_1+ \tq_1\,, \tinitialk_2 + \tq_2 \rightarrow 
\tinitialk_1\,, \tinitialk_2\,, k\,, \finalk_X)
\\&\qquad\qquad\times \Ampl(\tinitialk_1 + \txfer_1\,, \tinitialk_2 + \txfer_2\rightarrow 
\tinitialk_1\,, \tinitialk_2\,, k\,, \finalk_X)
\rangle\,
\,.
\end{aligned}
\end{equation}
As the $\txfer_\alpha$ implicitly carry a factor of $\hbar$, just as argued in \sect{subsec:Wavefunctions} for the momentum mismatch $q$, we may neglect the shift in the wavefunctions.  Dropping the tildes, and associating the $\xfer_\alpha$ integrals with $\Ampl$ and the $q_\alpha$ integrals with $\Ampl^*$, our expression is revealed as an integral over a perfect square,
\begin{equation}
\begin{aligned}
\Rad^\mu_\class 
& = \sum_X \prod_{\alpha=1,2} \Lexp \int \! \df(k) \, k_X^\mu
\biggl | \int \! \df(\xfer_\alpha + \initialk_\alpha)  \; 
\del^{(4)}( \xfer_1+  \xfer_2- k- \finalk_X)\\
& \hspace*{25mm} \times   e^{i b \cdot  \xfer_1/\hbar}  \,   
\Ampl( \initialk_1 +  \xfer_1,  \initialk_2 +  \xfer_2\rightarrow 
\initialk_1\,,  \initialk_2\,, k\,, \finalk_X) \biggr|^2 \Rexp
\,.
\label{eqn:radiatedMomentumClassicalAllOrder}
\end{aligned}
\end{equation}
The perfect-square structure allows us to define a \textit{radiation kernel\/},
\begin{equation}
\begin{aligned}
\RadKer(k, \finalk_X)
&\equiv \hbar^{3/2} \prod_{\alpha = 1, 2}  \int \! \df( \initialk_\alpha +  \xfer_\alpha)  \; 
\del^{(4)}( \xfer_1+  \xfer_2- k- \finalk_X) \\
& \qquad \qquad \times  e^{i b \cdot  \xfer_1/\hbar}  \,   
\Ampl( \initialk_1 +  \xfer_1,  \initialk_2 +  \xfer_2\rightarrow 
\initialk_1\,,  \initialk_2\,, k\,, \finalk_X), \\
= & \hbar^{3/2}\prod_{\alpha = 1, 2} \int \! \dd^4 \xfer_\alpha\; 
\del(2 p_\alpha \cdot \xfer_\alpha + \xfer_\alpha^2)\, \del^{(4)}( \xfer_1+  \xfer_2- k- \finalk_X) \\
& \quad\times \Theta(\initialk_\alpha^0+\xfer_\alpha^0)\, e^{i b \cdot  \xfer_1/\hbar}  \,   
\Ampl( \initialk_1 +  \xfer_1,  \initialk_2 +  \xfer_2\rightarrow 
\initialk_1\,,  \initialk_2\,, k\,, \finalk_X)\,,
\label{eqn:defOfR}
\end{aligned}
\end{equation}
so that
\begin{equation}
\begin{aligned}
\Rad^\mu_\class &=  \sum_X \hbar^{-3}\Lexp \int \! \df(k) \, k_X^\mu
\left |\RadKer(k, \finalk_X) \right|^2 \Rexp
\,.
\label{eqn:radiatedMomentumClassical}
\end{aligned}
\end{equation}
The prefactor along with the normalization of $\RadKer$ are again chosen so that the classical limit of the radiation kernel will be of $\Ord(\hbar^0)$. Let us now focus once more on the leading contribution, with $X=\emptyset$.  As usual, rescale $\xfer \rightarrow \hbar\xferb$, and remove an overall factor of $\tilde g^6$ and accompanying $\hbar$'s from the amplitude and its conjugate. Then the LO radiation kernel is
\begin{equation}
\begin{aligned}
\RadKerCl(\wn k) 
& \equiv \hbar^2 \prod_{\alpha = 1, 2} \int \! \dd^4 \xferb_\alpha \, \del(2p_\alpha \cdot \xferb_\alpha + \hbar\xferb_\alpha^2) \,
\del^{(4)}( \xferb_1+  \xferb_2- \wn k)
e^{i b \cdot  \xferb_1}  
\\& \hspace{70pt} \times  \AmplB^{(0)}( \initialk_1 +  \hbar \xferb_1,  \initialk_2 + \hbar \xferb_2\rightarrow 
\initialk_1\,,  \initialk_2\,, \hbar \wn k)\,,
\label{eqn:defOfRLO}
\end{aligned}
\end{equation}
ensuring that the leading-order momentum radiated is simply
\begin{equation}
\begin{aligned}
\Rad^{\mu, (0)}_\class &= \tilde g^6 \Lexp \int \! \df(\wn k) \, \wn k^\mu \left | \RadKerCl(\wn k) \right|^2 \Rexp\,.
\label{eqn:radiatedMomentumClassicalLO}
\end{aligned}
\end{equation}

\subsubsection{Conservation of momentum}
\label{sect:classicalConservation}

Conservation of momentum certainly holds to all orders, as we saw in \sect{sect:allOrderConservation}.  However, it is worth making sure that we have not spoiled this critical physical property in our previous discussion, or indeed in our discussion of the classical impulse in \sect{sec:classicalImpulse}. One might worry, for example, that there is a subtlety with the order of limits.

There is no issue at LO and NLO for the impulse, because
\begin{equation}
\DeltaPlo + \DeltaPloTwo = 0 ,\quad \DeltaPnlo + \DeltaPnloTwo = 0.
\end{equation}
These follow straightforwardly from the definitions of the observables, \eqn~\eqref{eqn:impulseGeneralTerm1classicalLO} and \eqn~\eqref{eqn:classicalLimitNLO}. The essential point is that the amplitudes entering into these orders in the impulse conserve momentum for two particles. At LO, for example, using \eqn~\eqref{eqn:impulseGeneralTerm1classicalLO} the impulse on particle 2 can be written as
\begin{multline}
\DeltaPloTwo=  \frac{i\tilde g^2}{4} \Lexp \hbar^2\! \int \!\dd^4 \qb_1  \dd^4 \qb_2 \; 
\del(\qb_1\cdot p_1) \del(\qb_1\cdot p_2)  \del^{(4)}(\qb_1 + \qb_2)
\\\times  
e^{-i b \cdot \qb_1} 
\, \qb_2^\mu  \, \AmplB^{(0)}(p_1,\,p_2 \rightarrow 
p_1 + \hbar\qb_1, p_2 + \hbar\qb_2)\,\Rexp.
\end{multline}
In this equation, conservation of momentum at the level of the four point amplitude $\AmplB^{(0)}(p_1,\,p_2 \rightarrow p_1 + \hbar\qb_1, p_2 + \hbar\qb_2)$ is expressed by the presence of the four-fold delta function $\del^{(4)}(\qb_1 + \qb_2)$. Using this delta function, we may replace $\qb_2^\mu$ with $- \qb_1^\mu$ and then integrate over $\qb_2$, once again using the delta function. The result is manifestly $-\DeltaPlo$,  \eqn~\eqref{eqn:impulseGeneralTerm1classicalLO}. A similar calculation goes through at NLO.

In this sense, the scattering is conservative at LO and at NLO. At NNLO, however, we must take radiative effects into account.  This backreaction is entirely described by the quadratic part of the impulse, $\ImpB$. As indicated in \eqn~\eqref{eqn:commutatorVanishes}, $\ImpA$ is always conservative. From our perspective here, this is because it involves only four-point amplitudes.  Thus to understand conservation of momentum we need to investigate $\ImpB$. The lowest order case in which a five point amplitude can enter $\ImpB$ is at NNLO. Let us restrict attention to this lowest order case, taking the additional state $X$ to be a messenger.

For $\ImpB$ the lowest order term inolving one messenger is, in the classical regime,
\begin{equation}
\hspace*{-3mm}\begin{aligned}
\ImpBclsup{(\textrm{rad})} =&\, \tilde g^6
\Lexp \hbar^{4}\!\int \! d\Phi(\wn k)  \prod_{\alpha = 1,2} \dd^4\xferb_\alpha\, 
\dd^4 \qb_1 \dd^4 \qb_2 \;\del(2 \xferb_\alpha\cdot p_\alpha + \xferb_\alpha^2)
\\&\times  \del(2 \qb_1\cdot p_1) \del(2 \qb_2\cdot p_2)\,
e^{-i b \cdot \qb_1}\,\xferb_1^\mu\,
\del^{(4)}(\xferb_1+\xferb_2 + \bar k)\, \del^{(4)}(\qb_1+\qb_2)
\\&\quad\times \AmplB^{(0)}(\initialk_1\,, \initialk_2 \rightarrow 
\initialk_1+\hbar\xferb_1\,, \initialk_2+\hbar\xferb_2, \hbar\kb)
\\&\qquad\times 
\AmplB^{(0)*}(\initialk_1+\hbar \qb_1\,, \initialk_2 + \hbar \qb_2 \rightarrow 
\initialk_1+\hbar\xferb_1\,, \initialk_2+\hbar\xferb_2, \hbar\kb)
\,\Rexp\,.
\end{aligned} 
\label{eqn:nnloImpulse}
\end{equation}
To see that this balances the radiated momentum, we use \eqn~\eqref{eqn:ExpectedMomentum2classicalLO}. The structure of the expressions are almost identical; conservation of momentum holds because the factor $\kb^\mu$ in \eqn~\eqref{eqn:ExpectedMomentum2classicalLO} is balanced by $\xferb_1^\mu$ in \eqn~\eqref{eqn:nnloImpulse} and $\xferb_2^\mu$ in the equivalent expression for particle 2.

Thus conservation of momentum continues to hold in our expressions once we have passed to the classical limit, at least through NNLO. At this order there is non-zero momentum
radiated, so momentum conservation is non-trivial from the classical point of view. We will see by explicit calculation in QED that our classical impulse correctly incorporates the impulse from the ALD force in addition to the Lorentz force.

\subsection{Perspectives from classical field theory}
\def\position{x}

Before jumping into examples, it is useful to reflect on the total radiated momentum, expressed in terms of amplitudes, by digressing into classical field theory. To do so we must classically describe the distribution and flux of energy and momentum in the radiation field itself. Although our final conclusions also hold in YM theory and gravity, let us work in electrodynamics for simplicity. Here the relevant stress-energy tensor is
\begin{equation}
T^{\mu\nu}(x) = F^{\mu\alpha}(x) F_\alpha{}^\nu(x) + \frac 14 \eta^{\mu\nu} F^{\alpha\beta}(x) F_{\alpha\beta}(x) \,.\label{eqn:EMfieldStrength}
\end{equation}
In particular, the (four-)momentum flux through a three dimensional surface $\partial \Omega$ with surface element $\d\Sigma_\nu$ is
\begin{equation}
K^\mu = \int_{\partial \Omega}\!\! \d \Sigma_\nu T^{\mu\nu}(x)\,.
\end{equation}
We are interested in the total momentum radiated as two particles scatter. At each time $t$, we therefore surround the two particles with a large sphere. The instantaneous flux of momentum is measured by integrating over the surface area of the sphere; the total momentum radiated is then the integral of this instantaneous flux over all times. It is straightforward to determine the momentum radiated by direct integration over these spheres using textbook methods --- see appendix D of \cite{Kosower:2018adc}.

A simpler but more indirect method is the following. We wish to use the Gauss theorem to write
\begin{equation}
K^\mu = \int_{\partial \Omega}\!\! \d \Sigma_\nu T^{\mu\nu}(x) =  \int \! \d^4x \, \partial_\nu T^{\mu\nu}(x)\,.
\end{equation}
However, the spheres surrounding our particle are not the boundary of all spacetime: they do not include the timelike future and past boundaries. To remedy this, we use a trick due to Dirac~\cite{Dirac:1938nz}. 

The radiation we have in mind is causal, so we solve the Maxwell equation with retarded boundary conditions. We denote these fields by $F^{\mu\nu}_\textrm{ret}(x)$.
We could equivalently solve the Maxwell equation using the advanced Green's function. If we wish to determine precisely the same fields $F^{\mu\nu}_\textrm{ret}(x)$ but using the advanced Green's function, we must add a homogeneous solution of the Maxwell equation. Fitting the boundary conditions in this way requires subtracting the incoming radiation field $F^{\mu\nu}_\textrm{in}(x)$ which is present in the advanced solution (but not in the retarded solution) and adding the outgoing radiation field (which is present in the retarded solution, but not the advanced solution.) In other words,
\begin{equation}
F^{\mu\nu}_\textrm{ret}(x) - F^{\mu\nu}_\textrm{adv}(x) = - F^{\mu\nu}_\textrm{in}(x) + F^{\mu\nu}_\textrm{out}(x)\,.
\end{equation}
Now, the radiated momentum $K^\mu$ in which we are interested is described by $F^{\mu\nu}_\textrm{out}(x)$. The field $F^{\mu\nu}_\textrm{in}(x)$ transports the same total amount of momentum in from infinity, ie it transports momentum $-K^\mu$ out. Therefore the difference between the momenta transported out to infinity by the retarded and by the advanced fields is simply $2 K^\mu$. This is useful, because the contributions of the point-particle sources cancel in this difference.

The relationship between the momentum transported by the retarded and advanced field is reflected at the level of the Green's functions themselves. 
The difference in the Green's function takes an instructive form:%
\begin{equation}
\begin{aligned}
\tilde G_\textrm{ret}(\kb) - \tilde G_\textrm{adv}(\kb) &= 
\frac{(-1)}{(\kb^0 + i \epsilon)^2 - \v{\kb}^2} 
- \frac{(-1)}{(\kb^0 - i \epsilon)^2 - \v{\kb}^2} 
\\&=  i     \left( \Theta(\kb^0) - \Theta(-\kb^0) \right) \del(\kb^2)\,.
\end{aligned}
\end{equation}
In this equation, $\v{\kb}$ denotes the spatial components of wavenumber four-vector $\kb$. This difference is a homogeneous solution of the wave equation since it is supported 
on $\kb^2 = 0$. The two terms correspond to positive and negative angular frequencies. As we will see, the relative sign ensures that the momenta transported to infinity add.

With this in mind, we return to the problem of computing the momentum radiated and write
\begin{equation}
2 K^\mu = \int_{\partial \Omega} \!\!\d \Sigma_\nu \Big(T^{\mu\nu}_\textrm{ret}(x) -T^{\mu\nu}_\textrm{adv}(x) \Big)\,.
\end{equation}
In this difference, the contribution of the sources at timelike infinity cancel, so we may regard the surface $\partial \Omega$ as the boundary of spacetime. Therefore,
\begin{equation}
2K^\mu = \int \! d^4 x \, \partial_\nu \!\left(T^{\mu\nu}_\textrm{ret}(x) -T^{\mu\nu}_\textrm{adv}(x) \right) =- \int \! d^4 x \left( F^{\mu\nu}_\textrm{ret}(x) - F^{\mu\nu}_\textrm{adv}(x)\right) J_\nu(x)\,,
\end{equation}
where the last equality follows from the equations of motion. We now pass to momentum space, noting that
\begin{equation}
F^{\mu\nu}(x) = -i\! \int \! \dd^4 \bar k \left( \bar k^\mu \tilde A^\nu(\bar k) - \bar k^\nu \tilde A^\mu(\bar k) \right) e^{-i \bar k \cdot x}\,.
\end{equation}
Using conservation of momentum, the radiated momentum becomes
\begin{equation}
\begin{aligned}
2K^\mu 
&= i\! \int \! \dd^4 \bar k \; \bar k^\mu \left( \tilde A^\nu_\textrm{ret}(\bar k) - \tilde A^\nu_\textrm{adv}(\bar k) \right) \tilde J_\nu^*(\bar k), \\
&= -\int \! \dd^4 \bar k \; \bar k^\mu \left(\Theta(\bar k^0) - \Theta(-\bar k^0)\right) \del(\bar k^2) \tilde J^\nu(\bar k) \tilde J_\nu^*(\bar k)\,.
\label{eqn:momentumMixed}
\end{aligned}
\end{equation}
The two different $\Theta$ functions arise from the outgoing and incoming radiation fields. Setting $k'^\mu = - k^\mu$ in the second term, and then dropping the prime, it is easy to see that the two terms add as anticipated. We arrive at a simple general result for the momentum radiated:
\begin{equation}
\begin{aligned}
K^\mu &= -\int \! \dd^4 \wn k \,  \Theta(\wn k^0)\del(\wn k^2) \, \wn k^\mu \, \tilde J^\nu(\wn k) \tilde J_\nu^*(\wn k) \\
&= -\int \! \df( \wn k) \, \bar k^\mu \, \tilde J^\nu(\bar k) \tilde J_\nu^*(\bar k) \,.
\label{eqn:classicalMomentumRadiated}
\end{aligned}
\end{equation}
It is now worth pausing to compare this general classical formula for the radiated momentum to the expression we derived previously in \eqn~\eqref{eqn:radiatedMomentumClassical}. Evidently the radiation kernel we defined in \eqn~\eqref{eqn:defOfR} is related to the classical current $\tilde J^\mu(\kb)$. This fact was anticipated in ref.~\cite{Luna:2017dtq}. Indeed, if we introduce a basis of polarisation vectors $\varepsilon^{h}_\mu(\kb)$ associated with the wavevector $\kb$ with helicity $h$, we may write the classical momentum radiated as
\begin{equation}
K^\mu = \sum_h \int \! \df(\kb) \, \kb^\mu \, 
\left| \varepsilon^{h} \cdot \tilde J(\kb) \right|^2\,,
\label{eqn:classicalMomentumRadiated1}
\end{equation}
where here we have written the sum over helicities explicitly. Similar expressions hold in classical YM theory and gravity \cite{Goldberger:2016iau}.

\section{Examples}
\label{sec:LOradiation}
\def\pol{\varepsilon}

At leading-order the amplitude appearing in the radiation kernel in equation~\eqref{eqn:defOfRLO} is a five-point, tree amplitude (figure~\ref{fig:5points}) that can be readily computed. In Yang--Mills theory,
\newcommand{\treeLa}{\begin{tikzpicture}[thick, baseline={([yshift=-\the\dimexpr\fontdimen22\textfont2\relax] current bounding box.center)}, decoration={markings,mark=at position 0.6 with {\arrow{Stealth}}}]
	\begin{feynman}
	\vertex (v1);
	\vertex [below = 0.3 of v1] (v2);
	\vertex [above right = .1 and .25 of v1] (o1);
	\vertex [above left = .1 and .25 of v1] (i1);
	\vertex [below right = .1 and .25 of v2] (o2);
	\vertex [below left = .1 and .25 of v2] (i2);
	\vertex [above right = .05 and .125 of v1] (g1);
	\vertex [below right = 0.125 and 0.125 of g1] (g2);
	\draw (i1) -- (v1);
	\draw (v1) -- (g1);
	\draw (g1) -- (o1);
	\draw (i2) -- (v2);
	\draw (v2) -- (o2);
	\draw (g1) -- (g2);
	\draw (v1) -- (v2);
	\end{feynman}	
	\end{tikzpicture}}
\newcommand{\treeLb}{\begin{tikzpicture}[thick, baseline={([yshift=-\the\dimexpr\fontdimen22\textfont2\relax] current bounding box.center)}, decoration={markings,mark=at position 0.6 with {\arrow{Stealth}}}]
	\begin{feynman}
	\vertex (v1);
	\vertex [below = 0.3 of v1] (v2);
	\vertex [above right = .1 and .25 of v1] (o1);
	\vertex [above left = .1 and .25 of v1] (i1);
	\vertex [below right = .1 and .25 of v2] (o2);
	\vertex [below left = .1 and .25 of v2] (i2);
	\vertex [above left = .05 and .125 of v1] (g1);
	\vertex [below left = 0.125 and 0.125 of g1] (g2);
	\draw (i1) -- (g1);
	\draw (g1) -- (v1);
	\draw (v1) -- (o1);
	\draw (i2) -- (v2);
	\draw (v2) -- (o2);
	\draw (g1) -- (g2);
	\draw (v1) -- (v2);
	\end{feynman}	
	\end{tikzpicture}}
\newcommand{\treeLc}{\begin{tikzpicture}[thick, baseline={([yshift=-\the\dimexpr\fontdimen22\textfont2\relax] current bounding box.center)}, decoration={markings,mark=at position 0.6 with {\arrow{Stealth}}}]
	\begin{feynman}
	\vertex (v1);
	\vertex [below = 0.3 of v1] (v2);
	\vertex [above right = .1 and .25 of v1] (o1);
	\vertex [above left = .1 and .25 of v1] (i1);
	\vertex [below right = .1 and .25 of v2] (o2);
	\vertex [below left = .1 and .25 of v2] (i2);
	\vertex [below right = 0.15 and 0.25 of v1] (g2);
	\draw (i1) -- (v1);
	\draw (v1) -- (o1);
	\draw (i2) -- (v2);
	\draw (v2) -- (o2);
	\draw (v1) -- (g2);
	\draw (v1) -- (v2);
	\end{feynman}	
	\end{tikzpicture}}
\newcommand{\treeYM}{\begin{tikzpicture}[thick, baseline={([yshift=-\the\dimexpr\fontdimen22\textfont2\relax] current bounding box.center)}, decoration={markings,mark=at position 0.6 with {\arrow{Stealth}}}]
	\begin{feynman}
	\vertex (v1);
	\vertex [below = 0.3 of v1] (v2);
	\vertex [above right = .1 and .25 of v1] (o1);
	\vertex [above left = .1 and .25 of v1] (i1);
	\vertex [below right = .1 and .25 of v2] (o2);
	\vertex [below left = .1 and .25 of v2] (i2);
	\vertex [below = .15 of v1] (g1);
	\vertex [right = 0.275 of g1] (g2);
	\draw (i1) -- (v1);
	\draw (v1) -- (o1);
	\draw (i2) -- (v2);
	\draw (v2) -- (o2);
	\draw (g1) -- (g2);
	\draw (v1) -- (v2);
	\end{feynman}	
	\end{tikzpicture}}
\[
\bar{\mathcal{A}}^{(0)}(\wn k^a) &=
\sum_D \colStructure^a(D) \bar{A}^{(0)}_D(p_1 + w_1, p_2 + w_2\rightarrow p_1, p_2; k, h) \\
&= \Big[\colStructure^a\!\left(\treeLa \right)\!A_{\scalebox{0.5}{\treeLa}} 
+ \colStructure^a\!\left(\treeLb \right)\! A_{\scalebox{0.5}{\treeLb}} 
\\ & \qquad\qquad + \colStructure^a\!\left(\treeLc \right)\! A_{\scalebox{0.5}{\treeLc}}
+ (1\leftrightarrow 2) \Big]
- i\,\colStructure^a\!\left(\treeYM \right)\! A_{\scalebox{0.5}{\treeYM}}\,.
\]
Explicitly, the colour factors are given by
\begin{equation}
\begin{gathered}
\colStructure^a\!\left(\treeLa \right) = (\newT_1^a \cdot \newT_1^b) \newT_2^b\,, \qquad
\colStructure^a\!\left(\treeLb \right) = (\newT_1^b \cdot \newT_1^a) \newT_2^b\,,\\
\colStructure^a\!\left(\treeLc \right) = \frac12\colStructure^a\!\left(\treeLa \right) + \frac12\colStructure^a\!\left(\treeLb \right), \qquad
\colStructure^a\!\left(\treeYM \right) = i\hbar f^{abc} \newT_1^b \newT_2^c\,,\label{eqn:radColourFactors}
\end{gathered}
\end{equation}
with the replacement $1\leftrightarrow2$ for diagrams with gluon emission from particle 2. Just as in the 4-point case at  1-loop, this is an overcomplete set for specifying a basis, because 
\begin{equation}
\begin{aligned}
\colStructure^a\!\left(\treeLb \right) = (\newT_1^a\cdot \newT_1^b) \newT_2^b +  i\hbar f^{bac} \newT_1^c \newT_2^b = \colStructure^a\!\left(\treeLa \right) +  \colStructure^a\!\left(\treeYM \right).\label{eqn:JacobiSetUp}
\end{aligned}
\end{equation}
\begin{figure}[t]
	\centering
	\begin{tikzpicture}[decoration={markings,mark=at position 0.6 with {\arrow{Stealth}}}]
	\begin{feynman}
	\vertex (v1);
	\vertex [above left=1 and 0.66 of v1] (i1) {$\initialk_1+\xfer_1$};
	\vertex [above right=1 and 0.8 of v1] (o1) {$\initialk_1$};
	\vertex [right=1.2 of v1] (k) {$k$};
	\vertex [below left=1 and 0.66 of v1] (i2) {$\initialk_2+\xfer_2$};
	\vertex [below right=1 and 0.8 of v1] (o2) {$\initialk_2$};
	\draw [postaction={decorate}] (i1) -- (v1);
	\draw [postaction={decorate}] (v1) -- (o1);
	\draw [postaction={decorate}] (i2) -- (v1);
	\draw [postaction={decorate}] (v1) -- (o2);
	\diagram*{(v1) -- [photon] (k)};
	\filldraw [color=white] (v1) circle [radius=10pt];
	\draw [pattern=north west lines, pattern color=patternBlue] (v1) circle [radius=10pt];
	\end{feynman}	
	\end{tikzpicture} 
	\caption[The amplitude appearing in the leading-order radiation kernel.]{The amplitude $\Ampl^{(0)}(\initialk_1+\xfer_1\,,\initialk_2+\xfer_2\rightarrow
		\initialk_1\,,\initialk_2\,,k)$ appearing in the radiation kernel at leading order.}
	\label{fig:5points}
\end{figure}
Hence the full basis of colour factors is only 3 dimensional, and the colour decomposition of the 5-point tree is
\begin{multline}
\bar{\mathcal{A}}^{(0)}(\wn k^a) = \colStructure^a\!\left(\treeLa \right)\Big(A_{\scalebox{0.5}{\treeLa}} + A_{\scalebox{0.5}{\treeLb}} + A_{\scalebox{0.5}{\treeLc}}\Big) \\ 
+ \frac12\colStructure^a\!\left(\treeYM \right)\Big(-iA_{\scalebox{0.5}{\treeYM}} + 2 A_{\scalebox{0.5}{\treeLb}} + A_{\scalebox{0.5}{\treeLc}}\Big) + (1\leftrightarrow2)\,.
\end{multline}
Given that the second structure is $\mathcal{O}(\hbar)$, it would appear that we could again neglect the second term as a quantum correction. However, this intuition is not quite correct, as calculating the associated partial amplitude shows:
\begin{multline}
-iA_{\scalebox{0.5}{\treeYM}} + 2  A_{\scalebox{0.5}{\treeLb}} + A_{\scalebox{0.5}{\treeLc}} = -\frac{4\,\varepsilon^{h}_\mu(\wn k)}{\hbar^2} \bigg[\frac{2p_1\cdot p_2}{{\wn w_2^2\, p_1\cdot\wn k}}\, \frac{p_1^\mu}{\hbar} + \frac{1}{\hbar\, \wn w_1^2 \wn w_2^2}\Big(2 p_2\cdot\wn k\, p_1^\mu \\- p_1\cdot p_2 \,(\wn w_1^\mu - \wn w_2^\mu) - 2p_1\cdot\wn k\, p_2^\mu\Big) + \mathcal{O}(\hbar^0) \bigg]\,,\label{eqn:radSingular}
\end{multline}
where we have used $p_1\cdot\wn w_2 = p_1\cdot\wn k + \hbar\wn w_1^2/2$ on the support of the on-shell delta functions in the radiation kernel~\eqref{eqn:defOfRLO}. The partial amplitude appears to be singular, as there is an extra power of $\hbar$ downstairs. However, this will cancel against the extra power in the colour structure, yielding a classical contribution. Meanwhile in the other partial amplitude the potentially singular terms cancel trivially, and the contribution is classical:
\begin{multline}
A_{\scalebox{0.5}{\treeLa}} + A_{\scalebox{0.5}{\treeLb}} + A_{\scalebox{0.5}{\treeLc}} = \frac{2}{\hbar^2} \frac{\varepsilon^{h}_\mu(\wn k)}{\wn w_2^2\, p_1\cdot\wn k}\bigg[ 2p_1\cdot p_2\,\wn w_2^\mu +  \frac{p_1\cdot p_2}{p_1\cdot\wn k} p_1^\mu(\wn w_1^2 - \wn w_2^2) \\ - 2p_1\cdot\wn k\, p_2^\mu + 2p_2\cdot \wn k\, p_1^\mu + \mathcal{O}(\hbar)\bigg]\,.
\end{multline}
Summing all colour factors and partial amplitudes, the classical part of the 5-point amplitude is
\begin{align}
&\bar{\mathcal{A}}^{(0)}(\wn k^a) = \sum_D \colStructure^a(D) \bar{A}^{(0)}_D(p_1 + w_1, p_2 + w_2\rightarrow p_1, p_2; k, h) \nonumber\\
&= - \frac{4\varepsilon_\mu^h(\wn k)}{\hbar^2} \bigg\{ \frac{\newT_1^a (\newT_1\cdot \newT_2)}{\wn w_2^2 \, \wn k \cdot p_1} \bigg[-(p_1\cdot p_2)\left(\wn w_2^\mu - \frac{\wn k\cdot\wn w_2}{\wn k\cdot p_1} p_1^\mu\right) + \wn k\cdot p_1 \, p_2^\mu  - \wn k\cdot p_2\, p_1^\mu\bigg] \nonumber \\ &\qquad + \frac{if^{abc}\,\newT_1^b \newT_2^c}{\wn w_1^2 \wn w_2^2}\bigg[2\wn k\cdot p_2\, p_1^\mu 
 - p_1\cdot p_2\, \wn w_1^\mu + p_1\cdot p_2 \frac{\wn w_1^2}{\wn k\cdot p_1}p_1^\mu\bigg] + (1\leftrightarrow 2)\bigg\}\,,
\end{align}
where we have used that $\wn w_1^2 - \wn w_2^2 = -2\wn k\cdot \wn w_2$ since the outgoing radiation is on-shell. Finally, we can substitute into the radiation kernel in \eqn~\eqref{eqn:defOfRLO} and take the classical limit. Averaging over the wavepackets sets $p_\alpha = m_\alpha u_\alpha$ and replaces quantum colour charges with their classical counterparts, yielding
\[
&\mathcal{R}^{a,(0)}_\text{YM}(\wn k) = -\int\!\dd^4\,\wn w_1 \dd^4 \wn w_2 \, \del^{(4)}(\wn k - \wn w_1 - \wn w_2) \del(u_1\cdot\wn w_1) \del(u_2\cdot\wn w_2)\, e^{ib\cdot\wn w_1} \varepsilon_\mu^{h}\\ 
&\times \bigg\{ \frac{c_1\cdot c_2}{m_1} \frac{c_1^a}{\wn w_2^2 \, \wn k \cdot u_1} \left[-(u_1\cdot u_2)\left(\wn w_2^\mu - \frac{\wn k\cdot\wn w_2}{\wn k\cdot u_1} u_1^\mu\right) + \wn k\cdot u_1 \, u_2^\mu - \wn k\cdot u_2\, u_1^\mu\right]\\
& \qquad + \frac{if^{abc}\,c_1^b c_2^c}{\wn w_1^2 \wn w_2^2}\left[2\wn k\cdot u_2\, u_1^\mu - u_1\cdot u_2\, \wn w_1^\mu + u_1\cdot u_2 \frac{\wn w_1^2}{\wn k\cdot u_1}\,u_1^\mu\right] + (1\leftrightarrow 2)\bigg\}\,.\label{eqn:LOradKernel}
\]
Our result is equal to the leading order current $\tilde{J}^{\mu,(0)}_a$ obtained in 
\cite{Goldberger:2016iau} by iteratively solving the Wong equations in \eqn~\eqref{eqn:Wong-momentum} and \eqn~\eqref{eqn:Wong-color} for timelike particle worldlines.

\subsection{Inelastic black hole scattering}
\label{sec:inelasticBHscatter}

Let us turn to an independent application of our LO Yang--Mills radiation kernel~\eqref{eqn:LOradKernel}. By returning to the colour-kinematics structure of the underlying amplitude, we can readily use the double copy to calculate results for gravitational wave emission from black hole scattering.

To apply the double copy we need the overcomplete set of colour factors in equation~\eqref{eqn:radColourFactors}. This is because the object of fundamental interest is now the Jacobi identity that follows from~\eqref{eqn:JacobiSetUp},
\begin{equation}
\colStructure^a\!\left(\treeLb \right) - \colStructure^a\!\left(\treeLa \right) = \colStructure^a\!\left(\treeYM \right),
\end{equation}
with an identical identity holding upon exchanging particles 1 and 2. Unlike our example in section~\ref{sec:LOimpulse}, this is a non-trivial relation, and we must manipulate the numerators of each topology into a colour-kinematics dual form. This can be readily achieved by splitting the topologies with four-point seagull vertices, and adding their kinematic information to diagrams with the same colour structure. It is simple to verify that, in the classical limit, a basis of colour-kinematics dual numerators for this amplitude is
\begin{align}
\sqrt{2}\, n_{\scalebox{0.5}{\treeLb}} &= 4(p_1\cdot p_2)\, p_1\cdot\varepsilon^{h}_k + 2\hbar\Big(p_1\cdot \wn k\, (p_1 + p_2)^\mu + p_1\cdot p_2\, (\wn w_1 - \wn w_2)^\mu\Big)\cdot\varepsilon^h_k + \mathcal{O}(\hbar^2)\nonumber\\
\hspace{-7mm}\sqrt{2}\, n_{\scalebox{0.5}{\treeLa}} &= 4(p_1\cdot p_2)\, p_1\cdot\varepsilon^{h}_k + 2\hbar\Big(p_1\cdot \wn k \, (p_1 - p_2)^\mu + 2 p_2\cdot\wn k\, p_1^\mu\Big)\cdot\varepsilon^h_k + \mathcal{O}(\hbar^2)\label{eqn:vecnums}\\
\sqrt{2}\, n_{\scalebox{0.5}{\treeYM}} &= 2\hbar\Big(2 p_1\cdot \wn k\, p_2^\mu - 2p_2\cdot\wn k\, p_1^\mu + p_1\cdot p_2\, (\wn w_1 - \wn w_2)^\mu\Big)\cdot\varepsilon^h_k +\mathcal{O}(\hbar^2)\,, \nonumber
\end{align}
where $\varepsilon^h_k \equiv \varepsilon^h_\mu(\wn k)$. It is crucial that we keep $\mathcal{O}(\hbar)$ terms, as we know from equation~\eqref{eqn:radSingular} that when the YM amplitude is not written on a minimal basis of colour factors there are terms which are apparently singular in the classical limit. The factors of $\sqrt{2}$ are to account for the proper normalisation of colour factors involved in the double copy --- see discussion around equation~\eqref{eqn:scalarYMamp}.

With a set of colour-kinematics dual numerators at hand, we can now double copy by replacing colour factors with these numerators, leading to
\begin{multline}
\hbar^{\frac{7}{2}}  {\mathcal{M}}^{(0)}_\textrm{JNW}(\wn k) = \left(\frac{\kappa}{2}\right)^3 \bigg[ \frac{1}{\hbar\,\wn w_2^2} \bigg(\frac{n^\mu_{\scalebox{0.5}{\treeLa}} n^\nu_{\scalebox{0.5}{\treeLa}}}{2p_1\cdot \wn k} - \frac{n^\mu_{\scalebox{0.5}{\treeLb}}n^\nu_{\scalebox{0.5}{\treeLb}}}{2p_1\cdot \wn k + \hbar \wn w_1^2 - \hbar \wn w_2^2}\bigg) + (1\leftrightarrow 2)\\  + \frac{1}{\hbar^2 \wn w_1^2 \wn w_2^2}n^\mu_{\scalebox{0.5}{\treeYM}} n^\nu_{\scalebox{0.5}{\treeYM}} \bigg] e_{\mu\nu}^h(\wn k)\,,\label{eqn:JNWamplitude}
\end{multline}
where we have used the outer product of the polarisation vectors (of momentum $\wn k$) from the numerators,
\begin{equation}
e_{\mu\nu}^{h} = \frac12\left(\varepsilon^{h}_\mu \varepsilon^{h}_\nu + \varepsilon^{h}_\nu \varepsilon^{h}_\mu - P_{\mu\nu}\right) + \frac12\left(\varepsilon^{h}_\mu \varepsilon^{h}_\nu - \varepsilon^{h}_\nu \varepsilon^{h}_\mu\right) + \frac12 P_{\mu\nu}\,.\label{eqn:polarisationOuterProd}
\end{equation}
Here $P_{\mu\nu} = \eta_{\mu\nu} - \left(\wn k_\mu \wn r_\nu + \wn k_\nu \wn r_\mu\right)/(\wn k\cdot \wn r)$ is a transverse projector with reference momentum $\wn r$. Since the amplitude is symmetric in its numerators, we can immediately restrict attention to the initial symmetric and traceless piece --- the polarisation tensor for a graviton. It now simply remains to Laurent expand in $\hbar$ to retrieve the parts of the amplitude which contribute to the classical radiation kernel for gravitational radiation. 

Rather than doing so at this point, it is more pertinent to note that our result is not yet an amplitude in Einstein gravity. As in our 4-point discussion in the previous chapter, this result is polluted by dilaton interactions. In particular, the corresponding current is for the scattering of JNW naked singularities in Einstein--dilaton gravity \cite{Goldberger:2016iau}. A convenient way to remove the dilaton states is to use a scalar ghost \cite{Johansson:2014zca}, as shown explicitly for the classical limit of this amplitude in ref.~\cite{Luna:2017dtq}. We introduce a new massles, adjoint-representation scalar $\chi$, minimally coupled to the YM gauge field, and use the double copy of its amplitude to remove the pollution. The ghost couples to the scalar fields in our action~\eqref{eqn:scalarAction} via the interaction term
\begin{equation}
\mathcal{L}_{\chi \textrm{int}} = -2g \sum_{\alpha=1,2} \Phi_\alpha^\dagger \chi \Phi_\alpha\,.\label{eqn:adjointScalar}
\end{equation}
On the same 5-point kinematics as our previous YM amplitude, the equivalent numerators for the topologies in~\eqref{eqn:vecnums} with interactions mediated by the new massless scalar are
\[
\sqrt{2}\, \tilde n_{\scalebox{0.5}{\treeLb}} &= -4 (p_1 - \hbar\wn w_2) \cdot\varepsilon^h_k \\
\sqrt{2}\, \tilde n_{\scalebox{0.5}{\treeLa}} &= -4 p_1 \cdot\varepsilon^h_k \\
\sqrt{2}\, \tilde n_{\scalebox{0.5}{\treeYM}} &= -2\hbar (\wn w_1 - \wn w_2)\cdot\varepsilon^h_k\,. \label{eqn:scalarnums}
\]
Since $\chi$ is in the adjoint representation, the appropriate colour factors are again those in equation~\eqref{eqn:radColourFactors} (there is now no seagull topology). The numerators hence trivially satisfy colour-kinematics duality. We can therefore double copy to yield the ghost amplitude,
\begin{multline}
\hbar^{\frac72} {\mathcal{M}}^{(0)}_\textrm{ghost}(\wn k) = \left(\frac{\kappa}{2}\right)^3 \bigg[\frac1{\hbar\,\wn w_2^2}\bigg(\frac{\tilde n^\mu_{\scalebox{0.5}{\treeLa}} \tilde n^\nu_{\scalebox{0.5}{\treeLa}}}{2p_1\cdot \wn k} - \frac{\tilde n^\mu_{\scalebox{0.5}{\treeLb}} \tilde n^\nu_{\scalebox{0.5}{\treeLb}}}{2p_1\cdot \wn k + \hbar \wn w_1^2 - \hbar\wn w_2^2}\bigg) + (1\leftrightarrow 2)\\  + \frac{1}{\hbar^2 \wn w_1^2 \wn w_2^2} \tilde n^\mu_{\scalebox{0.5}{\treeYM}} \tilde n^\nu_{\scalebox{0.5}{\treeYM}}\bigg] e^h_{\mu\nu}(\wn k)\,,\label{eqn:ghostAmp}
\end{multline}
This is an amplitude for a ghost in the sense that we can now write \cite{Luna:2017dtq}
\begin{equation}
\mathcal{M}^{(0)}_\textrm{Schwz}(\wn k) = \mathcal{M}^{(0)}_\textrm{JNW} - \frac1{D-2}\mathcal{M}^{(0)}_\textrm{ghost}\,,
\end{equation}
where the appearance of the spacetime dimension comes from matching the ghost to the dilaton propagator. Expanding the numerators, in $D=4$ one finds that
\begin{multline}
\hbar^2 \bar{\mathcal{M}}^{(0)}_\textrm{Schwz}(\wn k) =   \bigg[\frac{{P}_{12}^\mu {P}_{12}^\nu}{\wn w_1^2 \wn w_2^2} + \frac{p_1\cdot p_2}{2\wn w_1^2 \wn w_2^2}\left({Q}^\mu_{12} {P}_{12}^\nu + {Q}_{12}^\nu {P}_{12}^\mu\right) \\ + \frac14\left((p_1\cdot p_2)^2 - \frac12\right)\left(\frac{{Q}^\mu_{12} {Q}^\nu_{12}}{\wn w_1^2 \wn w_2^2} - \frac{{P}^\mu_{12} {P}^\nu_{12}}{(\wn k\cdot p_1)^2 (\wn k\cdot p_2)^2}\right)\bigg]e_{\mu\nu}^{h} + \mathcal{O}(\hbar)\,,\label{eqn:ampSchwarzschild}
\end{multline}
where
\begin{subequations}
	\begin{gather}
	P_{12}^\mu = (\wn k\cdot p_1)\, p_2^\mu - (\wn k\cdot p_2)\, p_1^\mu\,,\label{eqn:gaugeinvariant1}\\
	Q_{12}^\mu = (\wn w_1 - \wn w_2)^\mu - \frac{\wn w_1^2}{\wn k\cdot p_1} p_1^\mu + \frac{\wn w_2^2}{\wn k\cdot p_2} p_2^\mu\,,\label{eqn:gaugeinvariant2}
	\end{gather}\label{eqn:gaugeInvariants}
\end{subequations}
are two gauge invariant functions of the kinematics. Substituting the amplitude into the LO radiation kernel~\eqref{eqn:defOfRLO} yields the LO current for the scattering of two Schwarzschild black holes. Note that whereas we took the classical limit before double copying, this result was first obtained in ref.~\cite{Luna:2017dtq} by only taking the classical limit (via a large mass expansion) once the gravity amplitudes were at hand.

\subsubsection{Reissner--Nordstr\"{o}m black holes}

In our previous calculation, the adjoint massless scalar introduced in equation~\eqref{eqn:adjointScalar} merely acted as a useful computational trick to remove internal dilaton pollution. However, if we consider other black hole species it can be promoted to a far more fundamental role. For example, let us consider gravitational radiation emitted by the scattering of two Reissner--Nordstr\"{o}m black holes. RN black holes are solutions to Einstein--Maxwell theory rather than vacuum general relativity, and thus have non-zero electric (or magnetic) charges $Q_\alpha$. At leading order their gravitational interactions are the same as for Schwarzschild black holes, but the total current due to gravitational radiation will be different, precisely because of terms sourced from electromagnetic interactions, mediated by a massless vector field.

This is in contrast to our previous example, where we ``squared'' the vector numerators to obtain tensor interactions, in the sense that $A^\mu \otimes A^\mu \sim H_{\mu\nu}$. To obtain electromagnetic interactions in the gravity amplitude guaranteed by the double copy we need to use numerators in a vector and scalar representation respectively, such that we have $A^\mu \otimes \phi \sim \tilde{A}^\mu$. This is exactly what the sets of numerators in equations~\eqref{eqn:vecnums} and~\eqref{eqn:scalarnums} provide.

The double copy does not require that one square numerators, merely that colour data is replaced with numerators satisfying the same Jacobi identities. Thus the double copy construction
\begin{multline}
\hbar^{\frac72}  {\mathcal{M}}^{(0)}(\wn k) = \frac{\kappa}{2} e^2 Q_1 Q_2\bigg[ \frac{1}{\hbar\,\wn w_2^2} \bigg(\frac{n^\mu_{\scalebox{0.5}{\treeLa}} \tilde n^\nu_{\scalebox{0.5}{\treeLa}}}{2p_1\cdot \wn k} - \frac{n^\mu_{\scalebox{0.5}{\treeLb}}\tilde n^\nu_{\scalebox{0.5}{\treeLb}}}{2p_1\cdot \wn k + \hbar \wn w_1^2 - \hbar \wn w_2^2}\bigg) + (1\leftrightarrow 2)\\  + \frac{1}{\hbar^2 \wn w_1^2 \wn w_2^2}n^\mu_{\scalebox{0.5}{\treeYM}} \tilde n^\nu_{\scalebox{0.5}{\treeYM}} \bigg] e_{\mu\nu}^h\label{eqn:RNamp}
\end{multline}
is guaranteed to be a well-defined gravity amplitude. Note that we have altered the coupling replacement in the double copy appropriately.  

A consequence of using kinematic numerators from alternative sets is that the amplitude is asymmetric in its Lorentz indices; specifically, for the scalar diagrams there are no seagull vertex terms, while the scalar boson triple vertex term is manifestly different to the pure vector case. Since the graviton polarisation tensor is symmetric and traceless, the graviton amplitude is obtained by symmetrising over the Lorentz indices in~\eqref{eqn:RNamp}. Substituting the result into equation~\eqref{eqn:defOfRLO} yields the LO gravitational radiation kernel due to electromagnetic interactions,
\begin{multline}
\mathcal{R}_{\rm RN,grav}^{(0)} = \frac{e^2\kappa}{4} Q_1 Q_2 \!\int\! \dd^4\wn w_1 \dd^4\wn w_2\, \del(p_1\cdot\wn w_1) \del(p_2\cdot\wn w_2) \del^{(4)}(\wn w_1 + \wn w_2 - \wn k) e^{ib\cdot\wn w_1} \\\times \bigg[\frac{{Q}_{12}^\mu {P}_{12}^\nu + {Q}_{12}^\nu {P}_{12}^\mu}{\wn w_1^2 \wn w_2^2} + (p_1\cdot p_2) \bigg(\frac{{Q}_{12}^\mu {Q}_{12}^\nu}{\wn w_1^2 \wn w_2^2} - \frac{{P}_{12}^\mu {P}_{12}^\nu}{(\wn k\cdot p_1)^2 (\wn k\cdot p_2)^2} \bigg)\bigg] e_{(\mu\nu)}^{h}\,.\label{eqn:RNradKernel}
\end{multline}
We verify this result in appendix~\ref{app:worldlines}, by calculating the corresponding classical energy-momentum tensor from perturbative solutions to the field equations of Einstein--Maxwell theory.

To obtain an amplitude for graviton emission we symmetrised the result from the double copy. This rather conveniently restricted to a graviton amplitude. However, because we used double copy numerators from different theories there is also a non-zero contribution from the antisymmetric part of the polarisation tensor, $e_{[\mu\nu]}^{h}$. For gravity, this corresponds to the axion mode $B_{\mu\nu}$, and thus $\mathcal{M}^{[\mu\nu]}$ should correspond to axion emission. It is a simple matter to show that
\begin{equation}
\hbar^{\frac52} \mathcal{M}^{(0)}_{\rm RN, axion} = -\frac{e^2\kappa}{\wn w_1^2 \wn w_2^2} Q_1 Q_2 \left({Q}^{\nu}_{12} {P}^\mu_{12} - {Q}^\mu_{12} {P}^\nu_{12}\right) e_{[\mu\nu]}^{h} + \mathcal{O}(\hbar)\,.\label{eqn:axion}
\end{equation}
As well as being antisymmetric, for an on-shell state with momentum $\wn k^\mu$ the axion polarisation tensor must satisfy $e^h_{[\mu\nu]}(\wn k) \wn k^\mu = 0$ . Thus it has the explicit form $e^h_{[\mu\nu]}(\wn k) = \epsilon_{\mu\nu\rho\sigma} \wn k^\rho \xi^\sigma/(\wn k \cdot \xi)$, where $\xi^\sigma$ is an unspecified reference vector. We will find it convenient to take $\xi^\sigma = p_1^\sigma + p_2^\sigma$. Expanding the amplitude and then including this choice we have
\[
\hbar^{\frac72} \bar{\mathcal{M}}^{(0)}_{\rm RN,axion} &= -\frac{2e^2\kappa}{\wn w_1^2 \wn w_2^2} Q_1 Q_2 \Big[\left((\wn k\cdot p_1)\, p_2^\mu  - (\wn k\cdot p_2)\, p_1^\mu\right)(\wn w_1 - \wn w_2)^\nu \\ & \qquad\qquad\qquad\qquad\qquad\qquad - \left(\wn w_1^2 - \wn w_2^2\right) p_2^\mu p_1^\nu \Big] \epsilon_{\mu\nu\rho\sigma} \frac{\wn k^\rho \xi^\sigma}{\wn k \cdot \xi}\\
&= \frac{4 e^2\kappa}{\wn w_1^2 \wn w_2^2}Q_1 Q_2\, \epsilon_{\mu\nu\rho\sigma} p_1^\mu p_2^\nu \wn w_1^\rho \wn w_2^\sigma\,,
\]
leading to a LO axion radiation kernel
\begin{multline}
\mathcal{R}_{\rm RN,axion}^{(0)} = e^2\kappa Q_1 Q_2 \! \int\! \dd^4\wn w_1 \dd^4\wn w_2\, \del(p_1\cdot\wn w_1) \del(p_2\cdot\wn w_2) \del^{(4)}(\wn w_1 + \wn w_2 - \wn k) \\ \times \frac{e^{ib\cdot\wn w_1}}{\wn w_1^2 \wn w_2^2} \epsilon_{\mu\nu\rho\sigma} u_1^\mu u_2^\nu \wn w_1^\rho \wn w_2^\sigma\,.
\end{multline}
The electromagnetic scattering is only able to radiate axions due to their coupling with the vector bosons. It is not possible to couple axions to a scalar particle in the absence of spin, as to do so breaks diffeomorphism invariance and axion gauge symmetry \cite{Goldberger:2017ogt}. The amplitude of \eqn~(\ref{eqn:axion}) is therefore purely due to the antisymmetrisation of the triple gauge vertex; given that we have an isolated single vertex responsible, we can verify that the radiation is indeed axionic by using Einstein--Maxwell axion-dilaton coupled gravity. The action is
\begin{equation}
\hspace{-3mm} S=\frac1{2\kappa}\int\! \d^4x \,{\sqrt{g}}\left[R - \frac{1}{2}(\partial_\mu\phi_\textrm{d})^2 - \frac{1}{2}e^{4\phi_\textrm{d}}(\partial_\mu\zeta)^2 -  e^{-2\phi_\textrm{d}}F^2\! - \zeta F^{\mu\nu}{F}^*_{\mu\nu}\right].
\end{equation}
Here ${F}^*_{\mu\nu}$ is the dual electromagnetic field strength, $\phi_\textrm{d}$ is the dilaton field, $\zeta$ is the axion pseudoscalar defined in \eqn~\eqref{eqn:axionscalar}, and $g=-\det(g_{\mu\nu})$. Treating the axion-photon interaction in the final term perturbatively and integrating by parts gives
\begin{equation}
\mathcal{L}_{\rm int} \sim A_\lambda\partial_{[\alpha}A_{\beta]}H^{\lambda\alpha\beta}\,.
\end{equation}
Allocating momenta to the axion-photon vertex in the same way as in the scattering amplitudes, this interaction term corresponds to the Feynman rule
\hspace{-0.5cm}
\begin{tabular}[h!]{cccc}
	\begin{minipage}[c]{0.03\textwidth}
	\end{minipage}
	\begin{minipage}[c]{0.18\textwidth}
		\centering
		\scalebox{0.8}{
		\begin{tikzpicture}[thick, baseline={([yshift=-\the\dimexpr\fontdimen22\textfont2\relax] current bounding box.center)}, decoration={zigzag}]
		\tikzset{zigzag/.style={decorate, decoration=zigzag}}
		\begin{feynman}
		\vertex (v1);
		\vertex [above = 1.44 of v1] (v2) {$\mu \nu$};
		\vertex [below right = 1 and 1.5 of v1] (v3);
		\vertex [below left = 1 and 1.5 of v1] (v4);
		\vertex [left = 0.06 of v1] (a1) ;
		\vertex [right = 0.06 of v1] (a2) ;
		\vertex [above = 1.44 of a1] (a3) ;
		\vertex [above = 1.44 of a2] (a4) ;
		\vertex [right = 0.2 of v1] (a5) ;
		\vertex [above = 1.44 of a5] (a6) ;
		\vertex [below right = 1 and 1.5 of v1] (v33) {$\sigma$};
		\vertex [below left = 1 and 1.5 of v1] (v44) {$\rho$};
		\draw[zigzag] (a1) -- (a3);
		\draw[zigzag] (a2) -- (a4);
		\diagram*{(a5) -- [white, scalar, momentum' = {[arrow style=black,thick]\(k\)}] (a6)};
		\diagram*{(v3) -- [photon, momentum = \(q_2\)] (v1)};
		\diagram*{(v4) -- [photon, momentum = \(q_1\)] (v1)};
		\filldraw [color=black] (v1) circle [radius=2pt];
		\end{feynman}
		\end{tikzpicture}}
	\end{minipage}&
	\hspace{-0.7cm}
	\begin{minipage}[c]{0.8\textwidth}
		\begin{equation}
		\begin{aligned}
		=-2i\kappa\,\big\{(q_2-q_1)_\mu  k_{[\rho} \eta_{\sigma]\nu} -& (q_2-q_1)_\nu k_{[\rho}\eta_{\sigma]\mu} \\ &+ \eta_{\mu[\rho}\eta_{\sigma]\nu} \left[q_1\cdot k - q_2\cdot k\right]\big\}.
		\end{aligned}
		\end{equation}
	\end{minipage}
	\begin{minipage}[c]{0.03\textwidth}
	\end{minipage}
	\vspace{6pt}
\end{tabular}
\noindent Constructing a five-point amplitude for massive two external scalars with this axion emission vertex and taking the classical limit then precisely reproduces the results of \eqn~\eqref{eqn:axion}, verifying that the antisymmetric double copy artefact is indeed axionic.

\subsection{Momentum conservation and radiation reaction}
\label{sec:ALD}

Let us return to gauge theory, and in particular the YM radiation kernel in equation~\eqref{eqn:LOradKernel}. An immediate corollary of this result is the LO radiation in classical electrodynamics; replacing colour with electric charges and ignoring the structure constant terms yields
\begin{equation}
\begin{aligned}
\RadKerCl_\text{EM}(\kb)&=\int \! \dd^4 \xferb_1 \dd^4 \xferb_2 \;
\del(\ucl_1\cdot\xferb_1) \del(\ucl_2\cdot\xferb_2) 
\del^{(4)}(\kb - \xferb_1 - \xferb_2) \, e^{i\xferb_1 \cdot b} 
\\& \hphantom{\rightarrow}\times\biggl\{
\frac{1}{m_1} \frac{Q_1^2Q_2^{\vphantom{2}}}{\xferb_2^2} 
\biggl[-\ucl_2\cdot\pol^h_k + \frac{(\ucl_1\cdot \ucl_2)(\xferb_2\cdot\pol^h_k)}{\ucl_1\cdot\kb} 
+ \frac{(\ucl_2\cdot\kb)(\ucl_1\cdot\pol^h_k)}{\ucl_1\cdot\kb} 
\\&\hspace*{40mm} 
- \frac{(\kb\cdot\xferb_2)(\ucl_1\cdot \ucl_2)(\ucl_1\cdot\pol^h_k)}{(\ucl_1\cdot\kb)^2}\bigg] 
+ (1 \leftrightarrow 2)\biggr\} \,.
\label{eqn:Rcalculation}
\end{aligned}
\end{equation}
It is a simple calculation to see that the LO current which solves the Maxwell field equation has precisely the same expression, up to an overall sign~\cite{Kosower:2018adc}.

Now, we have already seen that conservation of momentum holds exactly (in \sect{sect:allOrderConservation}) and in our classical expressions (in \sect{sect:classicalConservation}). Let us ensure that there is no subtlety in these discussions by explicit calculation.

To do so, we calculate the part of the NNLO impulse $\ImpBclsup{(\textrm{rad})}$ which encodes radiation reaction, defined in \eqn~\eqref{eqn:nnloImpulse}. The two amplitudes appearing in equation~\eqref{eqn:nnloImpulse} are in common with the amplitudes relevant for the radiated momentum, equation~\eqref{eqn:defOfRLO}, though they are evaluated at slightly different kinematics. It will be convenient to change the sign of $\xferb_\alpha$ here; with that change, the amplitudes are
\begin{multline}
\AmplB^{(0)}(\initialk_1, \initialk_2 \rightarrow \initialk_1-\hbar\xferb_1\,, \initialk_2-\hbar\xferb_2 \, , \kb) = \frac{4Q_1^2 Q_2^{\vphantom{2}}}{\hbar^{2}\, \xferb_2^2} 
\bigg[-p_2\td\pol^h_k + \frac{(p_1\td p_2)(\xferb_2\td\pol^h_k)}{p_1\cdot\bar{k}} 
\\+ \frac{(p_2\cdot\kb)(p_1\td\pol^h_k)}{p_1\td\kb} 
- \frac{(\kb\td\xferb_2)(p_1\td p_2)(p_1\td\pol^h_k)}{(p_1\cdot\kb)^2}\bigg] + ( 1 \leftrightarrow 2)
\end{multline}
and
\begin{multline}
\AmplB^{(0)*}(\initialk_1+\hbar \qb_1\,, \initialk_2 + \hbar \qb_2 \rightarrow 
\initialk_1-\hbar\xferb_1\,, \initialk_2-\hbar\xferb_2 \,, \kb)
= \frac{4Q_1^2 Q_2^{\vphantom{2}}}{\hbar^{2}\, \xferb_2'^2} 
\bigg[\tm p_2\td\pol^{h*}_k
\\\tp \frac{(p_1\td p_2)(\xferb'_2\td\pol^{h*}_k)}{p_1\cdot\kb} 
\tp \frac{(p_2\td\kb)(p_1\td\pol_k^{h*})}{p_1\cdot\kb} 
\tm \frac{(\kb\td\xferb'_2)(p_1\td p_2)(p_1\td\pol^{h*}_k)}{(p_1\cdot\kb)^2}\bigg] 
 + ( 1 \leftrightarrow 2)\,,
\end{multline}
where we find it convenient to define $\xferb_\alpha' = \qb_\alpha + \xferb_\alpha$ (after the change of sign). 

We can now write the impulse contribution as
\begin{multline}
\ImpBclsup{(\textrm{rad})} = -e^6 \Lexp \int \! d\Phi(\kb) 
\prod_{\alpha = 1,2} \int \dd^4\xferb_\alpha\, \dd^4 \xferb'_\alpha \; \xferb_1^\mu \; 
\\ \times \mathcal{X}^*(\xferb'_1, \xferb'_2, \kb) \mathcal{X}(\xferb_1, \xferb_2, \kb)
\Rexp\,  ,
\label{eqn:impulseNNLO}
\end{multline}
where 
\begin{equation}
\begin{aligned}
\mathcal{X}(\xferb_1, \xferb_2, \kb) &= {4}  \, \del(2 \xferb_1\cdot p_1)
\del(2 \xferb_2\cdot p_2) \del^{(4)}(\bar k - \xferb_1 - \xferb_2) 
\, e^{i b \cdot  \xferb_1}
\\ 
& \quad \times 
\biggl\{Q_1^2 Q_2^{\vphantom{2}} \frac{\pol_{\mu}^h(\wn k) }{\xferb_2^2}
\bigg[-p_2^\mu + \frac{p_1\cdot p_2 \, \xferb_2^\mu}{p_1\cdot\kb} 
+ \frac{p_2\cdot\kb \, p_1^\mu}{p_1\cdot\kb} 
\\&\hspace*{35mm}
- \frac{(\kb\cdot\xferb_2)(p_1\cdot p_2) \, p_1^\mu}{(p_1\cdot\kb)^2}\bigg] 
+ (1 \leftrightarrow 2)\biggr\}\,.
\end{aligned}
\label{eqn:X1}  
\end{equation}
This expression is directly comparable to those for radiated momentum: \eqn~\eqref{eqn:impulseNNLO}, and the equivalent impulse contribution to particle 2, balance the radiated momentum \eqn~\eqref{eqn:radiatedMomentumClassicalLO} using $\xferb_1^\mu + \xferb_2^\mu = \bar k^\mu$, provided that the radiation kernel, \eqn~\eqref{eqn:Rcalculation}, is related to integrals over $\mathcal{X}$. Indeed this relationship holds: the integrations present in the radiation kernel are supplied by the $\xferb_\alpha$ and $\xferb'_\alpha$ integrals in \eqn~\eqref{eqn:impulseNNLO}; these integrations disentangle in the sum of impulses on particles 1 and 2 when we impose $\xferb_1^\mu + \xferb_2^\mu = \bar k^\mu$, and then form the square of the radiation kernel.

It is interesting to compare this radiated momentum with the situation in traditional formulations of classical physics, where one must include the ALD radiation reaction force 
by hand in order to enforce momentum conservation. Because the situation is simplest when only one particle is dynamical, let us take the mass $m_2$ to be very large compared to 
$m_1$ in the remainder of this section, and work in particle 2's rest frame.  In this frame, it does not radiate, and the only radiation reaction is on particle 1 --- the radiated momentum is precisely balanced by  the impulse on particle 1 due to the ALD force. We can therefore continue our discussion with reference to our expression for radiated momentum, \eqn~\eqref{eqn:radiatedMomentumClassicalLO}, and the radiation kernel, \eqn~\eqref{eqn:Rcalculation}. In this situation we may also simplify the kernels by dropping the $(1 \leftrightarrow 2)$ instruction: these terms will be dressed by an inverse power of $m_2$, and so are subdominant when $m_2 \gg m_1$. 

We will soon compute the impulse due to the ALD force directly from its classical expression in \eqn~\eqref{eqn:ALDclass}. But in preparation for that comparison there is one step which we must take. Classical expressions for the force---which involve only the particle's momentum and its derivatives---do not involve any photon phase space. So we must perform the integration over $\df(\wn k)$ which is present in \eqn~\eqref{eqn:radiatedMomentumClassicalLO}. 

To organise the calculation, we integrate over the $\qb_1$ variables in the radiation kernel, \eqn~\eqref{eqn:Rcalculation} using the four-fold delta function, so that we may write the radiated momentum as
\begin{align}
\hspace*{-3mm}\Rad^{\mu,(0)}_\class = -\frac{e^6 Q_1^4 Q_2^2}{m_1^2}\! \int\!\dd^4\qb\, \dd^4\qb'\;
e^{-ib\cdot(\qb - \qb')} \del(\ucl_1\cdot(\qb - \qb')) 
\frac{\del(\ucl_2\cdot\qb)}{\qb^2} \frac{\del(\ucl_2\cdot\qb')}{\qb'^2} \phInt\,,\label{eqn:rrmidstage}
\end{align}
where we renamed the remaining variables, $\xferb_2\rightarrow \qb$ and $\xferb'_2\rightarrow \qb'$, in order to match the notation used later. After some algebra we find
\begin{multline}
\phInt = \int \! \df (\kb) \, \del(\ucl_1\cdot \kb - \wn E) \, \kb^\mu \,
\left[ 1 + \frac{(\ucl_1\cdot \ucl_2)^2(\qb\cdot \qb')}{\wn E^2} 
+ \frac{(\ucl_2\cdot\kb)^2}{\wn E^2}  \right. \\ 
\left. - \frac{(\ucl_1\cdot \ucl_2)(\ucl_2\cdot\kb)\,\kb\cdot(\qb+\qb')}{\wn E^3} + \frac{(\ucl_1\cdot \ucl_2)^2(\kb\cdot\qb)(\kb\cdot\qb')}{\wn E^4}  \right].
\label{eqn:phaseSpaceIntegral}
\end{multline}
The quantity $\wn E$ is defined to be $\wn E = \ucl_1 \cdot \wn k$; in view of the delta function, the integral is constrained so that $\wn E = \ucl_1 \cdot \qb$. This quantity is the wavenumber of the photon in the rest frame of particle 1, and is fixed from the point of view of the phase space integration. As a result, the integrals are simple: there are two delta functions (one explicit, one in the phase space measure) which can be used to perform the $\kb^0$ integration and to fix the magnitude of the spatial wavevector. The remaining integrals are over angles. The relevant results were calculated in appendix~C of ref.~\cite{Kosower:2018adc}, and are
\begin{equation}
\begin{gathered}
\int \! \df (\kb) \, \del(\ucl_1\cdot \kb - \wn E) \, \kb^\mu = \frac{\wn E^2}{2\pi} u_1^\mu \Theta(\wn E)\,,\\
\int \! \df (\kb) \, \del(\ucl_1\cdot \kb - \wn E) \, \kb^\mu \kb^\nu \kb^\rho = \frac{\wn E^4}{\pi}\left(u_1^\mu u_1^\nu u_1^\rho - \frac12 u_1^{(\mu} \eta^{\nu\rho)}\right) \Theta(\wn E)\,.
\end{gathered}
\end{equation}
The radiated momentum then takes a remarkably simple form after the phase space integration:
\begin{equation}
\begin{aligned}
\Rad^{\mu,(0)}_\class = -\frac{e^6 Q_1^4 Q_2^2}{3\pi m_1^2} &\int\!\dd^4\qb \,
\dd^4\qb'\; e^{-ib\cdot(\qb - \qb')} \del(\ucl_1\cdot(\qb - \qb'))
\frac{\del(\ucl_2\cdot\qb)}{\qb^2} \frac{\del(\ucl_2\cdot\qb')}{\qb'^2}
\\ &\times\Theta(\ucl_1\cdot\qb)\,\left[(\ucl_1\cdot\qb)^2 
+ \qb\cdot\qb'(\ucl_1\cdot \ucl_2)^2\right] \ucl_1^\mu \,.
\label{eqn:radTheta}
\end{aligned}
\end{equation}

The $\Theta$ function is a remnant of the photon phase space volume, so it will be convenient to remove it. The delta functions in the integrand in \eqn~\eqref{eqn:radTheta} constrain the components of the vectors $\qb$ and $\qb'$ which lie in the two dimensional space spanned by $u_1$ and $u_2$. Let us call the components of $q$ and $q'$ in this plane to be $q_\parallel$ and $q'_\parallel$. Then the delta functions set $q_\parallel = q'_\parallel$. As a result, the integrand (ignoring the $\Theta$ function) is symmetric in $q_\parallel \rightarrow - q_\parallel$. Consequently we may symmetrise to find
\begin{equation}
\begin{aligned}
\Rad^{\mu,(0)}_\class = -\frac{e^6 Q_1^4 Q_2^2}{6\pi m_1^2} &\int\!\dd^4\qb\, \dd^4\qb'\;
e^{-ib\cdot(\qb - \qb')} \del(\ucl_1\cdot(\qb - \qb')) 
\frac{\del(\ucl_2\cdot\qb)}{\qb^2} \frac{\del(\ucl_2\cdot\qb')}{\qb'^2} 
\\ &\hspace*{10mm}\times\left[(\ucl_1\cdot\qb)^2 
+ \qb\cdot\qb'(\ucl_1\cdot \ucl_2)^2\right]\ucl_1^\mu \,.
\label{eqn:rrResult}
\end{aligned}
\end{equation}

It is now remarkably simple to see that this expression is equal but opposite to the impulse obtained from the classical ALD force in \eqn~\eqref{eqn:ALDclass}. Working in perturbation theory, the lowest order contribution to $dp_1 / d\tau$ is of order $e^2$, due to the (colour-stripped) LO Lorentz force~\eqref{eqn:Wong-momentum}. We can determine this explicitly using the methods of appendix~\ref{app:worldlines}: with particle 2 kept static, one finds
\begin{equation}
\frac{d p^{\mu,(0)}_1}{d \tau_1} =i e^2 Q_1 Q_2 \int \!\dd^4 \qb \, \del(\qb \cdot u_2) \, e^{- i \qb \cdot (b + u_1 \tau_1)} \,  \frac{\qb^\mu \, u_1 \cdot u_2 - u_2^\mu \, \qb \cdot u_1}{\qb^2}\,.
\label{eqn:LOforce}
\end{equation}
Therefore $\Delta {p^\mu_1}_{\rm ALD}$ is at least of order $e^4$. However, this potential contribution to the ALD impulse vanishes. To see this, observe that the acceleration due to the LO Lorentz force gives rise to an ALD impulse of
\begin{equation}
\Delta {p^\mu_1}_{\rm ALD} = \frac{e^4 Q_1^3 Q_2}{6\pi m_1}\int\!\dd^4\qb\, \del(\qb\cdot u_1) \del(\qb\cdot u_2) \, e^{-i\wn{q}\cdot b} \, \qb\cdot u_1 \, \big(\cdots\big) = 0\,.
\end{equation}
An alternative point of view on the same result is to perform the time integral in equation~\eqref{eqn:ALDclass}, noting that the second term in the ALD force is higher order. The impulse is then proportional to $f^\mu(+\infty) - f^\mu(-\infty)$, the difference in the asymptotic Lorentz forces on particle 1. But at asymptotically large times the two particles are infinitely far away, so the Lorentz forces must vanish. Since this second argument does not rely on perturbation theory we may ignore the first term in the ALD force law.

Thus, the first non-vanishing impulse due to radiation reaction is of order $e^6$. Since we only need the leading order Lorentz force to evaluate the ALD impulse, we can anticipate that the result will be very simple. Indeed, integrating the ALD force, we find that the impulse on particle 1 due to radiation reaction is
\begin{multline}
\Delta {p^\mu_1}_{\rm ALD} = \frac{e^6 Q_1^4 Q_2^2}{6\pi m_1^2} u_1^\mu \!\int\! \dd^4\qb\,\dd^4\qb'\, \del(\qb\cdot u_2) \del(\qb'\cdot u_2)  \del(u_1\cdot(\qb-\qb')) \, e^{-ib\cdot(\qb-\qb')} \\ \times\frac{1}{\qb^2 \qb'^2} \left[(\qb\cdot u_1)^2 + \qb\cdot\qb'(u_1\cdot u_2)^2 \right].
\label{eqn:classicalRadiationImpulse}
\end{multline}
This is precisely the expression~\eqref{eqn:rrResult} we found using our quantum mechanical approach.

\section{Discussion}
\label{sec:KMOCdiscussion}

In order to apply on-shell scattering amplitudes to the calculation of classically observable quantities for black holes, one needs a definition of the observables in the quantum theory.  One also needs a path and clear set of rules for taking the classical limit of the quantum observables.  In this first part of the thesis we have constructed one such path. Our underlying motivation is to understand the dynamics of classical general relativity through the double copy. In particular, we are interested in the relativistic two-body problem which is so central to the physics of the compact binary coalescence events observed by LIGO and Virgo. Consequently, we focused on observables in two-body events.

We have shown how to construct two observables relevant to this problem: the momentum transfer or impulse~(\ref{eqn:defl1}) on a particle; and the momentum emitted as radiation~(\ref{eqn:radiationTform}) during the scattering of two charged but spinless point particles.  We have shown how to restore $\hbar$'s and classify momenta in \sect{sec:RestoringHBar}; in \sect{sec:PointParticleLimit}, how to choose suitable wavefunctions for localised single particle states; and established in section~\ref{sec:classicalLimit} the conditions under which the classical limit is simple for point-particle scattering. With these formalities at hand we were able to further provide simplified leading and next-to-leading-order expressions in terms of on-shell scattering amplitudes for the impulse in \eqns{eqn:impulseGeneralTerm1classicalLO}{eqn:classicalLimitNLO}, and for the radiated momentum in \eqn~\eqref{eqn:radiatedMomentumClassical}.  These expressions apply directly to both gauge theory and gravity. In sections~\ref{sec:examples} and \ref{sec:LOradiation}, we used explicit expressions for amplitudes in QED, Yang--Mills theory and perturbative Einstein gravity to obtain classical results. We have been careful throughout to ensure that our methods correctly incorporate conservation of momentum, without the need to introduce an analogue of the Abraham--Lorentz--Dirac radiation reaction.

Other momentum observables should be readily accessible by similar derivations: for example the total radiated angular momentum is of particular current interest \cite{Damour:2020tta}, and is moreover accessible from worldline QFT \cite{Jakobsen:2021smu}; it would be very interesting to understand how this observable fits into our formalism. Higher-order corrections, to the extent they are unambiguously defined in the classical theory, require the harder work of computing two- and higher-loop amplitudes, but the formalism of these chapters will continue to apply.

Our setup has features in common with two related, but somewhat separate, areas of current interest. One area is the study of the potential between two massive bodies. The second is the study of particle scattering in the eikonal. Diagrammatically, the study of the potential is evidently closely related to the impulse of chapter~\ref{chap:impulse}. To some extent this is by design: we wished to construct an on-shell observable related to the potential. But we have also been able to construct an additional observable, the radiated momentum, which is related to the gravitational flux.

It is interesting that classical physics emerges in the study of the high-energy limit of quantum scattering~\cite{Amati:1987wq,tHooft:1987vrq,Muzinich:1987in,Amati:1987uf,Amati:1990xe}, see also refs.~\cite{Damour:2016gwp,Damour:2017zjx}. 
Indeed the classical centre-of-momentum scattering angle can be obtained from the eikonal function (see, for example ref.~\cite{DAppollonio:2010krb}). This latter function must therefore be related as well to the impulse, even though we have not taken any high-energy limit. Indeed, the impulse and the scattering angle are equivalent at LO and NLO, 
because no momentum is radiated at these orders. Therefore the scattering angle completely determines the change in momentum of the particles (and vice versa). The connection to the eikonal function should be interesting to explore.

At NNLO, on the other hand, the equivalence between the angle and the impulse fails. This is because of radiation: knowledge of the angle tells you where the particles went,
but not how fast. In this respect the impulse is more informative than the angle. Eikonal methods are still applicable in the radiative case~\cite{Amati:1990xe},
so they should reproduce the high-energy limit of the expectation value of the radiated momentum. Meanwhile at low energies, methods based on soft theorems could provide a bridge between the impulse and the radiated momentum~\cite{Laddha:2018rle,Laddha:2018myi,Sahoo:2018lxl}. Indeed, a first step in these directions was recently made in \cite{A:2020lub}. Radiation reaction physics can also be treated in this regime \cite{DiVecchia:2021ndb}, and we look forward to future progress in understanding how these references overlap with our formalism.

The NLO scattering angle is, in fact, somewhat simpler than the impulse: see ref.~\cite{Luna:2016idw} for example. Thanks to the exponentiation at play in the eikonal limit, it is the triangle diagram which is responsible for the NLO correction. But the impulse contains additional contributions, as we discussed in \sect{sec:nloQimpulse}. Perhaps this is because the impulse must satisfy an on-shell constraint, unlike the angle.

The focus of our study of inelastic scattering has been the radiation kernel introduced in~\eqref{eqn:defOfR}. Equivalent to a classical current, this object has proven to be especially versatile in the application of amplitudes methods to gravitational radiation. It has played a direct role in studies of the Braginsky--Thorne memory effect \cite{Bautista:2019tdr} and gravitational shock waves \cite{Cristofoli:2020hnk}; derivations of the connections between amplitudes, soft limits and classical soft theorems \cite{A:2020lub}; and calculations of Newman--Penrose spinors for long-range radiation in split-signature spacetimes \cite{Monteiro:2020plf}. The kernel is also closely related to progress in worldline QFT \cite{Mogull:2020sak,Jakobsen:2021smu}. However, the reader may object that we have not in fact calculated the total emitted radiation. This was recently achieved in ref.~\cite{Herrmann:2021lqe}, using integral evaluation techniques honed in $\mathcal{N}=8$ supergravity \cite{Parra-Martinez:2020dzs}. To calculate the full observable the authors of \cite{Herrmann:2021lqe} took a similar approach to our treatment of the non-linear impulse contributions at 1-loop, treating the product of radiation kernels as a cut of a 2-loop amplitude. Their application of the formalism presented here has led to state-of-the-art results for post--Minkowskian bremsstrahlung, already partially recovered from PM effective theory \cite{Mougiakakos:2021ckm}. 

In these two chapters we restricted attention to spinless scattering. In this context, for colourless particles such as astrophysical black holes the impulse (or equivalently, the angle) is the only physical observable at LO and NLO, and completely determines the interaction Hamiltonian between the two particles~\cite{Damour:2016gwp,Damour:2017zjx}. The situation is richer in the case of arbitrarily aligned spins --- then the change in spins of the particles is an observable which is not determined by the scattering angle. Fully specifying the dynamics of black holes therefore requires including spin in our formalism: this is the topic of the second part of the thesis.

%% file: chapter5/chapter5.tex
\part{Spinning black holes}
\label{part:spin}

\chapter{Observables for spinning particles}
\label{chap:spin}

\section{Introduction}

To begin this part of the thesis we will continue in the vein of previous chapters and use quantum field theory, now for particles with non-zero spin, to calculate observables for spinning point-particles. Our focus will be the leading-order scattering of black holes, however the formalism is applicable more widely \cite{Maybee:2019jus}. In chapter~\ref{chap:intro} we discussed at some length how scattering amplitudes have been applied to the dynamics of spinning Kerr black holes, and more fundamentally how minimally coupled amplitudes behave as the on-shell avatar of the no-hair theorem. Here we remove the restriction to the aligned-spin configuration in the final results of \cite{Guevara:2018wpp,Bautista:2019tdr}, and the restriction to the non-relativistic limit in the final results of \cite{Chung:2018kqs}.  We use on-shell amplitudes to directly compute relativistic classical observables for generic spinning-particle scattering, reproducing such results for black holes obtained by classical methods in \cite{Vines:2017hyw}, thereby providing more complete evidence for the correspondence between minimal coupling to gravity and classical black holes.

We will accomplish this by relaxing the restriction to scalars in previous chapters. In addition to the momentum impulse $\Delta p^\mu$, there is now another relevant on-shell observable, the change $\Delta s^\mu$ in the spin (pseudo-)vector $s^\mu$, which we will call the \textit{angular impulse}.  We introduce this quantity in \sect{sec:GRspin}, where we also review classical results from \cite{Vines:2017hyw} for binary black hole scattering at 1PM order. 
In \sect{sec:QFTspin} we consider the quantum analogue of the spin vector, the Pauli--Lubanski operator; manipulations of this operator allow us to write expressions for the angular impulse akin to those for the impulse of chapter~\ref{chap:impulse}. Obtaining the classical limit requires some care, which we discuss before constructing example gravity amplitudes in \sect{sec:amplitudes} from the double copy. 

Rather than working at this stage with massive spinor representations valid for any quantum spin $s$, in the vein of \cite{Arkani-Hamed:2017jhn}, we will ground our intuition in explicit field representations of the Poincar\'{e} group --- we will work with familiar spin 1/2 Dirac fermions and massive, spin 1 bosons. Explicit representations can indeed be chosen for any generic spin $s$ field; the applications of these representations to black hole physics was studied in ref.~\cite{Bern:2020buy}. However, for higher spins the details become extremely involved. In \sect{sec:KerrCalcs} we show that substituting familiar low spin examples into our general formalism exactly reproduces the leading terms of all-multipole order expressions for the impulse and angular impulse of spinning black holes \cite{Vines:2017hyw}. Finally, we discuss how our results further entwine Kerr black holes and scattering amplitudes in \sect{sec:angImpDiscussion}.

This chapter is based on work conducted in collaboration with Donal O'Connell and Justin Vines, published in \cite{Maybee:2019jus}.

\section{Spin and scattering observables in classical gravity}\label{sec:GRspin}

Before setting up our formalism for computing the angular impulse, let us briefly review aspects of this observable in relativistic classical physics. 

\subsection{Linear and angular momenta in asymptotic Minkowski space}

To describe the incoming and outgoing states for a weak scattering process in asymptotically flat spacetime we can use special relativistic physics, working as in Minkowski spacetime. There, any isolated body has a constant linear momentum vector $p^\mu$ and an antisymmetric tensor field $J^{\mu\nu}(x)$ giving its total angular momentum about the point $x$, with the $x$-dependence determined by $J^{\mu\nu}(x')=J^{\mu\nu}(x)+2p^{[\mu}(x'-x)^{\nu]}$, or equivalently $\nabla_\lambda J^{\mu\nu}=2p^{[\mu}\delta^{\nu]}{}_\lambda$.

Relativistically, centre of mass (CoM) position and intrinsic and orbital angular momenta are frame-dependent concepts, but a natural inertial frame is provided by the direction of the momentum $p^\mu$, giving the proper rest frame.  We define the body's proper CoM worldline to be the set of points $r$ such that $J^{\mu\nu}(r)p_\nu=0$, i.e.\ the proper rest-frame mass-dipole vector about $r$ vanishes, and we can then write
\begin{equation}\label{eqn:Jmunu}
J^{\mu\nu}(x)=2p^{[\mu}(x-r)^{\nu]}+S^{\mu\nu},
\end{equation}
where $r$ can be any point on the proper CoM worldline, and where $S^{\mu\nu}=J^{\mu\nu}(r)$ is the intrinsic spin tensor, satisfying
\begin{equation}
S^{\mu\nu} p_\nu=0.\label{eqn:SSC}
\end{equation}
Equation \eqref{eqn:SSC} is often called the ``covariant'' or Tulczyjew--Dixon spin supplementary condition (SSC) \cite{Fokker:1929,Tulczyjew:1959} in its (direct) generalization to curved spacetime in the context of the Mathisson--Papapetrou--Dixon equations \cite{Mathisson:1937zz,Mathisson:2010,Papapetrou:1951pa,Dixon1979,Dixon:2015vxa} for the motion of spinning extended test bodies. 
Given the condition \eqref{eqn:SSC}, the complete information of the spin tensor $S^{\mu\nu}$ is encoded in the momentum $p^\mu$ and the spin pseudovector \cite{Weinberg:1972kfs},
\begin{equation}
s_\mu = \frac{1}{2m}\epsilon_{\mu\nu\rho\sigma} p^\nu S^{\rho\sigma} = \frac{1}{2m}\epsilon_{\mu\nu\rho\sigma} p^\nu J^{\rho\sigma}(x),\label{eqn:GRspinVec}
\end{equation}
where $\epsilon_{0123} = +1$ and $p^2=m^2$.  Note that $s\cdot p=0$; $s^\mu$ is a spatial vector in the proper rest frame.
Given \eqref{eqn:SSC}, the inversion of the first equality of \eqref{eqn:GRspinVec} is
\begin{equation}
S_{\mu\nu} =\frac{1}{m} \epsilon_{\mu\nu\lambda\tau} p^\lambda s^\tau.\label{eqn:SSCintrinsicSpin}
\end{equation}
The total angular momentum tensor $J^{\mu\nu}(x)$ can be reconstructed from $p^\mu$, $s^\mu$, and a point $r$ on the proper CoM worldline, via \eqref{eqn:SSCintrinsicSpin} and \eqref{eqn:Jmunu}.


\subsection{Scattering of spinning black holes in linearised gravity}

Following the no-hair property emphasised by equation~\eqref{eqn:multipoles} of chapter~\ref{chap:intro}, the full tower of gravitational multipole moments of a spinning black hole, and thus also its (linearised) gravitational field, are uniquely determined by its monopole $p^\mu$ and dipole $J^{\mu\nu}$.  This is reflected in the scattering of two spinning black holes, in that the net changes in the holes' linear and angular momenta depend only on their incoming linear and angular momenta.  It has been argued in \cite{Vines:2017hyw} that the following results concerning two-spinning-black-hole scattering, in the  1PM approximation to GR, follow from the linearised Einstein equation and a minimal effective action description of spinning black hole motion, the form of which is uniquely fixed at 1PM order by general covariance and appropriate matching to the Kerr solution.



Consider two black holes with incoming momenta $p_1^\mu=m_1 u_1^\mu$ and $p_2^\mu=m_2 u_2^\mu$, defining the 4-velocities $u^\mu=p^\mu/m$ with $u^2=1$, and incoming spin vectors $s_1^\mu=m_1 a_1^\mu$ and $s_2^\mu=m_2 a_2^\mu$, defining the rescaled spins $a^\mu=s^\mu/m$ (with units of length, whose magnitudes measure the radii of the ring singularities).  Say the holes' zeroth-order incoming proper CoM worldlines are orthogonally separated at closest approach by a vectorial impact parameter $b^\mu$, pointing from 2 to 1, with $b\cdot u_1 =b\cdot u_2=0$.  Then, according to the analysis of \cite{Vines:2017hyw}, the net changes in the momentum and spin vectors of black hole 1 are given by
\begin{alignat}{3}
\begin{aligned}
\Delta p_1^{\mu} &= \textrm{Re}\{\mathcal Z^\mu\}+O(G^2),
\\
\Delta s_1^{\mu} &= - u_1^\mu a_1^\nu\, \textrm{Re}\{\mathcal Z_\nu\} - \epsilon^{\mu\nu\alpha\beta} u_{1\alpha} a_{1\beta}\, \textrm{Im}\{\mathcal Z_\nu\}+O(G^2),
\end{aligned}\label{eqn:KerrDeflections}
\end{alignat}
where
\begin{equation}
\mathcal Z_\mu = \frac{2G m_1 m_2}{\sqrt{\gamma^2 - 1}}\Big[(2\gamma^2 - 1)\eta_{\mu\nu} - 2i\gamma \epsilon_{\mu\nu\alpha\beta} u_1^\alpha u_2^\beta\Big]\frac{ b^\nu + i\Pi^\nu{ }_\rho (a_1+a_2)^\rho}{[b + i\Pi(a_1+a_2)]^2}\,,
\end{equation}
with $\gamma = u_1\cdot u_2$ the relative Lorentz factor, and with
\begin{equation}
\begin{aligned}
\Pi^\mu{ }_\nu &= \epsilon^{\mu\rho\alpha\beta} \epsilon_{\nu\rho\gamma\delta} \frac{{u_1}_\alpha {u_2}_\beta u_1^\gamma u_2^\delta}{\gamma^2 - 1}\\ &= \delta^\mu{ }_\nu +\frac1{\gamma^2 - 1}\bigg(u_1^\mu({u_1}_\nu - \gamma {u_2}_\nu) + u_2^\mu({u_2}_\nu - \gamma {u_1}_\nu)\bigg) \label{eqn:projector}
\end{aligned}
\end{equation}
the projector into the plane orthogonal to both incoming velocities.
The analogous results for black hole 2 are given by interchanging the identities $1\leftrightarrow 2$.

If we take black hole 2 to have zero spin, $a_2^\mu\to0$, and if we expand to quadratic order in the spin of black hole 1, corresponding to the quadrupole level in 1's multipole expansion, then we obtain the results shown in \eqref{eqn:JustinImpResult} and \eqref{eqn:JustinSpinResult} below.  In the remainder of this chapter, developing necessary tools along the way, we show how those results can be obtained from classical limits of scattering amplitudes. In particular, we will consider one-graviton exchange between a massive scalar particle and a massive spin $s$ particle, with minimal coupling to gravity, with $s=1/2$ to yield the dipole level, and with $s=1$ to yield the quadrupole level.

\section{Spin and scattering observables in quantum field theory}
\label{sec:QFTspin}

We have already established general formulae in quantum field theory for the impulse and radiated momentum; as the angular impulse is also on-shell similar methods should be applicable. A first task is to understand what quantum mechanical quantity corresponds to the classical spin pseudovector of equation~\eqref{eqn:GRspinVec}. This spin vector is a quantity associated with a single classical body, and we therefore momentarily return to discussing single-particle states. 

Particle states of spin $s$ are irreducible representations of the little group. For massive particles in 4 dimensions the little group is isomorphic to $SU(2)$, and thus we can adopt the simplest coherent states considered in section~\ref{sec:classicalSingleParticleColour}. The size of the representation is now determined by the spin quantum number $s$ associated with the states. For fractional spins the normalisation in \eqn~\eqref{eqn:ladderCommutator} of course generalises to an anticommutation relation.

For spinning particles we will thus adopt the wavepackets in equation~\eqref{eqn:InitialStateSimple}, with the important distinction that the representation states now refer to the little group, not a gauge group. To make this distinction clear we will denote the little group states by $|\xi\rangle$ rather than $|\chi\rangle$ (not to be confused with the parameter~\eqref{eqn:defOfXi}). The momentum space wavefunctions $\varphi(p)$ remain entirely unchanged.

\subsection{The Pauli--Lubanski spin pseudovector}

What operator in quantum field theory is related to the classical spin pseudovector of equation~\eqref{eqn:GRspinVec}? We propose that the correct quantum-mechanical
interpretation is that the spin is nothing but the expectation value of the \textit{Pauli--Lubanski} operator,
\begin{equation}
\W_\mu = \frac{1}{2}\epsilon_{\mu\nu\rho\sigma} \P^\nu \J^{\rho\sigma}\,,\label{eqn:PLvec}
\end{equation}
where $\P^\mu$ and $\mathbb{J}^{\rho\sigma}$ are translation and Lorentz generators respectively. In particular,
our claim is that the expectation value
\begin{equation}
\langle s^\mu \rangle = \frac1{m} \langle \mathbb{W}^\mu\rangle = \frac1{2m} \epsilon^{\mu\nu\rho\sigma} \langle \mathbb{P}_\nu \mathbb{J}_{\rho\sigma}\rangle
\end{equation}
of the Pauli--Lubanski operator on a single particle state~\eqref{eqn:InitialStateSimple} is the quantum-mechanical generalisation of the classical spin pseudovector. Indeed, a simple comparison of equations~\eqref{eqn:GRspinVec} and~\eqref{eqn:PLvec} indicates a connection between the two quantities. We will provide abundant evidence for this link in the remainder of this chapter --- it is shown in greater detail in appendix~B of ref.~\cite{Maybee:2019jus}.

The Pauli--Lubanski operator is a basic quantity in the classification of free particle states, although it receives less attention in introductory accounts of quantum field theory than it should. With the help of the Lorentz algebra
\[
[\J^{\mu\nu}, \P^\rho] &= i \hbar (\eta^{\mu\rho} \P^\nu - \eta^{\nu\rho} \P^\mu) \,, \\
[\J^{\mu\nu}, \J^{\rho\sigma}] &= i \hbar (\eta^{\nu\rho} \J^{\mu\sigma} - \eta^{\mu\rho} \J^{\nu\sigma} - \eta^{\nu\sigma} \J^{\mu\rho} + \eta^{\mu\rho} \J^{\mu\sigma}) \, ,
\]
it is easy to establish the important fact that the Pauli--Lubanski operator commutes with the momentum:
\[
[\P^\mu, \W^\nu] = 0\,.\label{eqn:PWcommute}
\] 
Furthermore, as $\W^\mu$ is a vector operator it satisfies
\[
[\J^{\mu\nu}, \W^\rho] = i\hbar (\eta^{\mu\rho} \W^\nu - \eta^{\nu\rho} \W^\mu) \,.
\]
It then follows that the commutation relations of $\W$ with itself are
\[
[\W^\mu, \W^\nu] = i\hbar \epsilon^{\mu\nu\rho\sigma} \W_\rho \P_\sigma\,.
\]
On single particle states this last commutation relation takes a particularly instructive form. Working in the rest frame of our massive particle state, evidently $W^0 = 0$. The remaining generators satisfy\footnote{We normalise $\epsilon^{123} = +1$, as usual.}
\[
[ \W^i, \W^j] = i \hbar m \,\epsilon^{ijk} \W^k \,,
\]
so that the Pauli--Lubanski operators are nothing but the generators of the little group. Not only is this the basis for their importance, but also we will find that these commutation relations are directly useful in our computation of the change in a particle's spin during scattering.

Because the $\mathbb{W}^\mu$ commutes with the momentum, we have
\begin{equation}
\langle p'\, j| \W^\mu |p\, i \rangle \propto \del_\Phi(p-p')\,.
\end{equation}
We define the matrix elements of $\W$ on the states of a given momentum to be
\[
\langle p'\, j| \W^\mu |p\, i \rangle \equiv m \s^\mu_{ij}(p)\, \del_\Phi(p-p') \,,\label{eqn:PLinnerProd}
\]
so that the expectation value of the spin vector over a single particle wavepacket is
\[
\langle \s^\mu \rangle = \sum_{i,j} \int d\Phi(p) \, | \varphi(p) |^2 \, \xi^*_i \s^\mu_{ij} \xi_j\,.
\]
The matrix $\s^\mu_{ij}(p)$, sometimes called the spin polarisation vector, will be important below. These matrices inherit the commutation relations of the Pauli--Lubanski vector, so that in particular
\[
[\s^\mu(p), \s^\nu(p) ] = \frac{i\hbar}{m} \, \epsilon^{\mu\nu\rho\sigma} \s_\rho(p) p_\sigma \,.
\]

Specialising now to a particle in a given representation, we may derive well known \cite{Ross:2007zza,Holstein:2008sx,Bjerrum-Bohr:2013bxa,Guevara:2017csg} explicit expressions for the spin polarisation $s^\mu_{ij}(p)$ by starting with the Noether current associated with angular momentum. Such derivations for the simple spin 1/2 and 1 cases were given in appendix B of~\cite{Maybee:2019jus} --- for a Dirac spin $1/2$ particle, the spin polarisation is
\[
s^\mu_{ab}(p) = \frac{\hbar}{4m} \bar{u}_a(p) \gamma^\mu \gamma^5 u_b(p)\,.\label{eqn:spinorSpinVec}
\]
Meanwhile, for massive vector bosons we have
\begin{equation}
s_{ij}^\mu(p) = -\frac{i\hbar}{m} \epsilon^{\mu\nu\rho\sigma} p_\nu \varepsilon{^*_i}_\rho(p) {\varepsilon_j}_\sigma(p)\,. \label{eqn:vectorSpinVec}
\end{equation}
We have these normalised quantities to be consistent with the algebraic properties of the Pauli--Lubanski operator.

\subsection{The change in spin during scattering}

Now that we have a quantum-mechanical understanding of the spin vector, we move on to discuss the dynamics of the spin vector in a scattering process. Following the set-up of chapter~\ref{chap:impulse} we consider the scattering of two stable, massive particles which are quanta of different fields, and are separated by an impact parameter $b^\mu$. We will explicitly consider scattering processes mediated by vector bosons and gravitons. The relevant incoming two-particle state is therefore that in equation~\eqref{eqn:inState}, but with little group states $\xi_\alpha$.

The initial spin vector of particle 1 is
\[
\langle s_1^\mu \rangle = \frac1{m_1} \langle \Psi |\W^\mu_1 |\Psi \rangle\,,
\]
where $\W^\mu_1$ is the Pauli--Lubanski operator of the field corresponding to particle 1. Since the $S$ matrix is the time evolution operator from the far past to the far future, the final spin vector of particle 1 is
\[
\langle s_1'^\mu \rangle = \frac1{m_1} \langle \Psi | S^\dagger \W^\mu_1 S| \Psi \rangle\,.
\]
We define the angular impulse on particle 1 as the difference between these quantities:
\begin{equation}
\langle \Delta s_1^\mu \rangle = \frac1{m_1}\langle\Psi|S^\dagger \mathbb{W}_1^\mu S|\Psi\rangle - \frac1{m_1}\langle\Psi|\mathbb{W}_1^\mu|\Psi\rangle\,.\label{eqn:defOfAngImp}
\end{equation}
Writing $S = 1 + iT$ and making use of the optical theorem yields
\begin{equation}
\langle \Delta s_1^\mu\rangle = \frac{i}{m_1}\langle\Psi|[\mathbb{W}_1^\mu,{T}]|\Psi\rangle + \frac{1}{m_1}\langle\Psi|{T}^\dagger[\mathbb{W}_1^\mu,{T}]|\Psi\rangle\,.\label{eqn:spinShift}
\end{equation}
Just as with equation~\eqref{eqn:defl1}, it is clear that the second of these terms will lead to twice as many powers of the coupling constant for a given interaction. Therefore only the first term is able to contribute at leading order. We will be exclusively considering tree level scattering $\mathcal{A}^{(0)}$, so the first term is the sole focus of our attention\footnote{The expansion of the second term is very similar to that of the colour impulse in ref.~\cite{delaCruz:2020bbn}.}.

Our goal now is to express the leading-order angular impulse in terms of amplitudes. To that end we substitute the incoming state in equation~\eqref{eqn:inState} into the first term of \eqn~\eqref{eqn:spinShift}, and the leading-order angular impulse is given by
\begin{multline}
\langle\Delta s^{\mu,(0)}_1\rangle = \frac{i}{m_1}\prod_{\alpha=1,2} \int\! \df(p_\alpha') \df(p_\alpha) \,\varphi_\alpha^*(p_\alpha') \varphi_\alpha(p_\alpha) e^{ib\cdot(p_1-p'_1)/\hbar} \\ \times \left\langle p_1'\,\xi'_1 ; p_2'\, \xi'_2\left|\W_1^\mu\, {T} -{T}\, \W_1^\mu \right|p_1 \, \xi_1; p_2\, \xi_2 \right\rangle.
\end{multline}
Scattering amplitudes can now be explicitly introduced by inserting a complete set of states between the spin and interaction operators, as in equation~\eqref{eqn:p1Expectation}. In their first appearance this yields
\begin{multline}
\int\! \df(r_1) \df(r_2) \d\mu(\zeta_1) \d\mu(\zeta_2)\langle p'_1\, \xi'_1; p'_2\, \xi'_2|\W_1^\mu|r_1\, \zeta_1; r_2, \zeta_2\rangle \langle r_1\, \zeta_1; r_2\, \zeta_2|T|p_1\, \zeta_1 ; p_2\, \zeta_2\rangle \\
 =  m_1 \langle \xi'_1\, \xi'_2 |{\s}^\mu_1(p'_1)\mathcal{A}(p_1, p_2 \rightarrow p'_1, p'_2)|\xi_1\,\xi_2\rangle\,  \del^{(4)}(p'_1 + p'_2 - p_1 - p_2)\,,
\end{multline}
where, along with the definition of the scattering amplitude, we have used the definition of the spin polarisation vector~\eqref{eqn:spinorSpinVec}. The result for the other ordering of $T$ and $\W^\mu_1$ is very similar. 

An essential point is that under the little group state inner product above, the spin polarisation vector and amplitude do not commute: they are both matrices in the little group representation, and we have simply suppressed the explicit indices. This novel feature of the angular impulse will become extremely important. Substituting into the full expression for $\langle\Delta s_1^{\mu,(0)}\rangle$ and integrating over the delta functions, we find that the observable is
\begin{equation}
\begin{aligned}
\langle\Delta s^{\mu,(0)}_1\rangle = i \prod_{\alpha = 1, 2} &\int \!\df(p_\alpha') \df(p_\alpha) \,\varphi_\alpha^*(p_\alpha') \varphi_\alpha (p_\alpha) \del^{(4)}(p_1' + p_2' - p_1 - p_2)  \\ & \times  e^{ib\cdot(p_1-p'_1)/\hbar} \langle\xi'_1\, \xi'_2| \s_{1}^\mu(p_1') \mathcal{A}^{(0)}(p_1, p_2 \rightarrow p'_1, p'_2) \\ &\hspace{30mm} - \mathcal{A}^{(0)}(p_1, p_2 \rightarrow p'_1, p'_2) \s_{1}^{\mu}(p_1)|\xi_1\, \xi_2\rangle\,.
\end{aligned}
\end{equation}
We now eliminate the delta function by introducing the familiar momentum mismatches $q_\alpha = p'_\alpha - p_\alpha$ and performing an integral. The leading-order angular impulse becomes
\begin{equation}
\begin{aligned}
\langle\Delta s^{\mu,(0)}_1\rangle = i & \int\! \df(p_1) \df(p_2)\, \dd^4q\,\del(2p_1\cdot q + q^2) \del(2p_2\cdot q - q^2) \Theta(p_1^0 + q^0) \\ \times& \Theta(p_2^0 - q^0) \varphi_1^*(p_1 + q) \varphi_2^*(p_2 - q) \varphi_1(p_1) \varphi_2(p_2) e^{-ib\cdot q/\hbar} \\\times& \langle \xi'_1\, \xi'_2|\s_{1}^{\mu}(p_1 + q) \mathcal{A}^{(0)}(p_1, p_2 \rightarrow p_1 + q, p_2 - q) \\ &\qquad\qquad - \mathcal{A}^{(0)}(p_1, p_2 \rightarrow p_1 + q, p_2 - q) \s^{\mu}(p_1)|\xi_1\, \xi_2\rangle\,.\label{eqn:exactAngImp}
\end{aligned}
\end{equation}

\subsection{Passing to the classical limit}
\label{sec:classicalLim}

The previous expression is an exact, quantum formula for the leading-order change in the spin vector during conservative two-body scattering. As a well-defined observable, we can extract the classical limit of the angular impulse by following the formalism introduced in part~\ref{part:observables}.

Recall that the basic idea is simple: the momentum space wavefunctions must localise the particles, without leading to a large uncertainty in their momenta. They therefore have a finite but small width $\Delta x = \ell_w$ in position space, and $\Delta p = \hbar/\ell_w$ in momentum space. This narrow width restricts the range of the integral over $q$ in equation~\eqref{eqn:exactAngImp} so that $q \lesssim \hbar /\ell_w$. We therefore introduce the wavenumber $\qb = q/\hbar$. We further found that our explicit choice of wavefunctions $\varphi_\alpha$ were very sharply peaked in momentum space around the value $\langle p_\alpha^\mu \rangle = m_\alpha u_\alpha^\mu$, where $u_\alpha^\mu$ is a classical proper velocity. We neglect the small shift $q = \hbar \qb$ in the wavefunctions present in equation~\eqref{eqn:exactAngImp}, and also the term $q^2$ compared to the dominant $2 p \cdot q$ in the delta functions, arriving at
\[
\langle\Delta s^{\mu,(0)}_1\rangle =&\, i \!  \int \! \df(p_1) \df(p_2)\, \dd^4q\,\del(2p_1\cdot q ) \del(2p_2\cdot q) |\varphi_1(p_1)|^2 |\varphi_2(p_2)|^2 \\&\times e^{-ib\cdot q/\hbar}\, \bigg\langle\s_{1}^{\mu}(p_1 + q) \mathcal{A}^{(0)}(p_1, p_2 \rightarrow p_1 + q, p_2 - q) \\ &\qquad\qquad\qquad\qquad - \mathcal{A}^{(0)}(p_1, p_2 \rightarrow p_1 + q, p_2 - q) \s_{1}^{\mu}(p_1)\bigg\rangle   \,.
\]
Notice we have also dropped the distinction between the little group states, simply writing an expectation value as in~\eqref{eqn:amplitudeDef}. This is permissible since we established coherent states suitable for the classical limit of $SU(2)$ states in section~\ref{sec:classicalSingleParticleColour}. Adopting the notation for the large angle brackets from \eqn~\eqref{eqn:angleBrackets}, the angular impulse takes the form
\[
\langle\Delta s^{\mu,(0)}_1\rangle =  &\, \Lexp i\! \int \!\dd^4 q\, \del (2p_1 \cdot q) \del (2p_2\cdot q) e^{-ib\cdot q/\hbar}  \\ & \times\bigg(\s^{\mu}(p_1 +  \hbar\qb) \mathcal{A}^{(0)}(p_1, p_2 \rightarrow p_1 + q, p_2 - q) \\ &\qquad\qquad - \mathcal{A}^{(0)}(p_1, p_2 \rightarrow p_1 + q, p_2 - q) \s_{1}^{\mu}(p_1)\bigg)\Rexp\,,
\label{eqn:intermediate}
\]

An important $\hbar$ shift remaining is that of the spin polarisation vector $\s_{1}^{\mu}(p_1 + \hbar\wn q)$. This object is a Lorentz boost of $\s_{1}^{\mu}(p_1)$. In the classical limit $q$ is small, so the Lorentz boost $\Lambda^{\mu}{ }_\nu p_1^\nu = p_1^\mu + \hbar\wn q^\mu$ is infinitesimal. In the vector representation an infinitesimal Lorentz transformation is $\Lambda^\mu\,_\nu=\delta^\mu_\nu + w^\mu{ }_\nu$, so for our boosted momenta $w^\mu{ }_\nu p_1^\nu = \hbar\wn q^\mu$. The appropriate generator is
\begin{equation}
w^{\mu\nu} = -\frac{\hbar}{m_1^2}\left(p_1^\mu \wn q^\nu - \wn q^\mu p_1^\nu\right)\,.\label{eqn:LorentzParameters}
\end{equation} 
This result is valid for particles of any spin as it is purely kinematic, and therefore can be universally applied in our general formula for the angular impulse. In particular, since $w_{\mu\nu}$ is explicitly $\mathcal{O}(\hbar)$ the spin polarisation vector transforms as
\begin{equation}
\s_{1\,ij}^{\mu}(p_1 + \hbar\wn q) = \s_{1\,ij}^{\mu}(p_1) - \frac\hbar{m^2} p^\mu \qb \cdot s_{ij}(p_1)\,.
\label{eqn:infinitesimalBoost}
\end{equation}
The angular impulse becomes
\begin{multline}
\langle \Delta s_1^{\mu,(0)} \rangle \rightarrow \Delta s_1^{\mu,(0)}\label{eqn:limAngImp}
= \Lexp i\! \int\!\dd^4\wn q\,\del(2p_1\cdot\wn q) \del(2p_2\cdot\wn q)e^{-ib\cdot\wn q} \\ \times \bigg(-\hbar^3 \frac{p_1^\mu}{m_1^2} \qb \cdot \s_{1} (p_1) \mathcal{A}^{(0)}(\wn q) + \hbar^2 \big[\s^\mu_1(p_1), \mathcal{A}^{(0)}(\wn q)\big] \bigg)\Rexp\,.
\end{multline}
The little group states have manifested themselves in the appearance of a commutator. The formula appears to be of a non-uniform order in $\hbar$, but fortunately this is not really the case: any terms in the amplitude with diagonal indices will trivially vanish under the commutator; alternatively, any term with a commutator will introduce a factor of $\hbar$ through the algebra of the Pauli--Lubanski vectors.
Therefore all terms have the same weight, $\hbar^3$, independently of factors appearing in the amplitude. 
The analogous formula for the leading order, classical momentum impulse was given in \eqn~\eqref{eqn:impulseGeneralTerm1classicalLO}.
We will make use of both the momentum and angular impulse formulae below.

There is a caveat regarding the uncertainty principle in the context of our spinning particles. In the following examples we restrict to low spins: spin 1/2 and spin 1. Consequently the expectation of the spin vector $\langle s^\mu \rangle$ is of order $\hbar$; indeed $\langle s^2 \rangle = s(s+1) \hbar^2$. This requires us to face the quantum-mechanical distinction between $\langle s^\mu s^\nu \rangle$ and $\langle s^\mu \rangle \langle s^\nu \rangle$. Because of the uncertainty principle, the uncertainty $\sigma_1^2$ associated with the operator $s^1$, for example, is of order $\hbar$, and therefore the difference between $\langle s_1^2 \rangle$ and $\langle s_1 \rangle^2$ is of order $\hbar^2$. Thus the difference $\langle s^\mu s^\nu \rangle - \langle s^\mu \rangle \langle s^\nu \rangle$ is of order $\langle s^\mu s^\nu \rangle$. We are therefore not entitled to replace $\langle s^\mu s^\nu \rangle$ by $\langle s^\mu \rangle \langle s^\nu \rangle$ for any arbitrary states, and will make the distinction between these quantities below. As we showed explicitly (for $SU(3)$ states) in chapter~\ref{chap:pointParticles}, this limitation can be overcome by studying very large spin representations. To elucidate the details of taking the classical limit of amplitudes with spin we will primarily work with the explicit spin 1/2 and spin 1 fields in this chapter; however, we will comment on the all--spin generalisation of our results, studied in \cite{Guevara:2019fsj}. The large spin limit will then play a crucial role in the next chapter.

The procedure for passing from amplitudes to a concrete expectation value is as follows. Once one has computed the amplitude, and evaluated any commutators, explicit powers of $\hbar$ must cancel. We then evaluate the integrals over the on-shell phase space of the incoming particles simply by evaluating the momenta $p_\alpha$ as $p_\alpha = m_\alpha u_\alpha$. An expectation value over the spin wave functions $\xi_\alpha$ remains; these are always of the form $\langle s^{\mu_1} \cdots s^{\mu_n}\rangle$ for various values of $s$. Only when the spin $s$ is large can we factorise this expectation value.

\section{Classical limits of amplitudes with finite spin}
\label{sec:amplitudes}

We have constructed a general formula for calculating the leading classical contribution to the angular impulse from scattering amplitudes. In the limit these amplitudes are Laurent expanded in $\hbar$, with only one term in the expansion providing a non-zero contribution. How this expansion works in the case for charged scalar amplitudes was established in part~\ref{part:observables}, but now we need to consider examples of amplitudes for particles with spin. The identification of the spin polarisation vector defined in \eqn~\eqref{eqn:PLinnerProd} will be crucial to this limit.

We will again look at the two lowest spin cases, considering tree level scattering of a spin $1/2$ or spin $1$ particle off a scalar in Yang--Mills theory and gravity. Tree level Yang--Mills amplitudes will now be denoted by $\mathcal{A}_{s_1-0}$, and those for Einstein gravity as $\mathcal{M}_{s_1-0}$. To ensure good UV behaviour of our amplitudes, we adopt minimally coupled interactions between the massive states and gauge fields. This has the effect of restricting the classical value of the gyromagnetic ratio to $g_L =2$, for all values of $s$ \cite{Chung:2018kqs,Ferrara:1992yc}.

\subsection{Gauge theory amplitudes}

We will continue to consider Yang--Mills theory minimally coupled to matter in some representation of the gauge group. The common Lagrangian is that in equation~\eqref{eqn:scalarAction}. Our calculations will be in the vein of section~\ref{sec:LOimpulse}, and we will always have the same $t$-channel colour factor. The amplitude for scalar-scalar, $\mathcal{A}_{0-0}$, scattering is of course that in equation~\eqref{eqn:treeamp}.

\subsubsection*{Spinor-scalar}

We can include massive Dirac spinors $\psi$ in the Yang--Mills amplitudes by using a Lagrangian $\mathcal{L} = \mathcal{L}_0 + \mathcal{L}_{\textrm{Dirac}}$, where the Dirac Lagrangian 
\begin{equation}
\mathcal{L}_\textrm{Dirac} = \bar{\psi}\left(i\slashed{D} - m\right)\psi\label{eqn:DiracL}
\end{equation}
includes a minimal coupling to the gauge field, and $\mathcal{L}_0$ is the scalar Langrangian in equation~\eqref{eqn:scalarAction}. The tree level amplitude for spinor-scalar scattering is then
\begin{equation}
i\mathcal{A}^{ab}_{1/2-0} = \frac{ig^2}{2 \hbar q^2}\bar{u}^a(p_1+q)\gamma^\mu u^{b}(p_1)(2p_2 - q)_\mu\, \tilde{\newT}_1\cdot \tilde{\newT}_2\,,
\end{equation}
where we have normalised the (dimensionful) colour factors consistent with the double copy. We are interested in the pieces of this amplitude that survive to the classical limit. To extract them we must set the momentum transfer as $q = \hbar\wn q$ and expand the amplitude in powers of $\hbar$. 

The subtlety here is the on-shell Dirac spinor product. In the limit, when $q$ is small, we can follow the logic of \eqn~\eqref{eqn:infinitesimalBoost} and interpret $\bar{u}^a(p_1 + \hbar\wn q) \sim \bar{u}^a(p_1) + \Delta\bar{u}^a(p_1)$ as being infinitesimally Lorentz boosted, see also~\cite{Lorce:2017isp}. One expects amplitudes for spin 1/2 particles to only be able to probe up to linear order in spin (i.e. the dipole of a spinning body) \cite{Vaidya:2014kza,Guevara:2017csg,Guevara:2018wpp}, so in deriving the infinitesimal form of the Lorentz transformation we expand to just one power in the spin. The infinitesimal parameters $w_{\mu\nu}$ are exactly those determined in \eqn~\eqref{eqn:LorentzParameters}, so in all the leading terms of the spinor product are
\begin{equation}
\bar{u}^a(p_1 + \hbar\wn q) \gamma_\mu u^b(p_1) = 2{p_1}_\mu \delta^{ab} + \frac{\hbar}{4m^2}\bar{u}^a(p_1) p{_1}^\rho \wn q^\sigma [\gamma_\rho,\gamma_\sigma] \gamma_\mu u^b(p_1) + \mathcal{O}(\hbar^2)\,.\label{eqn:spinorshift}
\end{equation}
Evaluating the product of gamma matrices via the identity
\begin{equation}
[\gamma_\mu, \gamma_\nu] \gamma_\rho = 2\eta_{\nu\rho}\gamma_\mu - 2\eta_{\mu\rho}\gamma_\nu -  2i\epsilon_{\mu\nu\rho\sigma} \gamma^\sigma \gamma^5\,,\label{eqn:3gammaCommutator}
\end{equation}
where $\epsilon_{0123} = +1$ and $\gamma^5 = i\gamma^0 \gamma^1 \gamma^2 \gamma^3$, the spinor product is just
\begin{multline}
\bar{u}^a(p_1 + \hbar\wn q) \gamma_\mu u^b(p_1) = 2{p_1}_\mu\delta^{ab}  + \frac{\hbar}{2m_1^2} \bar{u}^a(p_1) p_1^\rho \wn q^\sigma \left(\gamma_\sigma\eta_{\mu\rho} - \gamma_\rho\eta_{\mu\sigma}\right) u^b(p_1) \\ - \frac{i\hbar}{2m_1^2} \bar{u}^a(p_1) p{_1}^\rho \wn q^\sigma \epsilon_{\rho\sigma\mu\delta} \gamma^\delta\gamma^5 u^b(p_1) + \mathcal{O}(\hbar^2)\,.
\end{multline}
Comparing with our result from \eqn~\eqref{eqn:spinorSpinVec}, the third term clearly hides an expression for the spin 1/2 polarisation vector. Making this replacement and substituting the spinor product into the amplitude yields, for on-shell kinematics, only two terms at an order lower than $\mathcal{O}(\hbar^2)$:
\begin{equation}
\hbar^3\mathcal{A}^{ab}_{1/2-0} = \frac{2g^2}{\wn q^2} \left((p_1\cdot p_2)\delta^{ab} - \frac{i}{m_1}  \epsilon( p{_1}, \wn q, p_2, s^{ab}_1) + \mathcal{O}(\hbar^2)\right) \tilde{\newT}_1\cdot\tilde{\newT}_2\,,\label{eqn:spinorYMamp}
\end{equation}
where here and below we adopt the short-hand notation ${s_1}^{\mu}_{ab} = {s_1}^{\mu}_{ab}(p_1)$ and
\[
\epsilon(a,b,c,d) = \epsilon_{\mu\nu\rho\sigma} a^\mu b^\nu c^{\rho} d^{\sigma} \,,
\qquad \epsilon_\mu(a,b,c) = \epsilon_{\mu\nu\rho\sigma} a^\nu b^\rho c^\sigma\,.
\]
Upon substitution into the impulse in \eqn~\eqref{eqn:impulseGeneralTerm1classicalLO} or angular impulse in \eqn~\eqref{eqn:limAngImp} the apparently singular denominator in the $\hbar\rightarrow 0$ limit is cancelled. It is only these quantities, not the amplitudes, that are classically well defined and observable.

\subsubsection*{Vector-scalar}

Now consider scattering a massive vector rather than spinor. The minimally coupled gauge interaction can be obtained by applying the Higgs mechanism to the Yang--Mills Lagrangian\footnote{Regardless of minimal coupling, for vector states with masses generated in this way the classical value of $g_L=2$ \cite{Chung:2018kqs,Ferrara:1992yc}.}, which when added to the scalar Lagrangian $\mathcal{L}_0$ yields the tree-level amplitude
\begin{multline}
i\mathcal{A}^{ij}_{1-0} = -\frac{ig^2}{2\hbar q^2} \varepsilon{_i^*}^\mu(p_1+q) \varepsilon^\nu_j(p_1) \left(\eta_{\mu\nu}(2p_1+q)_\lambda - \eta_{\nu\lambda}(p_1-q)_\mu \right.\\\left. - \eta_{\lambda\mu}(2q+p_1)_\nu \right)(2p_2 - q)^\lambda\, \tilde{\newT}_1\cdot \tilde{\newT}_2\,.
\end{multline}
To obtain the classically significant pieces of this amplitude we must once more expand the product of on-shell tensors, in this case the polarisation vectors. In the classical limit we can again consider the outgoing polarisation vector as being infinitesimally boosted, so $\varepsilon{_i^*}^\mu(p_1 + \hbar\wn q) \sim \varepsilon{_i^*}^\mu(p_1) + \Delta\varepsilon{_i^*}^\mu(p_1)$. 

However, from spin 1 particles we expect to be able to probe $\mathcal{O}(s^2)$, or quadrupole, terms \cite{Vaidya:2014kza,Guevara:2017csg,Guevara:2018wpp}. Therefore it is salient to expand the Lorentz boost to two orders in the Lorentz parameters $w_{\mu\nu}$, so under infinitesimal transformations we take
\begin{equation}
\varepsilon_i^\mu(p) \mapsto \Lambda^\mu{ }_\nu\, \varepsilon_i^\nu(p) \simeq \left(\delta^\mu{ }_\nu - \frac{i}{2} w_{\rho\sigma} (\Sigma^{\rho\sigma})^\mu{ }_\nu - \frac18 \left((w_{\rho\sigma} \Sigma^{\rho\sigma})^2\right)^\mu{ }_\nu \right) \!\varepsilon_i^\nu(p)\,,
\end{equation}
where $(\Sigma^{\rho\sigma})^\mu{ }_\nu = i\left(\eta^{\rho\mu} \delta^\sigma{ }_\nu - \eta^{\sigma\mu} \delta^\rho{}_\nu\right)$. Since the kinematics are again identical to those used to derive \eqn~\eqref{eqn:LorentzParameters}, we get
\begin{equation}
\varepsilon{_i^*}^\mu(p_1 + \hbar\wn q)\, \varepsilon^\nu_j(p_1) = \varepsilon{_i^*}^\mu \varepsilon^\nu_j - \frac{\hbar}{m_1^2} (\wn q \cdot \varepsilon_i^*) p_1^\mu \varepsilon^\nu_j - \frac{\hbar^2}{2 m_1^2} (\wn q\cdot\varepsilon_i^*) \wn q^\mu \varepsilon^\nu_j + \mathcal{O}(\hbar^3)\,,\label{eqn:vectorS}
\end{equation}
where now $\varepsilon_i$ will always be a function of $p_1$, so in the classical limit $\varepsilon_i^*\cdot p_1 = \varepsilon_i\cdot p_1=0$. Using this expression in the full amplitude, the numerator becomes
\begin{multline}
n_{ij} = 2(p_1\cdot p_2)(\varepsilon_i^*\cdot\varepsilon_j) - 2\hbar (p_2\cdot\varepsilon_i^*)(\wn q\cdot\varepsilon_j) + 2\hbar (p_2\cdot\varepsilon_j)(\wn q\cdot\varepsilon_i^*) \\ + \frac{1}{m_1^2}\hbar^2 (p_1\cdot p_2)(\wn q\cdot\varepsilon_i^*)(\wn q\cdot\varepsilon_j) + \frac{\hbar^2}{2}\wn q^2 (\varepsilon_i^*\cdot \varepsilon_j) + \mathcal{O}(\hbar^3)\,.
\end{multline}
How the spin vector enters this expression is not immediately obvious, and relies on Levi--Civita tensor identities. At $\mathcal{O}(\hbar)$, $\epsilon^{\delta\rho\sigma\nu} \epsilon_{\delta\alpha\beta\gamma} = -3!\, \delta^{[\rho}{ }_\alpha \delta^{\sigma}{ }_\beta \delta^{\nu]}{ }_\gamma$ leads to
\begin{align}
\hbar (p_2\cdot\varepsilon{_i^*})(\wn q \cdot \varepsilon_j) - \hbar (p_2 \cdot \varepsilon_j)(\wn q \cdot \varepsilon_i^*) = \frac{\hbar}{m_1^2}&p_1^\rho \wn q^\sigma p_2^\lambda \epsilon_{\delta\rho\sigma\lambda}\epsilon^{\delta\alpha\beta\gamma}{\varepsilon_i^*}_\alpha {\varepsilon_j}_\beta {p_1}_\gamma \nonumber \\ &\equiv  -\frac{i}{m_1} \epsilon(p_1, \wn q, p_2, {s_1}_{ij})\,,
\end{align}
where again we are able to identify the spin 1 polarisation vector calculated in \eqn~\eqref{eqn:vectorSpinVec} and introduce it into the amplitude. There is also a spin vector squared contribution entering at $\mathcal{O}(\hbar^2)$; observing this is reliant on applying the identity $\epsilon^{\mu\nu\rho\sigma} \epsilon_{\alpha\beta\gamma\delta} = -4!\, \delta^{[\mu}{ }_\alpha \delta^{\nu}{ }_\beta \delta^{\rho}{ }_\gamma \delta^{\sigma]}{ }_\delta$ and the expression in \eqn~\eqref{eqn:vectorSpinVec} to calculate
\begin{equation}
\sum_k \left(\wn q \cdot s_1^{ik}\right) (\wn q \cdot s_1^{kj}) = -\hbar^2 (\wn q \cdot \varepsilon_i^*) (\wn q\cdot \varepsilon_j) - \hbar^2 \wn q^2 \delta_{ij} + \mathcal{O}(\hbar^3)\,.
\end{equation}
This particular relationship is dependent on the sum over helicities $\sum_h {\varepsilon^*_h}^\mu \varepsilon^\nu_h = -\eta^{\mu\nu} + \frac{p_1^\mu p_1^\nu}{m_1^2}$ for massive vector bosons, an additional consequence of which is that $\varepsilon_i^*\cdot \varepsilon_j = -\delta_{ij}$. Incorporating these rewritings of the numerator in terms of spin vectors, the full amplitude is
\begin{multline}
\hbar^3\mathcal{A}^{ij}_{1-0} = \frac{2g^2}{\wn q^2}\left((p_1\cdot p_2)\delta^{ij} - \frac{i}{m_1} \epsilon(p_1, \wn q, p_2, s_1^{ij}) + \frac{1}{2 m_1^2}(p_1\cdot p_2)(\wn q\cdot s_1^{ik}) (\wn q \cdot s_1^{kj}) \right.\\\left. + \frac{\hbar^2 \wn q^2}{4m_1^2}\left(2p_1\cdot p_2 + m_1^2\right) + \mathcal{O}(\hbar^3)\right) \tilde{\newT}_1\cdot \tilde{\newT}_2\,.
\end{multline}
The internal sum over spin indices in the $\mathcal{O}(s^2)$ term will now always be left implicit. In classical observables we can also drop the remaining $\mathcal{O}(\hbar^2)$ term, as this just corresponds to a quantum correction from contact interactions.

\subsection{Gravity amplitudes}

Rather than re-calculate these amplitudes in perturbative gravity, let us apply the double copy\footnote{One can easily verify that direct calculations with graviton vertex rules given in \cite{Holstein:2008sx} reproduce our results.}. Note that for massive states with spin this ability is reliant on our gauge theory choice of $g_L=2$, as was noted in \cite{Goldberger:2017ogt}. Only with this choice is the gravitational theory consistent with the low energy spectrum of string theory \cite{Goldberger:2017ogt,Chung:2018kqs}, of which the double copy is an intrinsic feature.

For amplitudes in the LO impulse the Jacobi identity is trivial, as we saw in section~\ref{sec:LOimpulse}. We can therefore simply replace colour factors with the desired numerator. In particular, if we replace the colour factor in the previous spin $s$--spin 0 Yang--Mills amplitudes with the scalar numerator from \eqn~\eqref{eqn:scalarYMamp} we will obtain a spin $s$--spin 0 gravity amplitude, as the composition of little group irreps is simply $(\mathbf{2s + 1})\otimes\mathbf{1}=\mathbf{2s + 1}$. Using the scalar numerator ensures that the spin index structure passes to the gravity theory unchanged. Thus we can immediately obtain that the classically significant part of the spin $1/2$--spin 0 gravity amplitude is
\begin{equation}
\hbar^3 \mathcal{M}^{ab} = -\left(\frac{\kappa}{2}\right)^2\frac{4}{\wn q^2} \bigg[(p_1\cdot p_2)^2\delta^{ab} - \frac{i}{m_1}(p_1\cdot p_2)\, \epsilon(p_1, \wn q, p_2, s_1^{ab}) + \mathcal{O}(\hbar^2)\bigg]\,,\label{eqn:spinorScalarGravAmp}
\end{equation}
while that for spin 1--spin 0 scattering is
\begin{multline}
\hbar^3\mathcal{M}^{ij} = -\left(\frac{\kappa}{2}\right)^2 \frac{4}{\wn q^2} \left[(p_1\cdot p_2)^2\delta^{ij} - \frac{i}{m_1}(p_1\cdot p_2)\, \epsilon(p_1, \wn q, p_2, s_1^{ij}) \right.\\\left. + \frac{1}{2 m_1^2}(p_1\cdot p_2)^2(\wn q\cdot s_1^{ik}) (\wn q \cdot s_1^{kj}) + \mathcal{O}(\hbar^2)\right]. \label{eqn:vectorScalarGravAmp}
\end{multline}
Notice that the $\mathcal{O}(s)$ parts of these amplitudes are exactly equal, up to the different spin indices. This is a manifestation of gravitational universality: the gravitational coupling to the spin dipole should be independent of the spin of the field, precisely as we observe.

We have deliberately not labelled these as Einstein gravity amplitudes, because the gravitational modes in our amplitudes contain both gravitons and scalar dilatons. To see this, let us re-examine the factorisation channels in the $t$ channel cut, but now for the vector amplitude:
\[
\lim\limits_{\wn q^2 \rightarrow 0} \left(\wn q^2 \hbar^3 \mathcal{M}^{ij}\right) &= -4\left(\frac{\kappa}{2}\right)^2\, \bigg(p_1^\mu p_1^{\tilde{\mu}} \delta^{ij} - \frac{i}{m_1}p_1^\mu \epsilon^{\tilde{\mu}\rho\sigma\delta} {p_1}_\rho \wn q_\sigma {s_1}_\delta^{ij} \\ & \qquad\qquad\qquad + \frac{1}{2m_1^2} (\wn q\cdot s_1^{ik})(\wn q\cdot s_1^{kj}) p_1^\mu p_1^{\tilde{\mu}}\bigg) \mathcal{P}^{(4)}_{\mu\tilde{\mu}\nu\tilde{\nu}}\, p_2^\nu p_2^{\tilde{\nu}} \\ & - 4\left(\frac{\kappa}{2}\right)^2\left(p_1^\mu p_1^{\tilde{\mu}} \delta^{ij} + \frac{(\wn q\cdot s_1^{ik})(\wn q\cdot s_1^{kj})}{2 m_1^2} p_1^\mu p_1^{\tilde{\mu}}\right) \mathcal{D}^{(4)}_{\mu\tilde{\mu}\nu\tilde{\nu}}\, p_2^\nu p_2^{\tilde{\nu}}\,,
\]
where we have utilised the de-Donder gauge graviton and dilaton projectors from equation~\eqref{eqn:gravityProjectors}. As for the scalar case, the pure Einstein gravity amplitude for classical spin 1--spin 0 scattering can just be read off as the part of the amplitude contracted with the graviton projector. We find that
\begin{multline}
\hbar^3\mathcal{M}^{ij}_{1-0} = -\left(\frac{\kappa}{2}\right)^2 \frac{4}{\wn q^2} \left[\left((p_1\cdot p_2)^2 - \frac12 m_1^2 m_2^2\right)\delta^{ij} - \frac{i}{m_1}(p_1\cdot p_2)\,\epsilon(p_1, \wn q, p_2, s_1^{ij}) \right.\\\left. + \frac{1}{2 m_1^2}\left((p_1\cdot p_2)^2 - \frac12 m_1^2 m_2^2\right)(\wn q\cdot s_1^{ik}) (\wn q \cdot s_1^{kj}) + \mathcal{O}(\hbar^2)\right]. \label{eqn:vectorGravAmp}
\end{multline}
The spinor--scalar Einstein gravity amplitude receives the same correction to the initial, scalar component of the amplitude. 

Note that dilaton modes are coupling to the scalar monopole and $\mathcal{O}(s^2)$ quadrapole terms in the gravity amplitudes, but not to the $\mathcal{O}(s)$ dipole component. We also do not find axion modes, as observed in applications of the classical double copy to spinning particles \cite{Li:2018qap,Goldberger:2017ogt}, because axions are unable to couple to the massive external scalar.

\section{Black hole scattering observables from amplitudes}
\label{sec:KerrCalcs}

We are now armed with a set of classical tree-level amplitudes and formulae for calculating the momentum impulse $\Delta p_1^\mu$ and angular impulse $\Delta s_1^\mu$ from them. We also already have a clear target where the analogous classical results are known: the results for 1PM scattering of spinning black holes found in \cite{Vines:2017hyw}. 

Given our amplitudes only reach the quadrupole level, we can only probe lower order terms in the expansion of \eqn~\eqref{eqn:KerrDeflections}. Expanding in the rescaled spin $a_1^\mu$, and setting $a_2^\mu\to0$, the momentum impulse is
\begin{multline}
\Delta p_1^{\mu} = \frac{2 G m_1m_2}{\sqrt{\gamma^2 - 1}} \left\{(2\gamma^2 - 1) \frac{{b}^\mu}{b^2} + \frac{2\gamma}{b^4} \Big( 2{b}^\mu {b}^\nu-b^2\Pi^{\mu\nu}\Big)\, \epsilon_{\nu\rho}(u_1, u_2)\,  a_1^\rho\right.
\\
\left. - \frac{2\gamma^2 - 1}{{b}^6} \Big(4b^\mu b^\nu b^\rho-3b^2 b^{(\mu}\Pi^{\nu\rho)}\Big)a_{1\nu}a_{1\rho} + \mathcal{O}(a^3) \right\}+\mathcal O(G^2)\,, \label{eqn:JustinImpResult}
\end{multline}
where $\Pi^\mu{}_\nu$ is the projector into the plane orthogonal to $u_1^\mu$ and $u_2^\mu$ from \eqref{eqn:projector}.
Meanwhile the angular impulse to the same order is
\begin{multline}
\Delta s_1^{\mu} =-u_1^\mu a_{1\nu}\Delta p_1^\nu -\frac{2G m_1m_2}{\sqrt{\gamma^2-1}}\left\{
2\gamma \epsilon^{\mu\nu\rho\sigma}  u_{1\rho}\, \epsilon_{\sigma} (u_1,u_2, b) \frac{a_{1\nu} }{b^2}\right.
\\
\left. - \frac{2\gamma^2 - 1}{b^4} \epsilon^{\mu\nu\kappa\lambda} u_{1\kappa}  \Big( 2{b}_\nu {b}_\rho-b^2\Pi_{\nu\rho} \Big)a_{1\lambda} a_1^\rho 
+ \mathcal{O}(a^3) \right\}+\mathcal O(G^2)\,. \label{eqn:JustinSpinResult}
\end{multline}
In this section we demonstrate that both of these results can be recovered by using the classical pieces of our Einstein gravity amplitudes.

\subsection{Momentum impulse}

To calculate the momentum impulse we substitute $\mathcal{M}_{1-0}$ into the general expression in \eqn~\eqref{eqn:impulseGeneralTerm1classicalLO}. Following the prescription in \sect{sec:classicalLimit}, the only effect of the momentum integrals in the expectation value is to set $p_\alpha \rightarrow m_\alpha u_\alpha$ in the classical limit. This then reduces the double angle bracket to the single expectation value over the spin states:
\begin{equation}
\begin{aligned}
\Delta p_1^{\mu,(0)} &= -i m_1 m_2 \left(\frac{\kappa}{2}\right)^2 \!\int\!\dd^4\wn q\, \del(u_1\cdot\wn q) \del(u_2\cdot\wn q) e^{-ib\cdot\wn q} \frac{\wn q^\mu}{\wn q^2} \\&\times \left\langle \frac12(2\gamma^2 - 1) - \frac{i \gamma}{m_1} \epsilon(u_1, \wn q, u_2, s_1) + \frac{2\gamma^2 - 1}{4m_1^2} (\wn q\cdot s_1) (\wn q \cdot s_1) \right\rangle\\
&\equiv -4i m_1 m_2 \pi G \bigg((2\gamma^2 - 1) I^\mu  -  2i\gamma u_1^\rho u_2^\nu \epsilon_{\rho\sigma\nu\delta}\, \big\langle a_1^\delta \big\rangle I^{\mu\sigma} \\ &\hspace{60mm} + \frac{2\gamma^2 - 1}{2} \big\langle {a_1}_\nu {a_1}_\rho \big\rangle I^{\mu\nu\rho} \bigg)\,,
\end{aligned}
\end{equation}
where we have rescaled $a^\mu = s^\mu/m$ and defined three integrals of the general form
\begin{equation}
I^{\mu_1\cdots \mu_n} = \int\!\dd^4\wn q\, \del(u_1\cdot\wn q) \del(u_2\cdot \wn q) \frac{e^{-ib\cdot\wn q}}{\wn q^2} \wn q^{\mu_1} \cdots \wn q^{\mu_n}\,.\label{eqn:defOfI}
\end{equation}

The lowest rank integral of this type was evaluated in chapter~\ref{chap:impulse}, with the result
\begin{equation}
I^\mu = \frac{i}{2\pi \sqrt{\gamma^2 - 1}} \frac{{b}^\mu}{b^2}\,.\label{eqn:I1result}
\end{equation}
To evaluate the higher rank examples, note that the results must lie in the plane orthogonal to the four velocities. This plane is spanned by the impact parameter $b^\mu$, and the projector $\Pi^\mu{ }_\nu$ defined in \eqn~\eqref{eqn:projector}. Thus, for example,
\begin{equation}
I^{\mu\nu} = \alpha_2 b^\mu b^\nu + \beta_2 \Pi^{\mu\nu}\,.
\end{equation}
Given that we are working away from the threshold value $b = 0$, the left hand side is traceless and $\beta_2 = - \alpha_2\, b^2 /2$. Then contracting both sides with $b_\nu$, one finds
\begin{equation}
\alpha_2 b^2\, b^\mu = 2\int\!\dd^4\wn q\, \del(u_1\cdot\wn q) \del(u_2\cdot \wn q) \frac{e^{-ib\cdot\wn q}}{\wn q^2} \wn q^{\mu} (b\cdot\wn q) = \frac{1}{\pi \sqrt{\gamma^2 - 1}} \frac{b^\mu}{b^2}\,,
\end{equation} 
where we have used the result of \eqn~\eqref{eqn:I1result}. Thus the coefficient $\alpha_2$ is uniquely specified, and we find
\begin{equation}
I^{\mu\nu} = \frac{1}{\pi b^4 \sqrt{\gamma^2 - 1}} \left(b^\mu b^\nu - \frac12 b^2 \Pi^{\mu\nu} \right)\label{eqn:I2result}.
\end{equation}
Following an identical procedure for $I^{\mu\nu\rho}$, we can then readily determine that
\begin{equation}
I^{\mu\nu\rho} = -\frac{4i}{\pi b^6 \sqrt{\gamma^2 - 1} } \left(b^\mu b^\nu b^\rho - \frac34 b^2 b^{(\mu} \Pi^{\nu\rho)} \right).\label{eqn:I3result}
\end{equation}

Substituting the integral results into the expression for the leading order classical impulse, and expanding the projectors from \eqn~\eqref{eqn:projector}, then leads to
\begin{multline}
\Delta p_1^{\mu,(0)} = \frac{2G m_1 m_2}{\sqrt{\gamma^2 - 1}}\left((2\gamma^2 - 1) \frac{{b}^\mu}{b^2} + \frac{2\gamma}{b^4} ( 2 b^\mu {b}^\alpha-b^2\Pi^{\mu\alpha}) \epsilon_{\alpha\rho}(u_1, u_2) \big\langle a_1^\rho \big\rangle  \right.
\\
\left.- \frac{2\gamma^2 - 1}{{b}^6} (4b^\mu b^\nu b^\rho-3b^2 b^{(\mu}\Pi^{\nu\rho)})\langle a_{1\nu}a_{1\rho}\rangle\right).\label{eqn:QuantumImpRes}
\end{multline}
Comparing with \eqn~\eqref{eqn:JustinImpResult} we observe an exact match, up to the appearance of spin state expectation values, between our result and the $\mathcal{O}(a^2)$ expansion of the result for spinning black holes from \cite{Vines:2017hyw}. 

\subsection{Angular impulse}

Our expression, equation~\eqref{eqn:limAngImp}, for the classical leading-order angular impulse naturally has two parts: one term has a commutator while the other term does not. For clarity we will handle these two parts separately, beginning with the term without a commutator, which we will call the direct term.

\subsubsection*{The direct term}

Substituting our $\mathcal{O}(s^2)$ Einstein gravity amplitude, equation~\eqref{eqn:vectorGravAmp}, into the direct part of the general angular impulse formula, we find
\begin{align}
&\Delta s_1^{\mu,(0)}\big|_{\textrm{direct}} \nonumber
\equiv
\Lexp i\! \int\!\dd^4\wn q\,\del(2p_1\cdot\wn q) \del(2p_2\cdot\wn q)e^{-ib\cdot\wn q} \bigg(-\hbar^3 \frac{p_1^\mu}{m_1^2} \qb \cdot \s_{1} (p_1) \mathcal{M}_{1-0} \bigg) \Rexp \\
&\;= \Lexp \frac{i \kappa^2}{m_1^2} \int\! \dd^4\wn q\,\del(2p_1\cdot\wn q) \del(2p_2\cdot\wn q) \frac{e^{-ib\cdot\wn q}}{\qb^2}\, p_1^\mu \wn q \cdot \s_1(p_1) \\ &\qquad\times\bigg(\bigg((p_1\cdot p_2)^2  - \frac{1}{2}m_1^2m_2^2\bigg)  - \frac{i}{m_1}(p_1\cdot p_2) \, \epsilon(p_1, \wn q, p_2,s_1) \bigg)+ \mathcal{O}(s^3) \Rexp\,.\nonumber
\end{align}
As with the momentum impulse, we can reduce the double angle brackets to single, spin state, angle brackets by replacing $p_\alpha \rightarrow m_\alpha u_\alpha$, so that
\begin{equation}
\hspace{-3pt}\Delta s^{\mu,(0)}_1\big|_{\textrm{direct}}\!\! = 4\pi G m_2\, u_1^\mu\!  \left(\! i  \left(2\gamma^2 - 1\right)\! \langle s_1^\nu \rangle I_\nu + \frac{2}{m_1} \gamma 
\, u_1^\alpha u_2^\gamma \epsilon_{\alpha\beta\gamma\delta} \langle s_{1\nu}  s_1^{\delta} \rangle \, I^{\nu\beta} \!\right),
\end{equation}
where the integrals are again defined by \eqn~\eqref{eqn:defOfI}. We can now just substitute our previous evaluations of these integrals, equations~\eqref{eqn:I1result} and~\eqref{eqn:I2result}, to learn that
\begin{multline}
\Delta a_1^{\mu,(0)}\big|_{\textrm{direct}} = -\frac{2G m_2}{\sqrt{\gamma^2 -1}} u_1^\mu \left((2\gamma^2 - 1)\frac{{b}_\nu}{b^2} \spinExp{a_1^\nu} \right. \\ \left. + \frac{2\gamma}{b^4} \left( 2{b}^\nu {b}^\alpha-{b^2}\Pi^{\nu\alpha}\right) \epsilon_{\alpha\rho} (u_1, u_2) \spinExp{{a_1}_\nu a_1^\rho} \right).\label{eqn:linPiece}
\end{multline}

\subsubsection*{The commutator term}

Now we turn to the commutator piece of \eqn~\eqref{eqn:limAngImp}. The scalar part of our Einstein gravity amplitude, equation~\eqref{eqn:vectorGravAmp}, has diagonal spin indices, so its commutator vanishes. We encounter two non-vanishing commutators:
\[
[\s_1^\mu, \s_1^\nu]&=\frac{i \hbar}{m_1} \, \epsilon^{\mu\nu}(s_{1} , p_{1}) \,,\\
[\s_1^\mu, \qb \cdot \s_1 \, \qb\cdot \s_1] &= \frac{2 i \hbar}{m_1} \, \qb \cdot s_1 \, \epsilon^{\mu} (\qb, s_{1}, p_1) + \mathcal{O}(\hbar^2) \,,
\]
omitting a term which is higher order. Using these expressions in the commutator term, the result is
\[
&\Delta s_1^{\mu, (0)}|_\textrm{com} = i\, \Lexp \int\! \dd^4 \qb \, \del(2p_1 \cdot \qb) \del(2p_2 \cdot \qb) e^{-ib \cdot \qb} \, \hbar^2 [s^\mu(p_1), \mathcal{M}_{1-0}] \Rexp \\
&= i \kappa^2 \Lexp \int\! \dd^4 \qb \, \del(2p_1 \cdot \qb) \del(2p_2 \cdot \qb) \frac{e^{-ib \cdot \qb}}{\qb^2} \bigg( (p_1 \cdot p_2) \, \epsilon(p_1, \qb, p_2)_\sigma\epsilon^{\mu\nu\rho\sigma} s_{1 \nu} \frac{p_{1 \rho}}{m_1^2}  \\
& \qquad\qquad\qquad\qquad- \frac{i}{m_1^3} \left( (p_1\cdot p_2)^2 - \frac12 m_1^2 m_2^2\right) \qb \cdot \s_1 \, \epsilon^{\mu\nu}(\qb, p_1) s_{1\nu} \bigg) \Rexp\,.
\]
As is familiar by now, we evaluate the integrals over the momentum-space wavefunctions by setting $p_\alpha = m_\alpha u_\alpha$, but expectation values over the spin-space wavefunctions remain. The result can be organised in terms of the integrals $I^\alpha$ and $I^{\alpha\beta}$ defined in equation~\eqref{eqn:defOfI}:
\begin{multline}
\Delta s_1^{\mu, (0)}|_\textrm{com} = 2\pi i \, G m_2 \bigg( 4\gamma \epsilon^{\mu\nu\rho\sigma} \langle s_{1\,\nu} \rangle u_{1\,\rho} \epsilon_{\sigma\alpha} (u_1, u_2) I^\alpha 
\\ - \frac{2 i}{m_1} (2\gamma^2 - 1) \epsilon^{\mu\nu\rho\sigma} u_{1\, \rho} \langle s_{1\, \sigma} {s_{1}}^\alpha \rangle I_{\alpha\nu}
\bigg) \, .
\end{multline}
Finally, we perform the integrals using equations~\eqref{eqn:I1result} and~\eqref{eqn:I2result}, rescale the spin vector to $a_1^\mu$ and combine the result with the direct contribution in \eqn~\eqref{eqn:linPiece}, to find that the angular impulse at $\mathcal{O}(a^2)$ is
\[
\hspace{-4pt}\Delta s_1^{\mu,(0)}
= -\frac{2Gm_1 m_2}{\sqrt{\gamma^2 -1}}& \bigg\{(2\gamma^2 - 1) u_1^\mu \frac{{b}_\nu}{b^2} \spinExp{a_1^\nu} + \frac{2\gamma}{b^2} \epsilon^{\mu\nu\rho\sigma} \spinExp{a_{1\,\nu}} u_{1\,\rho} \epsilon_{\sigma}(u_1, u_2, b) \\ &- \frac{2\gamma}{b^2} u_1^\mu \left(\eta^{\nu\alpha} - \frac{2{b}^\nu {b}^\alpha}{b^2}\right) \epsilon_{\alpha\rho} (u_1, u_2) \spinExp{{a_1}_\nu a_1^\rho} \\ & + \frac{(2\gamma^2 - 1)}{b^2} \epsilon^{\mu\nu\rho\sigma} {u_1}_\rho \spinExp{{a_1}_\sigma {a_1}_\lambda} \left(\Pi^\lambda{}_\nu - \frac{2{b}_\nu {b}^\lambda}{b^2}\right)  
 \bigg\}\,.
\label{eqn:QuantumAngImpRes}
\]
This final result agrees with the classical result of equation~\eqref{eqn:JustinSpinResult}, modulo the remaining spin expectation values.

\section{Discussion}
\label{sec:angImpDiscussion}

Starting from a quantum field theory for massive spinning particles with arbitrary long-range interactions (mediated e.g.\ by gauge bosons or gravitons), we have followed a careful analysis of the classical limit $(\hbar\to0)$ for long-range scattering of spatially localised wavepackets.   We have thereby arrived at fully relativistic expressions for the angular impulse, the net change in the intrinsic angular momentum of the massive particles, due to an elastic two-body scattering process.  This, our central result of the chapter, expressed in terms of on-shell scattering amplitudes, is given explicitly at leading order in the coupling by  \eqref{eqn:limAngImp}.  Our general formalism places no restrictions on the order in coupling, and the expression \eqref{eqn:spinShift} for the angular impulse, like its analogues for the momentum and colour impulse found in earlier chapters, should hold at all orders. Since the publication of this formalism in ref.~\cite{Maybee:2019jus}, general, fully classical formulae have been proposed in \cite{Bern:2020buy,Kosmopoulos:2021zoq} for the angular impulse, expressed in terms of commutators of a modified eikonal phase. It would be very interesting to establish a firm relationship between this proposal and our results. Our 1PM calculations of the spin-squared parts of the fully covariant momentum and angular impulses have also been confirmed using post--Minkowskian EFT methods, and extended to next-to-leading-order in the coupling \cite{Liu:2021zxr}, equivalent to a 1-loop computation.

In this chapter we applied our general results to the examples of a massive spin 1/2 or spin 1 particle (particle 1) exchanging gravitons with a massive spin 0 particle (particle 2), imposing minimal coupling.  The results for the linear and angular impulses for particle 1, $\Delta p_1^\mu$ and $\Delta s_1^\mu$, due to its scattering with the scalar particle 2, are given by \eqref{eqn:QuantumImpRes} and \eqref{eqn:QuantumAngImpRes}.  These expressions are valid to linear order in the gravitational constant $G$, or to 1PM order, having arisen from the tree level on-shell amplitude for the two-body scattering process.  By momentum conservation (in absence of radiative effects at this order), $\Delta p_2^\mu=-\Delta p_1^\mu$, and the scalar particle has no intrinsic angular momentum, $s_2^\mu=\Delta s_2^\mu=0$.  The spin 1/2 case provides the terms through linear order in the rescaled spin $a_1^\mu=s_1^\mu/m_1$, and the spin 1 case yields the same terms through linear order plus terms quadratic in $a_1^\mu$.

Our final results \eqref{eqn:QuantumImpRes} and \eqref{eqn:QuantumAngImpRes} from the quantum analysis are seen to be in precise agreement with the results \eqref{eqn:JustinImpResult} and \eqref{eqn:JustinSpinResult} from \cite{Vines:2017hyw} for the classical scattering of a spinning black hole with a non-spinning black hole, through quadratic order in the spin --- except for the appearance of spin-state expectation values $\langle a_1^\mu\rangle$ and $\langle a_1^\mu a_1^\nu\rangle$ in the quantum results replacing $a_1^\mu$ and $a_1^\mu a_1^\nu$ in the classical result.  For any quantum states of a finite-spin particle, these expectation values cannot satisfy the  appropriate properties of their classical counterparts, e.g., $\langle a^\mu a^\nu\rangle \ne \langle a^\mu\rangle \langle a^\nu\rangle$.  Furthermore, we know from section~\ref{sec:classicalSingleParticleColour} that the intrinsic angular momentum of a quantum spin-$s$ particle scales like $\langle s^\mu\rangle=m\langle a^\mu\rangle\sim s\hbar$, and we would thus actually expect any spin effects to vanish in a classical limit where we take $\hbar\to0$ at fixed spin quantum number $s$.

For a fully consistent classical limit yielding non-zero contributions from the intrinsic spin we of course would need to take the spin $s\to\infty$ as $\hbar\to0$, so as to keep $\langle s^\mu\rangle\sim s\hbar$ finite. However, the expansions in spin operators of the minimally coupled amplitudes and impulses, expressed in the forms we have derived here, are found to be universal, in the sense that going to higher spin quantum numbers $s$ continues to reproduce the same expressions at lower orders in the spin operators.  We have seen this explicitly here for the linear-in-spin level, up to spin 1. 
This pattern was confirmed to hold for minimally coupled gravity amplitudes for arbitrary spin $s$ in ref.~\cite{Guevara:2019fsj}. The authors showed that applying the formalism we have presented here to amplitudes in the limit $s\to\infty$ fully reproduces the full Kerr 1PM observables listed in \eqref{eqn:KerrDeflections}, up to spin state expectation values. Using the coherent states in section~\ref{sec:classicalSingleParticleColour}, one can indeed then take the limit where $\langle a^\mu a^\nu\rangle = \langle a^\mu\rangle \langle a^\nu\rangle$ and so forth. The precise forms of $1/s$ corrections to the higher-multipole couplings were discussed in \cite{Chung:2019duq}.

Our formalism provides a direct link between gauge-invariant quantities, on-shell amplitudes and classical asymptotic scattering observables, with generic incoming and outgoing states, for relativistic spinning particles.  It is tailored to be combined with powerful modern techniques for computing relevant amplitudes, such as unitarity methods and the double copy.  Already, with our examples at the spin 1/2 and spin 1 levels, we have explicitly seen that it produces new evidence (for generic spin orientations, and without taking the non-relativistic limit) for the beautiful correspondence between classical spinning black holes and massive spinning quantum particles minimally coupled to gravity, as advertised in chapter~\ref{chap:intro}. Let us now turn to a striking application of this on-shell relationship.

%% file: chapter6/chapter6.tex
\chapter{A worldsheet for Kerr}
\label{chap:worldsheet}

\section{NJ shifts from amplitude exponentiation}
\label{sec:NJintro}

The Newman--Janis (NJ) shift in equation~\eqref{eqn:NJshift} is a remarkable exact property of the Kerr solution, relating it to the simpler non-spinning Schwarzschild solution via means of a complex translation.
A partial understanding of this phenomenon is available in the context of minimally-coupled scattering amplitudes. 
Rather than consider field representations of a definite spin as in the last chapter, here it is more instructive to follow Arkani--Hamed, Huang and Huang~\cite{Arkani-Hamed:2017jhn}, and consider massive little group representations of arbitrary spin. Specifically, we will introduce spinors $\ket{p_I}$ and $|p_I]$ with $SU(2)$ little group indices $I=1,2$, such that that the momentum is written\footnote{We adopt the conventions of ref.~\cite{Chung:2018kqs}.}
\[
p^\mu = \frac12 \epsilon^{IJ} \bra{p_J} \sigma^\mu | p_I] = \frac12 \bra{p^I} \sigma^\mu | p_I] \,.
\]
We raise and lower the little group indices $I, J, \ldots$ with two-dimensional Levi--Civita tensors, as usual. The $\sigma^\mu$ matrices
are a basis of the Clifford algebra, and we use the common choice
\[
\sigma^\mu = (1, \sigma_x, \sigma_y, \sigma_z) \,.
\]
Minimally coupled three-point amplitudes take a particularly simple form when written in these spinor helicity representations. In particular, the amplitude for a spin $s$ particle of mass $m$ and charge $Q$ absorbing a photon with positive-helicity polarisation vector $\varepsilon_k^+$ is simply given by \cite{Chung:2018kqs,Arkani-Hamed:2019ymq}
\begin{equation}
\begin{tikzpicture}[scale=1.0, baseline={([yshift=-\the\dimexpr\fontdimen22\textfont2\relax] current bounding box.center)}, decoration={markings,mark=at position 0.6 with {\arrow{Stealth}}}]
\begin{feynman}
\vertex (v1);
\vertex [below left = 0.66 and 1 of v1] (i1);
\vertex [below left = 0.5 and 0.9 of v1] (i11) {$1^s$};
\vertex [below right= 0.66 and 1 of v1] (o1);
\vertex [below right = 0.5 and 0.9 of v1] (o11) {$2^s$};
\vertex [above = 1 of v1] (v2) ;
\vertex [above = 1 of v1] (v22) {$ {}_{+}$};
\draw [postaction={decorate}] (i1) -- (v1);
\draw [postaction={decorate}] (v1) -- (o1);
\diagram*{(v2) -- [photon,momentum=\(k\)] (v1)};
\filldraw [color=black] (v1) circle [radius=2pt];
\end{feynman}	
\end{tikzpicture} = - \frac{Q}{\sqrt{\hbar}}\, \varepsilon^{+}_k\cdot(p_1 + p_2)\,\frac{\langle p_1^I \, p_2^J\rangle^{\odot 2s}}{m^{2s}}\,,
\end{equation}
where the exponent includes a prescription to symmetrise over the little group indices. We can deduce the classical limit of this generic amplitude in the same manner as in section~\ref{sec:amplitudes} for specific field representations. The photon momentum $k$ behaves as a wavenumber in the limit, so the outgoing momentum $p_2$ can be viewed as an infinitesimal Lorentz boost, with generators~\eqref{eqn:LorentzParameters}. In terms of spinors,
\begin{equation}
|p_2^I\rangle = |p_1^I\rangle - \frac{\hbar}{2m_1} \wn k \cdot \sigma \,|p_1^I]\,.
\end{equation}
As in our earlier examples we expand the amplitude in the spin polarisation vector, which in these variables is in general given by \cite{Arkani-Hamed:2019ymq,Chung:2018kqs}
\begin{equation}
s^\mu_{IJ}(p) = \frac{1}{m} s\hbar \langle p_I| \sigma^\mu|p_J]\,.
\end{equation}
From our single particle wavefunctions in chapter~\ref{chap:pointParticles}, we know that as $\hbar\rightarrow0$ the spin $s$ must tend to infinity (since it is the size of the little group representation), such that the combination $s \hbar$ appearing in state expectation values is finite. Thus, in the classical limit, the three-point amplitude is
\begin{equation}
\mathcal{A}_{3,+} = - \frac{2Q}{\sqrt{\hbar}} (p_1\cdot\varepsilon_k^+) \lim_{s\rightarrow\infty} \left(\mathbb{I} - \frac{\bar k \cdot a}{2s}\right)^{2s} = - \frac{2Q}{\sqrt{\hbar}} (p_1\cdot\varepsilon_k^+) e^{-\bar k \cdot a}\,,
\end{equation}
where recall that $a^\mu = s^\mu/m$. The denominator $\hbar$ factor is a remnant of the correct normalisation for amplitude coupling constants from section~\ref{sec:RestoringHBar} --- such factors will be unimportant in this chapter and thus uniformly neglected. 

Since $-2Q(p_1\cdot\varepsilon_k^+) = \mathcal{A}_{3,+}^\textrm{Coulomb}$ is the usual QED amplitude for a scalar of charge $Q$ and momentum $p$ absorbing a positive-helicity photon, we can conclude that
\[
\mathcal{A}_{3,+}^{\sqrt{\text{Kerr}}} =
e^{-\wn k \cdot a} \mathcal{A}_{3,+}^{\text{Coulomb}} \,.
\label{eq:rootKerrAmp}
\]
Similarly, the gravitational three-point amplitude for a massive particle is
\[
\mathcal{M}_{3,+}^{{\text{Kerr}}} =
e^{-\wn k \cdot a} \mathcal{M}_{3,+}^{\text{Schwarzschild}} \,,
\]
in terms of the ``Schwarzschild'' amplitude for a scalar particle
interacting with a positive-helicity graviton of momentum $k$.
A straightforward way to establish
the connection of these amplitudes to spinning black holes
\cite{Guevara:2018wpp,Chung:2018kqs,Guevara:2019fsj}
is to then use them to compute the impulse on a scalar probe at leading-order in the Kerr background \cite{Arkani-Hamed:2019ymq}.
The calculation can be performed using the formalism of this thesis on the one hand, in particular equation~\eqref{eqn:impulseGeneralTerm1classicalLO}; or
using classical equations of motion on the other.
A direct comparison of the two approaches makes it evident
\cite{Guevara:2019fsj,Arkani-Hamed:2019ymq}
that the NJ shift of the background
is captured by the exponential factors $e^{\pm k\cdot a}$.

This connection between the NJ shift and scattering amplitudes
suggests that the NJ shift should extend beyond the exact Kerr solution
to the \emph{interactions} of spinning black holes.
Indeed, it is straightforward to scatter two Kerr particles
(by which we mean massive particles with classical spin lengths
$a_1$ and $a_2$)
off one another using amplitudes.
The purpose of this chapter is to investigate
the classical interpretation of this fact.
To do so, we turn to the classical effective theory describing the worldline interactions of a Kerr particle \cite{Porto:2005ac,Porto:2006bt,Porto:2008tb,Steinhoff:2015ksa,Levi:2015msa}.
We will see that the NJ property endows this worldline action
with a remarkable two-dimensional worldsheet structure.
The Newman--Janis story emerges via Stokes's
theorem on this worldsheet with boundary, and indeed persists
for at least the leading interactions.
We will see that novel equations of
motion, making use of the spinor-helicity formalism in a purely classical context, allow us to make the shift manifest in the leading interactions.

Our effective action is constructed only from the information in the three-point amplitudes. At higher orders, information from four-point
and higher amplitudes (or similar sources) is necessary to fully specify the effective action. Therefore our action is in principle supplemented
by an infinite tower of higher-order operators. We may hope, however, that the worldsheet structure may itself constrain the allowed higher-dimension operators. 

As further applications of our methods, we will use a generalisation of the Newman--Janis shift \cite{Talbot:1969bpa} to introduce magnetic charges (in electrodynamics) and NUT parameters (in gravity) for the particles described by our equations of motion.
As an example, we compute the leading impulse on a probe particle with mass,
spin and NUT charge moving in a Kerr--Taub--NUT background.
The charged generalisation of the NJ complex map can similarly be connected to the behaviour of three-point amplitudes in the classical limit \cite{Moynihan:2019bor,Huang:2019cja,Chung:2019yfs,Moynihan:2020gxj,Emond:2020lwi,Kim:2020cvf}, and we will reproduce results recently derived from this perspective~\cite{Emond:2020lwi},
furthermore calculating the leading angular impulse for the first time.

The material in this chapter is organised as follows. We begin our discussion in the context of electrodynamics, constructing the effective action for a $\rootKerr$~
probe in an arbitrary electromagnetic background. In this case it is rather easy to understand how the worldsheet emerges. We discuss key
properties of the worldsheet, including the origin of the Newman--Janis shift, in this context. It turns out to be useful to perform the matching
in a spacetime with ``split'' signature $(+,+,-,-)$, largely because the three-point amplitude does not exist on-shell in Minkowski space.
The structure of the worldsheet is particularly simple
in split-signature spacetimes.
In section~\ref{sec:gr} we turn to the gravitational case, showing that the worldsheet naturally describes the dynamics of a spinning Kerr particle.
We discuss equations of motion in section~\ref{sec:spinorEOM},
focussing on the leading-order interactions which are not sensitive to
terms in the effective action which we have not constrained. In this section, we will see how useful the methods of spinor-helicity are for capturing the
chiral dynamics associated with the NJ shift, as well as magnetic charges.

This chapter is based on material first published in \cite{Guevara:2020xjx}, in collaboration with Alfredo Guevara, Alexander Ochirov, Donal O'Connell and Justin Vines.

\section{From amplitude to action}
\label{sec:rootKerrEFT}

We begin by concentrating on the slightly simpler example of
the $\rootKerr$~particle in electromagnetism.
We wish to construct an effective action
for a massive, charged particle with spin angular momentum $S^{\mu\nu}$.
Building on the work of Porto, Rothstein, Levi and Steinhoff~\cite{Porto:2005ac,Porto:2006bt,Porto:2008tb,Levi:2014gsa,Levi:2015msa},
we write the worldline action as
\[
S = \int\!\d\tau \bigg\{ {-m}\sqrt{u^2} - \frac12 S_{\mu\nu} \Omega^{\mu\nu} - Q A \cdot u \bigg\} +  S_\text{EFT} \,,
\label{eq:fullSimpleAction}
\]
where $u^\mu$ and $\Omega^{\mu\nu}$ are the linear
and angular velocities,\footnote{We will be fixing $\tau$
to be the proper time,
so the velocity $u^\mu = \d r^\mu/\d\tau$ will satisfy $u^2=1  $.
The angular velocity can be defined through a body-fixed frame $e^a_\mu(\tau)$ on the worldline as
\[
\Omega^{\mu\nu}(\tau) = e^\mu_a(\tau) \frac{\d~}{\d\tau} e^{a\nu}(\tau) \,.
\label{eq:angMom}
\]
The tetrad allows us to pass from body-fixed frame indices $a, b, \ldots$
to Lorentz indices $\mu,\nu, \ldots $, as usual.
More details on spinning particles in effective theory
can be found in recent reviews~\cite{Porto:2016pyg,Levi:2018nxp}.}
and $S_\text{EFT}$ contains additional operators
coupling the spinning particle to the electromagnetic field. The momentum is defined as the canonical conjugate of the velocity,
\begin{equation}
p_\mu = -\frac{\partial L}{\partial u^\mu} = m u_\mu + {\cal O}(A)\,.
\end{equation}
We will continue to assume the spin tensor to be transverse
according to the Tulczyjew covariant spin supplementary condition in equation~\eqref{eqn:SSC}.
We can therefore relate the spin angular momentum to the spin pseudovector $a^\mu$ by
\[
a^{\mu} = \frac1{2p^2} \epsilon^{\mu\nu\rho\sigma} p_\nu S_{\rho\sigma} \equiv \frac{1}{2p^2} \epsilon^\mu(p,S)\qquad \Leftrightarrow \qquad
S_{\mu\nu} = \epsilon_{\mu\nu}(p, a) \,.
\label{eq:spinEquivs}
\]
The effective action~\eqref{eq:fullSimpleAction} can be written independently of the choice of SSC, at the expense of introducing an additional term from minimal coupling \cite{Yee:1993ya,Porto:2008tb,Steinhoff:2015ksa}. This has played an important role in recent work pushing the gravitational effective action beyond linear-in-curvature terms \cite{Levi:2020kvb,Levi:2020uwu,Levi:2020lfn},
but for our present purposes a fixed SSC will suffice.
Note that any differences in the choice of the spin tensor $S_{\mu\nu}$
are projected out from the pseudovector $a^\mu$ by definition,
and it is the latter that will be central to our discussion.

We will only consider the effective operators in $S_\text{EFT}$
that involve one power of the electromagnetic field $A_\mu$,
which can be fixed by the three-point amplitudes.
Since these amplitudes are parity-even, 
the possible single-photon operators are
\begin{multline}
S_\text{EFT} = Q\sum_{n=1}^\infty \int\!\d\tau \, u^\mu a^\nu
\Big[ B_n  (a\cdot \partial)^{2n-2} F_{\mu\nu}^*(x)  \\ + C_n (a \cdot \partial)^{2n-1} F_{\mu\nu}(x) \Big]_{x=r(\tau)} \,.
\label{eq:introEFT}
\end{multline}
Notice that an odd number of spin pseudovectors is accompanied
by the dual field strength
\[
F^*_{\mu\nu} = \frac12 \epsilon_{\mu\nu\rho\sigma} F^{\rho \sigma} \,,
\]
while the plain field strength goes together with an even power of $a$.
By dimensional analysis,
the unknown constant coefficients $B_n$ and $C_n$ are dimensionless. 

\subsection{Worldsheet from source}

To determine the unknown coefficients,
we choose to match our effective action
to a quantity that can be derived directly from the three-point
$\rootKerr$~amplitude~\eqref{eq:rootKerrAmp}.
A convenient choice is the classical Maxwell spinor given by the amplitude for an incoming photon,
which is~\cite{Monteiro:2020plf}
\[
\phi(x) = -\frac{\sqrt{2}}{m} \Re \int\!\d\Phi(\wn k) \, \del(\wn k \cdot u) \, \ket{\wn k} \bra{\wn k} \, e^{- i \wn k \cdot x } \mathcal{A}_{3,+} \,.
\]
In this expression the integration is over on-shell massless phase space, cast in the notation of equation~\eqref{eqn:dfDefinition}.
This Maxwell spinor is defined in (2,2) signature.
Indeed, in Minkowski space the only solution of the zero-energy condition $\wn k \cdot u$ for
a massless, on-shell momentum
is $\wn k^\mu=0$, so the three-point amplitude cannot exist on-shell for non-trivial kinematics.
However, there is no such issue in (2,2) signature, which motivates analytically continuing from Minkowski space. (The spinor $\ket{\wn k}$ is constructed from the on-shell null momentum $\wn k$ as usual in spinor-helicity.)

In fact, the Newman--Janis shift makes it extremely natural for us to analytically continue to split signature even in the classical sense, without any consideration of three-point amplitudes.
The Maxwell spinor for a static $\rootKerr$~particle is explicitly
\[
\phi^{\sqrt{\text{Kerr}}}(x) = -\frac{Q}{4\pi} \frac{1}{(x^2 + y^2 + (z + i a)^2)^{3/2}} (x, y, z + ia) \cdot \boldsymbol{\sigma} \,.
\label{eq:phiKerrMink}
\]
In preparation for the analytic continuation $z= -iz'$,
we may choose to order the Pauli matrices
as $\boldsymbol{\sigma}=(\sigma_z, \sigma_x, \sigma_y)$.
Then the spinor structure in equation~\eqref{eq:phiKerrMink} becomes real,
while the radial fall-off factor in the Maxwell spinor simplifies to
\[
\frac{1}{(x^2 + y^2 - (z - a)^2)^{3/2}} \,, \nonumber
\]
where we have dropped the prime sign of $z$.
In short, we have a real Maxwell spinor in (2,2) signature,
and the spin $a$ is now a real translation in the timelike $z$ direction.

We now analytically continue the action~\eqref{eq:introEFT}
by choosing the spin direction to become timelike.
In doing so, we also continue the component of the EM field in 
the spin direction, consistent with a covariant derivative
$\partial + i Q A$.
In split signature, it is convenient to rewrite the effective action
ansatz in terms of self- and anti-self-dual field strengths,
which we define as
\[
F^\pm_{\mu\nu}(x) = F_{\mu\nu}(x) \pm F^*_{\mu\nu}(x) \,.
\]
Our action then depends on a new set of unknown Wilson coefficients
$\tilde{B}_n$ and $\tilde{C}_n$:
\[
S_\text{EFT} = Q \sum_{n=0}^\infty \int\!\d\tau \, u^\mu a^\nu
\big[ \tilde{B}_n (a \cdot \partial)^n F_{\mu\nu}^+(x)
+ \tilde{C}_n (a \cdot \partial)^n F_{\mu\nu}^-(x)
\big]_{x=r(\tau)} \,.\label{eq:2,2actionUnmatched}
\]
To determine these coefficients we can match to the three-point amplitude by computing the Maxwell spinor for the radiation field sourced by the $\rootKerr$~particle, which we assume to have constant spin $a^\mu$ and constant proper velocity $u^\mu$. In (2,2) signature the exponential factor in equation~\eqref{eq:rootKerrAmp} also picks up a factor $-i$, so we match our action to
\[
\phi(x) = -\frac{\sqrt{2}}{m} \Re\! \int\! \d\Phi(\wn k) \, \del(\wn k \cdot u) \, \ket{\wn k} \bra{\wn k} \, e^{- i \wn k \cdot x } \mathcal{A}_{3,+}^\text{Coulomb} e^{i \wn k\cdot a} \,.\label{eq:rKthreePointSpinor}
\]
Our matching calculation hinges upon the field strength sourced by the particle, which is determined by the $\rootKerr$~worldline current $\tilde j^{\mu}(\wn k)$. In solving the Maxwell equation we impose retarded boundary conditions precisely as in~\cite{Monteiro:2020plf}, placing our observation point $x$ in the future with respect to one time coordinate $t^0$, but choosing the proper velocity $u$ to point along the orthogonal time direction. It is useful to make use of the result
\[
\frac{1}{\wn k^2_\text{ret}} = -i \sign \wn k^0 \del(\wn k^2) + \frac{1}{\wn k^2_\text{adv}} \,,
\]
where the $\text{ret}$ and $\text{adv}$ subscripts indicate retarded and advanced Green's functions respectively. Since the advanced Green's
function has support for $t^0 < 0$, we may simply replace
\[
\frac{1}{\wn k^2_\text{ret}} = -i \sign(\wn k^0) \del(\wn k^2)  \,.\label{eq:2,2retardedProp}
\]
The field strength sourced by the current in split signature is therefore
\[
F^{\mu\nu}(x)  &= 2\!\int\!\dd^4\wn k\, \sign(\wn k^0) \del(\wn k^2) \, \wn k^{[\mu} \tilde j^{\nu]}\,e^{-i\wn k\cdot x}\\
&= 2\!\int\!\dd^4\wn k\, \left(\Theta(\wn k^0) - \Theta(-\wn k^0)\right) \del(\wn k^2) \,  \wn k^{[\mu} \tilde j^{\nu]}\,e^{-\wn ik\cdot x}\,,
\]
Notice that the appropriate integral measure is now precisely the invariant phase-space measure~\eqref{eqn:dfDefinition}; substituting the worldline current for our $\rootKerr$~effective action in equation~\eqref{eq:2,2actionUnmatched}, evaluated on a leading-order trajectory, we thus have
\begin{multline}
F^{\mu\nu}(x) = 4Q \Re\! \int\!\d\Phi(\wn k)\,\del(\wn k\cdot u) \bigg\{\wn k^{[\mu} u^{\nu]}\Big[1 + ia\cdot \wn k \sum_{n=0}^\infty \left(\tilde B_n (ia\cdot \wn k)^{n} + \tilde C_n (ia\cdot \wn k)^{n}\right)\Big]\\
+ i\wn k^{[\mu} \epsilon^{\nu]}(\wn k,u,a) \sum_{n=0}^\infty\left(\tilde B_n(ia\cdot \wn k)^{n} - \tilde C_n (ia\cdot \wn k)^{n}\right)\bigg\} \, e^{-i\wn k\cdot x}\,.\label{eq:unmatchedField}
\end{multline}

To match to the three-point $\rootKerr$~amplitude, we need to compute the Maxwell spinor $\phi$ and its conjugate, $\tilde \phi$. To do so, we introduce a basis of positive and negative helicity polarisation vectors $\varepsilon_k^{\pm}$. On the support of the delta function in~\eqref{eq:unmatchedField}, manipulations using the spinorial form of the polarisation vectors given in \cite{Monteiro:2020plf} then lead to
\[
\wn k^{[\mu} u^{\nu]} \sigma_{\mu\nu} &= +\frac{\sqrt{2}}{2}
 \varepsilon_k^+\cdot u\, |\wn k\rangle \langle \wn k|\\
\wn k^{[\mu} \epsilon^{\nu]}(\wn k,u,a) \sigma_{\mu\nu} &= -\frac{\sqrt{2}}{2} a\cdot \wn k\ \varepsilon_k^+\cdot u \,|\wn k\rangle \langle \wn k|\,.
\]
The latter equality relies upon the identity $\wn k^{[\mu} \epsilon^{\nu\rho\sigma\lambda]} = 0$, and the fact that $\sigma_{\mu\nu}$ is self-dual in this signature. With these expressions in hand, it is easy to see that the Maxwell spinor has a common spinorial basis, and takes the simple form
\begin{multline}
\phi(x) = 2\sqrt{2} Q \Re\! \int\! \d\Phi(\wn k)\, \del(\wn k\cdot u) e^{-i\wn k\cdot x}\, |\wn k\rangle \langle \wn k| \,\varepsilon_k^+\cdot u  \\ \times \left(1 + \sum_{n=0}^\infty 2 \tilde C_n (ia\cdot\wn k)^{n+1}\right).\label{eq:unmatchedMaxwell}
\end{multline}
Recall from equation~\eqref{eq:2,2actionUnmatched} that the Wilson coefficients $\tilde B_n$ and $\tilde C_n$ were identified with self- and anti-self-dual field strengths, respectively. Since a positive-helicity wave is associated with an anti-self-dual field strength, it is no surprise that the Maxwell spinor should depend only on this part of the $\rootKerr$~effective action. Fixing the $\tilde B_n$ coefficients requires the dual spinor, which is given by
\begin{multline}
\tilde \phi(x) = -2\sqrt{2} Q \Re\! \int\! \d\Phi(\wn k)\, \del(\wn k\cdot u) e^{-i\wn k\cdot x}\, |\wn k]\, [\wn k|\, \varepsilon_k^-\cdot u \\ \times \left(1 + \sum_{n=0}^\infty 2 \tilde B_n (ia\cdot\wn k)^{n+1}\right)\,.
\label{eq:unmatchedMaxwellDual}
\end{multline}
Here we have used that
\[
\wn k^{[\mu} u^{\nu]} \tilde\sigma_{\mu\nu} &= -\frac{\sqrt{2}}{2} \varepsilon_k^-\cdot u\, |\wn k]\,[\wn k|\\
\wn k^{[\mu} \epsilon^{\nu]}(\wn k,u,a) \tilde\sigma_{\mu\nu} &= -\frac{\sqrt{2}}{2} a\cdot\wn k\ \varepsilon_k^-\cdot u \,|\wn k]\,[\wn k|\,,
\]
recalling that $\tilde \sigma$ is anti-self-dual in split signature spacetimes.

It now only remains to match to the Maxwell spinors for the three-point~amplitude, as given in equation~\eqref{eq:rKthreePointSpinor}. The scalar Coulomb amplitudes for photon absorption are just $\mathcal{A}_\pm = -2mQ\, u\cdot\varepsilon_k^\pm$, so for the $\rootKerr$~three-point amplitude
\[
\phi(x) &= + 2\sqrt{2}  Q \Re\! \int\! \d\Phi(\wn k)\, \del(\wn k\cdot u) e^{-i\wn k\cdot x}\, |\wn k\rangle \langle\wn k| \, \varepsilon_k^+\cdot u \, e^{i\wn k\cdot a}\,,\\
\tilde\phi(x) &= -2\sqrt{2}  Q \Re\! \int\! \d\Phi(\wn k)\, \del(\wn k\cdot u) e^{-i\wn k\cdot x}\,  |\wn k]\, [\wn k|\, \varepsilon_k^-\cdot u\,e^{-i \wn k\cdot a}\,.
\]
Expanding the exponentials and matching to eqs.~\eqref{eq:unmatchedMaxwell} and \eqref{eq:unmatchedMaxwellDual} identifies
\begin{equation}
\tilde B_n = \frac{(-1)^{n+1}}{2(n+1)!}\,, \qquad \quad
\tilde C_n = \frac{1}{2(n+1)!}\,,
\end{equation}
which upon substitution into equation~\eqref{eq:2,2actionUnmatched} finally yields the $\rootKerr$~effective action in split-signature:
\[
S_\text{EFT}  &= Q \sum_{n=0}^\infty \int\!\d\tau\, u^\mu a^\nu
\bigg[ {-\frac{(-a\cdot \partial)^n}{2(n+1)!}}  F_{\mu\nu}^+(x)
+ \frac{(a\cdot \partial)^n}{2(n+1)!} F_{\mu\nu}^-(x)
\bigg]_{x=r(\tau)} \\ 
=& -\frac{Q}{2} \int\!\d\tau\, u^\mu a^\nu
\bigg[ \left(\frac{e^{-a \cdot \partial} - 1}{-a \cdot \partial}\right) F_{\mu\nu}^+(x)
- \left(\frac{e^{ a \cdot \partial} - 1}{a \cdot \partial}\right) F_{\mu\nu}^-(x)
\bigg]_{x=r(\tau)} \,.\label{eq:2,2actionMatchedCoefficients}
\]

So far, the Newman--Janis structure is hinted at by the translation
operators $e^{\pm a \cdot \partial}$ appearing in the effective action.
We can make this structure more manifest by writing the effective action equivalently as
\[
S_\text{EFT} & = -\frac{Q}{2}
\int\!\d\tau\!\int_0^1\!\d\lambda\, u^\mu a^\nu
\big[ e^{-\lambda 
(a \cdot \partial)} F_{\mu\nu}^+(x)
- e^{\lambda (a \cdot \partial)} F_{\mu\nu}^-(x)
\big]_{x=r(\tau)} \\ &
= -\frac{Q}{2} \int\!\d\tau\!\int_0^1\!\d\lambda\, u^\mu a^\nu
\big[ F^+_{\mu\nu}(r - \lambda a) - F^-_{\mu\nu}(r + \lambda a) \big] \,.
\label{eq:seftStep}
\]
Our effective action is now an integral over a two-dimensional region
--- a worldsheet, rather than a worldline.

To see that this worldsheet is indeed connected to the Newman--Janis shift,
let us recover this shift for the Maxwell spinor.
First, we can read off the worldsheet current~$J^\mu$ from the action
$S_\text{EFT}-Q\!\int\!\d\tau A_\mu u^\mu = -\!\int\!\d^4x A_\mu J^\mu$.
Then, the gauge field $A_\mu$ set up at a point~$x$ by this source
may be written as an integral of a Green's function $G(x-y)$ over the worldsheet:
\begin{align}
A^\mu(x)  = \int\!\d^4y\,G(x-y) J^\mu(y) \hphantom{+ \frac{1}{2}\!\int_0^1\!\d\lambda}&\\ 
=  Q\!\int\!\d\tau
\bigg\{ u^\mu G(x-r)
+ \frac{1}{2}\!\int_0^1\!\d\lambda \Big(
&\big[ u^\mu (a \cdot \partial)
+ \epsilon^{\mu}(u, a, \partial)
\big] G(x-r+\lambda a) \nonumber \\ 
- &\big[ u^\mu (a \cdot \partial)
- \epsilon^{\mu} (u, a, \partial)
\big] G(x-r-\lambda a) \Big)\!
\bigg\} \,. \nonumber
\end{align}
The field strength follows by differentiation, after which contraction with $\sigma$ matrices yields the Maxwell spinor,
\[
\phi(x) = 2Q\!\int\!\d\tau\, \sigma_{\mu\nu}
u^{\nu} \partial^{\mu}\! \left[ G(x-r) - \int_0^1\!\d\lambda
\,(a \cdot \partial) G(x-r-\lambda a) \right] ,
\label{eq:NJbyintegration}
\]
where the first term comes from the non-spinning part of the action~\eqref{eq:fullSimpleAction}.
Now the $a\cdot\partial$ operator acting on the Green's function
can be understood as a derivative with respect to $\lambda$.
This produces a $\lambda$ integral of a total derivative,
which reduces to the boundary terms.
Cancelling the first term in equation~\eqref{eq:NJbyintegration}
against the boundary contribution at $\lambda = 0$, we find simply that
\[
\phi(x) = 2Q\!\int\!\d\tau \, \sigma_{\mu\nu} u^\nu \partial^\mu  G(x-r-a)  \,.
\]
The Maxwell spinor depends only on the anti-self-dual part of the effective action, shifted by the spin length. The real translation in $(2,2)$ signature
is a result of this real worldsheet structure. We will shortly see that this structure persists for interactions.

\subsection{Worldsheet for interactions}

Let us analytically continue
the action~\eqref{eq:seftStep} back to Minkowski space:
\[
S_\text{EFT} & = \frac{Q}{2} \int\!\d\tau\!\int_0^1\!\d\lambda\,
u^\mu a^\nu \big[ i F^+_{\mu\nu}(r + i \lambda a)
- i F^-_{\mu\nu}(r - i \lambda a) \big] \\ &
= Q \Re \int_\Sigma \! \d\tau \d \lambda \, iF^+_{\mu\nu}(r + i \lambda a) \, u^\mu a^\nu \,.
\]
Here the self-dual and anti-self-dual field strengths are
\[
F^\pm_{\mu\nu} = F_{\mu\nu} \pm i F^*_{\mu\nu}
= \pm \frac{i}{2} \epsilon_{\mu\nu\rho\sigma} F^{\pm\,\rho\sigma} \,,
\label{eq:rKwsaction}
\]
and $\Sigma$ is the worldsheet, with $\tau$ running over $(-\infty,\infty)$, and $\lambda$ over $[0,1]$.

\begin{figure}[t]
	\center
	\includegraphics[width = 0.5\textwidth]{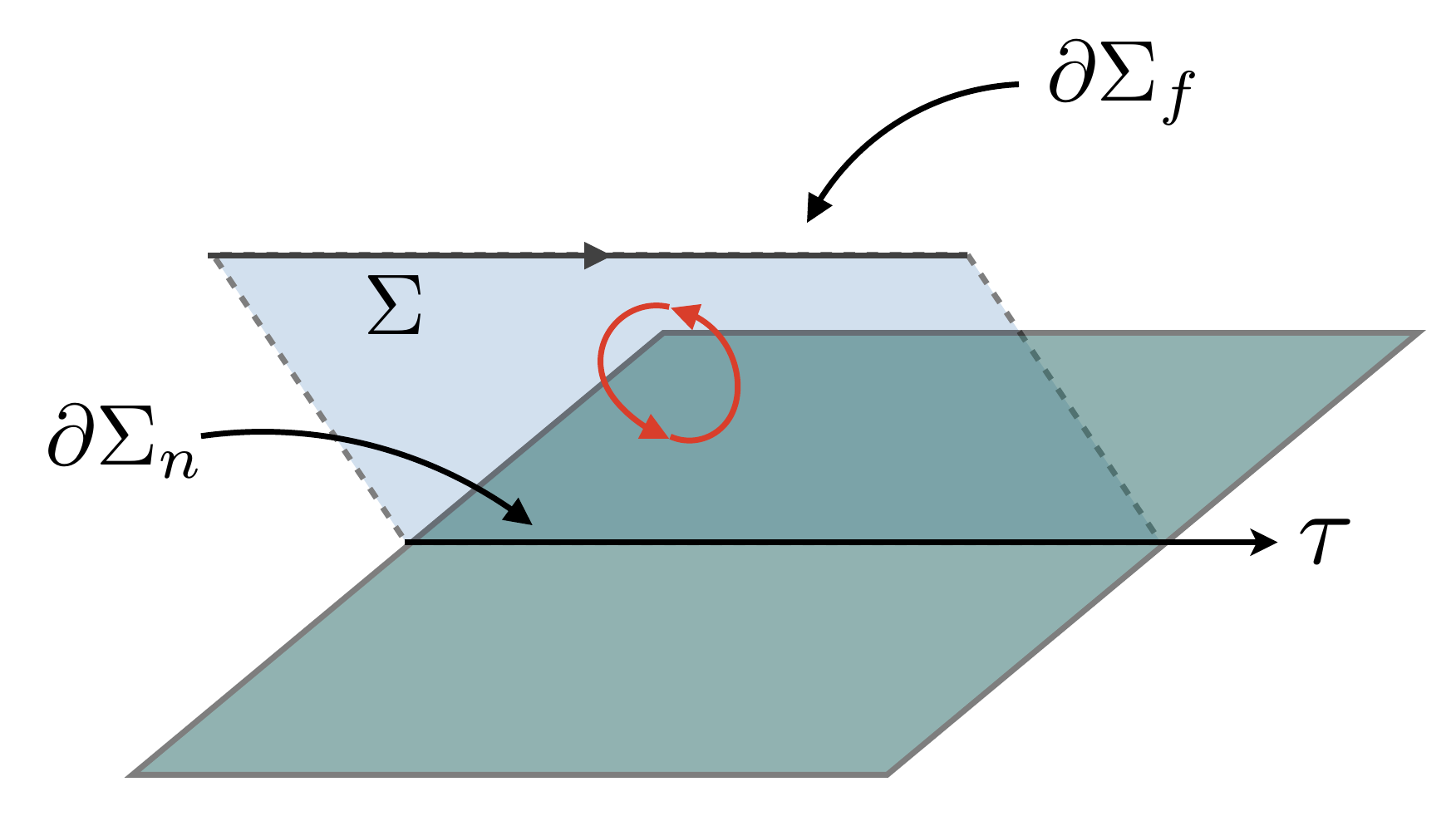}
	\vspace{-3pt}
	\caption[Geometry of the Kerr worldsheet effective action.]{Geometry of the effective action: boundary $\partial \Sigma_\text{n}$ of the  complex worldsheet (translucent plane) is  fixed to the particle worldline in real space (solid plane).\label{rootKerrIntegration}}
\end{figure}

Now that we are back in Minkowski space,
let us turn to the Newman--Janis structure of interactions.
Suppose that our spinning particle is moving
under the influence of an external electromagnetic field, generated by distant sources. The total interaction Lagrangian contains the worldsheet
term~\eqref{eq:rKwsaction} as well as the usual worldline minimal coupling:
\[
S_\text{int} = -Q \int_{\partial \Sigma_\text{n}}\!\!\!\d\tau A_\mu(r) u^\mu
+ Q \Re\! \int_\Sigma\!\d\tau \d \lambda \, iF^+_{\mu\nu}(r + i \lambda a) \, u^\mu a^\nu + \ldots \,,
\label{eq:sint}
\]
where $\partial \Sigma_\text{n}$ is the ``near'' boundary of the worldsheet, at $\lambda = 0$, as shown in figure~\ref{rootKerrIntegration}. We will similarly refer to the boundary at $\lambda = 1$ as the ``far'' boundary. The near boundary is the physical location of the object, while the far boundary is a timelike line embedded in the complexification of Minkowski space. We have also indicated the presence of unknown additional operators (involving at least two powers of the field strength) in the action by the ellipsis in equation~\eqref{eq:sint}.

It is convenient to introduce a complex coordinate $z = r + i \lambda a$ on the worldsheet. In terms of this coordinate, we may write the two-form
\[
F^+(z) =
\frac{1}{2} F^+_{\mu\nu}(z) \d z^\mu \wedge \d z^\nu = i F^+_{\mu\nu}(z) (u^\mu + i \lambda \dot a^\mu) a^\nu \, \d \tau \wedge \d \lambda \,,
\]
where $\dot a^\mu = \d a^\mu / \d \tau$. Since in the absence of interactions the spin is constant, $\dot a$ must be of order $F$.
Therefore, we may rewrite our interaction action as
\[
S_\text{int} = -Q \int_{\partial \Sigma_\text{n}} \!\!\! A_\mu(r) \d r^\mu
+ \frac{Q}{2} \Re\! \int_\Sigma F^+_{\mu\nu}(z) \, \d z^\mu \wedge \d z^\nu + \ldots \,.
\label{eq:sintNicer}
\]
In doing so, we have redefined the higher-order operators indicated by the ellipsis.

When the electromagnetic fields appearing in the action~\eqref{eq:sintNicer} are generated by external sources,
both $F$ and $F^*$ are closed two-forms,
so we may introduce potentials $A$ and $A^*$
such that $F = \d A$ and $F^* = \d A^*$.
The dual gauge potential $A^*$ is related to $A$ by duality,
but this relationship need not concern us here:
we only require that both potentials exist in the vicinity
of the $\rootKerr$~particle.
Hence we may also write $F^+ = \d A + i\, \d A^* = \d A^+$.
Then the action~\eqref{eq:sintNicer} becomes
\[
S_\text{int} &
=-Q \int_{\partial \Sigma_\text{n}}\!\!A
+ Q \Re\!\int_\Sigma\!\d A^+ + \ldots \\ &
=-Q \int_{\partial \Sigma_\text{n}}\!\!A
+ Q \Re\!\int_{\partial \Sigma_\text{n}}\!\!A^+
- Q \Re\!\int_{\partial \Sigma_\text{f}}\!\!A^+ + \ldots\,,
\]
where the boundary consists of two disconnected lines (the far and near boundaries).
The orientation of the integration contour was set by $F^+$,
as depicted in figure~\ref{rootKerrIntegration}.

Now, notice that on the near boundary $z = r(\tau)$ is real.
Hence $\Re A^+ = A$,
so we are only left with the far-boundary contribution in the action:
\[
S_\text{int} = -Q \Re\!\int_{\partial \Sigma_\text{f}}\!\!A^+ + \ldots
=-Q \Re\!\int\!\d \tau \, u^\mu A_\mu^+(r + i a) + \ldots \,.
\label{eq:rtKerrIntShift}
\]
Thus we explicitly see that the interactions
of a $\rootKerr$~particle can be described with a Newman--Janis shift.
We will exploit this fact explicitly in section~\ref{sec:spinorEOM}.
Before we do, we turn to gravitational interactions.

\section{Spin and gravitational interactions}
\label{sec:gr}

As a step towards a worldsheet action for a probe Kerr in a non-trivial background, it is helpful to understand how to make the electromagnetic
effective action~\eqref{eq:rKwsaction} generally covariant. In a curved spacetime, we cannot simply add a vector $\lambda a$ to a point $r$.
To see what to do, let us reintroduce translation operators as in equation~\eqref{eq:seftStep}. The worldsheet EFT term in Minkowski space is
\[
S_\text{EFT} &= Q \Re\! \int_\Sigma \! \d\tau \d\lambda\, 
i\, e^{i \lambda  \, a \cdot \partial} F^+_{\mu\nu}(x)\, u^\mu a^\nu \Big|_{x=r(\tau)}  \\
&= Q \Re \int_\Sigma \! \d\tau \d\lambda \, i
\sum_{n=0}^\infty \frac{1}{n!} (i \lambda \, a \cdot \partial)^n  F^+_{\mu\nu}(x) \, u^\mu a^\nu\Big|_{x=r(\tau)} \,.
\]
Now, it is clear that a minimal way to make this term generally covariant is to replace the partial derivatives $\partial$ with covariant 
derivatives $\nabla$, so in curved space we have
\[
S_\text{EFT} &= Q \Re \!\int_\Sigma \! \d\tau \d\lambda \,
\sum_{n=0}^\infty \frac{1}{n!} (i \lambda \, a \cdot \nabla)^n F^+_{\mu\nu}(x) \, i u^\mu a^\nu  \Big|_{x=r(\tau)} \,.
\label{eq:rKeftCurvedStep}
\]

It is therefore natural for us to consider a covariant translation operator
\[
e^{i\lambda\,a \cdot \nabla} \equiv \sum_{n=0}^\infty \frac{1}{n!} (i \lambda \, a \cdot \nabla)^n \,.
\label{eq:explicitTranslation}
\]
This operator generates translations along geodesics in the direction $a$.
To see
why, note that the perturbative expansion of such a geodesic beginning at a point
$x_0$ in the direction $a$ with parameter $\ell$ is
\[
x^\mu(\ell) = x_0^\mu + \ell \, a^\mu - \frac{\ell^2}{2} \Gamma^\mu_{\nu\rho}(x_0) a^\nu a^\rho + \ldots \,.
\]
Now consider the perturbative expansion of a scalar function $f(x)$ along such a
geodesic. We have
\begin{align}
f(x(\ell)) &= f(x_0) + \ell \, a^\mu \partial_\mu f(x_0) + \frac{\ell^2}2 a^\mu a^\nu 
\left(
\partial_\mu \partial_\nu f(x_0) - \Gamma^\alpha_{\mu\nu}(x_0) \partial_\alpha f(x_0)
\right) 
+ \ldots \nonumber\\
&=
f(x_0) + \ell (a \cdot \nabla) f(x_0) + \frac{\ell^2}2 (a \cdot \nabla) (a \cdot \nabla) f(x_0)
+ \ldots \\
&= e^{\ell \, a\cdot \nabla} f(x_0) \,.\nonumber
\end{align}

A traditional point of view on equation~\eqref{eq:rKeftCurvedStep} is that the
operators only act on the two-form $F^+_{\mu\nu}$. However, we can alternatively
think of the operator acting on a scalar function $F^+_{\mu\nu} u^\mu a^\nu$, 
provided we extend the definitions of the velocity $u$ and the spin $a$
so that they become fields on the domain of the translation operator. We can simply
do this by parallel-transporting $u(r(\tau))$ 
and $a(r(\tau))$ along the geodesic beginning at $r(\tau)$ in the direction $a(\tau)$
(using the Levi--Civita connection). We denote these geodesics by $z(\tau, \lambda)$; 
explicitly,
\[
z^\mu(\tau, \lambda) = r^\mu(\tau) + i \lambda a^\mu(\tau) + \frac{\lambda^2}{2} \Gamma^\mu_{\nu\rho}(r(\tau)) a^\nu a^\rho + \ldots \,.
\]
(Notice that the translation operator~\eqref{eq:explicitTranslation} has parameter
$i \lambda$.)
The parallel-transported vectors, with initial conditions $a(z(\tau, 0)) = a(\tau)$ 
and $u(z(\tau, 0)) = u(\tau)$,
have the similar perturbative expansions
\[
u^\mu(z(\tau, \lambda)) &= u^\mu(\tau) - i \lambda \Gamma^\mu_{\nu\rho} (r(\tau)) 
a^\nu(\tau) u^\rho(\tau) + \ldots \,,\\
a^\mu(z(\tau, \lambda)) &= a^\mu(\tau) - i \lambda \Gamma^\mu_{\nu\rho} (r(\tau)) 
a^\nu(\tau) a^\rho(\tau) + \ldots \,.
\]
We now view the translation operator in equation~\eqref{eq:rKeftCurvedStep}
as acting on the scalar quantity $F^+_{\mu\nu} u^\mu a^\nu$:
\[
e^{i \lambda a \cdot \nabla} F_{\mu\nu}(r(\tau)) a^\mu(\tau) u^\nu(\tau) 
= F_{\mu\nu}(z(\tau,\lambda)) a^\mu(z(\tau,\lambda)) u^\nu(z(\tau, \lambda)) \,.
\]
Expanding perturbatively to first order in $\lambda$, we have
\begin{align}
e^{i \lambda a \cdot \nabla} F_{\mu\nu}(r(\tau)) a^\mu(\tau) u^\nu(\tau)
&= \left(F_{\mu\nu}(r(\tau)) + i \lambda a^\rho \left(
\partial_\rho F_{\mu\nu} - \Gamma^\alpha_{\mu\rho} F_{\alpha\nu} 
-\Gamma^\alpha_{\nu\rho} F_{\mu \alpha} 
\right)\right) a^\mu u^\nu \nonumber \\
&= \left( F_{\mu\nu}(r(\tau)) + i \lambda a^\rho \nabla_\rho F_{\mu\nu}\right)a^\mu u^\nu \,.
\end{align}
The final expression is precisely the same as the picture in which 
the derivatives act only on the field strength: these are equivalent points of
view.

The worldsheet arises from interpreting the translation operators as genuine
translations. In curved space, the operators replace the straight-line sum 
$r+i a \lambda$ appearing in our action~\eqref{eq:rKwsaction} with the 
natural generalisation --- a geodesic in the direction $a$.\footnote{In general, 
these geodesics may become singular. We assume that such singularities do
not arise. If they were to arise, there would also be a divergence in the
interpretation of the EFT as an infinite sum of operators.}
We can express the curved-space effective action as
\[
S_\text{EFT} = Q \Re \!\int_\Sigma \! \d\tau \d \lambda \, iF^+_{\mu\nu}(z) \, u^\mu(z) a^\nu(z) \Big|_{z=z(\tau,\lambda)} \,.
\label{eq:rkCovActionVelocity}
\]
The surface $\Sigma$ is built up from the worldline of the particle, augmented by the geodesics in the direction $a$ for each $\tau$.

Note that, since we neglect higher-order interactions,
we may replace the velocity vector field $u(\tau,\lambda)$ 
in the action~\eqref{eq:rkCovActionVelocity}
with the similarly defined momentum field $p(\tau, \lambda)$.
Indeed, at $\lambda = 0$ the difference adds another order in the gauge field, 
and this persists for $\lambda \neq 0$ after parallel translation along the geodesics.
Therefore, up to $F^2$ operators that we are neglecting,
the $\rootKerr$~action may be written as
\[
S_\text{EFT} = \frac{Q}{m} \Re\!\int_\Sigma\!\d\tau \d\lambda \, iF^+_{\mu\nu}(z) \, p^\mu(z) a^\nu(z) \Big|_{z=z(\tau,\lambda)} \,.
\label{eq:rkCovActionMomentum}
\]

We are now ready for the fully gravitational Kerr worldsheet action,
which is naturally motivated as a classical double copy of this covariantised worldsheet action.
Recalling that we should double-copy from non-Abelian gauge theory rather than electrodynamics, we promote the field strength to the Yang--Mills case:
\[
Q F^+_{\mu\nu}(z) ~\rightarrow~ c^A(z) F^{A+}_{\mu\nu}(z) \,,
\]
where $c^A(z(\tau,\lambda))$ is a vector in the colour space (generated by parallel transport from the classical colour vector of a particle,
as described by the Yang--Mills--Wong equations~\eqref{eqn:classicalWong}).
The double copy replaces colour by kinematics, so we anticipate a replacement of the form $c^A \rightarrow u^\mu$.
Moreover, to replace $F^{A}_{\mu\nu}$ we need an object with three indices,
antisymmetric in two of them,
for which the spin connection
\[
\omega_\mu{}^{ab} = e^b_\nu\, \nabla_\mu e^{a\nu} =
e^b_\sigma \big( \partial_\mu e^{a\sigma} + \Gamma^\sigma_{\mu\nu} e^{a\nu} \big)\label{eq:spinConnection}
\]
is the natural candidate.
Since it is defined via a derivative of the (body-fixed) spacetime tetrad~$e_a^\mu$, which is
a dimensionless quantity, on dimensional grounds the replacement should be
of the form $F^{A}_{\mu\nu} \to m \omega_\mu{}^{ab}$.
Indeed, we find that the correct worldsheet action for Kerr is
\[
S_\text{EFT} = \Re\! \int_\Sigma \d\tau \d \lambda \, i\,u_\mu(z) \, \omega^{+\mu}{}_{ab}(z) \, p^a(z) a^b(z) \Big|_{z=z(\tau,\lambda)} \,,
\label{eq:kerrws}
\]
where $\omega^+$ is a self-dual part of the spin connection,
defined explicitly by
\[
\omega^{+\mu}{}_{ab}(x) = \omega^{\mu}{}_{ab}(x) + i\,\omega^{*\mu}{}_{ab}(x) \,, \qquad
\omega^{*\mu}{}_{ab}(x) = \frac12 \epsilon_{abcd}\,\omega^{\mu\,cd}(x) \,.
\label{eq:omegaDefs}
\]
In writing these equations, we have extended the body-fixed frame $e_a = e_a^\mu \partial_\mu$ of vectors to every point of the complex worldsheet. We do so by parallel transport.
As usual, the frame indices $a, b, \cdots$ take values from 0 to 3,
and $\epsilon_{abcd}$ is the flat-space Levi--Civita tensor,
with $\epsilon_{0123} = + 1$.

\subsection{Flat-space limit}

We will shortly prove that the worldsheet term~\eqref{eq:kerrws} reproduces all single-curvature terms in the known effective action for a Kerr black hole in an arbitrary background~\cite{Porto:2006bt,Porto:2008tb,Levi:2015msa}. But first we wish to show that the term is non-trivial even in
flat space, and is in fact the standard kinetic term for a spinning particle in Minkowski space~\cite{Porto:2005ac} in that context.

In flat space and Cartesian coordinates, the worldsheet effective term~\eqref{eq:kerrws} is
\[
S_\text{EFT} = \Re\! \int\!\d\tau\!\int_0^1\!\d\lambda\,
i\,u_\mu(\tau)\,\omega^{+\mu}{}_{ab}(r + i \lambda a)\, p^a(\tau) a^b(\tau) \,, 
\label{eq:recoverSpinKineticStep}
\]
since the parallel transport of the vectors $u, p$ and $a$ is now trivial, and the geodesics reduce to straight lines. In flat space, 
the frame $e^a_\mu(\tau, \lambda)$ is also independent of $\lambda$, since it is
generated by parallel transport. 
Thus, the spin connection is $\lambda$-independent and the $\lambda$ integral in equation~\eqref{eq:recoverSpinKineticStep} becomes trivial.

Given the $\lambda$ independence of the spin connection, we may write
\[
S_\text{EFT} & = \Re\! \int\!\d\tau\!\int_0^1\!\d\lambda\,i\,
u_\mu(\tau)\,\omega^{+\mu}{}_{ab}(r(\tau))\, p^a(\tau) a^b(\tau) \\ &
=-\!\int\!\d\tau\,
u_\mu(\tau)\,\omega^{*\mu}{}_{ab}(r(\tau))\, p^a(\tau) a^b(\tau) \,.
\]
Recalling the definitions of the dual spin connection~$\omega^*$ and the spin pseudovector~$a$, eqs.~\eqref{eq:omegaDefs} and \eqref{eq:spinEquivs}, we equivalently have
\[
S_\text{EFT} = - \frac12 \int\!\d\tau\,
u_\mu(\tau)\, \omega^{\mu}{}^{ab}(r(\tau))\, S_{ab}(\tau)
=- \frac12 \int\!\d\tau\, \Omega{}^{ab}(r(\tau))\, S_{ab}(\tau) \,.\label{eq:wsSpinKinematic}
\]
This is nothing but the spin kinetic term written in equation~\eqref{eq:fullSimpleAction}.
In this way, we see that the worldsheet expression~\eqref{eq:kerrws} already describes the basic dynamics of spin.

\subsection{Single-Riemann effective operators}

It is now straightforward to recover the full tower of single-Riemann operators in the Kerr effective action. Returning to the full curved-space case, we may write our action~\eqref{eq:kerrws} as
\[
S_\text{EFT} & = \Re \!\int\!\d\tau\!\int_0^1\!\d\lambda\,
e^{i \lambda \, a \cdot \nabla} i\, u_\mu(\tau) \, \omega^{+\mu}{}_{ab}(r(\tau)) \, p^a(\tau) a^b(\tau) \\ &
= \Re \!\int\!\d\tau\, i\left(\frac{e^{i a \cdot \nabla}-1}{i a \cdot \nabla}\right) u_\mu(\tau) \, \omega^{+\mu}{}_{ab}(r(\tau)) \, p^a(\tau) a^b(\tau) \\ &
= \sum_{n=0}^\infty
\Re\!\int\!\d\tau\, \frac{(i a \cdot \nabla)^n}{(n+1)!} i\,u_\mu(\tau) \, \omega^{+\mu}{}_{ab}(r(\tau)) \, p^a(\tau) a^b(\tau) \,,
\]
where we performed the $\lambda$ integral and expanded the translation operator $e^{i a \cdot \nabla}$.
The leading contribution is again the spin kinetic term,
as a short computation demonstrates.
We also encounter an infinite series of higher-derivative contributions for $n \geq 1$. To express them in terms of the Riemann tensor,
we recall that it satisfies
\[
R_{ab\,\mu\nu} = e^\alpha_a e^\beta_b R_{\alpha\beta\,\mu\nu}
= -\nabla_\mu \omega_{\nu ab} + \nabla_\nu \omega_{\mu ab}
+ \omega_{\mu ac}\,\omega_{\nu}{}^c_{~\,b}
- \omega_{\nu ac}\,\omega_{\mu}{}^c_{~\,b} \,.
\label{Connection2Riemann}
\]
Consistently omitting the quadratic in $\omega$ terms from the equation above, as well as the higher-order interaction contributions due to the difference between $p^a$ and $mu^a$,
we rewrite a typical effective operator as
\begin{align}
& -\frac{1}{(n+1)!} \Re\!\int\!\d\tau\, u^\mu p^a a^b
(i a \cdot \nabla)^{n-1} a^\nu \nabla_\nu \omega^+_{\mu ab}
\big|_{x = r(\tau)} \nonumber \\ & \quad
=-\frac{1}{(n+1)!} \Re\!\int\!\d\tau\, p^a a^b u^\mu a^\nu
(i a \cdot \nabla)^{n-1}
\big[ R^+_{ab\,\mu\nu} + \nabla_\mu \omega^+_{\nu ab}
\big]_{x = r(\tau)} + \ldots \label{eq:EffOperatorSteps}\\ & \quad
=-\frac{m}{(n+1)!} \Re\!\int\!\d\tau\, u^a a^b a^\nu
(i a \cdot \nabla)^{n-1}
\bigg[ u^\mu R^+_{ab\,\mu\nu} + \frac{D~}{d\tau} \omega^+_{\nu ab}
\bigg]_{x = r(\tau)} + \ldots \,.\nonumber
\end{align}
Here $R^+_{ab\,\mu\nu}=R_{ab\,\mu\nu}+iR^*_{ab\,\mu\nu}$ is defined via the dualisation of the first two indices. Notice that in equation~\eqref{eq:EffOperatorSteps} we treat 
the velocity $u$, momentum $p$ and spin $a$ as fields on the worldline, so that they
commute with the covariant derivative.

We may proceed by integrating the $D/\d \tau$ term by parts, after which it acts on factors of velocity and spin.
This generates curvature-squared (and higher) operators that we again neglect.
In this way, we arrive at the form of the leading interaction Lagrangian
\[
S_\text{int} =-m\!\int\!\d\tau\,u^a a^b u^\mu a^\nu \Re \sum_{n=1}^\infty
\frac{(i a \cdot \nabla)^{n-1}\!}{(n+1)!}
\bigg[ R_{ab\,\mu\nu} + i R^*_{ab\,\mu\nu}
\bigg]_{x = r(\tau)} + \ldots \,.
\]
Finally, separating the even and odd values of $n$ into two distinct sums
\[
S_\text{int} = m\!\int\!\d\tau
\bigg[ & \sum_{n=1}^\infty \frac{(-1)^n}{(2n)!} (a \cdot \nabla)^{2n-2}
R_{\alpha\beta\,\mu\nu} u^\alpha a^\beta u^\mu a^\nu \\ &
-  \sum_{n=1}^\infty \frac{(-1)^n}{(2n+1)!}
(a \cdot \nabla)^{2n-1} R^*_{\alpha\beta\,\mu\nu} u^\alpha a^\beta u^\mu a^\nu
\bigg]_{x = r(\tau)} + \ldots \,,\label{eq:LSinteractions}
\]
one can verify that this reproduces the leading interactions of a Kerr black hole, as discussed in detail by Levi and Steinhoff~\cite{Levi:2015msa}. 

It is interesting that the
worldsheet structure unifies the spin kinetic term with the
leading interactions of Kerr. The same phenomenon was observed
directly at the level of amplitudes in ref.~\cite{Chung:2018kqs}.

\section{Spinorial equations of motion}
\label{sec:spinorEOM}

Equation~\eqref{eq:rtKerrIntShift} explicitly displays a Newman--Janis shift for the leading interactions of the $\rootKerr$~solution. Now we take
a first look at the structure of the equations of motion encoding this shift. Since the Newman--Janis shift is chiral, we will find that it is very
convenient to describe the dynamics using the method of spinor-helicity, even in a fully classical setting. Our focus here will be to extract
expressions for observables from the equations of motion at leading order. Thus we are free to make field redefinitions, dropping total
derivatives which do not contribute to observables. We will also extend our work to magnetically charged objects, such as spinning dyons
and the gravitational Kerr--Taub--NUT analogue at the level of equations of motion.

We may write the leading order action for a $\rootKerr$~particle with trajectory $r(\tau)$ and spin $a(\tau)$ as
\[
S = - \int \d \tau \left( p \cdot \dot r(\tau) + \frac12 \epsilon(p, a, \Omega) + Q \Re u^\mu A_\mu^+(r + i a) \right) + \ldots \,.
\label{eq:rootKerrLeadingAction}
\]
By varying with respect to the position $r(\tau)$ it is easy to determine that
\[
\frac{\d p^\mu}{\d \tau} &= Q \Re F^{+\mu\nu}(r+ia) u_\nu + \ldots
= \frac{Q}{m} \Re F^{+\mu\nu}(r+ia) p_\nu + \ldots \,.
\label{eq:rtKerrEOMmomentum}
\]
In the second equality, we replaced the velocity $u = \dot r$ with the momentum $p/m$, noting that the difference between the momentum and $mu$ is of 
order $F$. To obtain a similar differential equation for the spin $a^\mu$, it is helpful to begin by differentiating $a \cdot p = 0$, finding
\[
p_\mu \frac{\d a^\mu}{\d \tau} = p_\mu \Re \frac{Q}{m} F^{+\mu\nu}(r+ia) a_\nu \,.
\]
Based on this simple result, it is easy to guess that the spin satisfies
\[
\frac{\d a^\mu}{\d \tau} = \frac{Q}{m} \Re  F^{+\mu\nu}(r+ia) a_\nu + \ldots \,,
\label{eq:rtKerrEOMspin}
\]
and indeed a more lengthy calculation using the Lagrangian~\eqref{eq:rootKerrLeadingAction} confirms this guess.

Our expressions~\eqref{eq:rtKerrEOMmomentum} and~\eqref{eq:rtKerrEOMspin} for the momentum and spin have the same basic structure,
and are consistent with the requirements that $p^2$ and $a^2$ are constant while $a \cdot p = 0$. 
As discussed in section~\ref{sec:NJintro}, in the context of scattering amplitudes it has proven to be very convenient to introduce spinor variables describing similar momenta; this logic also applies to spins.
Notice that there is nothing quantum about using spinor variables for momenta and spin: the momenta of particles in amplitudes need not be
small, and the spin can be arbitrarily large. We are simply taking advantage of the availability of spinorial representations of the Lorentz group.
A key motivation for introducing spinors in the present context is the chirality structure of eqs.~\eqref{eq:rtKerrEOMmomentum}
and~\eqref{eq:rtKerrEOMspin}, which hint at a more basic description using an intrinsically chiral formalism.

The little group of a massive momentum is ${SO}(3)$, so to construct the spin vector $a^\mu$ in terms of spinors we need only
form a little-group vector representation from little group spinors. The vector representation of ${SO}(3)$ is the symmetric tensor product of two
spinors, so we will need to symmetrise little group indices. Let $a^{IJ}$ be a constant symmetric two-by-two matrix; then
\[
a^\mu = \frac12 a^{IJ}\! \bra{p_J} \sigma^\mu | p_I] 
\]
is the spin vector. To understand how these expressions work, it may be helpful to work in a Lorentz frame $p^\mu = (\sqrt{p^2}, 0, 0, 0)$. Then
the spin is a purely spatial vector, so it is a linear combination of components in the $x$, $y$ and $z$ directions. Thus there is a basis of
three possible spins.
This is reflected in the three independent components
of the symmetric two-by-two matrix $a^{IJ}$. The algebra of the spinors
immediately guarantees that the spin $a$ and the momentum $p$ are orthogonal.

Given that we can always reconstruct the momentum and spin from the spinors, all we now need are dynamical equations for the spinors 
themselves. The leading-order spinorial equations of motion for $\rootKerr$~are
\[
\frac{\d}{\d \tau} \ket{p_I} &= \frac{Q}{2m} \maxwell(r(\tau) + i a(\tau)) \ket{p_I}\,, \\
\frac{\d}{\d \tau} |p_I] &=  \frac{Q}{2m} \tilde \maxwell(r(\tau) - i a(\tau)) |p_I] \,.
\label{eq:spinorEOMrtKerr}
\]
Notice that the evolution of the spinors is directly determined by the Maxwell spinor of whatever background the particle is moving in. The
NJ shifts indicated explicitly in equation~\eqref{eq:spinorEOMrtKerr} are an explicit consequence of the shift~\eqref{eq:rtKerrIntShift} at the
level of the effective action. It is straightforward to recover the vectorial equations~\eqref{eq:rtKerrEOMmomentum} 
and~\eqref{eq:rtKerrEOMspin} from our spinorial equation; for example, for the momentum,
\[
\frac{dp^\mu}{d\tau} &= \frac{Q}{2m}\epsilon^{IJ}\, \textrm{Re}\, \phi(r(\tau) + ia(\tau)) \ket{p_I} |p_J] \sigma^\mu\\
&= -\frac{Q}{2m} \textrm{Re}\, F_{\rho\sigma}(r(\tau) + ia(\tau)) p_\nu\, \textrm{tr}\left(\sigma^{\rho\sigma} \sigma^\mu \tilde \sigma^\nu \right).
\]
The trace can be evaluated using standard techniques, which yield the projector of a two-form onto its self-dual part,
\begin{equation}
\textrm{tr}\left(\sigma^\mu \tilde \sigma^\nu \sigma^{\rho\sigma} \right) =  \eta^{\nu\rho}\eta^{\mu\sigma} - \eta^{\mu\rho}\eta^{\nu\sigma} + i\epsilon^{\nu\mu\rho\sigma}\,.\label{eqn:sigmaTrace}
\end{equation}
The vector algebra now easily leads to~\eqref{eq:spinorEOMrtKerr}.

To illustrate the use of spinorial methods, consider scattering two $\rootKerr$~particles off one another. We will compute both the leading impulse 
$\Delta p_1$ and the leading angular impulse $\Delta a_1$ on one of the two particles during the scattering event. The primary goal of this thesis has been to obtain these observables using the methods
of scattering amplitudes; here, spinorial equations of motion render the computations even simpler. We denote
the spinor variables for particle 1 by $\ket{1, \tau}$ and $|1, \tau]$, and 
similarly for particle 2; these spinors are explicitly functions of proper time. In a scattering event we denote the initial spinors
as $\ket{1} \equiv \ket{1, -\infty}$ (and similarly for $|1]$.) The
final outgoing spinors are then $\ket{1'} \equiv \ket{1, + \infty}$.

The impulses on particle 1 are given in terms of a leading order kick of the spinor $\ket{\Delta1} \equiv \ket{1, +\infty} - \ket{1, -\infty}$ as\footnote{Notice that we are representing the impulses here as bispinors.}
\[
\Delta p_1 &= 2 \epsilon^{IJ} \Re \ket{\Delta 1_{J}} [1_{I}| \,,\\
\Delta a_1 &= 2 a_1^{IJ} \Re \ket{\Delta 1_{J}} [1_{I}|  \,.
\label{eq:impulsesFromSpinors}
\]
Thus we simply need to compute the kick suffered by the spinor of particle 1 to
determine \emph{both} impulses, in contrast to other methods available (including using amplitudes.) By direct
integration of the spinorial equation~\eqref{eq:spinorEOMrtKerr} we see that this spinorial kick is
\[
\ket{\Delta 1_I} = \frac{Q_1}{2m_1} \int_{-\infty}^\infty\! \d \tau \,\phi(r_1 + i a_1) \, \ket {1_I} \,.
\label{eq:deltaSpinorStep}
\]
At this level of approximation, we may take the trajectory $r_1$ to be a straight line with constant velocity, and take the spin $a_1$ to 
be constant, under the integral. Notice that we evaluate the Maxwell spinor at the shifted position $r_1 + i a_1$ because of the Newman--Janis
shift property at the level of interactions.

To perform the integration we need the Maxwell spinor influencing the motion of particle 1. This is the field of the second of our two particles.
It is easy to obtain this field --- indeed, by the standard Newman--Janis shift of the field set up by particle 2, we need only shift the Coulomb
field of a point-like charge. The field is
\[
\phi(x) = 2iQ_2 \!\int\! \dd^4\wn k \, \del(\wn k \cdot u_2) \frac{e^{-i \wn k \cdot (x+i a_2)}}{\wn k^2} \sigma_{\mu\nu} \wn k^\mu u_2^\nu \,.
\label{eq:explicitMaxwell}
\]
Note the explicit NJ shift by the spin $a_2$: this is the shift of the background, in contrast to the shift through $a_1$ of 
equation~\eqref{eq:deltaSpinorStep}. Of course, there is a pleasing symmetry between these shifts. Using the field~\eqref{eq:explicitMaxwell}
in our expression~\eqref{eq:deltaSpinorStep} for the change in the spinors of particle 1, we arrive at an integral expression for the spinor kick,
\[
\ket{\Delta 1_I} =  \frac{i Q_1 Q_2}{2m_1}\int \dd^4\wn k \, \del(\wn k \cdot u_1)\del(\wn k \cdot u_2) \frac{e^{-i\wn k \cdot (b+i a_1+i a_2)}}{\wn k^2} \wn k^\mu u_2^\nu 
\sigma_{\mu\nu} \ket {1_I} \,,
\]
where $b$ is the impact parameter. 
This expression contains complete information about both the linear and angular impulses. For example, substituting into 
equation~\eqref{eq:impulsesFromSpinors} and applying~\eqref{eqn:sigmaTrace} we find that the angular impulse is
\begin{multline}
\Delta a_1^\mu = \frac{Q_1 Q_2}{m_1} \Re\! \int\! \dd^4 \wn k \, \del(\wn k \cdot u_1) \del(\wn k \cdot u_2) \frac{e^{-i\wn k \cdot (b+i a_1+i a_2)}}{\wn k^2}
\\
\times
\left(
ia_1 \cdot u_2 \, \wn k^\mu - i\wn k \cdot a_1 \, u_2^\mu + \epsilon^\mu(\wn k, a_1, u_2) 
\right) \,.
\end{multline}

Spinorial equations of motion are also available for the leading order interactions of Kerr moving in a gravitational background. They are
\[
\frac{\d}{\d \tau} \ket{p_I} &= -\frac{1}{2} u^\mu \omega_\mu (r + i a) \ket{p_I}\,, \\
\frac{\d}{\d \tau} |p_I] &= - \frac{1}{2} u^\mu \tilde \omega_\mu (r - i a) |p_I] \,,
\label{eq:spinorEOMKerr}
\]
where the spin connection is written in terms of spinors:
\[
\omega_\mu \ket{p} = \omega_\mu{}^{ab} \sigma_{ab} \ket{p} \,.
\]

Using these spinorial equations of motion and a brief calculation in exact analogy with our $\rootKerr$~discussion above, it is remarkably straightforward
to recover the 1PM linear and angular impulse due to Kerr/Kerr scattering in equation~\eqref{eqn:KerrDeflections}~\cite{Vines:2017hyw}. 

In fact we can go further and consider the generalisation of Kerr with NUT charge, corresponding in the stationary case to the exact Kerr--Taub--NUT
solution. It is known that NUT charge can be introduced by performing the gravitational analogue of electric/magnetic
duality \cite{Talbot:1969bpa}. Working at linearised level, this deforms the linearised spinorial equation of motion to
\[
\frac{\d}{\d \tau} \ket{p_I} &= -  \frac{e^{-i \theta}}{2} u^\mu \omega_\mu (r + i a) \ket{p_I}\,, \\
\frac{\d}{\d \tau} |p_I] &= -  \frac{e^{+i \theta}}{2} u^\mu \tilde \omega_\mu (r - i a) |p_I] \,,
\]
where $\theta$ is a magnetic angle. The particle described by these equations has mass $m \cos \theta$ and NUT parameter $m \sin \theta$.
Using these equations, and defining the rapidity $w$ by $\cosh w = u_1 \cdot u_2$, we find that the leading order impulse in a Kerr--Taub--NUT/Kerr--Taub--NUT scattering event is given by
\begin{multline}
\Delta p_1^\mu = - 4\pi G m_1 m_2 \Re\! \int\! \dd^4\wn k  \, \del(\wn k \cdot u_1)\del(\wn k \cdot u_2) \frac{e^{-i \wn k \cdot (b+i a_1+i a_2)}}{\wn k^2} e^{i (\theta_2 -\theta_1)}
\\
\times
\left(
i \cosh 2w \,\wn k^\mu + 2 \cosh w \, \epsilon^\mu(\wn k, u_1, u_2)
\right) \,,
\end{multline}
in agreement with a previous computation performed using scattering amplitudes~\cite{Emond:2020lwi}. 
It is also straightforward to compute
the angular impulse using these methods; we find that
\[
\Delta a_1^\mu = 4\pi &G m_2 \Re\! \int\! \dd^4 \wn k  \, \del(\wn k \cdot u_1)\del(\wn k \cdot u_2) \frac{e^{-i\wn k \cdot (b+i a_1+i a_2)}}{\wn k^2} e^{i (\theta_2 -\theta_1)}
\\
\times&
\Big(
\cosh 2w \, \epsilon^\mu(\wn k, u_1, a_1) - 2 \cosh w \, u_1^\mu \epsilon(\wn k, u_1, u_2, a_1) \\& \qquad - 2i a_1\cdot u_2 \cosh w\, \wn k^\mu + i\wn k\cdot a_1 \left(2 \cosh w\, u_2^\mu - u_1^\mu\right)  
\Big) \,.
\]
These results can be integrated by means of the generalisation of~\eqref{eqn:I1result},
\begin{equation*}
\int\! \dd^4\wn k  \, \del(\wn k \cdot u_1)\del(\wn k \cdot u_2) \frac{e^{-i \wn k \cdot (b+i a_1+i a_2)}}{\wn k^2} \wn k^\mu = \frac{i}{2\pi \sqrt{\gamma^2 - 1}} \frac{{b}^\mu + i\Pi^\mu{ }_\nu(a_1 + a_2)^\nu}{[b + i\Pi(a_1 + a_2)]^2}\,,
\end{equation*}
where the projector $\Pi^\mu{ }_\nu$ was defined in equation~\eqref{eqn:projector}. A little algebra is enough to see that, when the magnetic angles are zero, our results precisely match the 1PM impulses for spinning black holes in equation~\eqref{eqn:KerrDeflections} --- the observables for Kerr--Taub--NUT scattering are simply phase rotations of their Kerr counterparts.

\section{Discussion}
\label{sec:discussion}

The Newman--Janis shift is often dismissed as a trick, without any underlying geometric justification. The central theme of this chapter is that we
should rather view Newman and Janis's work as an important insight. The Kerr solution is simpler than it first seems, and correspondingly
the leading interactions of Kerr are simpler than they might otherwise be. It seems appropriate to place the NJ shift at the heart of our
formalism for describing the dynamics of Kerr black holes, thereby taking maximum advantage of this leading order simplicity.

Our spinorial approach to the classical dynamics of Kerr (and its electromagnetic single-copy, $\rootKerr\,$) makes it trivial to include the
spin (to all orders in $a$) in scattering processes. Computing the evolution of the spinors, rather than the momenta and spin separately,
reduces the workload in performing these computations, and is even more efficient in some examples than computing with the help of
scattering amplitudes. However, we only developed these equations at leading order. At higher orders, spinor equations of motion will certainly exist and be worthy of study.

We found that the effective action for Kerr has the surprising property that it can be formulated in terms of a two-dimensional worldsheet
integral instead of the usual one-dimensional worldline effective theory. This remarkable fact provides some kind of geometric basis for
the Newman--Janis shift, where it emerges using Stokes's theorem. Our worldsheet actions contain terms integrated over some boundaries,
and other terms integrated over the ``bulk'' two-dimensional worldsheet. This structure is also familiar from brane world scenarios, but is
obviously surprising in the context of Kerr black holes. In Minkowski space, this worldsheet is embedded in a complexification of spacetime,
in a manner somewhat reminiscent of other work on complexified worldlines; see ref.~\cite{Adamo:2009ru}, for example. However, 
our worldsheet seems to be a bit of a different beast: it is not a complex line, but rather a strip with two boundaries.

The worldsheet emerged in our work, built up from the physical boundary worldline and geodesics in the direction of spin. This construction
is very different from the sigma models familiar from string theory. The dynamical variables in our action are the ``near'' worldsheet coordinates,
the spin, and body-fixed frame. But perhaps these dynamical variables emerge from a geometric description more reminiscent of the picture for strings.

It is remains to be seen whether the worldsheet structure persists when higher-order operators, involving two or more powers of the Riemann
curvature (or electromagnetic field strength), are included. But we can certainly hope that the surprising simplicity of Kerr persists to higher
orders --- the computation of observables, at finite spin, from both loop-level amplitudes \cite{Guevara:2018wpp,Chung:2018kqs,Bern:2020buy,Aoude:2020ygw,Kosmopoulos:2021zoq} and post--Minkowskian EFT methods \cite{Liu:2021zxr} certainly indicates that further progress can be made. The latter reference is particularly inspiring in this regard, due to the fully covariant nature of the results. Meanwhile, precision calculations reliant on the effective action in \eqref{eq:fullSimpleAction} require including higher-dimension operators in the action  \cite{Levi:2020kvb,Levi:2020uwu,Levi:2020lfn}. It would be particularly interesting to investigate the symmetry structure of the Kerr worldsheet, with an eye towards placing symmetry constraints on the tower of possible higher-dimension operators.


%% file: chapter7/chapter7.tex
\chapter{Conclusions}
\label{chap:conclusions}

On-shell scattering amplitudes are the quantum backbone of particle physics. Black holes are archetypes of classical general relativity. The central aim of this thesis has been to show that despite the apparent dissimilarities, the two are intimately connected. Furthermore, we have advocated that utilising purely on-shell data offers a powerful window into black hole physics.

In the first part of the thesis we gathered the technology needed to compute on-shell observables relevant for gravitational wave astronomy, basing our formalism in the humble study of explicit single particle wavepackets. We introduced our first scattering observable in chapter~\ref{chap:impulse}: the impulse, or total change in momentum of a scattering point-particle. We constructed explicit expressions for the impulse, valid in any quantum field theory, in terms of on-shell amplitudes. With our explicit wavepackets we were able to rigorously determine the classical regime of this observable in section~\ref{sec:classicalLimit}, encountering the Goldilocks inequalities, which we showed held the key to calculating the classical limit of scattering amplitudes.

Armed with a full understanding of the classical limit, and in particular a practical knowledge of crucial $\hbar$ factors, we were able to provide expressions for the classical limit of the impulse observable. Explicit examples at LO and NLO in section~\ref{sec:examples} led to expressions which agree with classical worldline perturbation theory --- a highly non-trivial constraint. Beyond these orders the impulse does not fully capture the dynamics of a spinless particle, due to the emission of radiation. In chapter~\ref{chap:radiation} we therefore included the total radiated momentum in our formalism, eventually finding that for inelastic scattering, amplitudes, via the radiation kernel~\ref{eqn:defOfR}, specify a point-particle's classical worldline current (as first noted in ref.~\cite{Luna:2017dtq}). We showed the efficacy of the double copy for computing this current, a fact that is crucial for gravitational wave astronomy. We also showed that basing our treatment of radiation and the impulse in quantum mechanics has the enormous advantage of ameliorating the conceptual difficulties inherent in treatments of radiating point-particles in classical field theory, explicitly recovering the LO predictions of the Abraham--Lorentz--Dirac force in electrodynamics from amplitudes.

Although we only considered unbound (scattering) events, it is in fact possible to determine the physics of bound states from our observables. This can be done concretely using effective theories \cite{Cheung:2018wkq}. It should also be possible to connect our observables more directly to bound states using analytic continuation, in a manner similar to the work of K\"{a}lin and Porto \cite{Kalin:2019rwq,Kalin:2019inp}. As in any application of traditional scattering amplitudes, however, time-dependent phenomena are not readily accessible. This reflects the fact that amplitudes are the matrix elements of a time evolution operator from the far past to the far future. For a direct application of our methods to the time-dependent gravitational waveform, we must overcome this limitation. One possible path of future investigation for upgrading the formalism presented here would start from the fact that the observables we have discussed are essentially expectation values. They are therefore most naturally discussed using the time-dependent in-in formalism, which has a well-known Schwinger--Keldysh diagrammatic formulation. Whether the double copy applies in this context remains to be explored.

The double copy offers an avenue, rooted in scattering amplitudes, to simpler calculations in gravity. Amplitudes methods are also especially potent when applied to the physics of spin, motivating the focus of the second part of the thesis. For a spinning body the impulse and radiation are not enough to uniquely specify the dynamics of a scattering event for two bodies with unaligned spins. For a black hole, uniquely constrained by the no-hair theorem, the full data is gained from knowledge of the angular impulse, or change in the spin vector. Rooting our intuition once again in the physics of single particle states, we identified the Pauli--Lubanski operator as the crucial quantum actor corresponding to the classical spin vector. We used this operator to obtain explicit expressions for the LO angular impulse in terms of amplitudes, finding notably more complex expressions than for the momentum impulse. We did not consider the higher order corrections, but they are very similar to those of the colour impulse in ref.~\cite{delaCruz:2020bbn} --- these would be necessary to access the results of \cite{Liu:2021zxr}, for example.

By carefully studying the classical limit of amplitudes for finite spin, and computing the momentum and angular impulses, we were able to precisely reproduce the leading spin, 1PM scattering data for Kerr black holes. The complete, all-spin classical expressions,~\eqref{eqn:KerrDeflections}, can be calculated from amplitudes by applying our methods to massive spinor helicity representations in the large spin limit \cite{Guevara:2019fsj}. These results are one artefact of a beautiful relationship between black holes and amplitudes: minimally coupled graviton amplitudes are the on-shell avatar of the no-hair theorem \cite{Vaidya:2014kza,Chung:2018kqs}. This provocative idea can provide non-geometric insights into intriguing results in general relativity, such as the Newman--Janis complex map between the Schwarzschild and Kerr solutions, which is naturally explained by the exponentiation of the minimally coupled 3--point amplitude in the large spin limit \cite{Arkani-Hamed:2019ymq}.

We extended this on-shell insight into black hole physics in chapter~\ref{chap:worldsheet}, where we showed that the Newman--Janis shift could be interpreted in terms of a worldsheet effective action. This holds both in gravity, and for the single-copy $\rootKerr$~solution in electrodynamics. Moreover, at the level of equations of motion we showed that the NJ shift holds also for the leading interactions of the Kerr black hole. These leading interactions were conveniently described using chiral classical equations of motion with the help of the spinor-helicity method familiar from scattering amplitudes. These spinor equations of motion are extremely powerful tools --- they offered remarkably efficient derivations of the on-shell observables calculated from amplitudes earlier in the thesis, encapsulating the full on-shell data for non-aligned black hole scattering in the kick of the holomorphic spinor. It was also a trivial matter to extend the scope of this technology to the magnetically charged Kerr--Taub--NUT black hole, facilitating the first calculation of the angular impulse for this solution. It would be interesting to see if our methods can be generalised to include the full family of parameters of the famous Plebanski--Demianski family \cite{Debever:1971,Plebanski:1976gy}.

The full geometry of our worldsheet remains to be explored, in particular its applicability when higher-order curvature terms are accounted for. Such corrections are crucial for the extending the power of the chiral spinor equations. A major motivation for tackling this problem comes from an obstacle towards progress in amplitudes computations of importance for gravitational wave astronomy: a well defined, tree-level expression for the four-point Compton amplitude, valid for any generic spin $s$, is not known. Application of BCFW recursion with arbitrary spin representations leads to spurious poles \cite{Arkani-Hamed:2017jhn,Aoude:2020onz}; these poles can only be removed by hand ambiguously \cite{Chung:2018kqs}. As we have seen in chapter~\ref{chap:impulse}, triangle diagrams play a crucial role in the computation of either the NLO impulse or potential. The Compton amplitude is a key component to the computation of these topologies using generalised unitarity, and thus progress in computing precision post--Minkowksian corrections for particles with spin is impeded by a crucial tree-level input into the calculation.

The all-spin Compton amplitude should have a well defined, unambiguous classical limit. Physically the amplitude corresponds to absorption and re-emission of a messenger boson, a problem which is tractable in classical field theory. Our worldsheet construction shows that the Newman--Janis shift holds for the leading interactions of Kerr black holes, and thus the results of chapter~\ref{chap:worldsheet} should offer a convenient way to approach this problem, provided that higher order contributions can be included. We hope that, in this manner, our work will be only the beginning of a programme to exploit the Newman--Janis structure of Kerr black holes to simplify their dynamics.

%% file: appendices/appendix1.tex
\chapter{Wavefunction integrals}
\label{app:wavefunctions}

In this appendix we detail the evaluation of the explicit momentum space wavefunction integrals that played a key role in our disccussion of the classical limit in chapters~\ref{chap:pointParticles} and~\ref{chap:impulse}.

We begin with the single particle momentum space wavefunction normalisation in \eqn~\eqref{eqn:LinearExponential}. To determine this we must compute the following integral:
\begin{equation}
m^{-2}\int \df(p)\;\exp\biggl[-\frac{2 p\cdot u}{m\xi}\biggr]\,.
\end{equation}
Let us parametrise the on-shell phase space in a similar manner to equation~\eqref{eqn:qbRapidityParametrisation}, but appropriate for timelike momenta, writing
\begin{equation}
p^\mu = E_p\bigl(\cosh\zeta,\,\sinh\zeta\sin\theta\cos\phi,\,\sinh\zeta\sin\theta\sin\phi,
\sinh\zeta\cos\theta\bigr)\
\label{eqn:Parametrization}               
\end{equation}
so that
\begin{equation}
\begin{aligned}
\df(p) &= (2\pi)^{-3}\dd E_p d\zeta d\Omega_2\,\del(E_p^2-m^2)\Theta(E_p)\,
E_p^3 \sinh^2\zeta
\\&=
(2\pi)^{-3}\dd E_p d\zeta d\theta d\phi \,\del(E_p^2-m^2)\Theta(E_p)\,
E_p^3 \sinh^2\zeta\sin\theta\,.
\end{aligned}
\end{equation}
Performing the $E_p$ integration, we obtain
\begin{equation}
\df(p)  \rightarrow \frac{m^2}{2(2\pi)^{3}} 
d\zeta d\theta d\phi\,\sinh^2\!\zeta\sin\theta\,,
\label{Measure}
\end{equation}
along with $E_p=m$ in the integrand.  The integral must be a Lorentz-invariant
function of $u$; as the only available Lorentz invariant is $u^2=1$, we conclude
that the result must be a function of $\xi$ alone.  We can compute it in the rest
frame of $u$, where our desired integral is
\begin{equation}
\frac12 \! \int_0^\infty \!\dd\zeta\,\sinh^2\!\zeta\!\int_0^\pi\! \dd\theta\,\sin\theta\!
\int_0^{2\pi} \!\dd\phi\, \exp\biggl[-\frac{2\cosh\zeta}{\xi}\biggr]
=\frac1{2(2\pi)^2} \xi\, K_1(2/\xi)\,,\label{eqn:normalisationIntegral}
\end{equation}
where $K_1$ is a modified Bessel function of the second kind.  The normalisation
condition~(\ref{eqn:WavefunctionNormalization}) then yields
\begin{equation}
\frac{2\sqrt2\pi}{\xi^{1/2} K_1^{1/2}(2/\xi)}
\end{equation}
for the wavefunction's normalisation.

Next, we compute $\langle p^\mu\rangle$.  Lorentz invariance implies that the expectation
value must be proportional to $u^\mu$;
again computing in the rest frame, we find that
\begin{equation}
\langle p^\mu\rangle = m u^\mu \,\frac{K_2(2/\xi)}{K_1(2/\xi)}\,.
\end{equation}
The phase-space measure fixes $\langle p^2\rangle = m^2$, so we conclude that
\begin{equation}
\begin{aligned}
\frac{\sigma^2(p)}{\langle p^2\rangle} =
1-\frac{\langle p\rangle^2}{\langle p^2\rangle} &= 1-\frac{K_2^2(2/\xi)}{K_1^2(2/\xi)}
\\&=-\frac32\xi+\Ord(\xi^2)\,,
\end{aligned}
\end{equation}
where as $\xi\rightarrow 0$ we have used the asymptotic approximations
\[
K_1(2/\xi) &= \sqrt{\frac{\pi \xi}{4}}\, \exp\left[-\frac{2}{\xi}\right]\left(1 + \frac{3\xi}{16} + \mathcal{O}(\xi^2)\right),\\ K_2(2/\xi) &= \sqrt{\frac{\pi \xi}{4}}\, \exp\left[-\frac{2}{\xi}\right] \left(1 + \frac{15\xi}{16} + \mathcal{O}(\xi^2)\right).
\]
For the single particle case it remains to compute the double expectation value in equation~\eqref{eqn:doubleMomExp}. We can again apply Lorentz invariance, which dictates that
\begin{equation}
\langle p^\mu p^\nu \rangle = A u^\mu u^\nu + B \eta^{\mu\nu}\,.\label{eqn:doubleExpVal}
\end{equation}
Contracting with the velocity leads to
\begin{equation}
A + B = \frac{\mathcal{N}^2}{m^2}\! \int\! \df(p)\, (u\cdot p)^2\, \exp\left[-\frac{2 u\cdot p}{m \xi}\right] = m^2 + \frac32 m^2 \xi \frac{K_2(2/\xi)}{K_1(2/\xi)}\,,
\end{equation}
where we used the result in equation~\eqref{eqn:normalisationIntegral}. Combining this constraint with the trace of~\eqref{eqn:doubleExpVal} is then enough to determine the coefficients as
\begin{equation}
A = m^2 + 2 m^2 \xi \frac{K_2(2/\xi)}{K_1(2/\xi)}\,, \qquad B = -\frac{m^2}{2} \xi \frac{K_2(2/\xi)}{K_1(2/\xi)}\,,
\end{equation}
yielding the result in equation~\eqref{eqn:doubleMomExp}.

Our discussion of point-particle scattering in section~\ref{sec:classicalLimit} hinged upon the integral 
\begin{equation}
T(\qb) = \frac{\Norm^2}{\hbar m^3}\exp\biggl[-\frac{\hbar \qb\cdot u}{m\xi}\biggr]\int \df(p)\;
\del(2 p\cdot\qb/m+\hbar\qb^2/m)\,\exp\biggl[-\frac{2 p\cdot u}{m\xi}\biggr]\,.
\end{equation}
This integral is dimensionless, and can depend only on two Lorentz invariants,
$\qb\cdot u$ and $\qb^2$, along with $\xi$.  
It is convenient to write it as a function of two
dimensionless variables built out of these invariants,
\begin{equation}
\omega \equiv \frac{\qb\cdot u}{\sqrt{-\qb^2}}\,,
\qquad \tau \equiv \frac{\hbar\sqrt{-\qb^2}}{2m}\,.
\end{equation}
We again work in the rest frame of $u^\mu$, and without loss of generality,
choose the $z$-axis of the $p$ integration to lie along the direction of
$\v{\qb}$.  The only components that appear in the integral are then
$\qb^0$ and $\qb^z$; after integration, we can obtain the dependence on
$\tau$ and $\omega$ via the replacements
\begin{equation}
\qb^0\rightarrow \frac{2 m\omega\tau}{\hbar}\,,
\qquad \qb^z \rightarrow \frac{2m\sqrt{1+\omega^2}\tau}{\hbar}\,;
\end{equation}
hence,
\begin{equation}
-\!\qb^2 \rightarrow \frac{4m^2\tau^2}{\hbar^2}\,.
\end{equation}
Using the measure~(\ref{Measure}), we find that
\begin{equation}
\begin{aligned}
\IntOne(\wn q) &= \frac1{2\pi\,\hbar m\,\xi K_1(2/\xi)} \exp\biggl[-\frac{\hbar \qb^0}{m\xi}\biggr]
\\&\qquad\times \int_0^\infty \!\d\zeta\,\sinh^2\!\zeta\! \int_0^\pi\! \d\theta\,\sin\theta\!
\int_0^{2\pi}\! \d\phi\, \exp\biggl[-\frac{2\cosh\zeta}{\xi}\biggr]
\\& \qquad\qquad\qquad \times \del(2 \qb^0 \cosh\zeta - 2\qb^z \sinh\zeta \cos\theta
+\hbar\qb^2/m)\,.
\end{aligned}
\end{equation}
The $\phi$ integral is trivial, and we can use the delta function to do the
$\theta$ integral:
\newcommand{\cosech}{\textrm{cosech}}
\begin{equation}
\begin{aligned}
\IntOne(\wn q) &= \frac1{2\hbar m\,\qb^z\,\xi K_1(2/\xi)} \exp\biggl[-\frac{\hbar \qb^0}{m\xi}\biggr]
\int_0^\infty d\zeta\;\sinh\zeta\;
\exp\biggl[-2\frac{\cosh\zeta}{\xi}\biggr]
\\& \qquad\qquad\qquad\times 
\Theta\bigl(1+ \qb^0 \coth\zeta/\qb^z +\hbar\qb^2\cosech\,\zeta/(2 m\qb^z)\bigr)\,
\\& \qquad\qquad\qquad\times 
\Theta\bigl(1-\qb^0 \coth\zeta/\qb^z -\hbar\qb^2\cosech\,\zeta/(2 m\qb^z)\bigr)\,.
\end{aligned}
\end{equation}
In the $\hbar\rightarrow 0$ limit, the first theta function will have no effect,
even with $\qb^2<0$.  Changing variables to $w=\cosh\zeta$, 
the second theta function will impose the constraint
\begin{equation}
w \ge \frac{\qb^z\sqrt{1-\hbar^2\qb^2/(4m^2)}}{\sqrt{-\qb^2}} -\frac{\hbar\qb^0}{2m}\,.
\end{equation}
In terms of $\omega$ and $\tau$,
this constraint is
\begin{equation}
w \ge \sqrt{1+\omega^2}\sqrt{1+\tau^2}-\omega\tau\,.
\end{equation}
Up to corrections of $\Ord(\hbar)$, the right-hand side is greater than 1,
and so becomes the lower limit of integration.  The result for the integral is then
\begin{equation}
\begin{aligned}
T(\qb) &=\frac1{8 m^2\sqrt{1+\omega^2}\tau\,K_1(2/\xi)} 
\exp\biggl[-\frac{2\omega\tau}{\xi}\biggr]\\
&\qquad\qquad\qquad \times\exp\biggl[-\frac2{\xi}\Bigl(\sqrt{1+\omega^2}\sqrt{1+\tau^2}-\omega\tau\Bigr)\biggr]
\\&=
\frac1{4 \hbar m\sqrt{(\qb\cdot u)^2-\qb^2}\,K_1(2/\xi)} \\
&\qquad\qquad\qquad \times \exp\biggl[-\frac2{\xi}\frac{\sqrt{(\qb\cdot u)^2-\qb^2}}{\sqrt{-\qb^2}}
\sqrt{1-\hbar^2\qb^2/(4m^2)}\biggr]
\,,
\end{aligned}
\end{equation}
which is the result listed in equation~\eqref{eqn:TIntegral}.

%% file: appendices/appendix2.tex
\chapter{Worldline perturbation theory}
\label{app:worldlines}

Throughout part~\ref{part:observables} we computed on-shell classical observables from scattering amplitudes. We justified our results by their agreement with iterative solutions to the appropriate classical equations of motion, as shown in detail in refs.~\cite{Kosower:2018adc,delaCruz:2020bbn}. To provide a flavour of these methods, let us calculate the LO current for gravitational radiation due to the electromagnetic scattering of Reissner--Nordstr\"{o}m black holes, which we studied using the double copy in section~\ref{sec:inelasticBHscatter}. 

Classically, the emitted gravitational radiation will satisfy the linearised Einstein--Maxwell field equation, which for the trace-reversed perturbation in de--Donder gauge is
\begin{equation}
\partial^2\bar h^{\mu\nu}(x) = T^{\mu\nu}(x) = \frac{\kappa}{2}\left(T_{\rm pp}^{\mu\nu}(x) + T_{\rm EM}^{\mu\nu}(x)\right).
\end{equation}
The electromagnetic field strength tensor $T^{\rm EM}_{\mu\nu}$ is that in \eqn~\eqref{eqn:EMfieldStrength}, while in gravity
\begin{equation}
T_{\rm pp}^{\mu\nu}(x) = \sum_{\alpha=1}^n m_\alpha \int\! \frac{d\tau_\alpha}{\sqrt{g}}\, u^\mu_\alpha(\tau)u^\nu_\alpha(\tau)\, \delta^{(D)}(x-x_\alpha(\tau))\label{eqn:ppGravTensor}
\end{equation}
for $n$ particles, where $g = -\det(g_{\mu\nu})$.  We are seeking perturbative solutions, and therefore expand worldline quantities in the coupling:
\[
x_\alpha^\mu(\tau_\alpha) &= b_\alpha^\mu+u_\alpha\tau_\alpha + \Delta^{(1)}x_\alpha^\mu(\tau_\alpha) + \Delta^{(2)} x_\alpha^\mu(\tau_\alpha) +\cdots\,,\\
v_\alpha^\mu(\tau_\alpha) &= u_\alpha + \Delta^{(1)}v_\alpha^\mu(\tau_\alpha) + \Delta^{(2)} v_\alpha^\mu(\tau_\alpha) +\cdots \,.
\]
Here $\Delta^{(i)}x_\alpha^\mu$ will indicate quantities entering at $\mathcal{O}(\tilde g^{2i})$, for couplings $\tilde g = e$ or $\kappa/2$. We also work with boundary conditions $x^\mu_\alpha(\tau_\alpha\rightarrow-\infty) = b^\mu_\alpha + u^\mu_\alpha\tau_\alpha$, so $v_\alpha^\mu(\tau_\alpha\rightarrow-\infty) = u^\mu_\alpha$. At leading order we can treat the contributions to the field equation separately, splitting the problem into gravitational radiation sourced by either the particle worldlines or the electromagnetic field. 

Of course we are interested in electromagnetic interactions of two particles, and thus require the leading order solution to the Maxwell equation in Lorenz gauge, $\partial^2A^\mu(x) = eJ^\mu_\textrm{pp}(x)$, where the (colour-dressed) current appears in~\eqref{eqn:YangMillsEOM} This is a very simple calculation, and once quickly finds that the field sourced by particle 2, say, is
\begin{equation}
F^{\mu\nu}_2(x) = ie \! \int\! \dd^4\wn q\, \del(u_2\cdot\wn q) e^{i\wn q\cdot b_2} e^{-i\wn q\cdot x} \frac{\wn q^\mu u_2^\nu - u_2^\mu \wn q^\nu}{\wn q^2}\,;\label{eqn:LOfieldStrength}
\end{equation}
see \cite{Kosower:2018adc} for details. The key point is that we can now obtain the leading deflections simply by solving the classical equation of motion, the Lorentz force, which yields
\begin{equation}
\frac{d \Delta^{(1)} v^{\mu}_1}{d\tau} = \frac{ie^2}{m_1}\! \int \! \dd^4 \wn q\, \del(u_2\cdot \wn q)  e^{-i\wn q \cdot (b_1 - b_2)} \frac{e^{-i\wn q\cdot u_1 \tau}}{\wn q^2} \Big(\wn q^\mu\, u_1\cdot u_2 - u_2^\mu\, u_1\cdot \wn q\Big)\,.
\end{equation}
We now have all the data required to determine the full LO gravitational current. Focusing on the radiation sourced from the point-particle worldlines, we can expand~\eqref{eqn:ppGravTensor} to find
\begin{multline}
T^{\mu\nu}_{\textrm{pp},(1)}(\wn k) = -2e^2\!\int\!\dd^4\wn q\, \del((\wn k - \wn q)\cdot u_1)\del(\wn q\cdot u_2) \frac{e^{i(\wn k - \wn q)\cdot b_1}e^{i\wn q\cdot b_2}}{\wn q^2 (\wn q\cdot u_1)} \bigg[u_1\cdot u_2\, \wn q^{(\mu} u_1^{\nu)} \\ - \frac12 u_1\cdot u_2\, \frac{\wn k\cdot \wn q \, u_1^\mu u_1^\nu}{\wn q\cdot u_1}  - \wn q\cdot u_1\, u_1^{(\mu} u_2^{\nu)} + \frac12 \wn k\cdot u_2 \, u_1^\mu u_1^\nu \bigg] +(1\leftrightarrow 2)\,.
\end{multline}
It is then convenient to relabel $\wn q = \wn w_2$, and introduce a new momentum $\wn w_1 = \wn k - \wn w_2$. In these variables, the analogous contribution from the EM stress tensor in~\eqref{eqn:LOfieldStrength} to the gravitational radiation is
\begin{multline}
T^{\mu\nu}_{\textrm{pp},(1)}(x) = -e^2\!\int\!\dd^4\wn w_1\dd^4\wn w_2\, \del(\wn w_1\cdot u_1)\del(\wn w_2\cdot u_2) \frac{e^{i\wn w_1\cdot b_1} e^{i\wn w_2\cdot b_2}}{\wn w_1^2 \wn w_2^2} e^{-i(\wn w_1 + \wn w_2)\cdot x}\\ \times\bigg[2 \wn w_1\cdot u_2\, u_1^{(\mu} \wn w_2^{\nu)}  - \wn w_1\cdot \wn w_2\, u_1^\mu u_2^\nu - u_1\cdot u_2 \, \wn w_1^\mu \wn w_2^\nu + (1\leftrightarrow 2)\bigg]\,. 
\end{multline}
Summing the two pieces and restoring classical momenta $p_\alpha = m_\alpha u_\alpha$ yields
\begin{multline}
T^{\mu\nu}_{(1)}(\wn k) = -\frac{e^2\kappa}{4} \!\int\!\dd^4\wn w_1 \dd^4\wn w_2\, \del(\wn w_1\cdot p_1) \del(\wn w_2\cdot p_2) \del^{(4)}(\wn k- \wn w_1 - \wn w_2) \, e^{i\wn w_1\cdot b_1} e^{i\wn w_2\cdot b_2}\\ \times \bigg[\frac{Q_{12}^\mu P_{12}^\nu + Q_{12}^\nu P_{12}^\mu}{\wn w_1^2 \wn w_2^2} + (p_1\cdot p_2)\left(\frac{Q_{12}^\mu Q_{12}^\nu}{\wn w_1^2 \wn w_2^2} - \frac{P_{12}^\mu P_{12}^\nu}{(\wn k\cdot p_1)^2 (\wn k\cdot p_2)^2}\right)\bigg]\,,\label{eqn:classicalRN}
\end{multline}
where we have adopted the gauge invariant functions defined in~\eqref{eqn:gaugeInvariants}. Upon contraction with a graviton field strength tensor this current then agrees (for  $b_2 = 0$, and up to an overall sign) with equation~\eqref{eqn:RNradKernel}, the LO gravitational radiation kernel for electromagnetic scattering of Reissner--Nordstr\"{o}m black holes.